\documentclass[usenatbib]{mn2e}
\def\hb{H$\beta$}

\def\oiii{[O~{\small III}]}
\def\kms{$\rm km\;s^{-1}$}
\def\deg{$^{\circ}$}
\def\smin{$\sigma_{_{{\rm MIN}}}/\sigma_{_{1.0}}$}
\def\slast{$\sigma_{_{{\rm LAST}}}$}
\def\snorm{$\sigma_{_{1.0}}$}
\def\fre{$1.0R_e$}
\def\altfre{$0.1R_e$}
\def\altfree{$0.5R_e$}
\def\vel{\rm v}
\usepackage{graphicx}
\usepackage{epsfig}
\usepackage{txfonts}
\usepackage{rotating}
\usepackage{lscape}
\usepackage{natbib}

\title[Kinematic properties of early-type galaxy haloes using
  PNe.]{Kinematic properties of early-type galaxy haloes using
  planetary nebulae.\thanks{Based in part on observations made with the
  William Herschel Telescope operated by the Isaac Newton Group in the
  Spanish Observatorio del Roque de los Muchachos on the island of La
  Palma, of the Instituto de Astrof\'isica de Canarias, and on
  observations collected at the European Southern Observatory, Chile,
  Program: 76.B-0788(A).}}

\author[Coccato et al.]{L. Coccato$^{1}$\thanks{E-mail: lcoccato@mpe.mpg.de}, O. Gerhard$^1$, M. Arnaboldi$^{2,3}$, P. Das$^1$, N. G. Douglas$^4$, K. Kuijken$^5$, \and M. R. Merrifield$^6$, N. R. Napolitano$^{7}$, E. Noordermeer$^6$, A. J. Romanowsky$^{8,9}$,  \and  M. Capaccioli$^{10,11}$, 
A. Cortesi$^6$, F. De Lorenzi$^1$, K. C. Freeman$^{12}$ \\ 
\smallskip \\
$^1$Max-Plank-Institut f\"ur Extraterrestrische Physik, Giessenbachstra$\beta$e, D-85741 Garching bei M\"unchen, Germany;  
$^2$European Southern Observatory,\\ Karl-Schwarzschild-Stra$\beta$e 2, D-85748 Garching bei M\"unchen, Germany;  
$^3$INAF, Osservatorio Astronomico di Pino Torinese, I-10025 Pino Torinese, Italy; \\
$^4$Kapteyn Astronomical Institute, Postbus 800, 9700 AV Groningen, The Netherlands;  
$^5$Leiden Observatory, Leiden University, PO Box 9513, 2300RA Leiden,\\ The Netherlands;  
$^6$School of Physics and Astronomy, University of Nottingham, University Park, Nottingham NG7 2RD, UK; 
$^7$INAF-Observatory of \\  Capodimonte, Salita Moiariello, 16, 80131, Naples, Italy; 
$^{8}$UCO/Lick Observatory, University of California, Santa Cruz, CA 95064, USA; 
$^{9}$Departamento de F\'isica, \\ Universidad de Concepci\'on, Casilla 160-C, Concepci\'on, Chile;
$^{10}$Dipartimento di Scienze Fisiche, Universit\'a Federico II, Via Cinthia, 80126, Naples, Italy; \\ 
$^{11}$INAF - VSTceN,  Salita Moiariello, 16, 80131, Naples, Italy; 
$^{12}$Research School of  Astronomy \& Astrophysics, ANU, Canberra, Australia.
}

\begin{document}

\date{today}

%\pagerange{\pageref{firstpage}--\pageref{lastpage}} \pubyear{2007}

\maketitle

\label{firstpage}

\begin{abstract}

We present new planetary nebulae (PNe) positions, radial velocities,
and magnitudes for 6 early-type galaxies obtained with the Planetary
Nebulae Spectrograph (PN.S), along with derived two-dimensional
velocity and velocity dispersion fields, and the $\alpha$
parameters (i.e. the number of PNe per unit luminosity). We also
present new deep absorption-line long-slit kinematics for 3 galaxies
in the sample, obtained with the VLT/FORS2 spectrograph.

We extend this study to include an additional 10 early-type
galaxies with PNe radial velocity measurements available from the
literature, including previous PN.S studies, in order to obtain a
broader description of the outer-halo kinematics in early-type
galaxies. These data extend the information derived from stellar
absorption-line kinematics to typically several and up to $\sim8$ effective
radii.

The combination of photometry, absorption-line and PNe kinematics
shows: i) a good agreement between the PNe number density distribution
and the stellar surface brightness in the region where the two 
  data sets overlap; ii) a good agreement between PNe and
absorption-line kinematics; iii) that the mean rms velocity profiles
fall into two groups, with part of the galaxies characterized by
slowly decreasing profiles and the remainder having steeply falling
profiles; iv) a larger variety of velocity dispersion radial profiles;
v) that twists and misalignments in the velocity fields are more
frequent at large radii, including some fast rotator galaxies; vi)
that outer haloes are characterised by more complex radial profiles of
the specific angular momentum-related $\lambda_R$ parameter than
observed within 1 $R_e$; vii) that many objects are more rotationally
dominated at large radii than in their central parts; and viii) that
the halo kinematics are correlated with other galaxy properties, such
as total B-band and X-ray luminosity,  isophotal shape, total
stellar mass, $V/\sigma$, and $\alpha$ parameter, with a clear
separation between fast and slow rotators.

\end{abstract}


\section{Introduction}
\label{sec:introduction}

The dynamics of galaxies provide fundamental information on their origin
and evolution.
Our knowledge of the kinematics and dynamics of early-type galaxies is
mainly based on measurements of the first 4 moments of the
line-of-sight velocity distribution (LOSVD) from stellar absorption
lines. These measurements are generally confined to within 1--2
effective radii ($R_e$) and can be obtained with long slits or
integral-field units.
Detailed LOSVD measurements for early-type galaxies (e.g.,  
 \citealt{Bender+94,Fisher97,Pinkney+03,Emsellem+04}) have 
provided us with a general picture of their dynamics. The higher-order
moments are needed  to obtain good constraints on the anisotropy and
mass distribution \citep{Gerhard93, Merritt93}.  Most of the studied
objects appear to be isotropic or slightly radially anisotropic
systems \citep{Kronawitter+00, Cappellari+06}. Some galaxies have
revealed the presence of cold stellar discs whose kinematics stand
out  only after $1 - 1.5$ $R_e$ \citep{Rix+99}.

Kinematics of early-type galaxies are related to their isophotal
shape. Deviations of the isophotes from a perfect ellipse are
parameterised by the amplitude $a_4$ of the $\cos 4\theta$ term in a
Fourier expansion of the isophote radius in polar coordinates
(e.g., \citealt{Bender+88}). Observational evidence
(e.g., \citealt{Bender88, Kormendy+89}) implies that
{\it discy} ellipticals ($a_4>0$) have significant rotation with
$V/\sigma$ $\geq 1$ and may generally be axisymmetric;
{\it boxy} ellipticals ($a_4<0$) exhibit no rotation, have a range of
values of $V/\sigma$ including strongly anisotropic systems ($V/\sigma
<< 1$), may be triaxial, and are in general more massive than discy
ellipticals.
These relations between isophotal shape and galaxy kinematics motivated
\citet{Kormendy+96} to revise the Hubble classification scheme for
early-type galaxies using the $a_4$ parameter, which is related to
intrinsic galaxy properties rather than the apparent ellipticity,
which is related to the galaxy's orientation on the sky.

A related modification to this scheme has been recently proposed
by \citet{Emsellem+07}, taking advantage of the advent of
integral-field units. These allow for two-dimensional maps of the
LOSVD moments typically out to $1R_e$
\citep{Emsellem+04}. Early-type galaxies have been divided into two
distinct classes: {\it slow} and {\it fast} rotators, according to the
stellar angular momentum they possess per unit of mass. The two
classes have different dynamical properties: slow rotators appear to
be more massive systems, nearly round with a significant kinematic
misalignment, implying a moderate degree of triaxiality, and span a
moderately large range of anisotropies; fast rotators appear to be
rather flattened systems, without significant kinematic misalignments,
nearly axisymmetric and span a larger range of anisotropies
\citep{Cappellari+07}.

In parallel,  the arrival of large photometric surveys such as the
Sloan Digital Sky Survey has consolidated findings of a bimodal
colour distribution in local galaxies (e.g., \citealt{Baldry+06}),
differentiating between a {\it blue cloud} of mostly starforming
spiral galaxies and a {\it red sequence} of mostly non-starforming
early-type galaxies.

Some of the main aims of galaxy formation studies are to understand
the mechanisms that allow galaxies to evolve from the blue cloud to
the red sequence, and to differentiate between processes that form
early-type galaxies that are fast rotators  or slow rotators.
Numerical simulations suggest that red-sequence galaxies are formed by
mergers of galaxies in the blue cloud, followed by a quenching of the
star formation (e.g., \citealt{Cattaneo+06, Faber+07, kang+07,
    Romeo+08}). The less-luminous fast rotators with discy isophotes
are preferentially formed through a series of minor mergers with less
massive companions. On the other hand, the more-luminous slow rotators
with boxy isophotes, are thought to form through a violent major
merger between galaxies of similar mass \citep[e.g.,][]{Naab+99,
  Naab+03}, or through multiple or hierarchical mergers
\citep{Weil+96,Burkert+07}.

These formation mechanisms are complicated further by the presence or
absence of gas during the merger event, which plays a fundamental role
in the final kinematic structure of the merger remnant
(e.g., \citealt{Barnes+96, Naab+06, Ciotti+07}). Mergers producing
red-sequence galaxies dominated by rotation appear to be gas-rich
({\it wet}), while gas-poor ({\it dry}) mergers produce red-sequence
galaxies dominated by random motions (e.g., \citealt{Bournaud+05,
  Cox+06, Naab+06}).

However, observations show that many of the most massive early-type
galaxies were already in place by $z \sim 2$ \citep{vanDokkum+04,
  Treu+05}, and the evolution of the galaxy luminosity function since
redshifts of $z\sim1$ argues against a significant contribution of
recent dissipationless dry mergers to the formation of the most
massive early-type galaxies in the red sequence
\citep{Scarlata+07}. Therefore it is clear that the processes
forming elliptical galaxies are not yet completely understood.

Numerical simulations of galaxy formation in a cosmological context
predict particular radial profiles for the total and dark matter
distributions (e.g., \citealt{Dekel+05, Naab+07}), for the $V/\sigma$
ratio (e.g., \citealt{Abadi+06}), the angular momentum, orbital
distribution and isophotal shape (e.g., \citealt{Naab+06}), depending
on which cosmology or merger type is assumed. The evaluation of these
quantities from observations through dynamical models is fundamental
to probe galaxy formation theories, numerical simulations and
cosmological scenarios. Unfortunately the picture is complicated by
the presence of unknown variables such as the three-dimensional shape
of the galaxy and the orbital distribution. Therefore only detailed
information on the LOSVD out to large radii can help to disentangle
possible scenarios in galaxy formation.

In spiral galaxies (or generally in gas-rich systems), the
distribution of dark matter can be determined by measuring the
kinematics of neutral or ionised gas, which can be easily observed at
large distances from the centre. This allowed for instance the
discovery of the presence of dark matter in galaxies in the 1970s.
Only a few early-type galaxies contain large gas rings with which the
mass distribution can be traced out to several $R_e$ (e.g.
\citealt{Bertola+93, Oosterloo+02, Weijmans+08}).  In the large
majority of ellipticals or lenticulars, the mass determination is more
difficult because absorption lines from a stellar spectrum can be
measured with a sufficient signal-to-noise ratio only up to 2
effective radii. Studies based on integrated light spectra provide
evidence for dark matter only in a fraction of ellipticals, with the
inferred mass profiles being nearly isothermal to the limit of the
data and the dark matter contributing $\sim10-50\%$ of the mass within
$R_e$ (e.g. \citealt{Kronawitter+00,Gerhard+01, Thomas+07}). For other
systems absorption-line data are not extended enough to provide
  conclusive evidence for dark matter \citep[see, e.g., the detailed
  recent analysis of][]{DeLorenzi+08a,DeLorenzi+08b}. The apparent
dichotomy may be related to the fact that the logarithmic gradient of
the mass-to-light ratio correlates with luminosity
\citep{Napolitano+05}, suggesting a link between the structural
parameters and the total mass of ellipticals.

In order to extend the kinematic information for early-type galaxies
to larger radii, alternative kinematic tracers have been identified
 to compensate for the rapid fall-off in the stellar light at
radii larger than $2R_e$, such as globular clusters and planetary
nebulae (PNe).
In particular, a lot of effort has been focused on PNe in the last
decade: they are generally believed to trace the main stellar
population of elliptical galaxies, and their relative bright \oiii\
emission line allows them to be easily detected at large radii from
the centre, making them an ideal tool for kinematic studies.

A dedicated instrument was installed at the William Herschel
Telescope: the Planetary Nebulae Spectrograph (PN.S,
\citealt{Douglas+02}). In the last seven years, our team has
undertaken a long-term observational campaign aimed at measuring the
kinematics of a selected sample of ellipticals and S0s, with the
principal aim to quantify the dark matter content in those systems. An
initial sample of round (E0--E2) and bright ($B \leq 12.5$) galaxies
has been selected, and later enlarged to include more
flattened objects ($\leq$ E5) and S0s.  A series of papers has
been published, presenting the initial results of the PN.S survey on
early-type galaxies and their dynamical analysis \citep{Romanowsky+03,
Douglas+07, Noordermeer+08, DeLorenzi+08b, Napolitano+08}).

To provide a general overview of the outer kinematics ($R \geq
1.5-2$ $R_e$) of early-type galaxies and fundamental constraints on
formation scenarios, we have used the literature to collect all the
existing PNe kinematics in ellipticals and S0s, amounting to 10
objects in total.  This includes galaxies previously observed with the
PN.S as well as galaxies observed by other authors using different
telescopes and techniques. 

In this paper we also present new PNe measurements for 6 galaxies
obtained with the PN.S. We will refer to them as {\it sample A}, while
we will  divide the other 10 galaxies from the literature 
  into {\it sample B} (6 galaxies with more than 80 PNe radial
  velocity measurements) and {\it sample C} (4 galaxies with less than
  80 PNe radial velocity measurements) in the rest of the paper.  The
total sample consists of 16 galaxies, and their principal
characteristics are summarised in Table \ref{tab:sample}.  In
  Section \ref{sec:observations} we present the PN.S. observations,
  in Section \ref{sec:spatial_distribution} we compare the PNe
spatial distribution to the stellar surface brightness.  In Section
\ref{sec:pne_kinematics} the smoothed two-dimensional velocity and
velocity dispersion fields are derived and compared to the
corresponding stellar kinematic data. In Section
\ref{sec:individual_galaxies} we provide notes on the individual
galaxies. In Section \ref{sec:discussion} we discuss the general
kinematic properties of early-type galaxies at large radii obtained
from the combination of stellar and PNe kinematics for all the
galaxies in the sample, and we close with a summary of our work in
Section \ref{sec:summary}.

\begin{table*}
\centering
\caption{Sample galaxies}
\begin{tabular}{l c c c c c c c c c c}
\hline
\hline
\noalign{\smallskip}
Name   &  Type   &      $D$     &$cz$     & $B_T$ & $R_e$& PA  &$N_{PNe}$&$R_{LAST}$& Log $L_X$  &    Reference               \\
\noalign{\smallskip}   
       &         &     (Mpc)    & (\kms)  &       &($''$)&(\deg)&       & (')  &Log (erg s$^{-1}$)&                                \\
 (1)   &  (2)    &      (3)     & (4)     & (5)   &(6)   & (7) &(8)     &(9)   &    (10)      &  (11)                  \\
\noalign{\smallskip}
\hline
\noalign{\smallskip}    
NGC  821$^{*}$  &  E6  &22.4 $\pm$     1.8&     1735   &11.72  &39    & 25  & 123(4) & 6.8  & 40.33       &                \\
NGC 3377$^{*}$  &  E5  &10.4 $\pm$    0.4 &     665    &11.07  &41    & 35  & 151(3) & 10.0 &$<$ 39.60    &                \\
NGC 3608$^{*}$  &  E2  &21.3 $\pm$     1.4&    1253    &11.69  &40    & 75  &  87(5) & 6.8  & 40.01       &                \\ 
NGC 4374$^{*}$  &  E1  &17.1 $\pm$    0.9&    1060    &10.01  &53    & 135 & 450(7)  & 6.9  & 40.83       &                \\
NGC 4564$^{*}$  &  E6  &13.9 $\pm$     1.1&    1142    &12.05  &22    & 47  & 49(1)  & 7.5  &$<$ 39.85    &                \\   
NGC 5846$^{*}$  &  E0  &23.1 $\pm$     2.1&    1714    &10.91  &53    & 70  & 123(1) & 6.0  & 41.65       &                \\    
%\noalign{\smallskip}                           
\hline                                          
%\noalign{\smallskip}                             
NGC 1023$^{*}$  &  S0  &10.6 $\pm$     0.8&    637     &10.08  &46    & 87  & 183(20)& 10.8 & 39.6        & \citet{Noordermeer+08}   \\
NGC 1344        &  E5  &18.4 $\pm$     2.5&     1169   & 11.24 &46    &165  & 194(3) & 6.7  &$<$ 39.48    & \citet{Teodorescu+05}          \\
NGC 3379$^{*}$  &  E1  &9.8  $\pm$     0.5&     889    &10.18  & 47   & 70  & 186(5) & 7.2  & 39.54       & \citet{Douglas+07}       \\
NGC 4494$^{*}$  &  E1  &15.8 $\pm$    0.8&    1344    &10.55  & 53 &  0  & 255(12) & 7.6  & 40.10       & \citet{Napolitano+08} \\
NGC 4697        &  E6  &10.9 $\pm$    0.7&     1236   &10.07  & 66 & 70  & 535(0)  & 6.6  & 40.12       & \citet{Mendez+01}              \\
NGC 5128        &  S0  &4.2 $\pm$ 0.3    &   547       & 7.30 & 255  & 35  & 780     & 8.4  &  40.10      & \citet{Peng+04} \\ 
\hline                                          
NGC 1316        &  S0  &20.0 $\pm$     1.6&     1793   & 9.40  &109   & 50  &  43    &  4.3 & 40.87       & \citet{Arnaboldi+98}           \\
NGC 1399        &  E1  &18.5 $\pm$     1.4&     1447   & 10.44 & 42   &110  &  37    & 8.3  & 41.63       & \citet{Saglia+00}              \\
NGC 3384        &  S0  &10.8 $\pm$    0.7 &     704    &10.75  &50    & 53  &  68    & 2.4  &$<$ 39.52    & \citet{Tremblay+95}            \\
NGC 4406        &  E3  &16.0 $\pm$     1.0&     -244   & 9.74  &91    & 130 &  16    & 4.0  & 42.05       & \citet{Arnaboldi+96}           \\
\noalign{\smallskip}
\hline
\end{tabular}
\label{tab:sample}
\begin{minipage}{17.5cm}
  Notes -- Col.1: Name of galaxy. Galaxies marked with $^*$ have been
  observed with the PN.S.  Col.2: Morphological type of galaxy
  according to the NASA/IPAC Extragalactic Database (NED).  Col.3:
  Distance of galaxy from \citet{Tonry+01} derived using the surface brightness fluctuations
  method. Distance moduli are shifted by $-0.16$
  mag to take into account the new Cepheid zero point of
  \citet{Freedman+01} as done by \citet{Jensen+03}.  Col.4: Galaxies'
  heliocentric systemic velocity listed in NED.  Col.5: Total $B$
  magnitude corrected for extinction and redshift listed in \citet[RC3
    hereafter]{rc3}.  Col.6: Effective radii given by
  \citet{Blakeslee+01}.  Col.7: Galaxy photometric position angle as
  given by RC3, except for NGC 1399 \citep{Saglia+00}, NGC 3379
  \citep{Capaccioli+90} and NGC 4494 \citep{Napolitano+08}.  Col.8:
  Total number of PNe with measured radial velocities. In parentheses
  we give the number of sources excluded by the friendless algorithm
  (not included in the listed total PNe number), for those galaxies
  for which we applied our adaptive kernel smoothing technique (see
  Section \ref{sec:smoothed_fields}).  Col.9: Maximum distance of the
  PNe detections from the galaxy centre.  Col.10: Logarithmic X-ray
  bolometric luminosity according to \citet{OSullivan+01}, except for
  NGC 1023, whose X-ray luminosity was converted from 0.5-2 keV
  measurements of \citet{David+06} to bolometric luminosity following
  the prescription of \citet{OSullivan+01}.  Col.11: Reference for the
  PNe data. The horizontal lines separate the galaxies with new
    PNe catalogues presented in this paper ({\it sample A}), from the
    other galaxies whose PNe data are available in the literature, with
    more than 80 PNe radial velocities ({\it sample B}) and with less
    than 80 PNe radial velocity measurements ({\it sample C}).
\end{minipage}
\end{table*}

\section{Observations and data reduction}
\label{sec:observations}

For galaxies in {\it sample A}, we present new PNe positions, radial
velocities and magnitudes, obtained as part of the PN.S elliptical
galaxy survey. For these {\it sample A} galaxies, data reduction, PNe
identification, radial velocity and magnitude measurements are
performed using the procedures described in \citet{Douglas+07}.
Catalogues with positions, velocities and magnitudes of PNe in galaxies
of {\it sample A} are shown in Table \ref{tab:PNe.catalog}. 
  Heliocentric systemic velocities derived from PNe are in agreement
  within the errorbars with the values tabulated in either the RC3 or
  NED catalogues.  Instrumental magnitudes are shifted in order to
match the bright cut off of the luminosity function with the expected
apparent magnitude $m'=M^*_{5007}+25+5 \log D$, where $D$ is the
distance in Mpc given in Table \ref{tab:sample} and $M^*_{5007}=-4.48$
is the absolute magnitude of the cut-off according to
\citet{Ciardullo+89}.

The number of PNe detected using the PN.S is over 100 for the majority
of the galaxies, with a remarkable number of 457 objects for NGC
4374. This is a major improvement when compared to the results of
previous techniques and instruments (typically $<$50 detections).
In the following sections, we use only PNe whose velocities are
within $3\sigma$ of the mean velocity of their neighbours, according
to the ``friendless algorithm'' described in \citet{Merrett+06}.

\section{PNe spatial distribution}
\label{sec:spatial_distribution}

\begin{figure*}
 \vbox{
   \hbox{
     \psfig{file=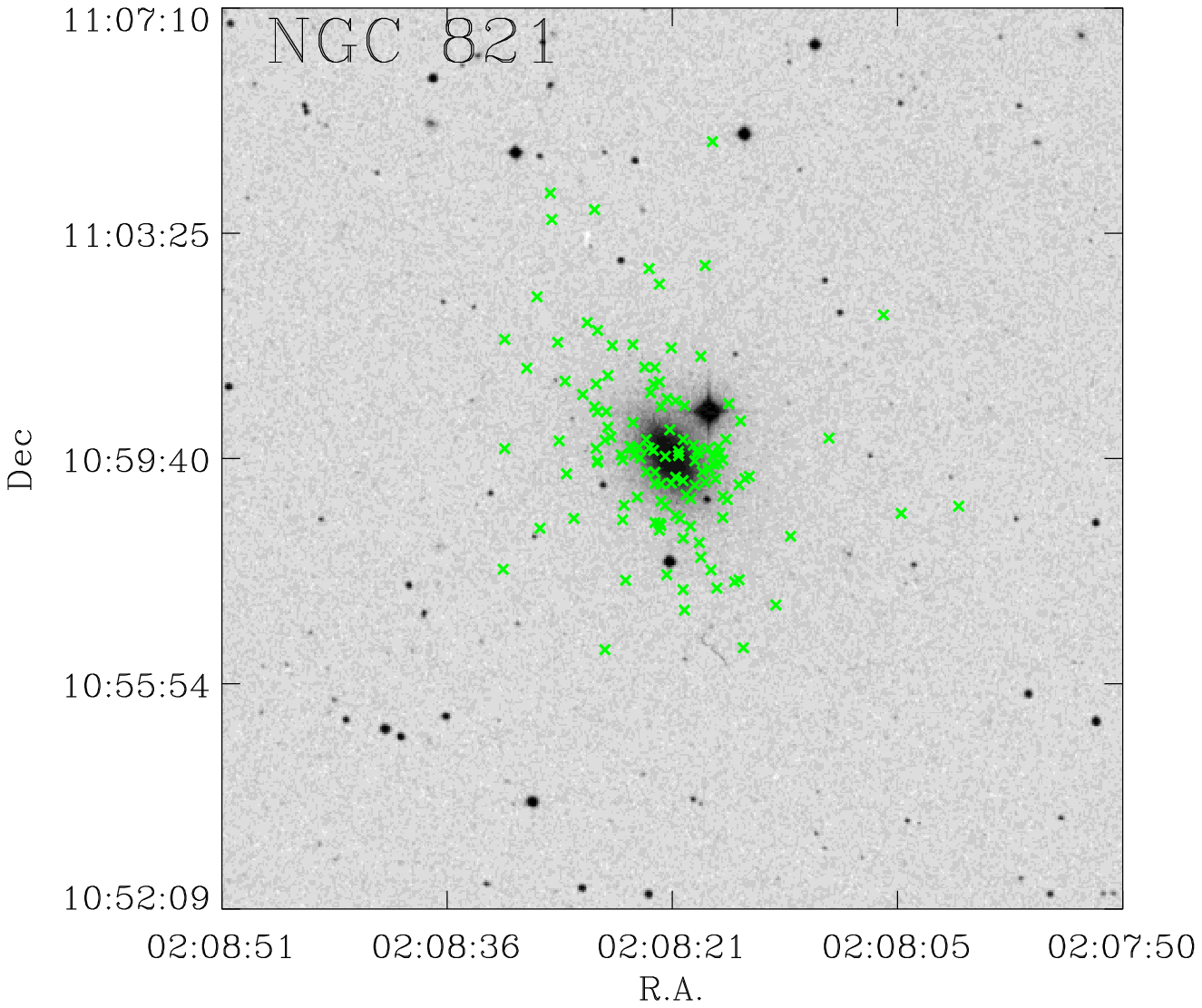,clip=,width=7.5cm}
     \psfig{file=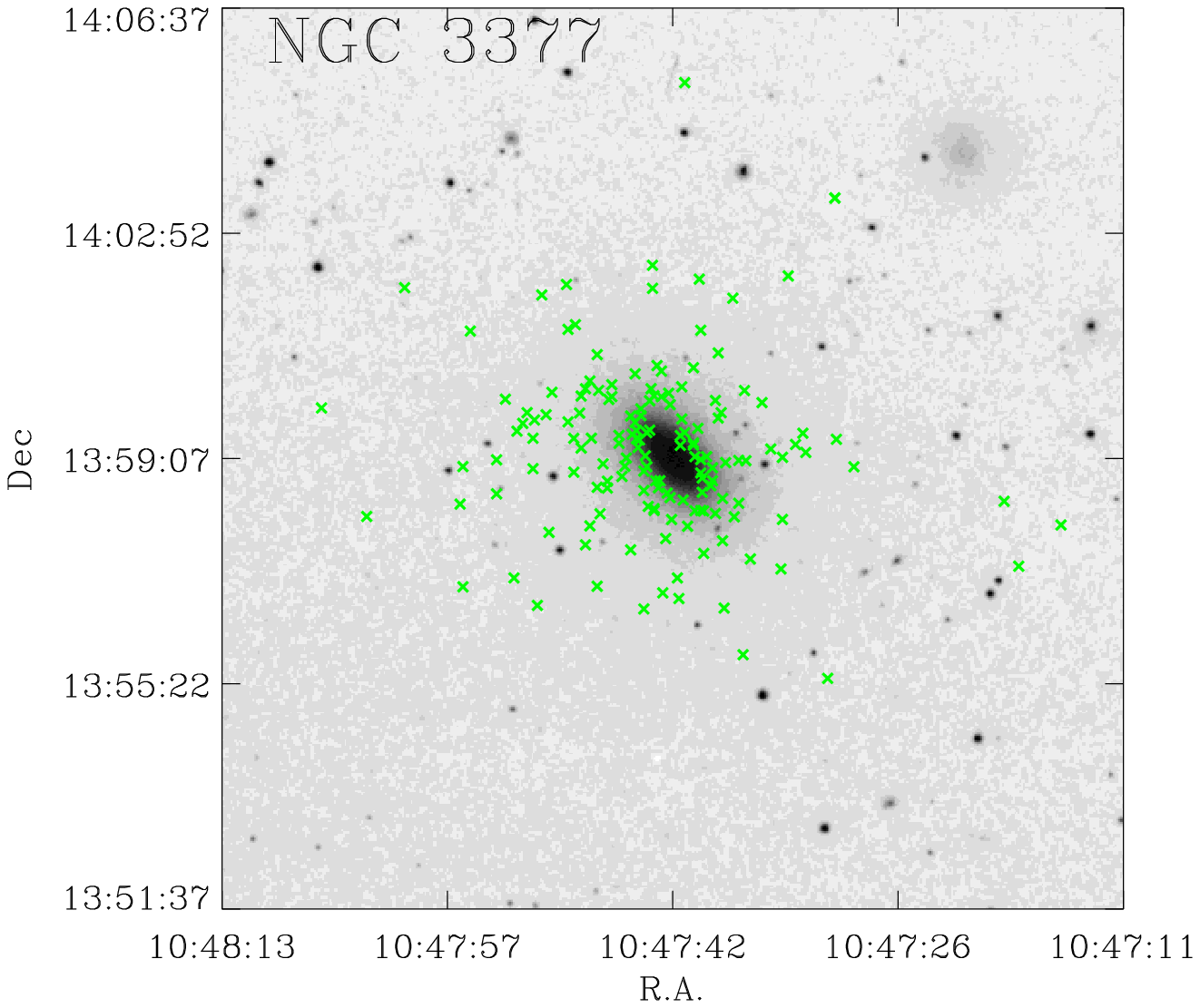,clip=,width=7.5cm}
   }
   \hbox{
     \psfig{file=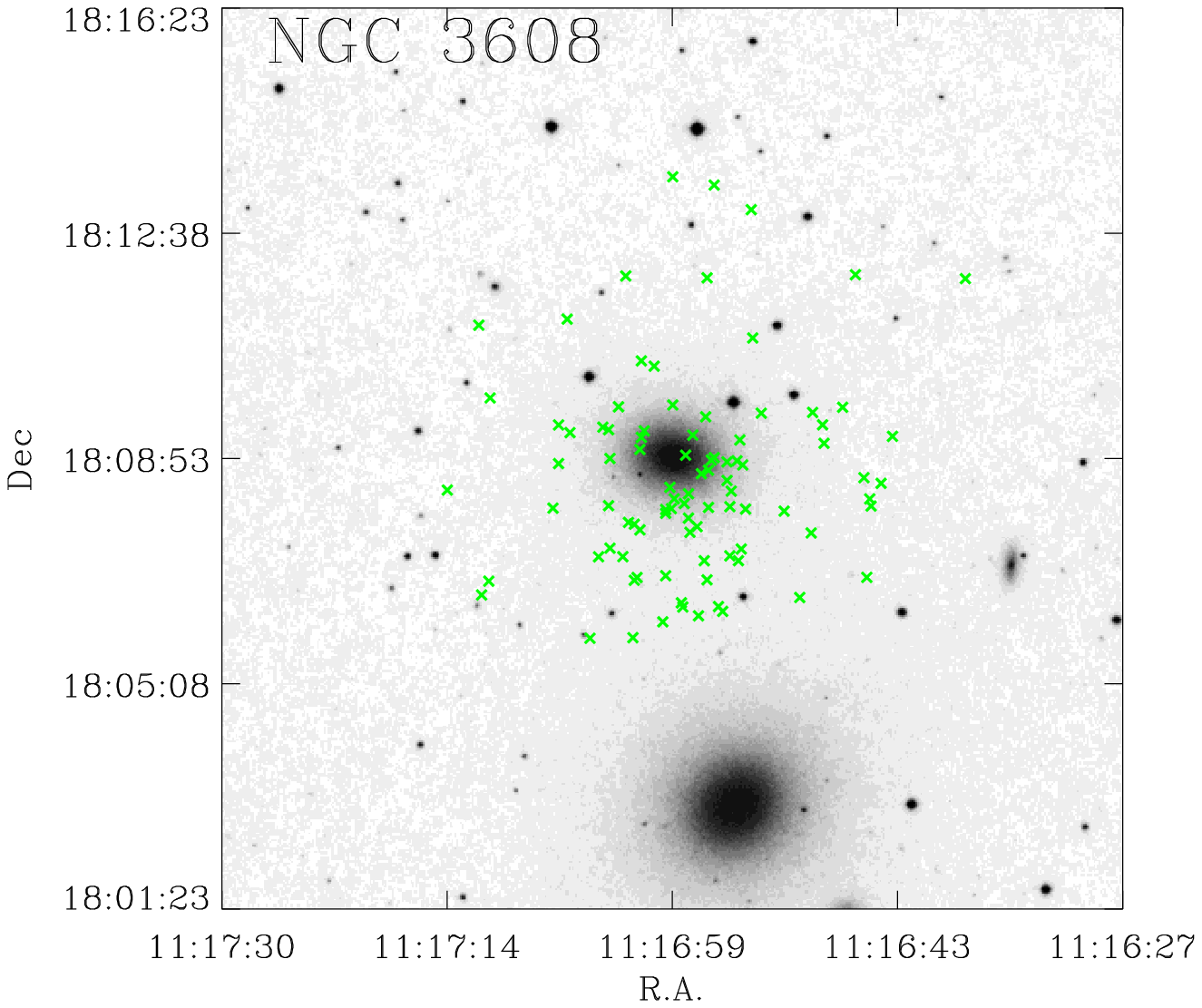,clip=,width=7.5cm}
     \psfig{file=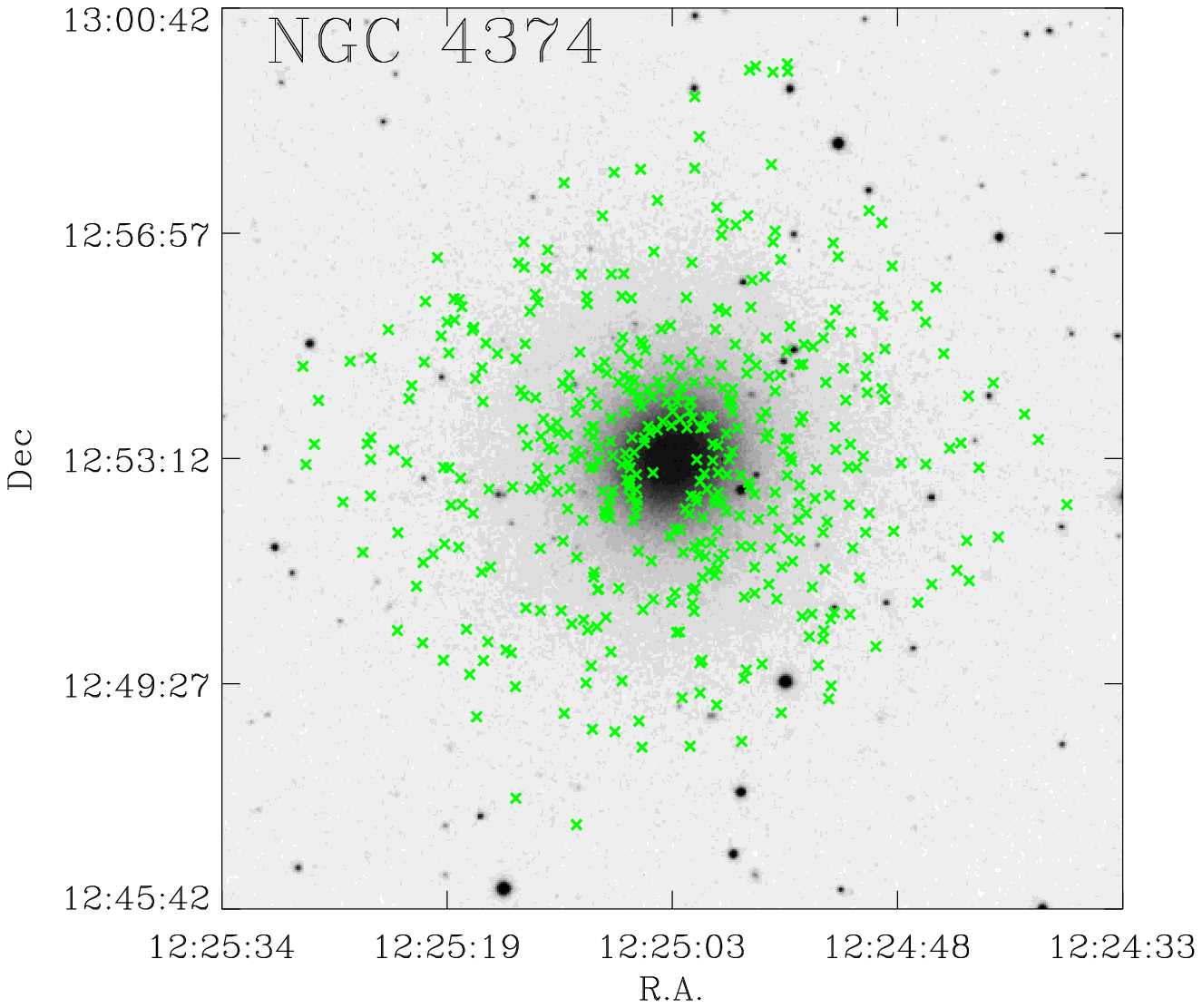,clip=,width=7.5cm}
   }
   \hbox{
     \psfig{file=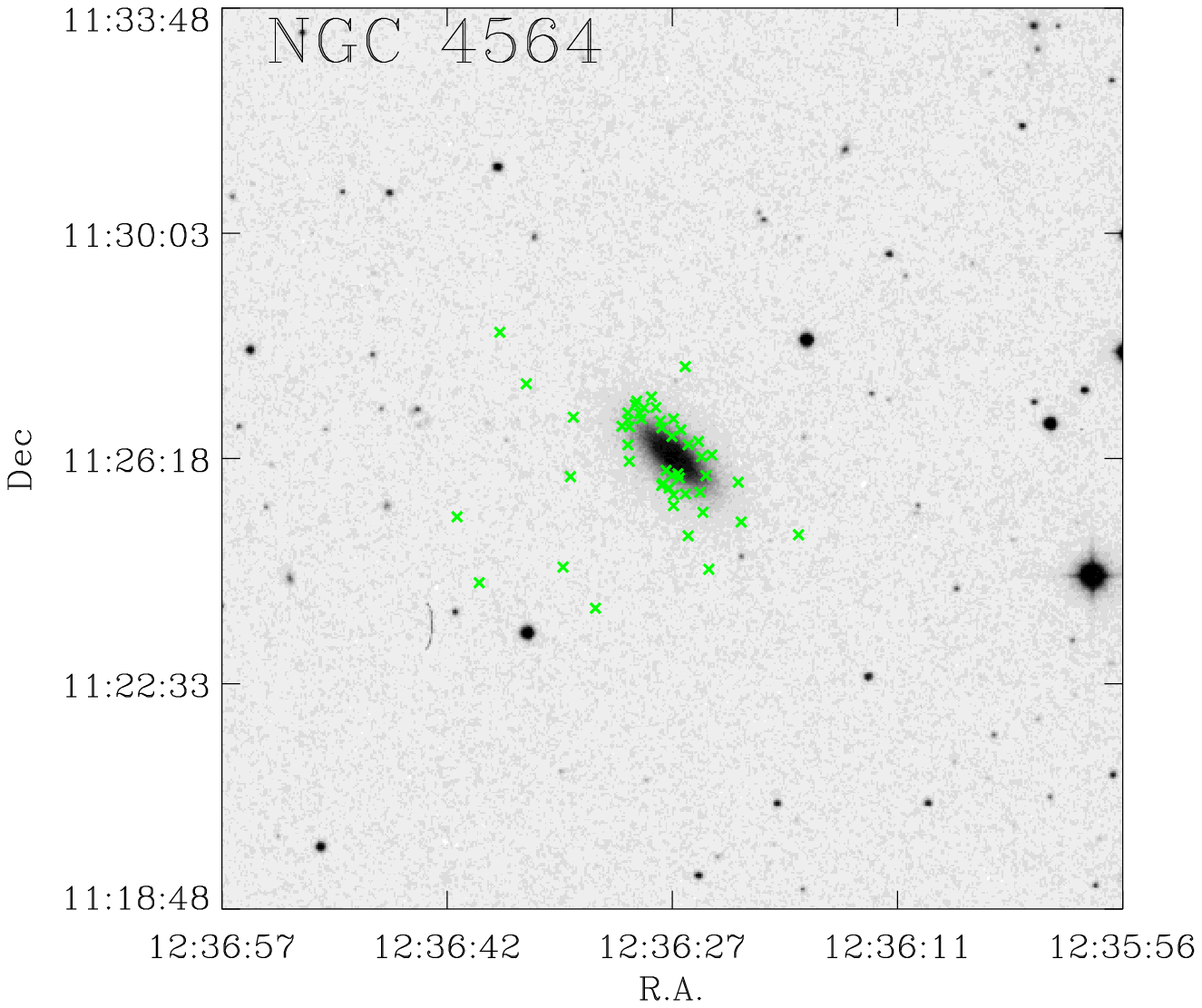,clip=,width=7.5cm}
     \psfig{file=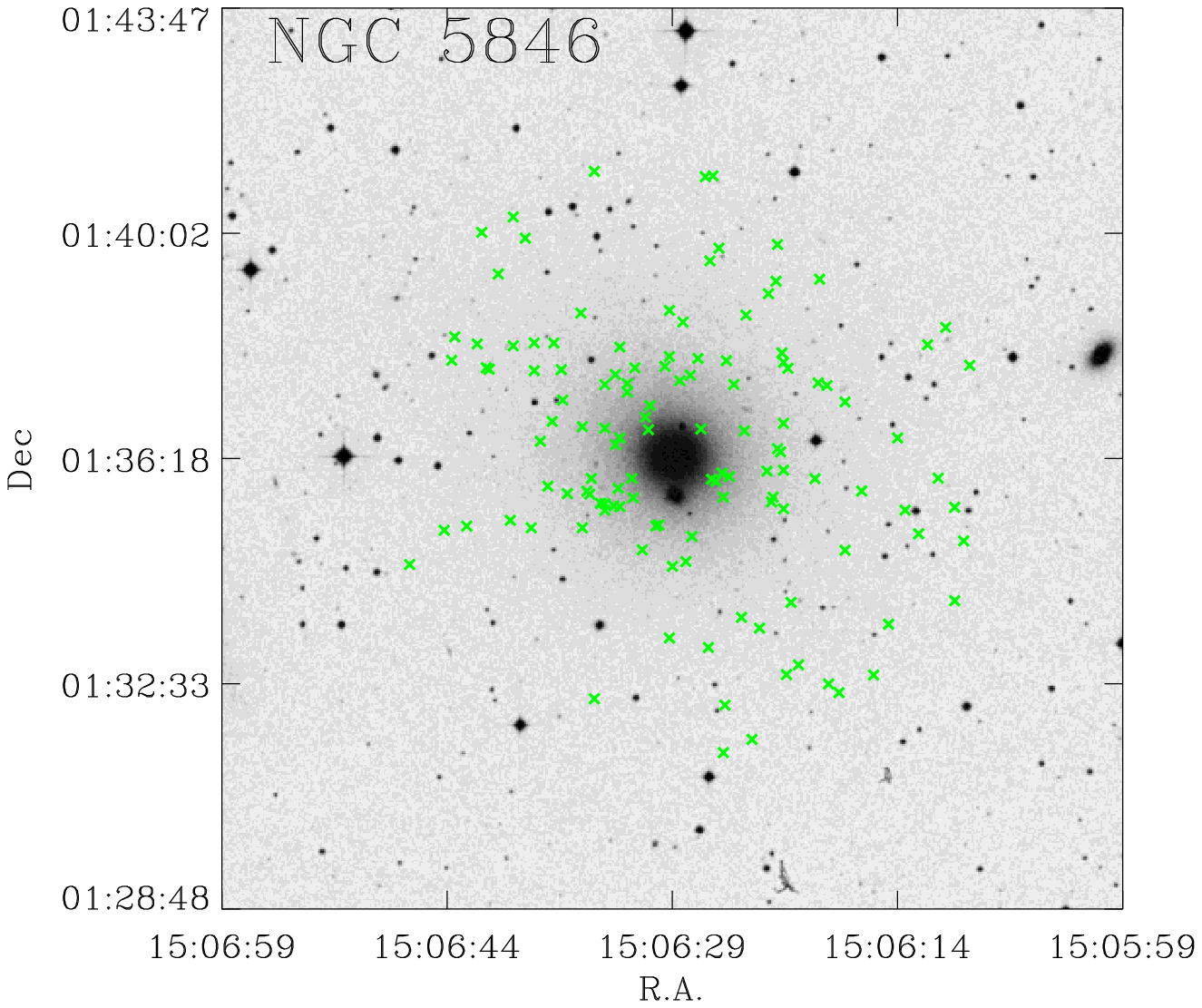,clip=,width=7.5cm}
   }
}
\caption{ DSS images of galaxies in {\it sample A} with the PNe
    positions marked as {\it green crosses}. The field-of-view is
    $13\times 13$ arcminutes. North is up and East is left.}
\label{fig:spatial_distribution}
\end{figure*}

Figure \ref{fig:spatial_distribution} shows the spatial distribution
of the PNe in {\it sample A} galaxies. For the PNe positions in the
galaxies of {\it samples B}  and {\it C}, we refer to the
references listed in Table \ref{tab:sample}.

The first property of the PNe spatial distribution we would like to
investigate is whether or not the number density of the PNe follows
the stellar surface brightness.
This has already been tested for many galaxies in the sample (see
reference list in Table \ref{tab:sample}), finding generally a good
agreement between the surface brightness and PNe counts.
\footnote{In some cases this could be complicated by population
effects. For example, \citet{Sambhus+06} found a subpopulation of
bright PNe in NGC 4697, which are geometrically and kinematically
peculiar. To discover such effects in the PN distribution requires
both a large PN sample and accurate PN magnitudes, depending on
how clear the signature is. Here we only consider how well the
entire PN population agrees with the surface photometry.} 

From Figure \ref{fig:spatial_distribution} we see that to a first
approximation the two-dimensional distribution of PNe is
symmetric. NGC 821 and NGC 4564 are possible exceptions in that they
may show an excess of PNe at large radii on their SE
side. Unfortunately, the number of PNe in those regions is too small
to determine whether the asymmetry in the distribution is
statistically significant.

In the following sections we present a comparative study between PNe
density number density and stellar surface brightness for the galaxies
in {\it sample A}  and {\it B}. For the galaxies in {\it sample
    C}, the PNe samples are too small to perform this analysis.

In Section \ref{sec:photometry} we describe the stellar surface
brightness and in Section \ref{sec:PNe_vs_sb} we compare it to
the radial PNe number density.

\subsection{Photometric data of the stellar component}
\label{sec:photometry}

Radial profiles of the stellar surface brightness, position angle,
ellipticity $\epsilon$, and $a_4$ shape parameter were obtained from
the combination of different datasets in the literature.  Mean values
are shown in Table \ref{tab:photometry} together with the sources of
the photometry.

Usually we considered HST photometry in the innermost $4-5''$, ground
based observations for $r>10-11''$ and their average values in
between.

From the $\epsilon$ and $a_4$ radial profile we derived their
characteristic values by averaging the values within the range $2\cdot
FWHM_{seeing} < R < 1.5 \cdot R_e$ as done in \citet{Bender+88},
\citet{Peletier+90} and \citet{Hao+06}. As weights in the average
procedure, we use the ratios $F/E$ where $F$ is the flux measured at
that position and $E$ is the error in the ($\epsilon$ or $a_4$)
measurement. Errors in $<\epsilon>$ and $<a_4>$ are computed
by measuring the scatter of the measurements in the same
spatial region.

The surface brightness radial profiles $\mu$ were fitted with i) a
S\'ersic profile and ii) a $R^{1/4}$ plus exponential disc profile,
using the robust Levenberg-Marquardt method implemented by
\citet{More+80}. The actual computation was done using the {\tt mpfit}
algorithm implemented by C. Markwardt under the {\tt IDL} environment
\footnote{The most recent version of the {\tt mpfit} algorithm
can be found at
http://cow.physics.wisc.edu/$\sim$craigm/idl/fitting.html}. We would
like to note that the aim is not to study in detail the
photometric components of the galaxies, but instead to obtain a
general description of the mean radial profile of the stellar light.
In the cases where HST data were not available, we excluded the
innermost $1''$ or $2''$ from the fit, to reduce possible
contamination due to seeing effects. Results of the fit are shown in
Table \ref{tab:photometry}.

\begin{table}
\centering
\caption{PNe catalogue for galaxies in {\it Sample A}}
\begin{tabular}{l c c c c}
\hline
\hline
\noalign{\smallskip}
Name   &  R.A.   &    Dec     &$V_{\odot}$&  $m_{5007}$ \\
\noalign{\smallskip}   
PNS-EPN-& hh:mm:ss  & deg:mm:ss  & \kms  & mag  \\
 (1)   &  (2)    &      (3)   & (4)    & (5)  \\
\noalign{\smallskip}
\hline
\noalign{\smallskip}    
NGC821-001 & 02:08:01.6 & +10:58:52.4 & $1670\pm21$ & 28.8 \\   
NGC821-002 & 02:08:05.5 & +10:58:45.2 & $1703\pm21$ & 28.1 \\  
NGC821-003 & 02:08:06.7 & +11:02:03.4 & $1742\pm20$ & 27.7 \\  
NGC821-004 & 02:08:10.4 & +11:00:00.4 & $1725\pm21$ & 28.6 \\  
NGC821-005 & 02:08:13.0 & +10:58:22.4 & $1823\pm22$ & 29.7 \\  
NGC821-006 & 02:08:14.0 & +10:57:13.7 & $1781\pm21$ & 28.7 \\  
NGC821-007 & 02:08:15.8 & +10:59:22.3 & $1669\pm21$ & 28.3 \\  
NGC821-008 & 02:08:16.1 & +10:59:20.0 & $1832\pm21$ & 28.3 \\  
NGC821-009 & 02:08:16.2 & +10:56:31.0 & $1629\pm21$ & 28.7 \\  
NGC821-010 & 02:08:16.4 & +11:00:17.5 & $1681\pm21$ & 28.8 \\  
..         &            &            &            &      \\
NGC5846-124 & 15:06:46.9 & +01:34:32.2 &$1971\pm22$ & 28.1 \\   
\noalign{\smallskip}
\hline
\end{tabular}
\label{tab:PNe.catalog}
\begin{minipage}{8.1cm}
  Notes -- The complete table is published as Supplementary Material in the
           online version of this article. PNe catalogues are also available at 
           http://www.strw.leidenuniv.nl/pns.
   Col.1:  PNe name according to the IAU standards 
           ({\tt PNS-EPN-<galaxy name>-<sequential number>}).
   Col.2:  Right ascension (J2000).
   Col.3:  Declination (J2000).
   Col.4:  Radial velocity in the heliocentric reference system.
   Col.5:  Apparent $m_{5007}$ magnitude. Magnitudes have an error of $\pm0.4$ mag.
\end{minipage}
\end{table}

\subsection{Comparison between PNe spatial distribution and surface brightness profiles}
\label{sec:PNe_vs_sb}

In order to compare the stellar surface brightness and the PNe number
density, we have to take into account the incompleteness in the PNe
detections, e.g. the number of undetected PNe because of their
low signal-to-noise ratio. The noise in the field has four main
contributions, the sky surface brightness (constant all over the
field), the detector readout noise (constant all over the field), the
galaxy background (variable over the field) and the presence of
foreground stars (variable over the field).

To measure how these four noise sources affect the PNe detection and
correct the PN number counts we proceed as follows:

\begin{figure*}
 \vbox{ \hbox{
     \psfig{file=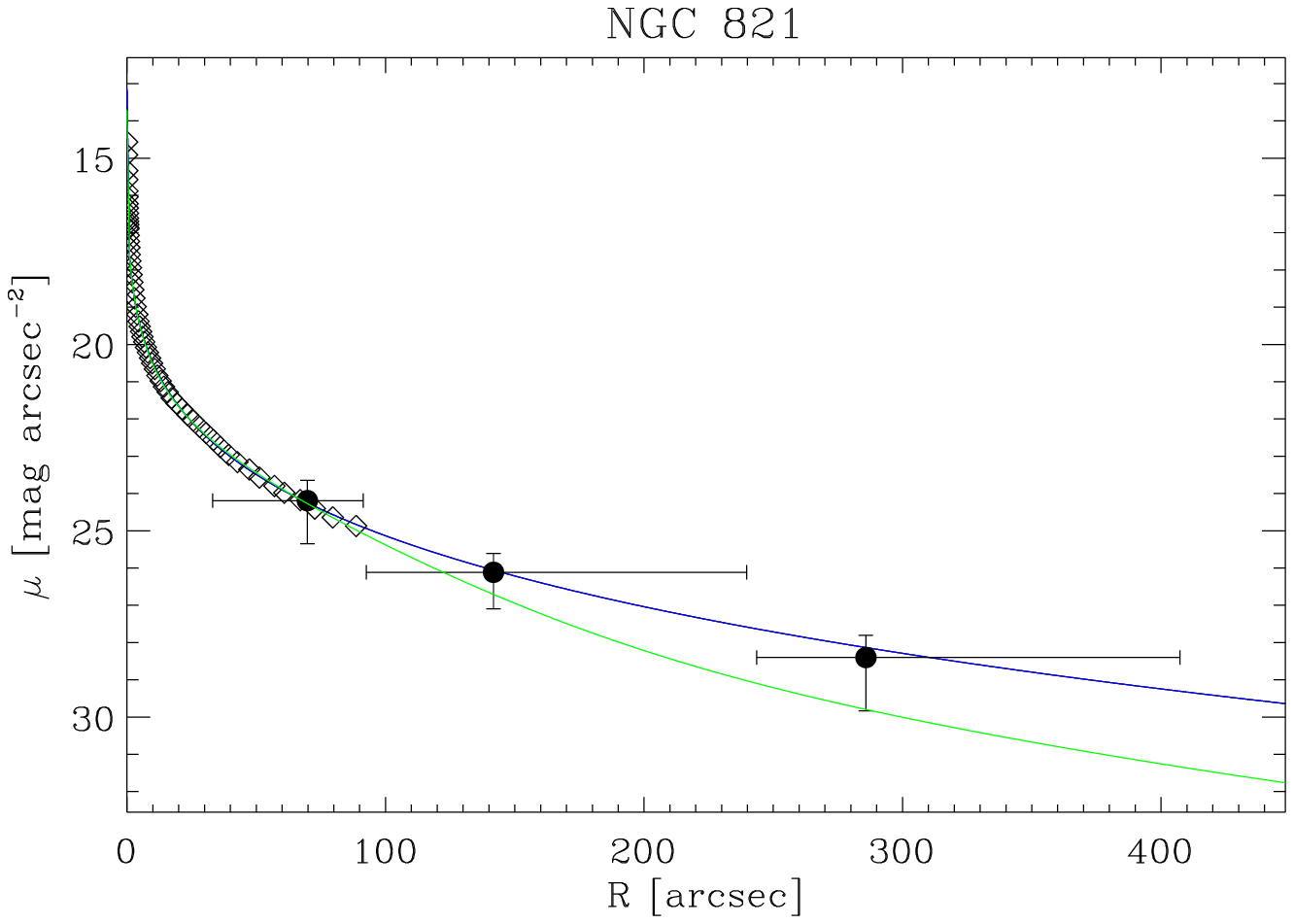,clip=,width=5.7cm}
     \psfig{file=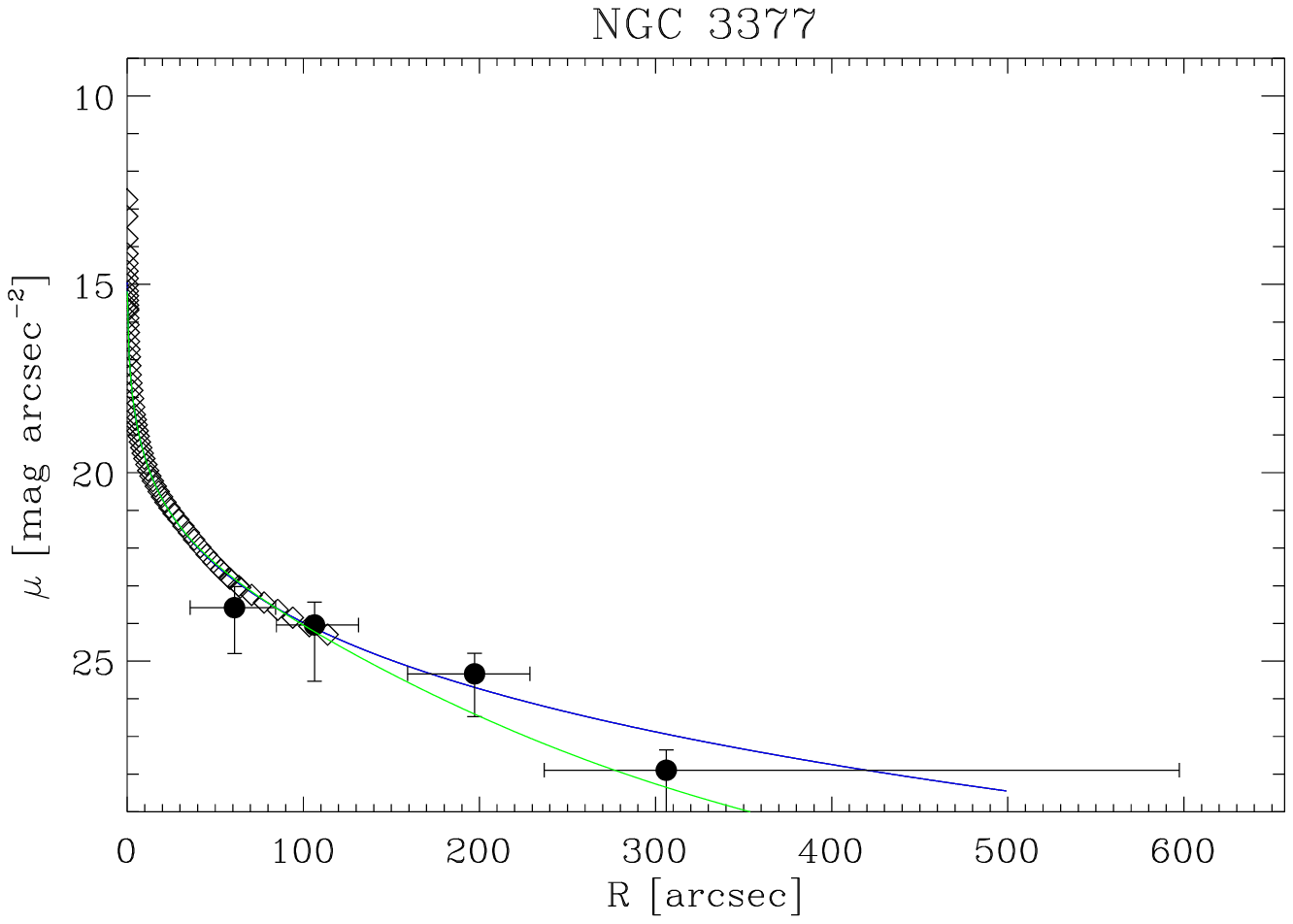,clip=,width=5.7cm}
     \psfig{file=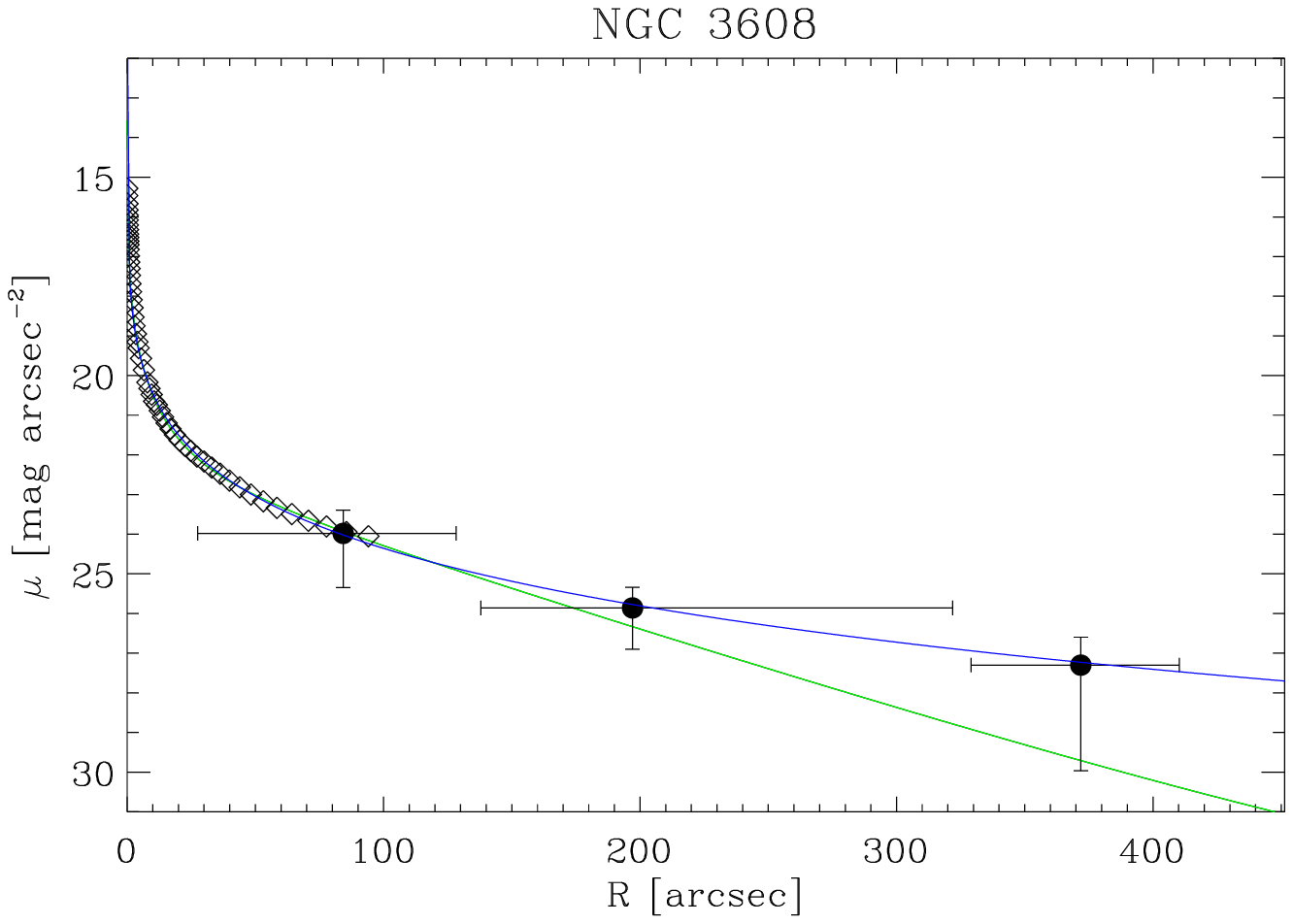,clip=,width=5.7cm}} 
\hbox{
     \psfig{file=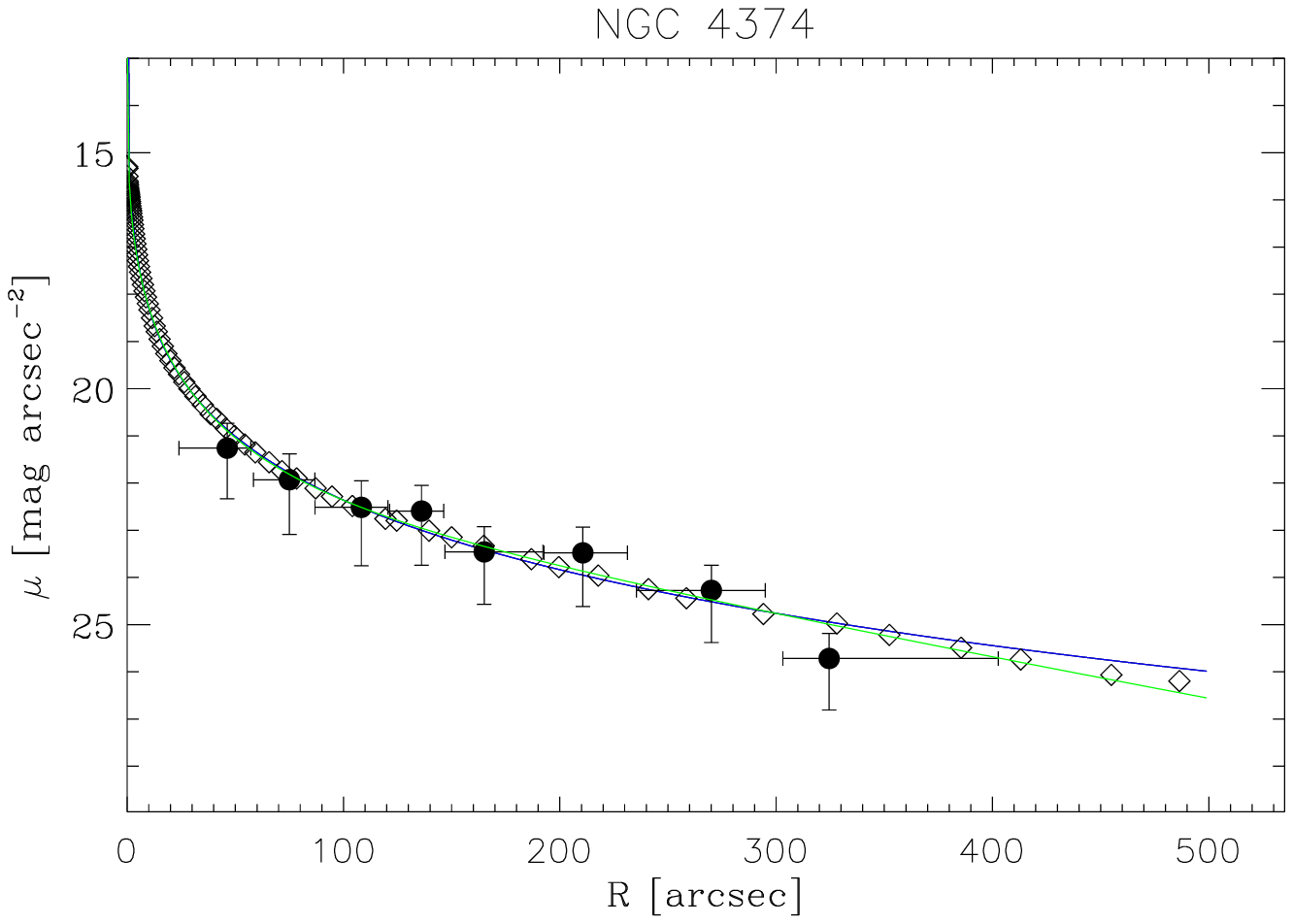,clip=,width=5.7cm}
     \psfig{file=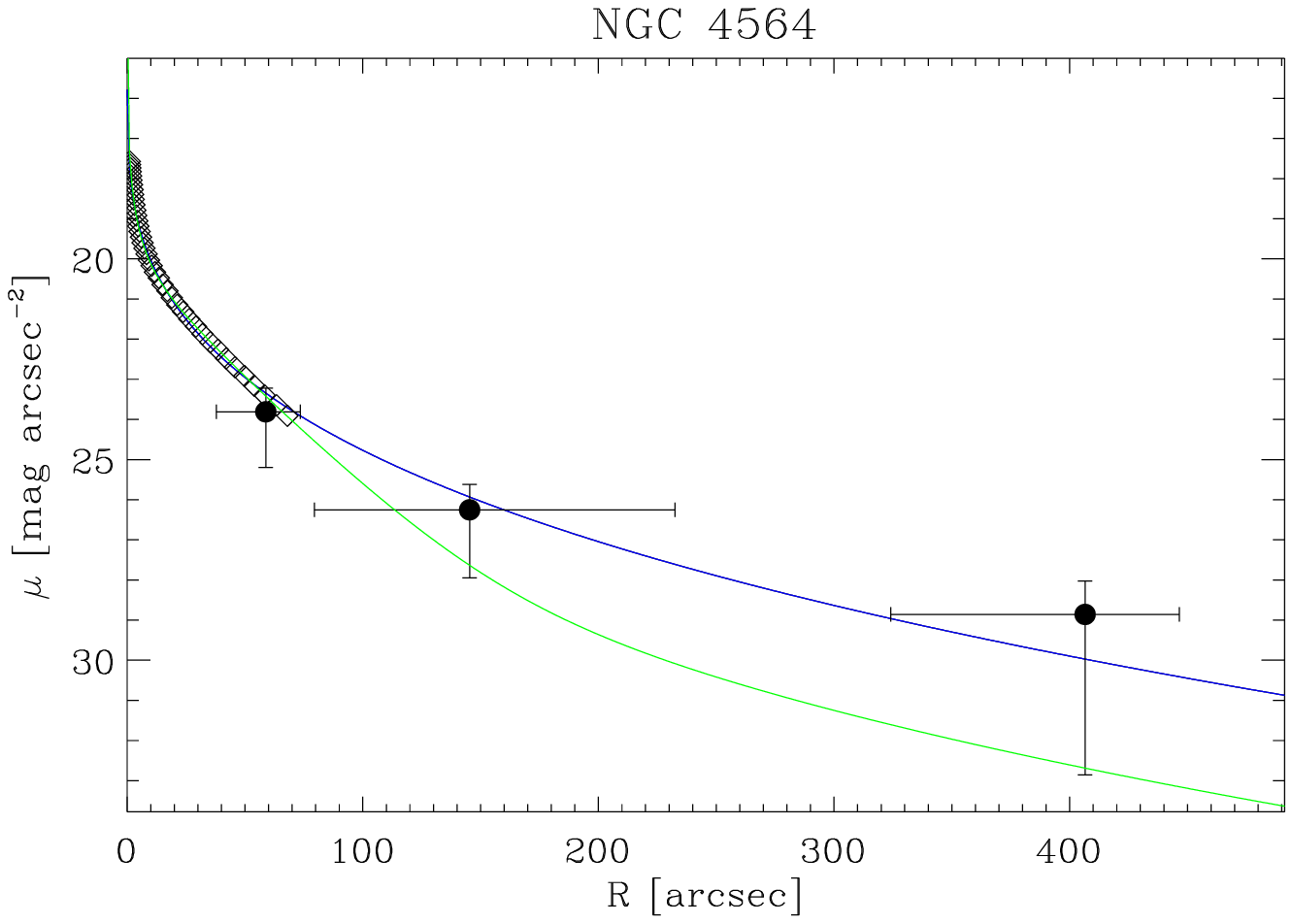,clip=,width=5.7cm}
     \psfig{file=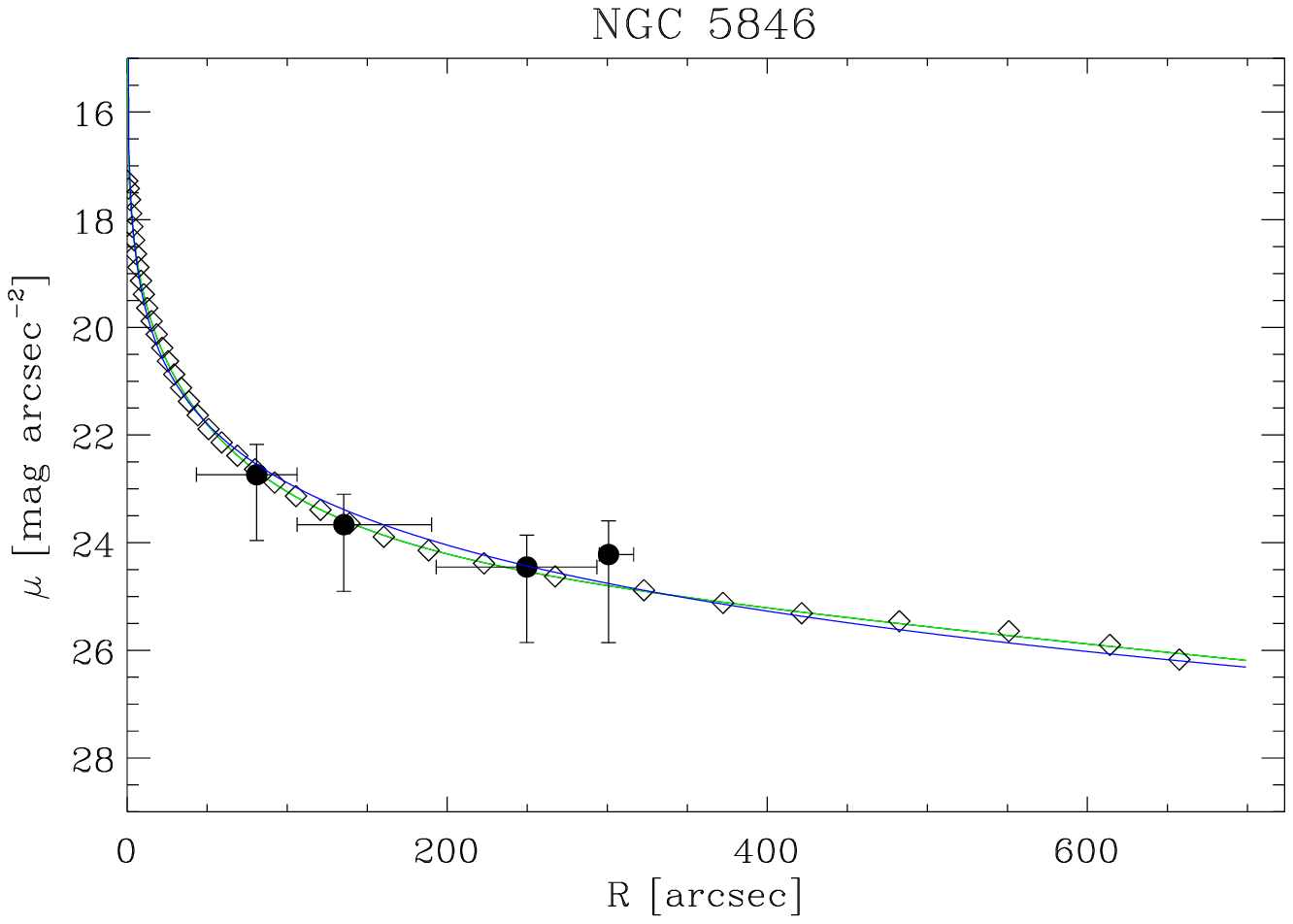,clip=,width=5.7cm}} 
\hbox{
     \psfig{file=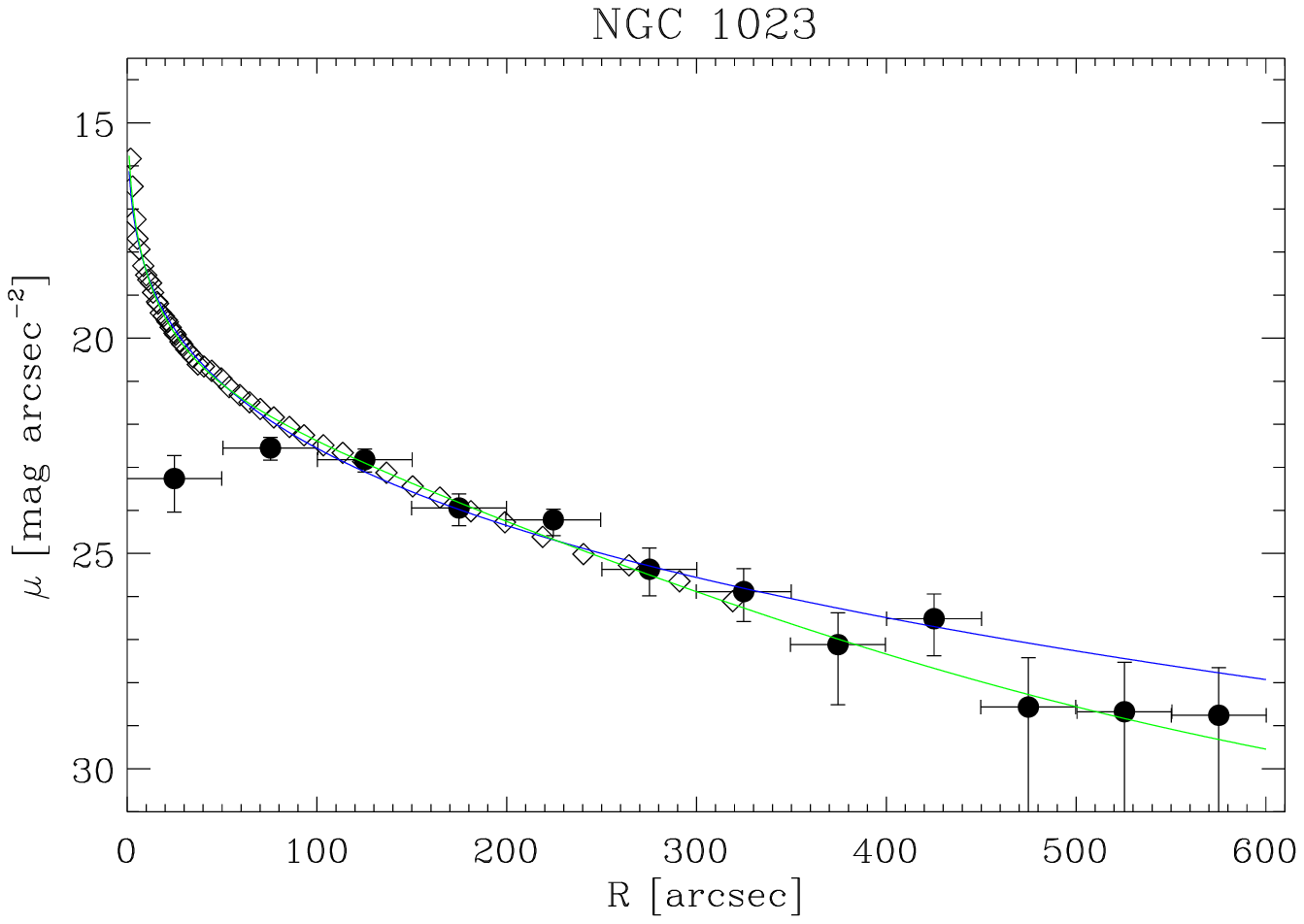,clip=,width=5.7cm}
     \psfig{file=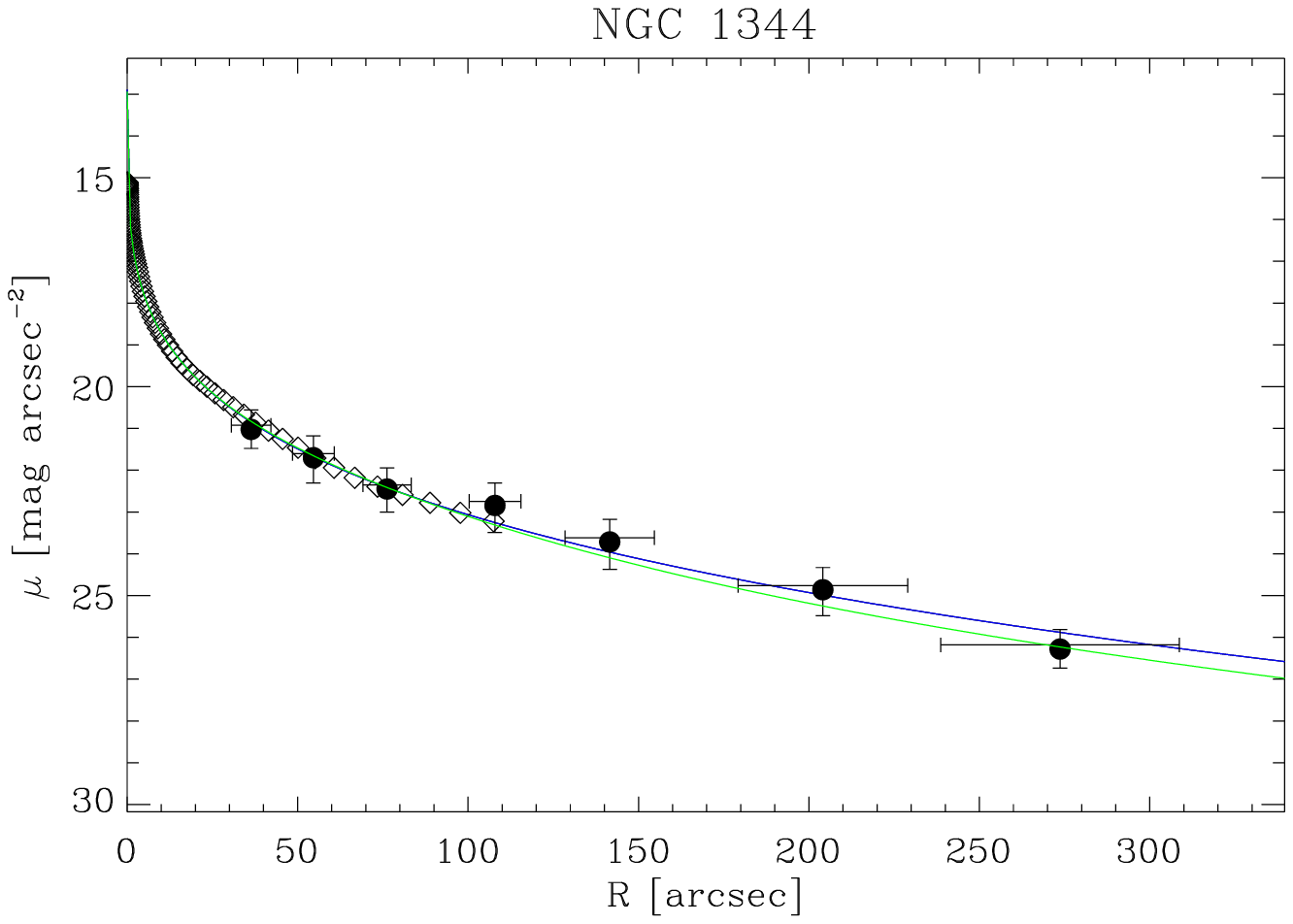,clip=,width=5.7cm}
     \psfig{file=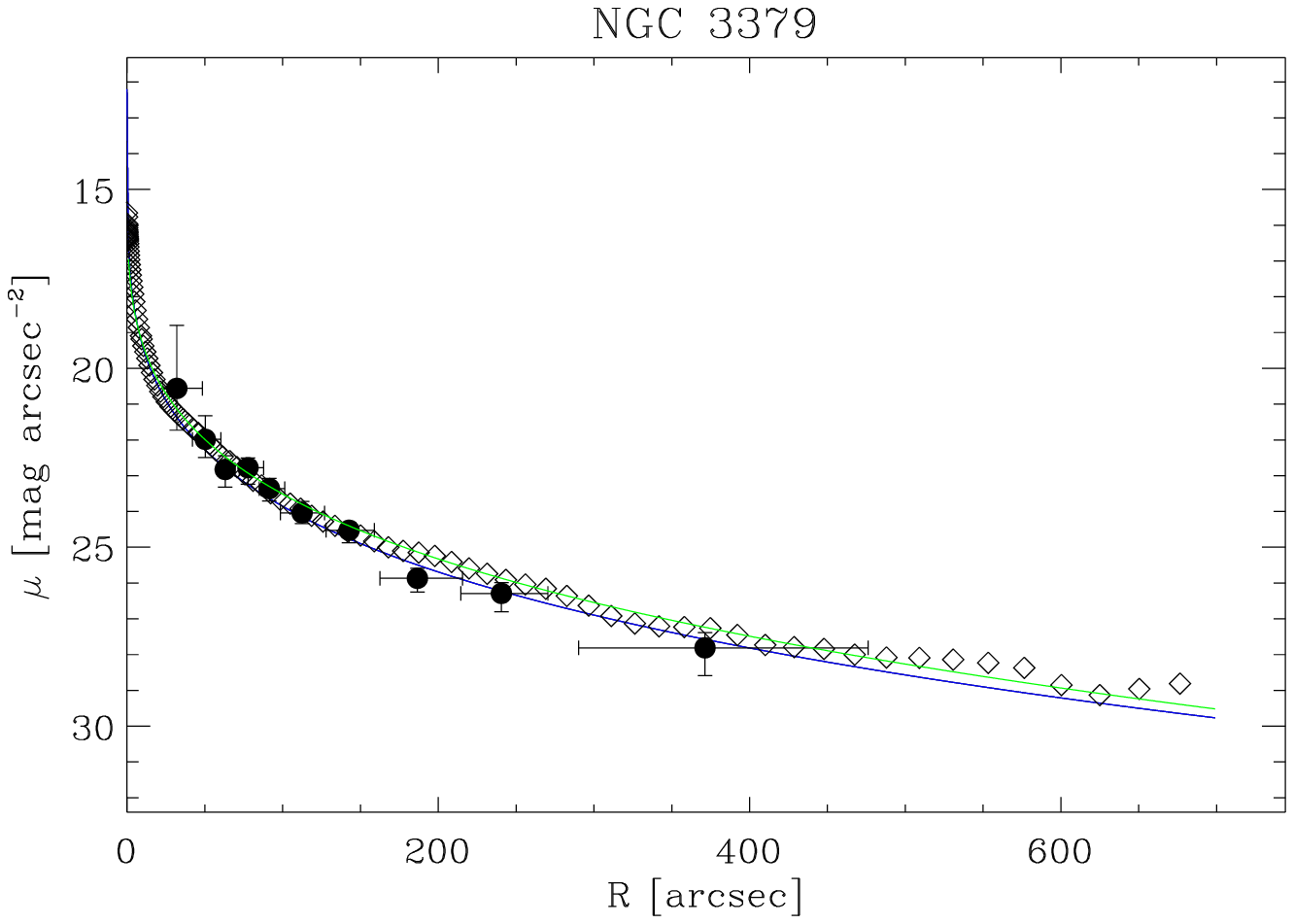,clip=,width=5.7cm}} 
\hbox{
     \psfig{file=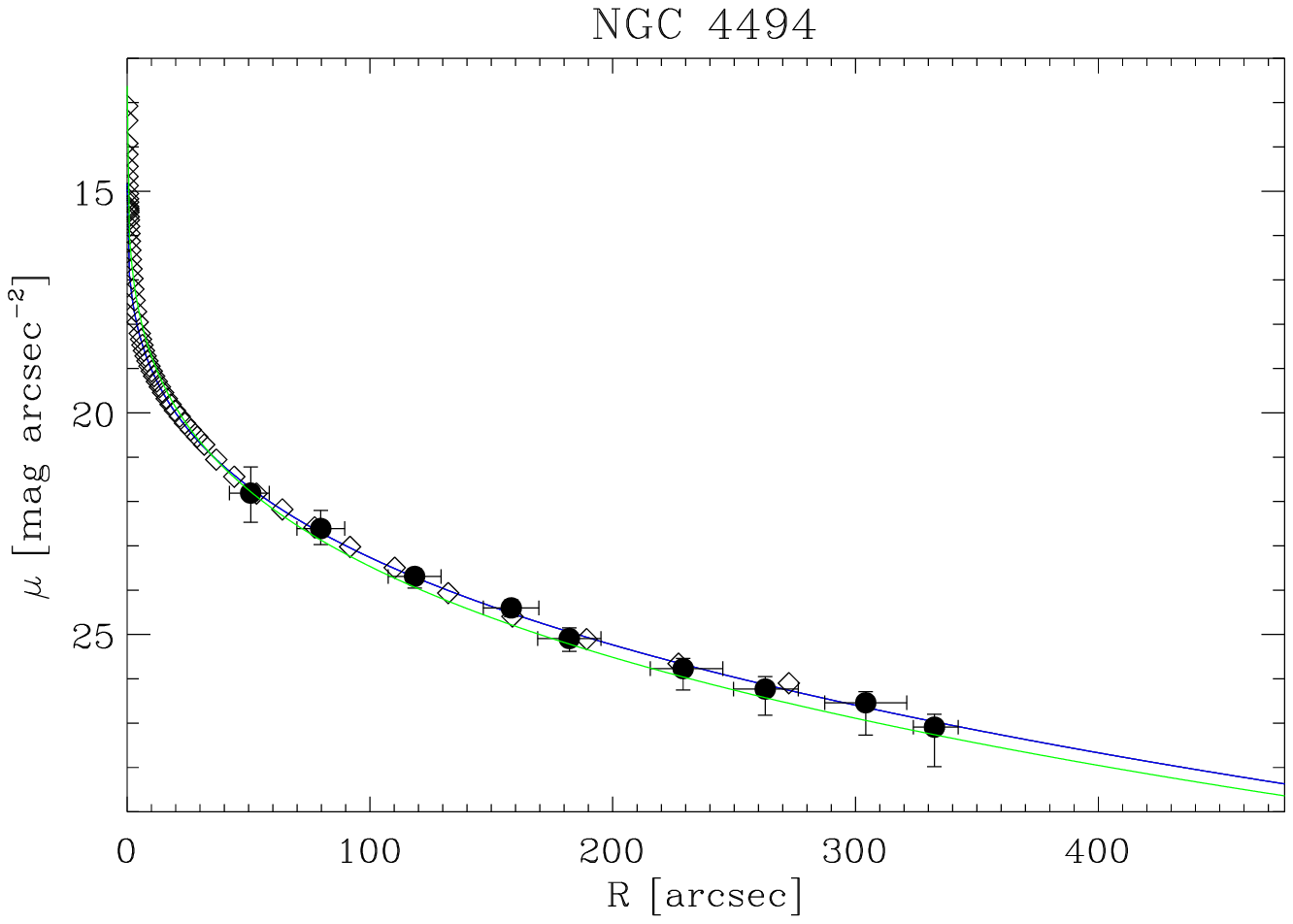,clip=,width=5.7cm}
     \psfig{file=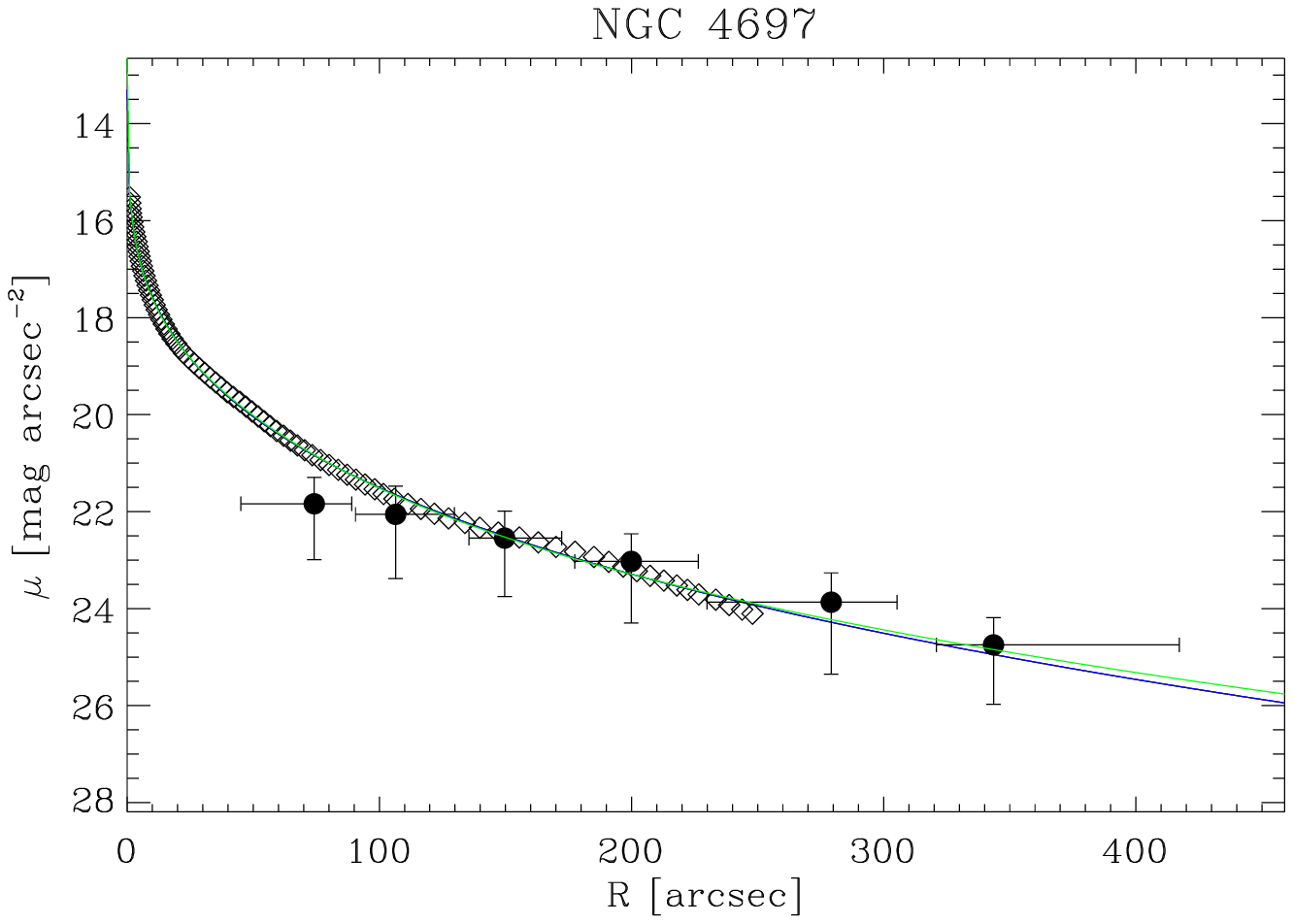,clip=,width=5.7cm}}

 }

 \caption{Surface density profiles of PNe and stars. {\it Open
     diamonds}: Surface brightness of the stellar component as function
   of semi-major axis, according to the references in Table \ref{tab:sample}.
   {\it Solid lines:} Extrapolation of the stellar surface brightness
   profile with the {\it blue line} representing the S\'ersic law and
   {\it green line} representing the $R^{1/4}$ plus exponential
   decomposition. {\it Filled circles with error bars}: scaled
   logarithmic PNe number density (see text for details).}
\label{fig:photom_comparison}
\end{figure*}

\begin{itemize}

\item{} We first generate artificial images, which mimic the sky
  and CCD readout noise in our observations and then we populate them
  with simulated point-like sources of different magnitudes.

\item{} We compute the planetary nebulae luminosity function (PNLF) of
  the simulated objects in the artificial images. In particular, we
  evaluate the magnitude $m_{80\%}$ at which the loss of objects is
  significant. $m_{80\%}$ is defined in such a way that for
  $m<m_{80\%}$, more than 80\% of PNe are recovered.

\item{} Now we consider the real, observed images and we insert a set
  of artificial point-like sources in them with a distribution in
  magnitude given by the observed PNLF, but brighter than
  $m_{80\%}$. In this way we are sure (at a 80\% confidence level)
  that our artificial sources would not be undetected because of the
  background noise (sky plus CCD readout noise).

\item{} We detect the artificial sources in the real images again, and
  since we know their positions, we can compute the radial
  completeness factor $c_{R}$, i.e., the inverse of the fraction of
  simulated objects recovered at different distances from the galaxy
  centre\footnote{Distance from the centre is computed taking into
    account the mean galaxy ellipticity: $R^2=X^2/(1-e)^2+Y^2$, where
    $Y$ is aligned along the major axis.}. Since we inserted only
  sources with $m<m_{80\%}$, the loss of objects is attributed only to
  the galaxy continuum and the stellar trails contamination rather
  than the sky background or detector noises.

\item{} Finally, we group the observed PNe in elliptical annuli, oriented
  along the galaxy photometric major axis. We compute $N_c/A$, the
  number of PNe brighter than $m_{80\%}$ in each annulus (corrected
  for the radial completeness factor $c_{R}$ evaluated at the annuli
  centres), divided by the area $A$ of the annulus. We then
  compute the logarithmic PNe number density
  $\tilde\rho_{PNe}(R)=-2.5\log_{10} \left( N_c/A \right)$.

\end{itemize}

In Figure \ref{fig:photom_comparison} we show the comparison between
the major axis stellar surface brightness profile and
$\tilde\rho_{PNe}(R)$ for galaxies in {\it samples A}  and {\it
  B}\footnote{NGC 5128 is not included because of the lack of an extended surface brightness
  radial profile.}.  $\tilde\rho_{PNe}(R)$ is shifted by an arbitrary
constant (different for each galaxy) to match the stellar profile.

For NGC 1344 and NGC 4697, which are not part of the PN.S dataset, we
could not evaluate the completeness correction with simulated sources
because the reduced images were not available. This affects mostly the
innermost regions of NGC 4697, in which the inner PNe density profile
in Figure \ref{fig:photom_comparison} falls slightly below the stellar
surface brightness profile.

\begin{table*}
\begin{scriptsize} 
\centering
\caption{Photometric properties of 11 sample galaxies.}
\begin{tabular}{l c c c c c c c c c c c c}
\hline
\hline
\noalign{\smallskip}
Name      &Band&  $\mu_S$        &    $R_S$       &  n            & $\mu_B$   & $R_{eB}$    &$\mu_0$    & $h$    &$< \epsilon >$  & $<a_4\cdot100/a>$  & Reference \\
\noalign{\smallskip}                                                                                                       %                                                      
 NGC      &    &[mag arcsec$^{-2}$]&[$''$]         &               &  [mag arcsec$^{-2}$] &   [$''$] & [mag arcsec$^{-2}$] &  [$''$]       &              &        [\%]       &             \\ 
 (1)      & (2)&  (3)          &    (4)            & (5)           &  (6)      &   (7)      &(8)         & (9)     &(10)           &        (11)       &   (12)    \\ 
\noalign{\smallskip}                                                                                                              %                                                      
\hline                                                                                                                             %                                                      
\noalign{\smallskip}                                                                                                             %                                                      
     821  & B  & $23.0 \pm 0.1  $  & $39.8 \pm 2.0$ & $4.7 \pm 0.2$ &$22.0\pm0.1$&$20\pm2$   &$22.4\pm0.2$  &$29\pm1$    &$0.38 \pm 0.01$ & $0.70 \pm 0.1$     & (1),(2)   \\
    3377  & B  & $22.6 \pm 0.6$    & $54 \pm 4 $    & $5.2 \pm 0.3$ &$21.5\pm0.2$&$26\pm3$   &$22.1\pm0.4$  &$40\pm3$    &$0.50 \pm 0.01$ & $0.9 \pm 0.9$     & (1),(2),(3)   \\
    3608  & B  & $25.28\pm 0.09$   & $157\pm10$     & $7.0\pm0.1$   &$21.9\pm0.1$&$19\pm1$   &$22.5\pm0.1$&$53\pm3$&$0.19 \pm 0.02$ & $-0.15\pm0.09$    & (2),(3)          \\  
    4374  & V  & $23.1 \pm 0.2 $   & $142 \pm 16$   & $8.0 \pm 0.6 $&$20.62\pm0.05$&$37\pm1$ &$22.4\pm0.08$&$123\pm3$&$0.13 \pm 0.02$ & $-0.4 \pm 0.1 $   &   (4)          \\ 
    4564  & B  & $22.1 \pm 0.1$    & $33.8 \pm 2$   & $3.1 \pm 0.2$ &$21.42\pm0.05$&$13.3\pm0.5$&$20.33\pm0.02$&$12.90\pm0.08$&$0.39 \pm 0.11$ & $0.7 \pm 0.9$ & (1)       \\
    5846  & V  & $29  \pm1$        & $2903\pm192$   &$12\pm2$       &$22.52\pm0.04$&$68\pm2$ &$24.26\pm0.07$&$379\pm16$&$0.07 \pm 0.05$ & $0.0 \pm 0.1$     & (5)          \\  
\hline	       						    
    1023  & B  & $21.4\pm0.1$      & $60\pm2$       & $3.9\pm0.2$   &$20.6\pm0.1$  &$31\pm3$ &$21.2\pm0.1$  &$62\pm1$      & $0.39\pm 0.05$ & $0.54 \pm 0.1$ $^{(*)}$& (6)    \\
    1344  & V  & $21.5 \pm 0.1$    & $50 \pm 2$     & $4.1\pm0.1$   &$21.27\pm0.07$&$42\pm2$ &$21.9\pm0.5$  &$34\pm3$      & $0.31\pm0.03$  & $0.1\pm0.2$   & (7)    \\
    3379  & B  & $22.1 \pm 0.1$    & $47.0\pm0.2$   &  $4.7\pm0.2$  &$22.25\pm0.08$&$57\pm1$ & $17.2\pm0.5$ & $0.1\pm0.5$  & $0.09\pm 0.02$ & $0.2 \pm 0.1$ & (8),(9)    \\
    4494  & V  & $21.63 \pm 0.07$  & $49\pm 1$      & $3.3 \pm 0.1$ &$21.74\pm0.06$&$49\pm2$ &$22.1\pm0.5$  &$30\pm7$      & $0.14\pm0.01$  &  $0.2\pm 0.2$ & (10)    \\ 
    4697  & B  & $20.60 \pm 0.04$  & $66\pm1$       &$3.53\pm0.06$  &$20.97\pm0.04$&$74\pm1$ &$20.1\pm0.3$  & $26\pm3$     & $0.42\pm 0.04$ & $1.4\pm 0.2$  & (2),(11)\\ 
\noalign{\smallskip}
\hline
\end{tabular}
\label{tab:photometry}
\end{scriptsize} 
\begin{minipage}{17.5cm}
Notes -- 
         Col.1: Galaxy name. Galaxies in {\it sample A} are listed in
         the upper part of the table.  The lower part of the table
         contains the {\it sample B} galaxies except NGC 5128.
         Col.2: Photometric band.
         Col.3: Surface brightness at $R_S$. 
         Col.4: Scale radius $R_S$ determined from the S\'ersic fit. 
         Col.5: S\'ersic index.
         Col.6: Bulge  brightness at $R_{eB}$ for the $R^{1/4}+$ exponential disc fit.
         Col.7: Bulge effective radius for the $R^{1/4}+$ exponential disc fit.
         Col.8: Central surface brightness of the exponential disc for the $R^{1/4}+$ exponential disc fit.
         Col.9: Scale radius of the exponential disc component fit. 
         Col.10: Weighted mean value for ellipticity in the range $2\cdot FWHM_{seeing} < R < 1.5 \cdot R_e$. 
         Col.11: Weighted mean value for the $a_4$ parameter in the
         	range $2\cdot FWHM_{seeing} < R < 1.5 \cdot R_e$.  Positive
         	values correspond to discy isophotes, negative values
         	to boxy isophotes. $^{(*)}$ From \citet{Emsellem+07}. 
                In that paper the error on $a_4$ is not given, therefore 
                we assumed an arbitrary error of 0.1.
         Col.12: Reference for the photometric datasets: 
	 	(1) \citet{Goudfrooij+94}; (2) \citet{Lauer+05}; 
		(3) \citet{Jedrzejewski87}; (4) \citet{Kormendy+08}; 
		(5) \citet{Kronawitter+00}; (6) \citet{Noordermeer+08};
	 	(7) \citet{Sikkema+07}; (8) \citet{Capaccioli+90}; 
		(9) \citet{Gebhardt+00}; (10) \citet{Napolitano+08}; 
		(11) \citet{DeLorenzi+08a}.
\end{minipage}
\end{table*}

The conclusion we can derive from this comparison is that the stellar
surface brightness and PNe counts agree well for the galaxies where
the two sets of data overlap in radius. In the cases where the
stellar surface brightness measurements are not extended enough in
radius to ensure an overlap with the PNe data, the PNe number density
still follows the extrapolation of the surface brightness fits at
larger radii.

\section{PNe kinematics}
\label{sec:pne_kinematics}

In this section we analyse the kinematics of the sample galaxies using
their PNe radial velocities and compare them to long-slit
absorption-line kinematics.

In Section \ref{sec:folded_catalogue} we define the ``folded
catalogue'', which we will use in Section \ref{sec:smoothed_fields},
\ref{sec:testing} and \ref{sec:results_fields} to measure the global
velocity and velocity dispersion fields of the PNe system and their
associated errors. In Section \ref{sec:comp_with_sk} we extract
velocity and velocity dispersion radial profiles from the PNe data and
compare them to stellar profiles, to test the consistency of the two
sources of kinematic information.

 This analysis is done only for galaxies in samples A and B for
  which sufficient data are available.

\subsection{Folded catalogue}
\label{sec:folded_catalogue}

To decrease the statistical noise resulting from a low detection
number, we assume that the galaxy is point-symmetric in phase space.
Therefore each point in phase space, $(x,y,\vel)$ has a {\it mirror}
counterpart $(-x,-y,-\vel)$. Here $x$ and $y$ give the positions of the
PNe on the sky, and are centred on the galaxy with $y$ aligned along
the galaxy photometric major axis, and $\vel$ gives the radial velocities
of the PNe corrected for the galaxy's systemic velocity (therefore
$\vel=0$ at $x=0$, $y=0$).
By the union of the original data set and the mirrored one, we {\it
  virtually} double the number of data points. This technique has
often been adopted in the past (e.g., \citealt{Arnaboldi+98,
  Peng+04}). Hereafter, the term ``folded catalogue'' refers to the
PNe catalogue obtained by the union of the original one and the
mirrored counterpart, and we use this one to map the 2-D kinematics of
the PNe system. The reliability of the point symmetry assumption is
tested {\it a posteriori} by repeating the calculations using only the
original catalogue and comparing the two results. We note that in
almost all cases the computed quantities based on the original and
folded PNe samples agree well; only in the case of NGC 4374, the
rotation field shows asymmetries of the order of 25 km s$^{-1}$
(Section \ref{sec:individual_4374}).

\subsection{Construction of smoothed two-dimensional velocity and velocity dispersion fields}
\label{sec:smoothed_fields}

The computation of the two-dimensional velocity and velocity
dispersion fields from the folded catalogue was done using an adaptive
kernel smoothing technique, which improves on the method described in
\citet{Peng+04}.  For comparison, two examples of unsmoothed
  velocity fields are shown in Appendix \ref{app:unsmoothed}.

As stated in Section
\ref{sec:spatial_distribution}, we use only PNe whose velocities are
within $3\sigma$ of the mean velocity of their neighbours. Generally,
the $3\sigma$ outliers we rejected are few, and they make no
difference to the output velocity and velocity dispersions fields. The
only exceptions are the two galaxies NGC 3377 and NGC 3608 in which
small variations in the resulting two dimensional fields are seen when
outliers are included. However, the outliers in these two cases were
more than $5\sigma$ away from the galaxy systemic velocity and
therefore we are confident that their exclusion is justified.

At every position $(x_P,y_P)$ on the sky we computed the velocity and
velocity dispersion by:

\begin{equation}
\tilde{V}(x_P,y_P)\ = \frac{\sum_i \vel_i \cdot w_{i,P}}{\sum_i w_{i,P}}
\label{eqn:v}
\end{equation}

\begin{eqnarray}
\tilde{\sigma}(x_P,y_P)\ &=&\left(<V^2>-<V>^2 -\Delta V^2 \right)^{1/2}  \nonumber \\
&=& \left[ \frac{\sum_i \vel_i^2 \cdot w_{i,P}}{\sum_i w_{i,P}}-\tilde{V}(x_P,y_P)^2 -\Delta V^2 \right]^{1/2}  \label{eqn:sigma} 
\end{eqnarray}
where $\vel_i$ is the $i-$th PN velocity
and $\Delta V^2$ is the measurement error ($\sim$20 \kms), which is a
combination of instrumental error related to the PN.S ($\sim$17 \kms,
\citealt{Merrett+06}) and the accuracy in determining the planetary
nebula position (a few \kms); $w_{i,P}$ is the distance-dependent
weight for the $i-$th PN, defined using the Gaussian kernel:

\begin{equation}
w_{i,P}=\exp \frac{-D_i^2}{2k(x_P,y_P)^2}
\end{equation}
where $k$ is the kernel amplitude and
$D_i=\sqrt{(x_i-x_P)^2+(y_i-y_P)^2}$ is the distance of the $i-$th PN
from $(x_P,y_P)$.

Errors on $\tilde{V}$ and $\tilde{\sigma}$ are obtained using Monte Carlo
simulations, which will be discussed in Section \ref{sec:testing}.

The weighting procedure depends on the distance $D$ from $(x_P,y_P)$
and on the amplitude $k$ of the kernel.  The latter is a measure of
the spatial resolution at which we would like to investigate the
kinematics. Large values of $k$ will lead to smoother profiles of the
velocity and velocity dispersion fields, highlighting the general
trend but suppressing kinematic structures at small scales. 
  Conversely, smaller values of $k$ will allow a better spatial
resolution, with the risk of amplifying the noise pattern due to the
low number of PNe and the measurement errors.
The optimal $k$ must therefore be chosen in order to find the best
compromise between spatial resolution and noise smoothing. It can be
different at each point in the observed field, since it depends on the
number density of points, the velocity gradient and the velocity
dispersion we want to resolve.

To determine $k$, we proceeded as follows. At each position on the
fields, we define $k$ to be linearly dependent on the distance $R_M$
of the $M-th$ closest PN. We arbitrarily chose $M=20$ but
tested in a range $10 < M < 60$ finding no significant differences in
the final results, even for those galaxies with a relatively low
number of PNe.

\begin{equation}
k(x,y)= A \cdot R_M(x,y) +B = A \sqrt{(x-x_M)^2+(y-y_M)^2} +B
\label{eqn:k}
\end{equation}
where $x_M$, $y_M$ are the coordinates of the $M-$th closest PN to $(x,y)$.
Our definition of $R_M$ can be used also to define the local PNe
number density $\rho(x,y)$:

\begin{equation}
  \rho(x,y)=\frac{M}{\pi R_M^2}
  \label{eqn:density}
\end{equation}
which can be used to relate $k$ to the local PNe density $\rho$, by
combining Equation \ref{eqn:k} and Equation \ref{eqn:density}:

\begin{equation}
k(x,y) = A \sqrt{\frac{M}{\pi \rho}} +B
\label{eqn:k2}
\end{equation}

The way $k$ is defined through Equation \ref{eqn:k} (and its
equivalent Equation \ref{eqn:k2}) allows it to be smaller in the
innermost regions where the spatial density is bigger and larger in
the outer regions, where the PNe density is smaller.

In order to select the most appropriate constants $A$ and $B$, we
built simulated sets of PNe (with the number density resembling the
observed one) with radial velocities distributed according to a
  chosen velocity gradient and velocity dispersion, which mimic
  the observations. The artificial sets were processed with the
adaptive kernel procedure using different values of $A$ and $B$ until
the simulated input velocity field was recovered. Therefore, for a
given density, velocity gradient and velocity dispersion we have the
most reliable values for $A$ and $B$ to use which give the best
compromise between kinematic resolution and smoothing (see Table
\ref{tab:k}).

We show in Figure \ref{fig:2dfields} the smoothed two-dimensional
velocity and velocity fields for galaxies in {\it sample A} derived
with this technique.  The two-dimensional fields of the {\it sample B}
galaxies are presented in Appendix B.

\begin{table}
\centering
\caption{Typical parameters and typical errors for the smoothed,
  two-dimensional velocity and velocity dispersion fields}
\begin{tabular}{l c c c c}
\hline
\hline
\noalign{\smallskip}
Name      & $A$& $B$ &  $<\Delta V>$   &   $<\Delta \sigma>$  \\
\noalign{\smallskip}                   
          &    & arcsec & \kms     & \kms   \\
 (1)      & (2)&     (3) &   (4)    &   (5)  \\
\noalign{\smallskip}                                      
\hline        
\noalign{\smallskip}  
NGC 0821  & 0.24 &  10.68  & 30& 20 \\
NGC 3377  & 0.57 &  -0.08  & 25& 20 \\
NGC 3608  & 0.08 &  33.56  & 40& 30 \\
NGC 4374  & 0.97 & 22.72   & 25& 20 \\
NGC 4564  & 0.24 & 10.68   & 20& 20 \\
NGC 5846  & 0.00 &  60.00  & 50& 30 \\
\noalign{\smallskip}
\hline
\end{tabular}
\begin{minipage}{8cm}
  Notes -- Cols. 2 -- 3: Values of $A$ and $B$ used in the kernel
  smoothing procedure (see Equation \ref{eqn:k}) as determined from
  the simulations. Cols. 4 -- 5 typical error on the two-dimensional
  velocity and velocity dispersion fields as determined from Monte
  Carlo simulations (see Section \ref{sec:testing} for details).
\end{minipage}
\label{tab:k}
\end{table}

\begin{figure*}
 \vbox{
   \hbox{
     \psfig{file=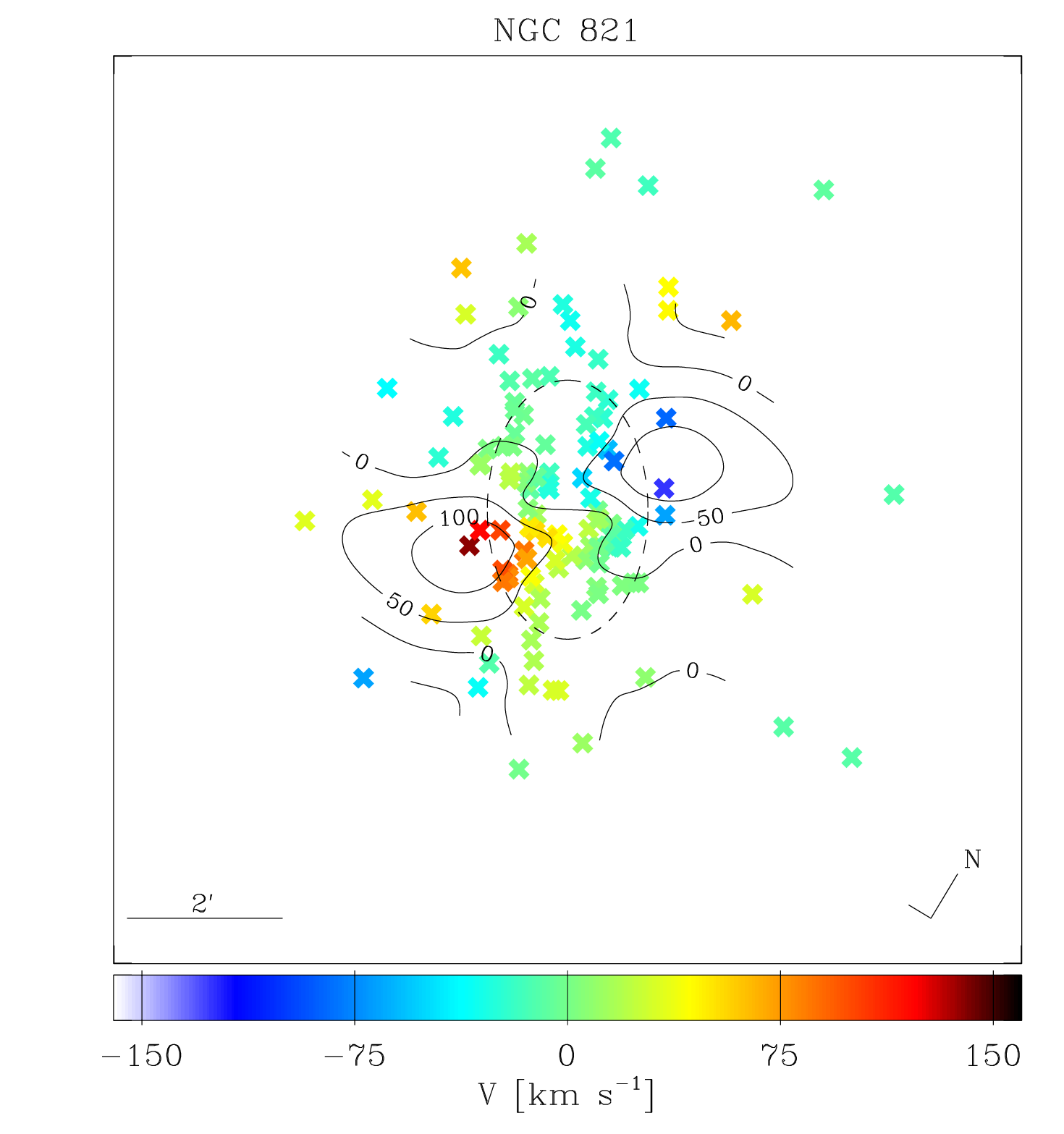,clip=,width=7.8cm}
     \psfig{file=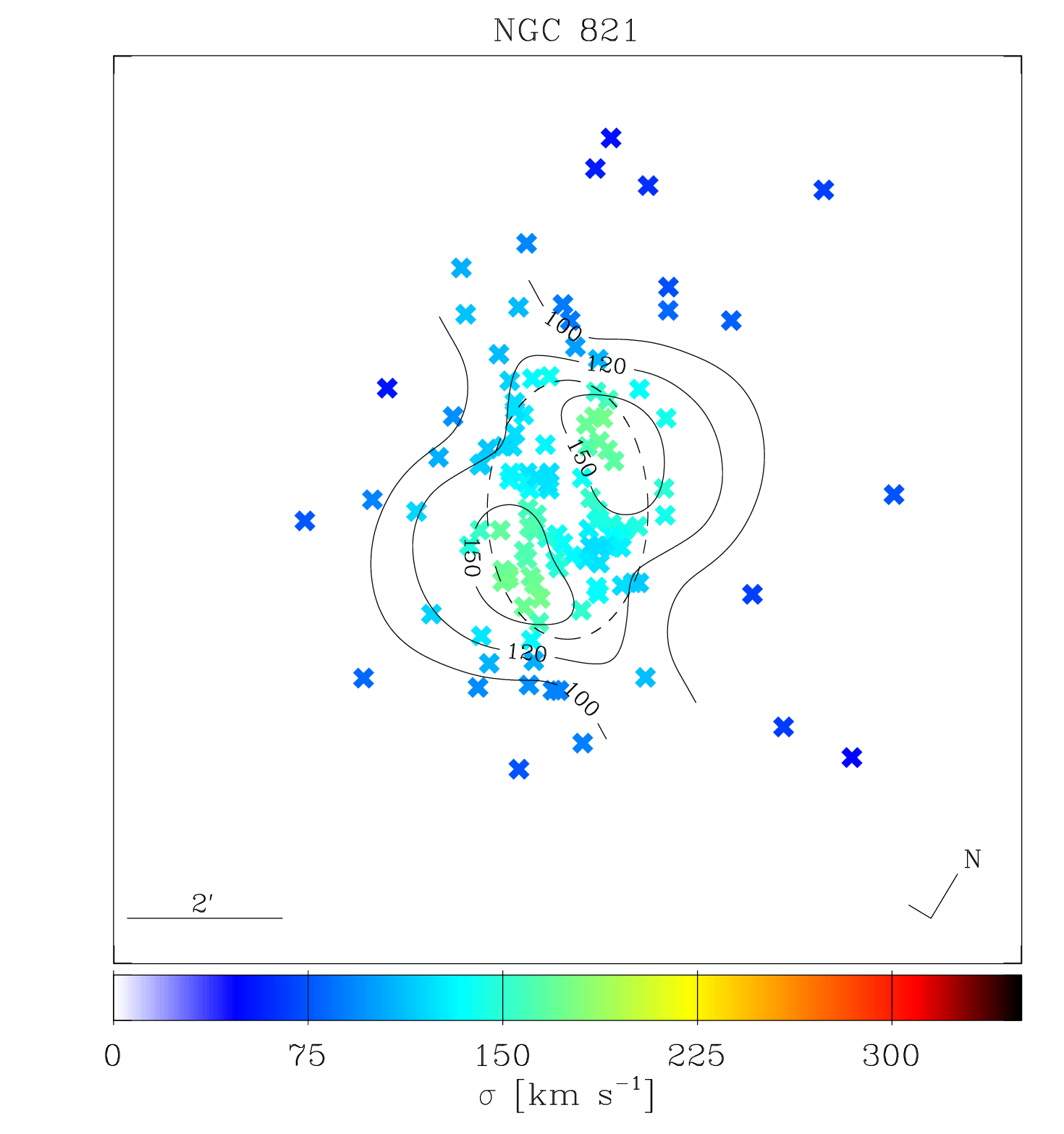,clip=,width=7.8cm}}
   \hbox{
     \psfig{file=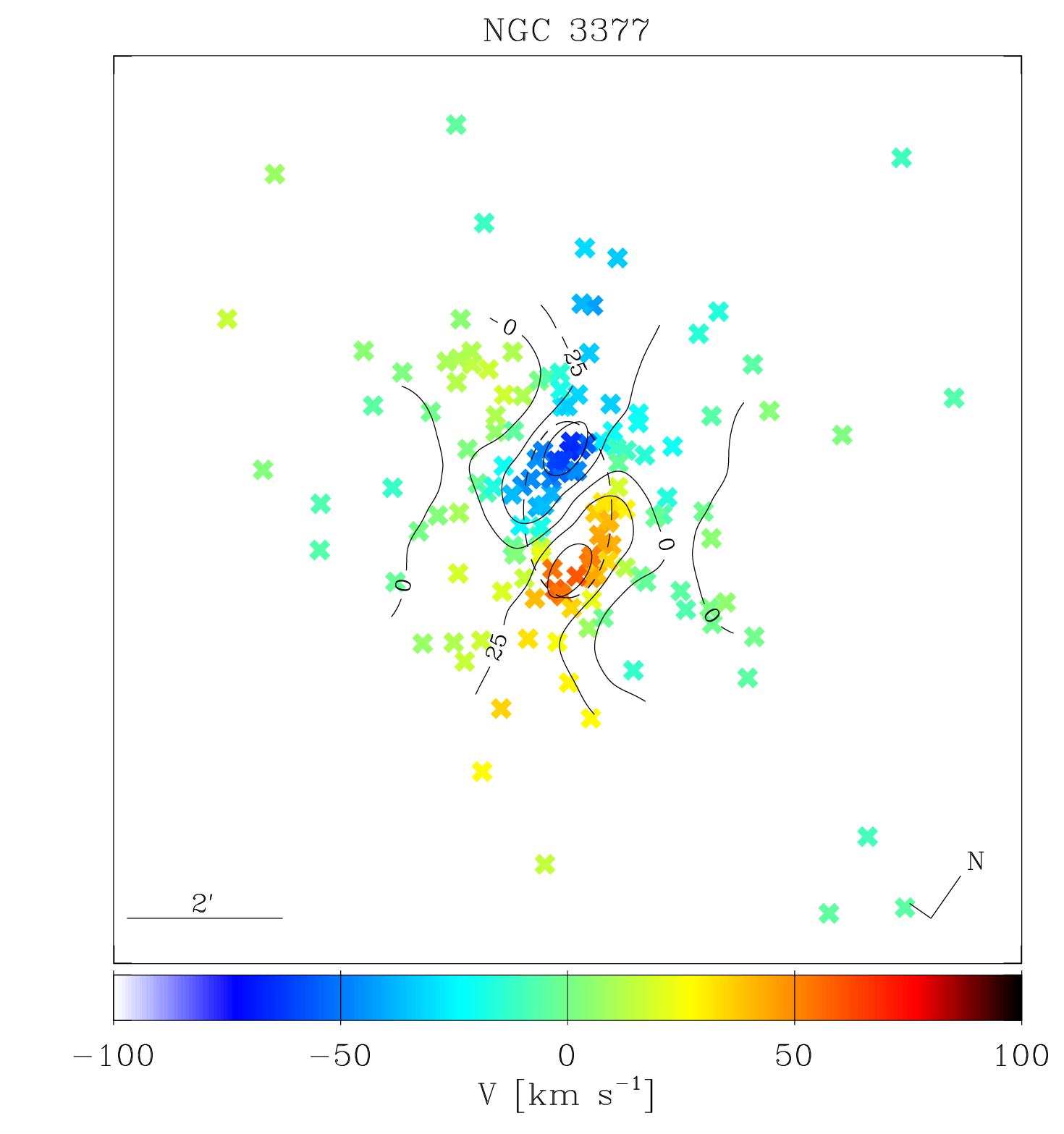,clip=,width=7.8cm}
     \psfig{file=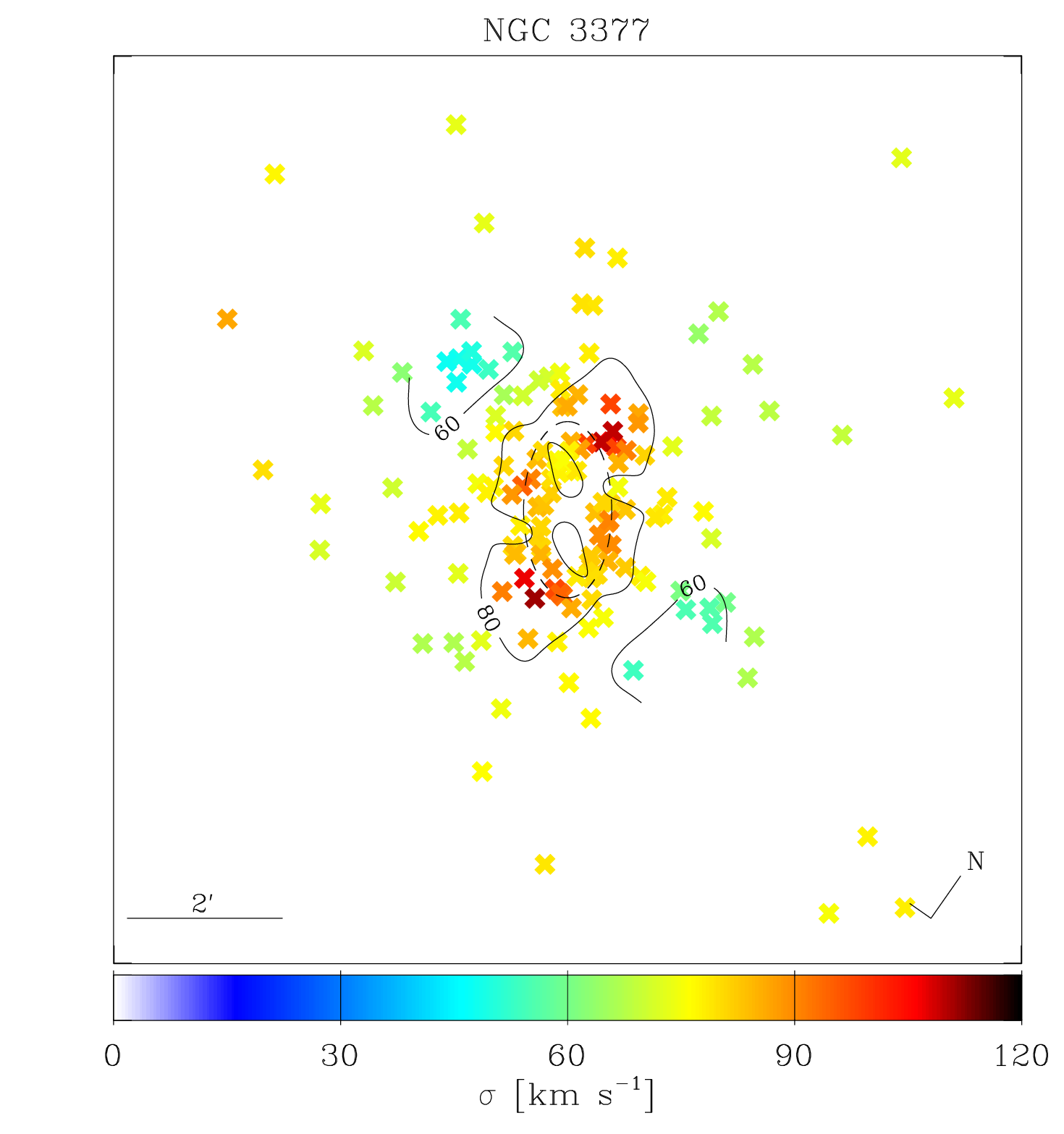,clip=,width=7.8cm}}
  }
  \caption{Smoothed two-dimensional velocity ({\it left panels})
    and velocity dispersion ({\it right panels}) fields of galaxies in
    {\it sample A} from PNe data. Spatial scale and orientation are given in the
    panels.  The photometric major axis as given in Table
    \ref{tab:sample} is aligned along the vertical axis. Crosses
    represent the locations of the PNe, while the colours represent
    the values of the smoothed velocity (or velocity dispersion) field
    at those points. The colour scale is given at the bottom of each
    panel. The {\it dashed ellipses} are located at 2 effective radii,
    which are listed in Table \ref{tab:sample}. Two-dimensional fields
    of the {\it sample B} galaxies are presented in Appendix
    \ref{sec:2d_literature}.}
  \label{fig:2dfields}
\end{figure*}

\addtocounter{figure}{-1}
\begin{figure*}
 \vbox{
   \hbox{
     \psfig{file=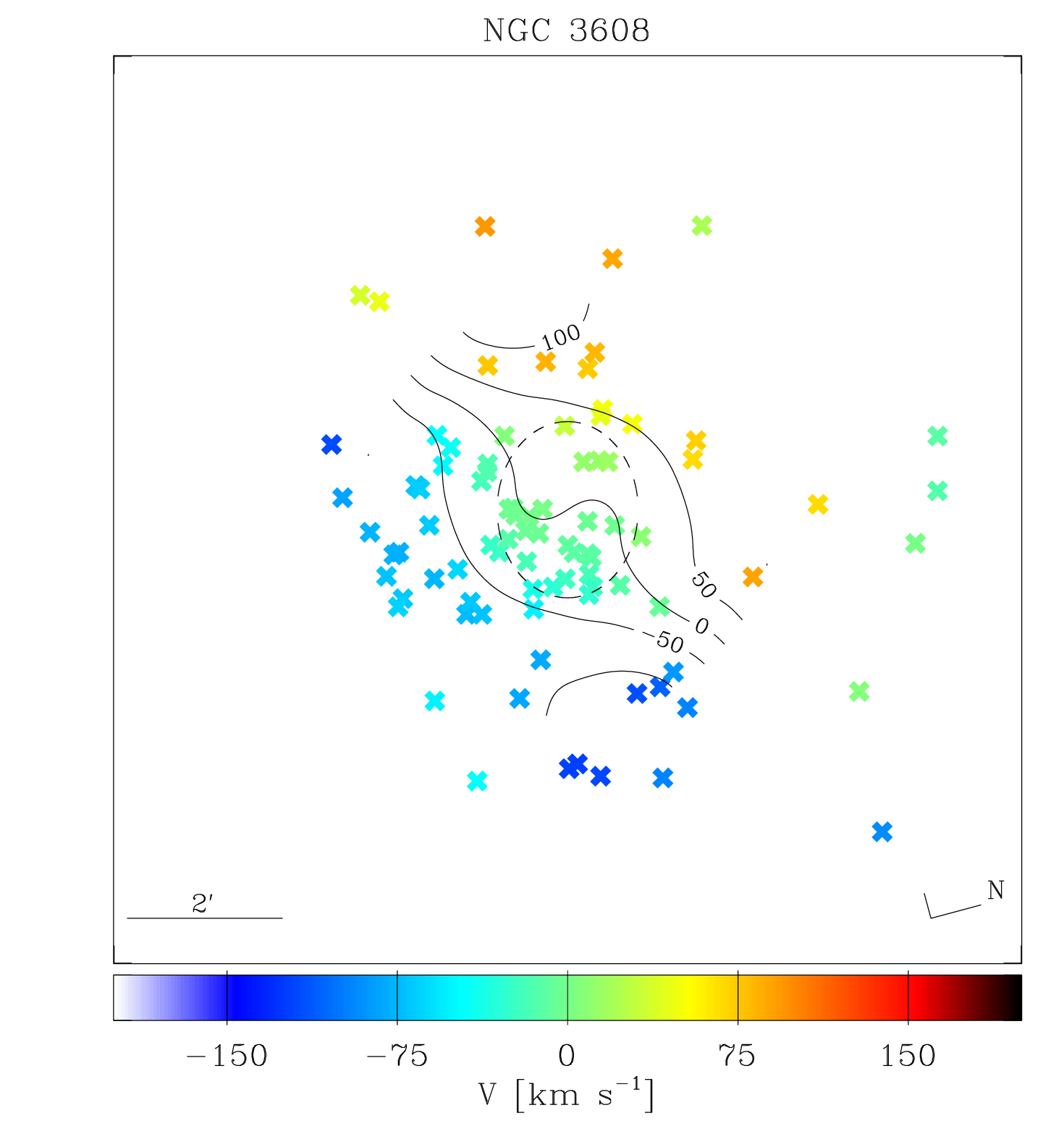,clip=,width=7.8cm}
     \psfig{file=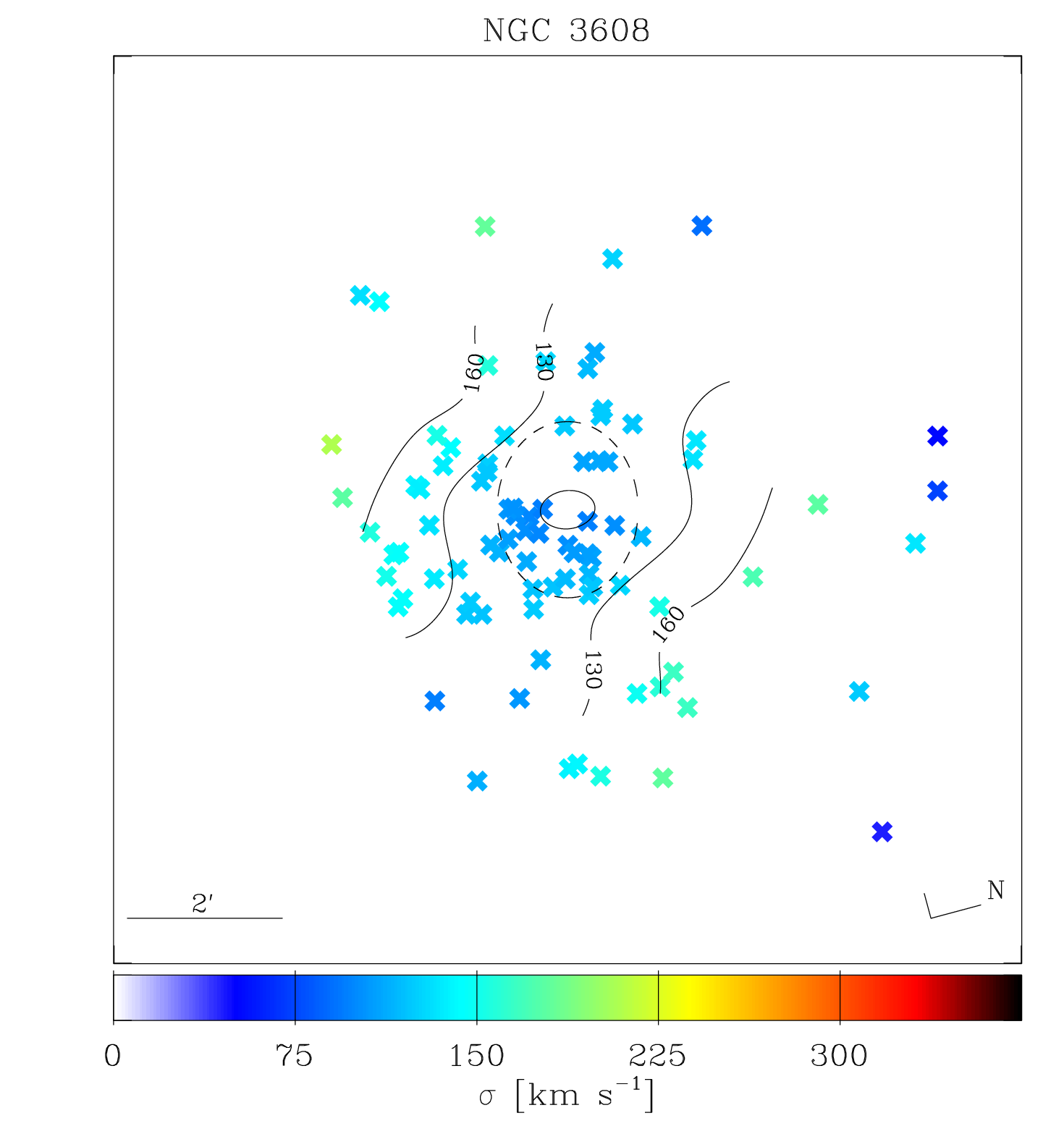,clip=,width=7.8cm}}
   \hbox{
     \psfig{file=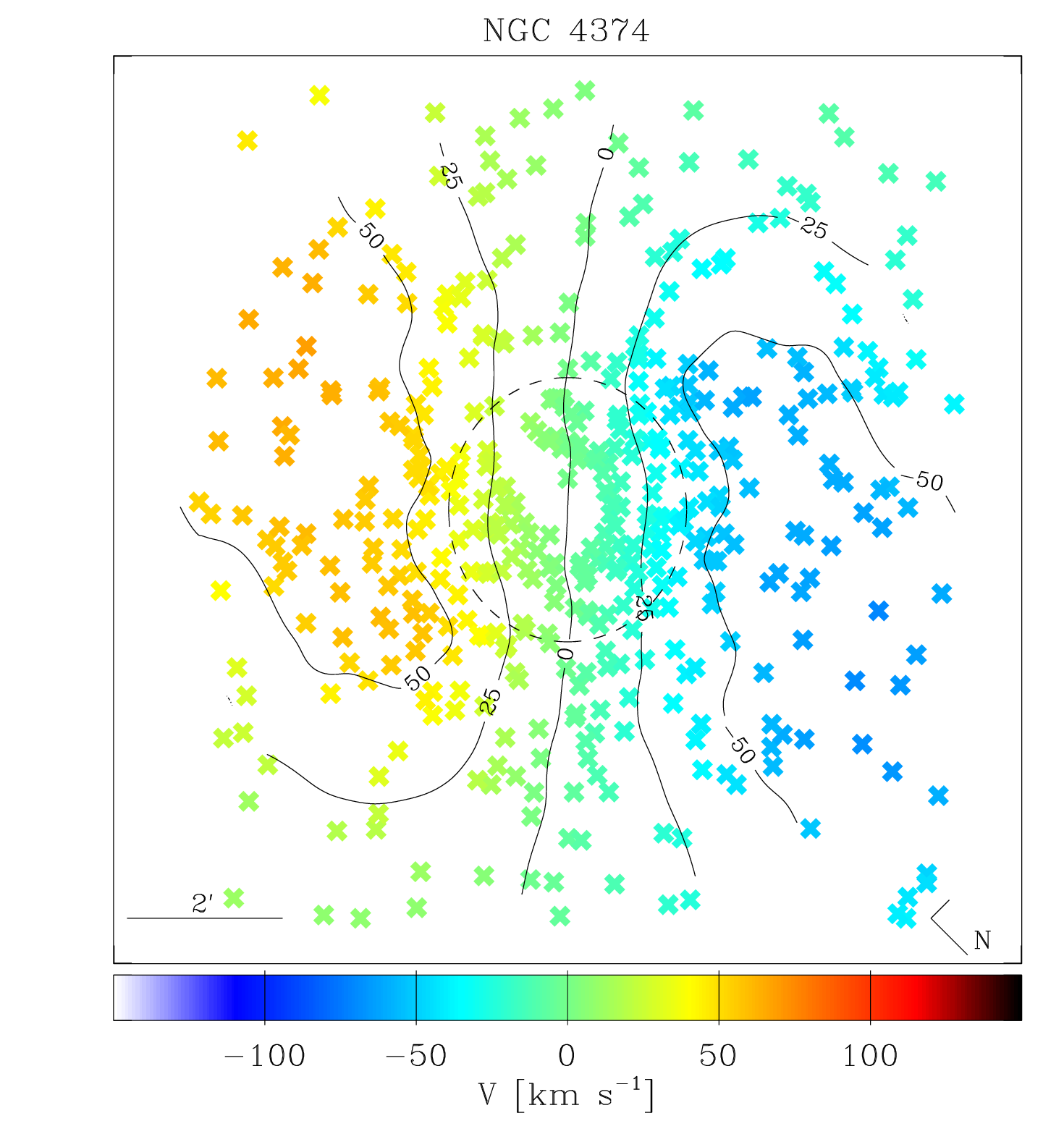,clip=,width=7.8cm}
     \psfig{file=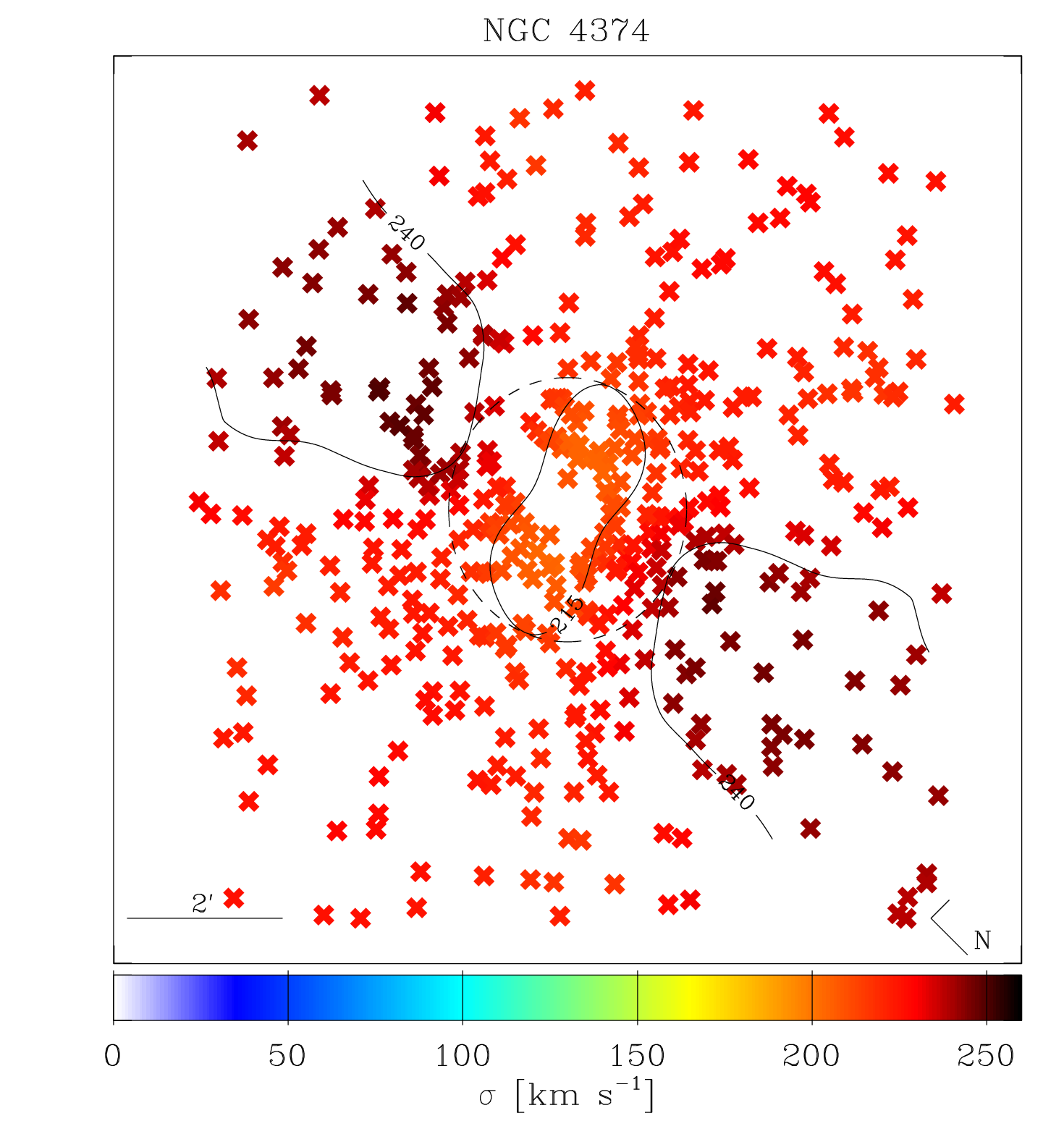,clip=,width=7.8cm}}
  }
  \caption{Continued}
\end{figure*}
\addtocounter{figure}{-1}
\begin{figure*}
   \hbox{
     \psfig{file=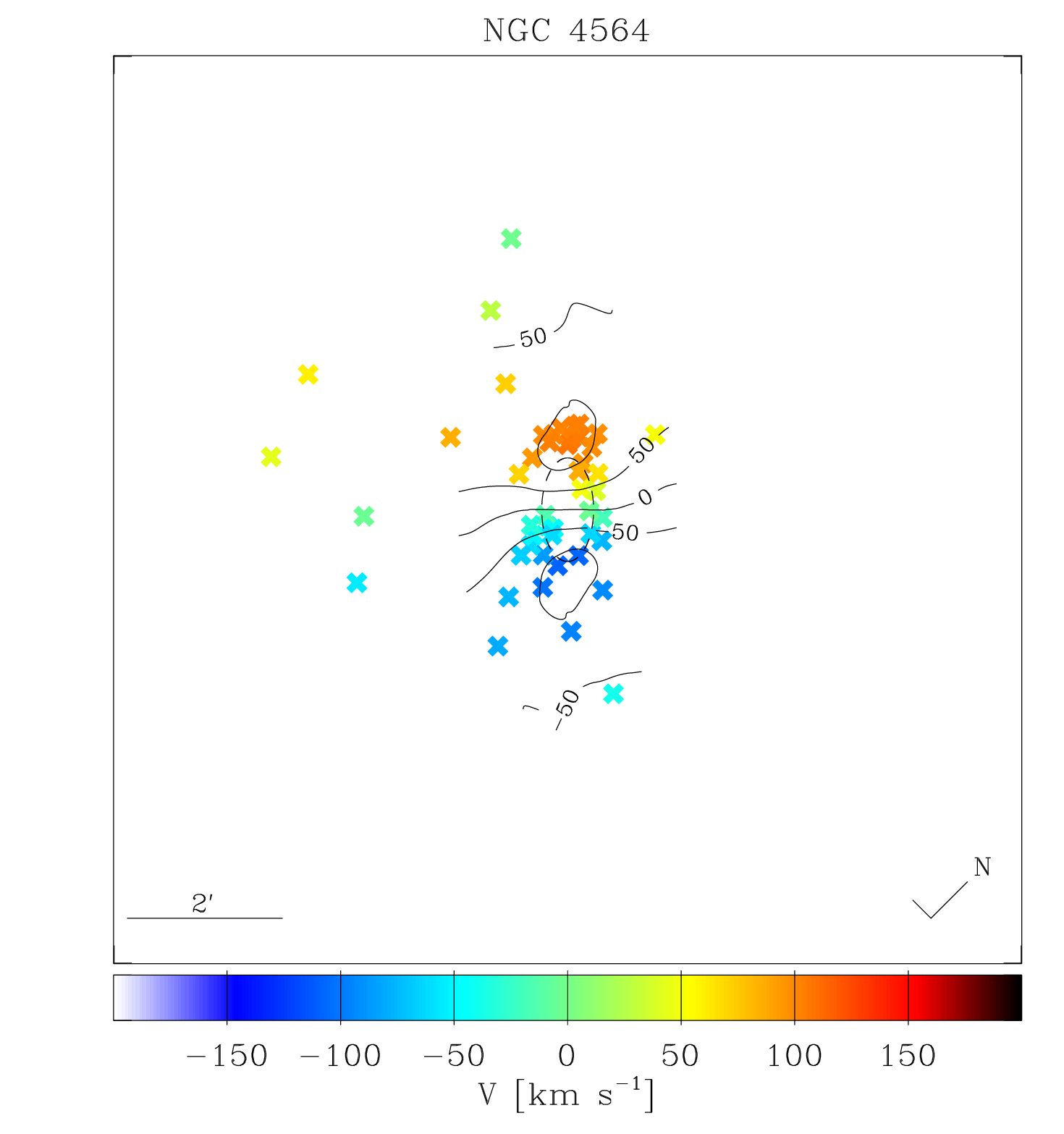,clip=,width=8.2cm}
     \psfig{file=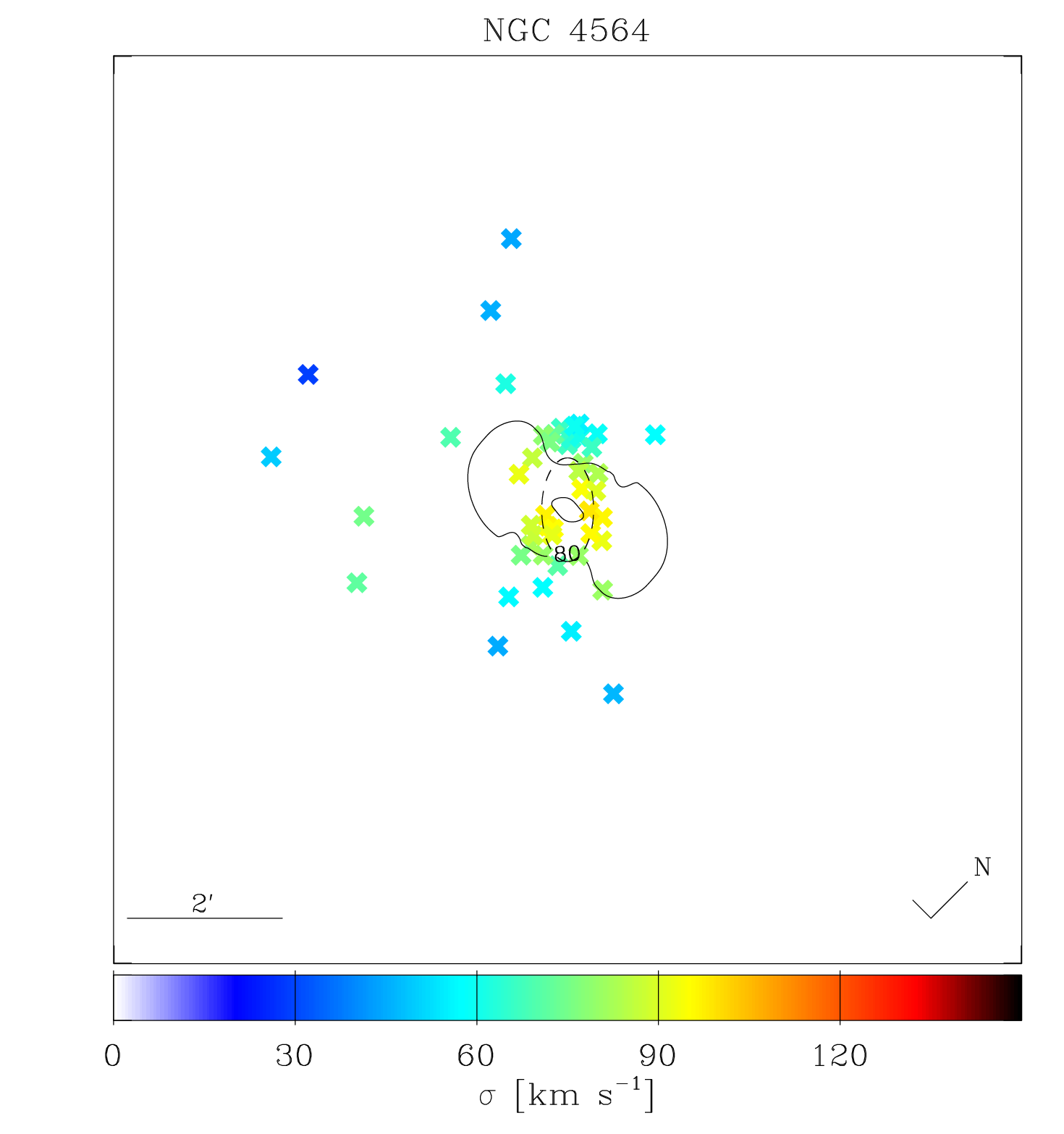,clip=,width=8.2cm}}
   \hbox{
     \psfig{file=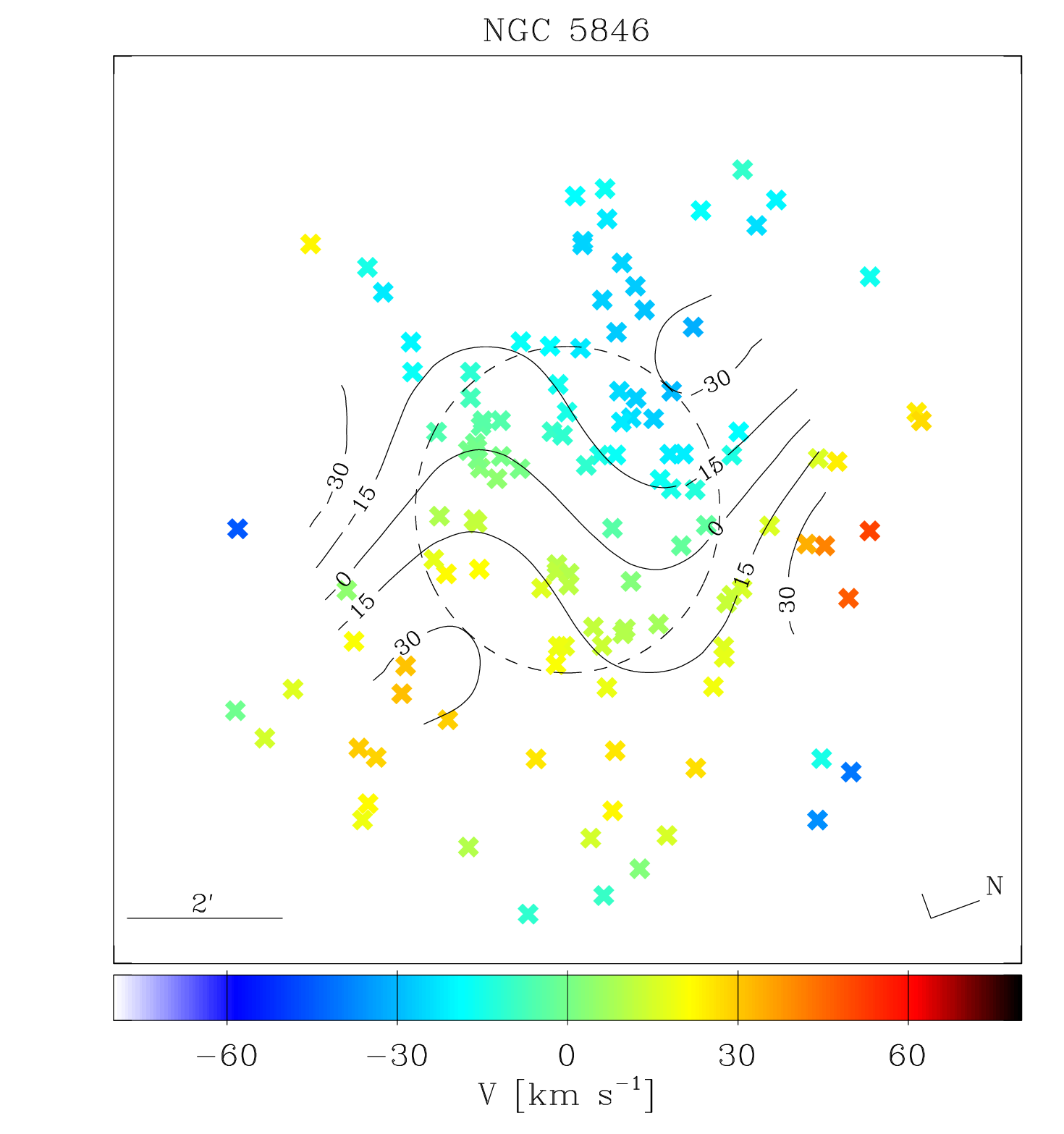,clip=,width=8.2cm}
     \psfig{file=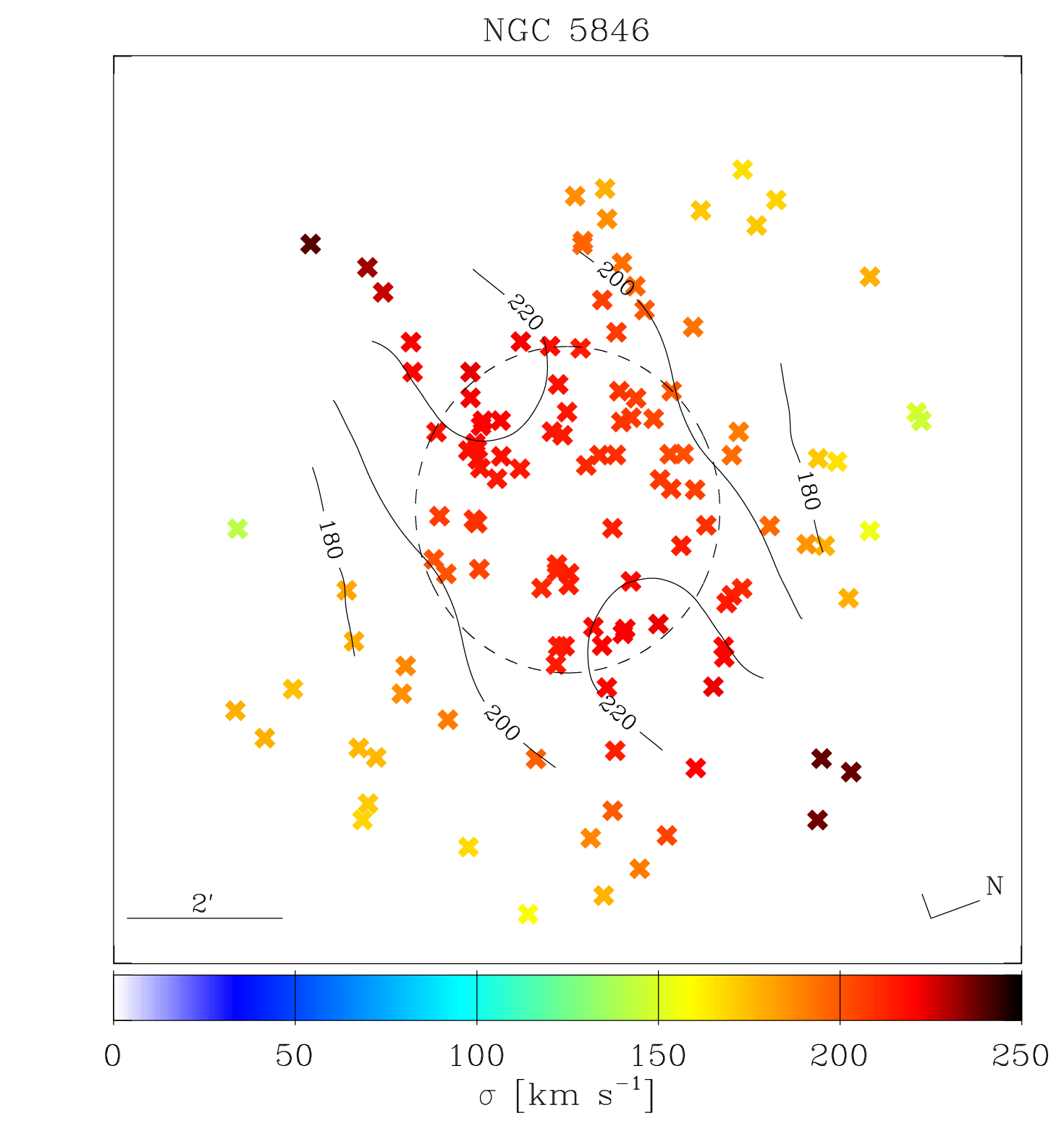,clip=,width=8.2cm}}
%  }
  \caption{Continued}
\end{figure*}

\subsection{Errors in the velocity and velocity dispersion fields}
\label{sec:testing}

Errors on the derived smoothed velocity and velocity dispersion fields
are computed by means of Monte Carlo simulations.

For each galaxy, we built 100 datasets of PNe with simulated radial
velocities at the same positions as in the observed (folded) dataset,
to mimic the observations. The radial velocity for each simulated
object was calculated from the observed two-dimensional smoothed
velocity field by the addition of a random value. The random value
must resemble the observed velocity dispersions and associated
measurement errors and therefore it was chosen randomly from a
Gaussian distribution centred at 0 and  with dispersion equal to

\begin{equation}
 \sigma=\sqrt{\tilde\sigma^2 + \Delta V^2}
 \label{eqn:simulations_sigma}
\end{equation}
where $\tilde\sigma$ is the velocity dispersion measured at that position
and $\Delta V$ is the velocity error.

These simulated datasets were processed with the same reduction script
and parameters as the observed ones. The statistics of the simulated
velocity and velocity dispersion fields give us the error associated
with any position in the observed fields. Average values are given in
Table \ref{tab:k}.

Two-dimensional error fields are not shown, but errors derived from
them will be shown in the plots of the kinematics extracted along the
kinematic major axis (Section \ref{sec:test_rotation}) and along the
photometric major and minor axes (Section
\ref{sec:pne_vs_star_kinematics}).

These simulations are used also to investigate two main aspects: i) if
the measured rotation is significant; and ii) if the misalignment
between kinematic ($PA_{KIN}$) and photometric ($PA_{PHOT}$) major
axes or the twisting of the velocity fields are significant.

\subsubsection{Testing the rotation}
\label{sec:test_rotation}

We extracted the velocity curve $V_{major}$ along the kinematic major
axis of the PNe system from the smoothed two-dimensional velocity
field.  To find its direction, we fitted the PNe velocities with a
simple rotation model:

\begin{equation}
 V_{PN}(\phi_{PN}) = V_{max} \cdot \cos (\phi_{PN} - PA_{KIN})   
\label{eqn:rotation}
\end{equation}
where $\phi_{PN}$ is the position angle of the PN on the sky, the
constant $V_{max}$ measures the amplitude of rotation and $PA_{KIN}$
is the position of the kinematic major axis. Angles are measured on
the sky plane, starting from North going counterclockwise.

Then we created 100 simulated data sets of PNe at the same positions
as the observed ones, with velocity equal to zero plus a random error.
The random value was generated as in Section \ref{sec:testing}, from a
Gaussian distribution with mean zero and dispersion depending on the
observed velocity dispersion and the measurement error (see Equation
\ref{eqn:simulations_sigma}).

We built the two-dimensional velocity field from the simulated
catalogues using the same procedures and the same parameters adopted
for the real catalogues, and we extracted the simulated rotation curve
along the direction of the kinematic major axis.

Since the simulated catalogues were artificially generated to have
their PNe velocities consistent with 0 \kms, the rotation we observed on
the simulated two-dimensional fields is only an artifact of the noise,
as specified by the limited number of data points, the measurement
errors and the intrinsic velocity dispersion of the galaxy.

\begin{figure*}
  \vbox{ \hbox{ \psfig{file=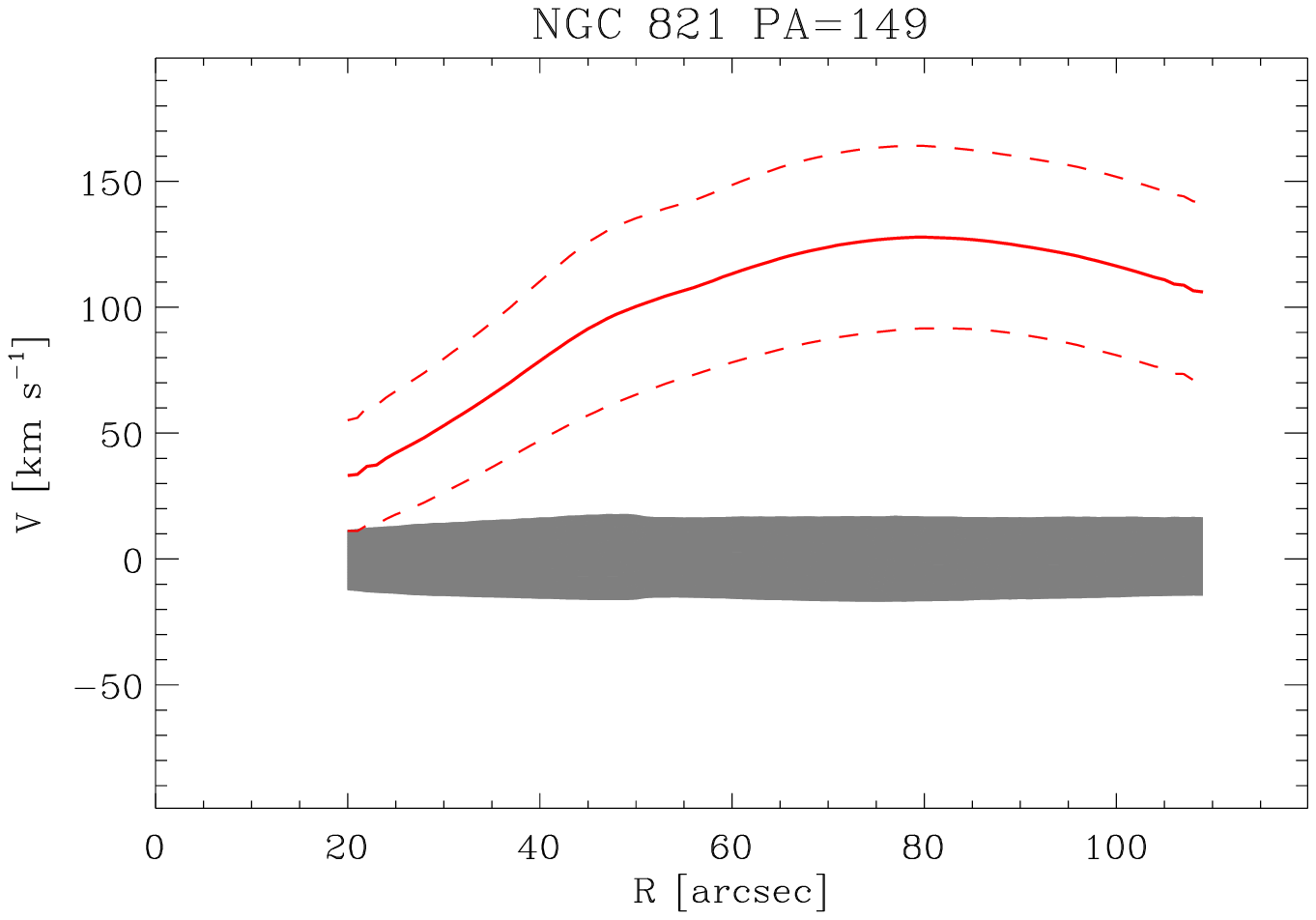,clip=,width=5.7cm}
  \psfig{file=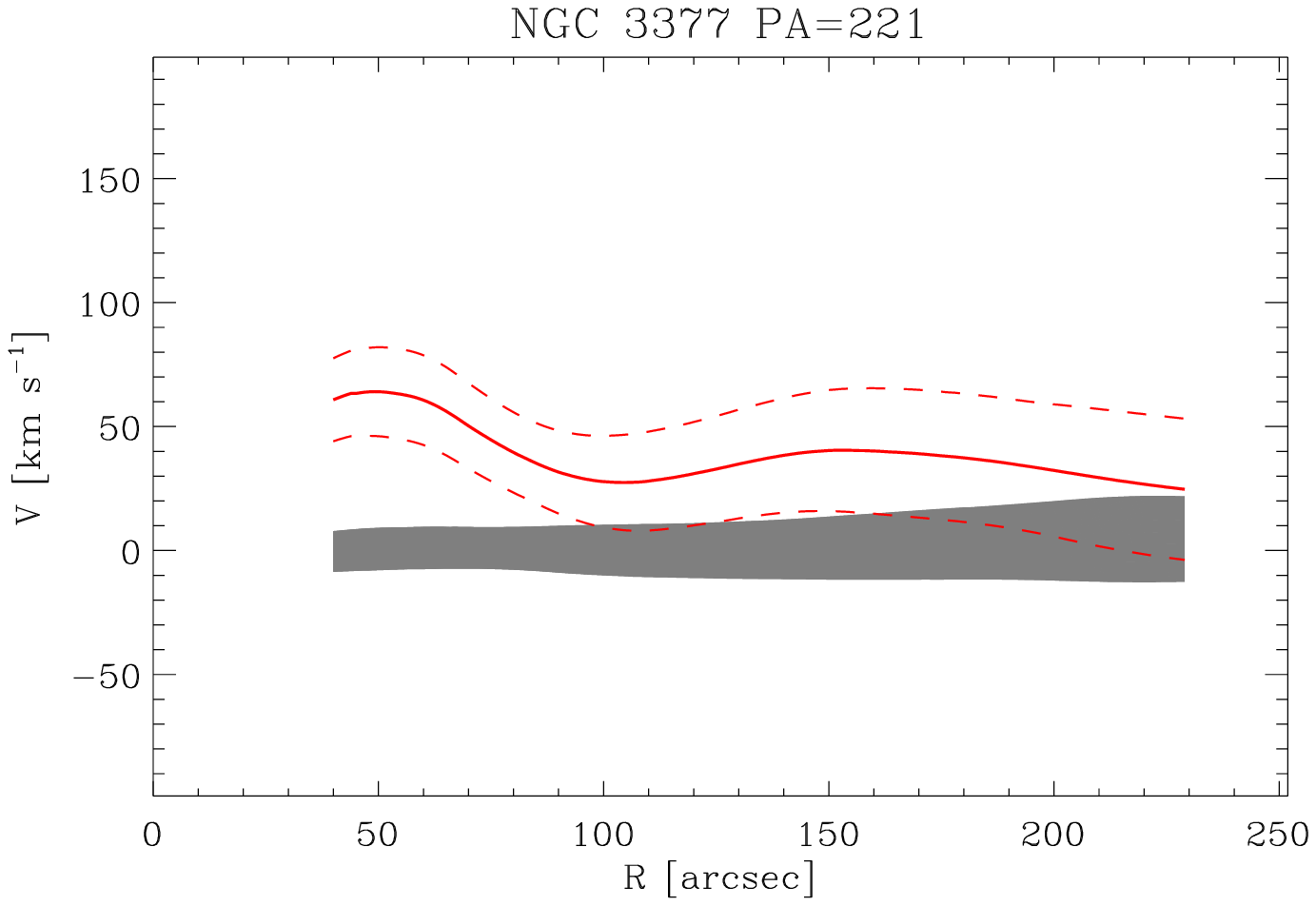,clip=,width=5.7cm}
  \psfig{file=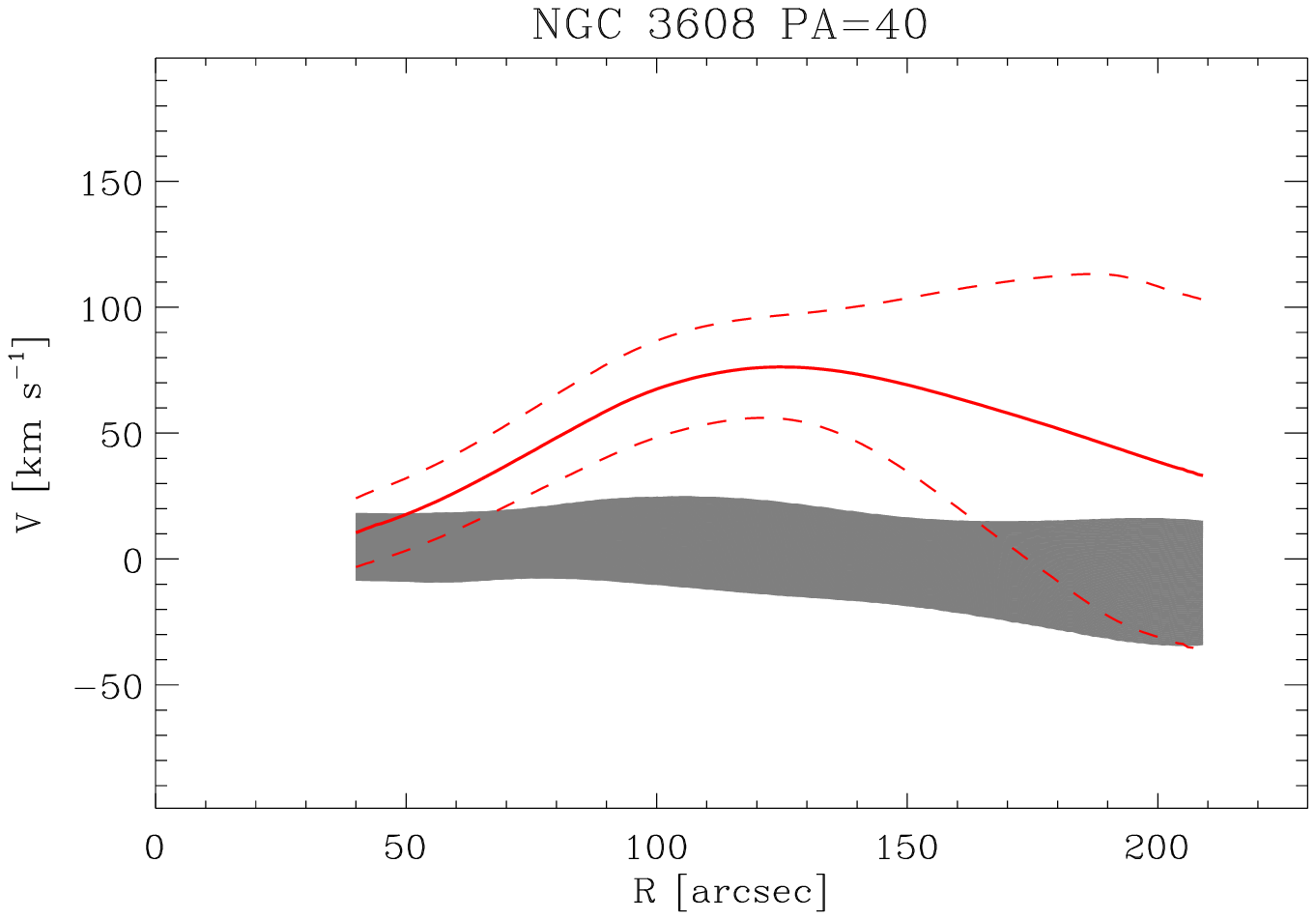,clip=,width=5.7cm}} \hbox{
  \psfig{file=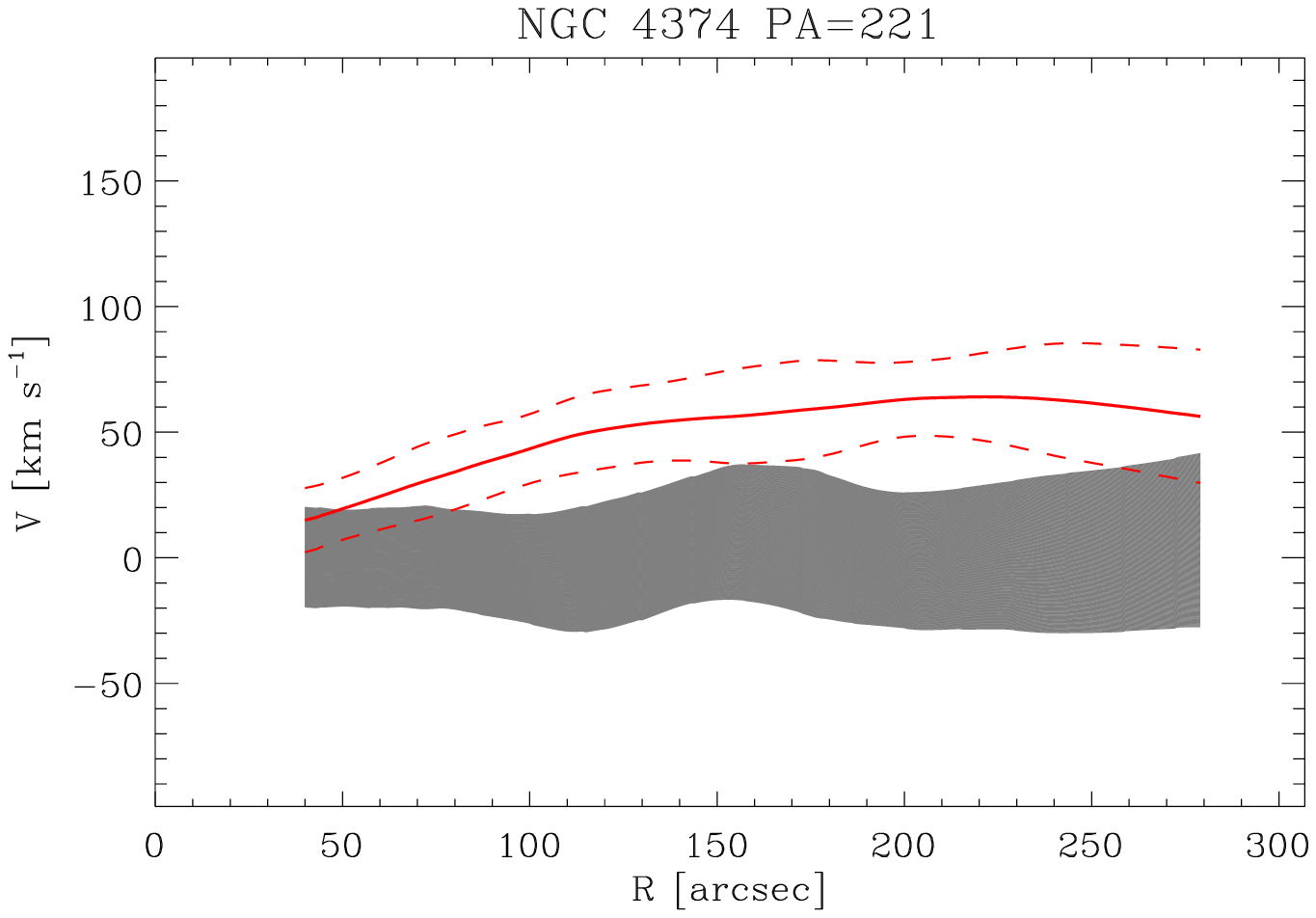,clip=,width=5.7cm}
  \psfig{file=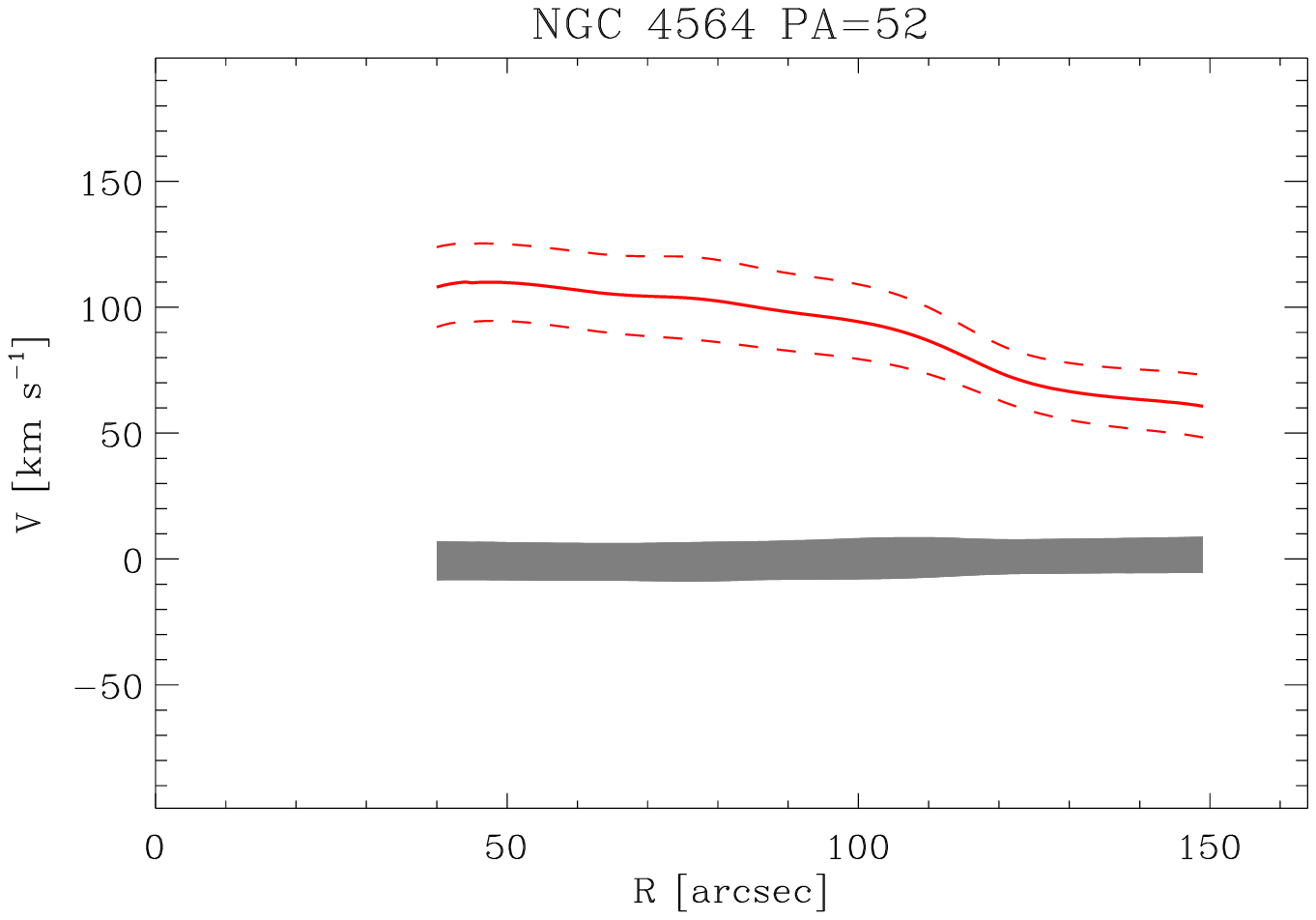,clip=,width=5.7cm}
  \psfig{file=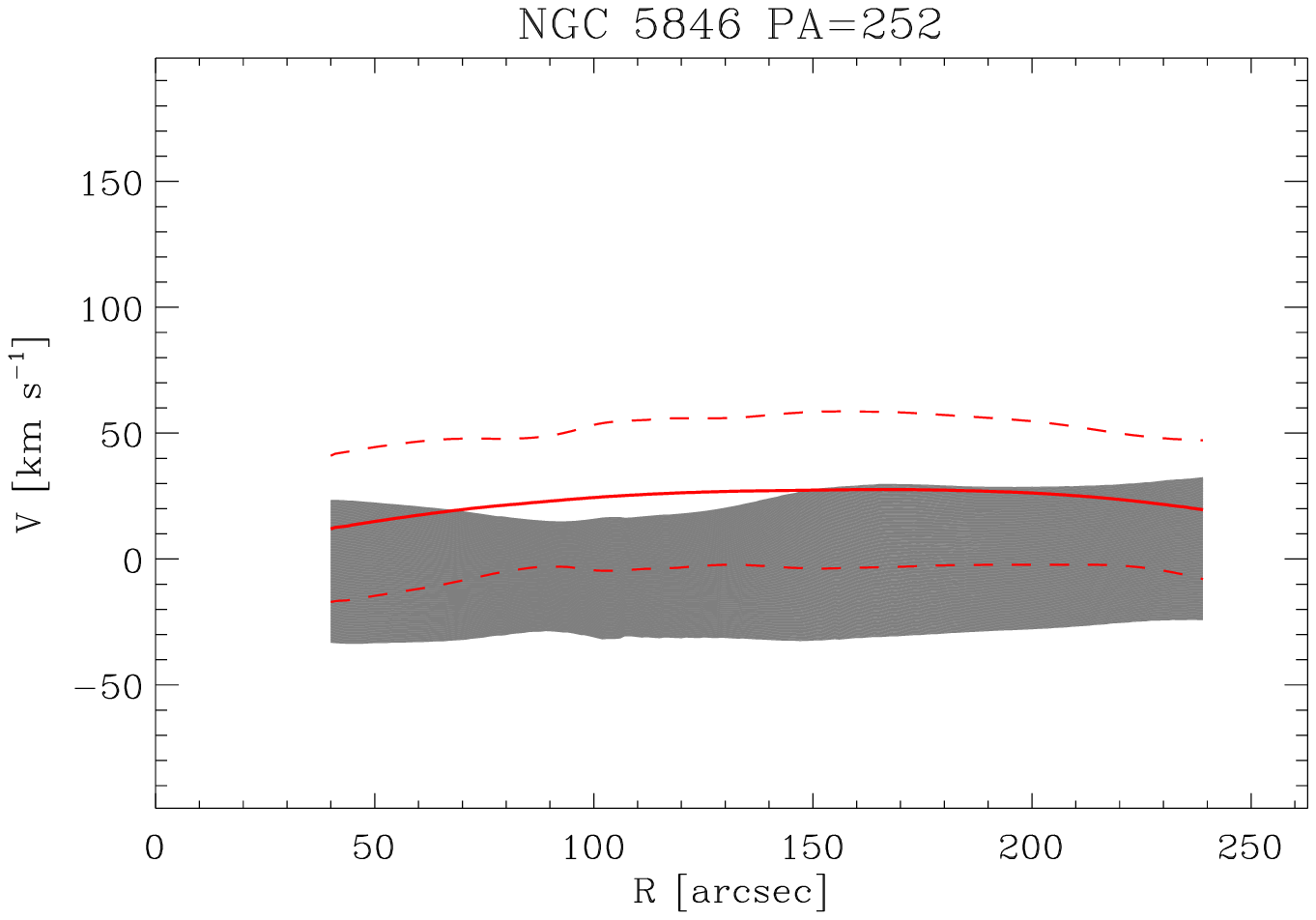,clip=,width=5.7cm}} }
  \caption{{\it Red line:} Radial profile of observed mean velocity
    for the PNe system extracted along the kinematic major axis of the
    two-dimensional field of Figure \ref{fig:2dfields}. {\it Red
      dashed lines}: $1\sigma$ error on the observed mean
      velocity. The {\it grey area} spans the region for which the
    rotation is consistent with 0 (within 1$\sigma$ level), as
    measured from 100 simulated datasets with zero intrinsic rotation
    (see Section \ref{sec:test_rotation} for details).}
\label{fig:simulations1}
\end{figure*}

By looking at the distribution of the 100 simulated rotation curves,
we determined the region in position-velocity space where the rotation
is consistent with 0 \kms\ at the $1\sigma$ level. In Figure
\ref{fig:simulations1} we show the velocity radial profiles measured
along the kinematic major axis of the observed two-dimensional field
together with the $1\sigma$ confidence level of zero rotation.

The general result is that the PNe rotation we measure is real (with
the exception of NGC 5846). The typical 1$\sigma$ range in which the
velocity is consistent with zero is around $20-30$ \kms, depending
mostly on the number of detections and the velocity dispersion value.

\subsubsection{Testing  misalignment and twisting}
\label{sec:test_misalig}

In this section we investigate how precisely we are able to identify
the kinematic major axis of the PNe system (and thereby ascertain
whether the misalignment with the photometric axis is consistent with
zero or not). We also investigate the reliability of distortions in
the velocity field, which may be interpreted as an effect of
triaxiality.

To do that, we measured the kinematic position angle of all the
simulated fields computed in Section \ref{sec:testing}.  The standard
deviation of these values gives us an estimate of the error in
the kinematic major axis determination. 
These errors typically range between 10 and 25 degrees, depending on
the number of PNe, and on the amount of rotation and velocity
dispersion: errors are smaller in galaxies with larger number of
  PNe and higher $V/\sigma$ (see also \citealt{Napolitano+01}).
Results for $PA_{KIN}$ and errors are given in Table
\ref{tab:kin_angles}, as well as the misalignment between kinematic
and photometric major axes.  The kinematic and photometric major axes
are aligned in NGC 3377 (on average) and NGC 4564, but misaligned in
NGC 821, NGC 3608 (marginally) and NGC 4374.

\begin{figure}
 \psfig{file=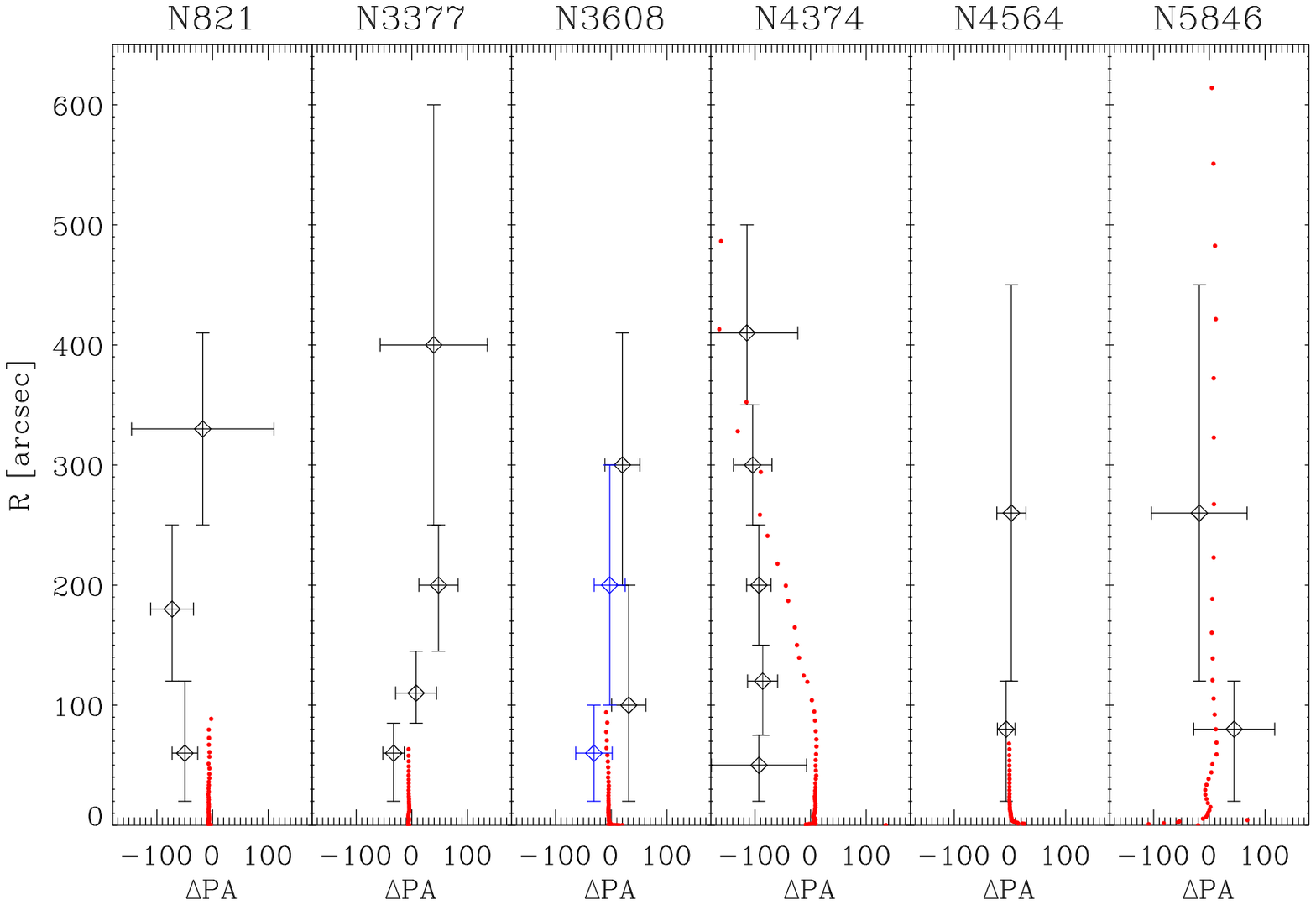,clip=,width=8.5cm}
 \caption{Radial dependence of the misalignment between the
   photometric and kinematic major axes (as derived from the PNe
     kinematics) $\Delta PA = PA_{PHOT}-PA_{KIN}$ ($\pm 180$ deg) for
   galaxies in {\it sample A}  ({\it black open diamonds}). 
     $PA_{PHOT}$ is the constant position angle given in Table
     \ref{tab:sample}.  {\it Blue open diamonds} in the NGC 3608
     panel represent the kinematic position angles calculated
     considering only the PNe on the north side of the galaxy. This
     sub-sample is defined and discussed in Section
     \ref{sec:individual_3608}. Significant misalignment is observed
   in NGC 821 and NGC 4374, while a twist in the direction of rotation
   is observed in NGC 3377. For NGC 3608 and NGC 5846 no definitive
   conclusions can be derived within the errors. {\it Red dots} show
   $PA_{PHOT}-PA(R)$, where $PA(R)$ is the photometric major axis
   radial profile from the referenced papers listed in Table
   \ref{tab:photometry}.}
\label{fig:radial_twisting}
\end{figure}

\begin{figure}
 \psfig{file=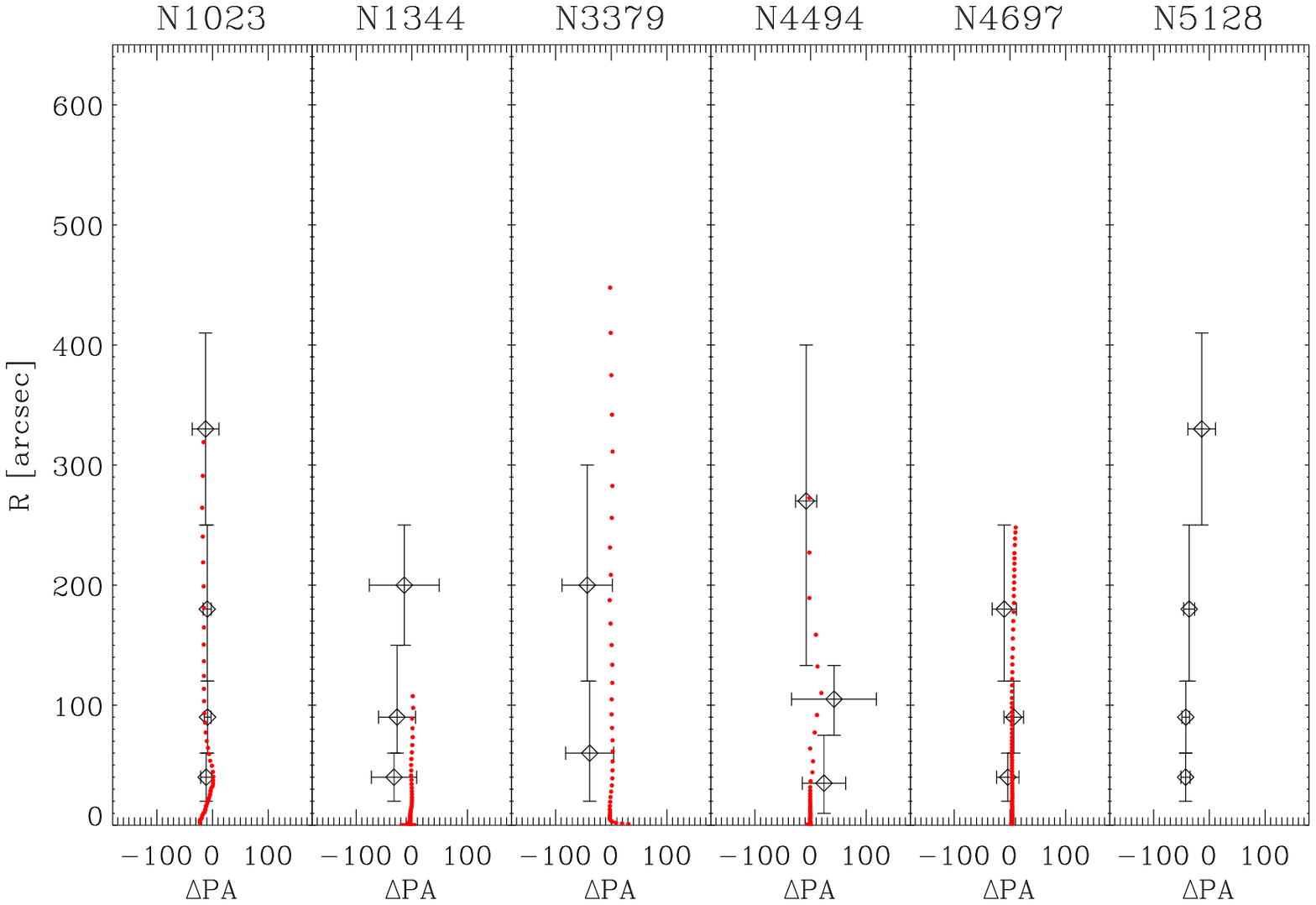,clip=,width=8.5cm}
 \caption{Same as Figure \ref{fig:radial_twisting} but for galaxies in
   {\it Sample B}.}
\label{fig:radial_twisting2}
\end{figure}

The kinematic major axis $PA_{KIN}$ might also not be constant with
radius. To test that, for each galaxy we divided the PNe dataset into
elliptical annuli (the ellipticity of each annulus is the galaxy's
mean ellipticity, as given in Table \ref{tab:photometry}) and calculated the
position angle at each radial position using Equation
\ref{eqn:rotation} and Monte Carlo simulations to compute the errors.
This is shown in Figures \ref{fig:radial_twisting} and
\ref{fig:radial_twisting2} for galaxies of {\it Sample A} and {\it
Sample B} respectively.

A significant twist is observed in NGC 3377. In the inner region
($R<80''$) the kinematic position angle (on the receding side) is
$PA_{KIN}=248\pm20$, while in the outer regions it decreases linearly
to a value of $PA_{KIN}=167\pm20$ at $R=200''$. In the last measured
bin, the rotation is very low (see Section
\ref{sec:individual_3377}) and therefore errors in $PA_{KIN}$ are
too large to derive definitive conclusions. For other galaxies, the
twist is consistent with 0 within the errors.

Together with the kinematical twisting, Figures
\ref{fig:radial_twisting} and \ref{fig:radial_twisting2} also show the
difference between kinematic and photometric major axis, using
$PA_{PHOT}$ as zero point. An interesting case is NGC 4374: the mean
$PA_{KIN}$ is misaligned by $86^{\circ}$ with respect to $PA_{PHOT}$
(see discussion in Section \ref{sec:individual_4374}) and the
photometric position angle radial profile changes by $100^{\circ}$, reaching
$PA_{KIN}$ in the halo regions.

\begin{table}
\centering
\caption{Kinematic position angles and misalignments.}
\begin{tabular}{l c c c c c}
\hline
\hline
\noalign{\smallskip}
Name      & $PA_{KIN}$  & $\mid \Delta PA \mid _{PNe}$ & Twist?  & $PA_{KIN}^{STARS}$ &$\mid \Delta PA \mid _{STARS}$ \\
\noalign{\smallskip}                   
   (NGC)  &  (deg)     & (deg)       &      &  (deg) &(deg)\\
\noalign{\smallskip}                                       
    (1)   &  (2)       &      (3)    & (4)  &  (5)   & (6) \\   
\hline        
\noalign{\smallskip}  
0821  & $149\pm25$ & $56\pm25$  &NO    & 31      &  6  \\
3377  & $221\pm20$  & $6\pm18$   &YES  & 226     & 11  \\
3608  & $40\pm29$  & $35\pm29$  &NO    &  $-95$  & 10  \\
4374  & $221\pm18$  & $86\pm18$  &NO   & 141     &  6  \\
4564  & $52\pm10$  & $5\pm10$   &NO    &  49     &  2  \\
5846  & $237\pm86$  & $13\pm86$  &NO   & 306     & 56  \\
\hline        
1023  & $93\pm10$  & $6\pm10$   &NO    &  89     &  2  \\
1344  & $197\pm27$  & $32\pm27$  &NO   &  --     & --  \\
3379  & $-71\pm40$ & $39\pm40$  & NO   &  $-108$ &  2  \\
4494  & $180\pm17$   & $0\pm17$   & NO &  --     & --  \\
4697  & $254\pm14$  & $4\pm14$   & NO  &  --     & --  \\
5128  & $252\pm6$   & $37\pm6$   & NO  &  --     & --  \\
\noalign{\smallskip}
\hline
\end{tabular}
\begin{minipage}{8.5cm}
Notes -- Col.1: Galaxy name. The horizontal line separates
galaxies of {\it Sample A}  and {\it Sample B}.  Col.2: Kinematic
position angle of the PNe system (receding side), measured from North
towards East.  Col.3: Absolute difference between $PA_{KIN}$ and
$PA_{PHOT}\pm 180$. $PA_{PHOT}$ is given in Table \ref{tab:sample}. In
the case of NGC 5846, the measured rotation is consistent with 0 \kms;
therefore the error on $PA_{KIN}$ is very large and the measurement of
$PA_{KIN}$ itself is not useful.  
Col.4: Twisting of the  PNe kinematic major axis, if present.
Col.5: Kinematic position angle of the stars (receding side), measured
from two dimensional velocity maps, if available \citep{Cappellari+07}.
Col.6: Absolute difference between $PA_{KIN}^{STARS}$ and
$PA_{PHOT}\pm 180$.
\end{minipage}
\label{tab:kin_angles}
\end{table}

\subsection{Results for the velocity and velocity dispersion fields}
\label{sec:results_fields}

Smoothed two-dimensional velocity and velocity dispersion fields 
are shown in Figure \ref{fig:2dfields} and Appendix B for the galaxies
in {\it sample A} and {\it sample B}. They will be used in the
computation of the $\lambda_R$ (proxy for angular momentum per
unit mass) (Section \ref{sec:lambda_profile}), the outer $<V/\sigma>$
ratio (Section \ref{sec:v_over_sigma}), and in the comparison with
stellar kinematics (Section \ref{sec:pne_vs_star_kinematics}).

A general result emerging from  inspection of the velocity fields
in Figure \ref{fig:2dfields} and the quantitative analysis in
Figure \ref{fig:simulations1} is that rotation is present in all {\it
sample A} galaxies, except in NGC 5846 in which the rotation is
consistent with 0 (see Section \ref{sec:test_rotation}).

In the galaxies with significant rotation, the kinematic major axis
may or may not be aligned with the photometric major axis, within the
error bars (see Table \ref{tab:kin_angles}). Both are aligned in NGC
3377 (on average) and NGC 4564, but misaligned in NGC 821, NGC 3608
(marginally) and NGC 4374. In the {\it sample B} galaxies,
misalignment is seen in Cen A (NGC 5128) and (marginally) in NGC 1344
and NGC 3379. Individual cases will be discussed separately in
Section \ref{sec:individual_galaxies}.

Twisting of the kinematic major axis is significant only in NGC 3377
(Figure \ref{fig:radial_twisting}).

\subsection{Comparison with stellar kinematics}
\label{sec:comp_with_sk}

An important aspect of our analysis is to check whether the PNe
kinematics is in agreement with the stellar absorption-line
kinematics.  To do that, we retrieved major and minor (where
available) stellar kinematic data from the literature and we compared
it to the PNe kinematics extracted along the same axis.

\subsubsection{Long-slit kinematics}
%\label{sec:long_slit}
Stellar kinematics from long-slit or integral-field spectroscopy is
available in the literature for the majority of the galaxies. We give
the list of references for each galaxy in Section
\ref{sec:individual_galaxies}.
For two galaxies in {\it sample A} (NGC 3377 and NGC 4374), we
obtained new deep long-slit observations.  Also the E1 galaxy NGC
4494, which is part of the PN.S galaxy sample, was observed during this
run. Its PNe kinematics and properties together with a new dynamical
analysis are discussed in a separate paper \citep{Napolitano+08}.

Appendix \ref{sec:long_slit} gives the kinematic data derived from the 
long-slit spectra.

\subsubsection{PNe radial profiles and comparison with stellar kinematics}
\label{sec:pne_vs_star_kinematics}

Radial kinematic profiles have been extracted from the PNe data in two
different ways.

The first method uses the interpolated smoothed two-dimensional fields
and their errors presented in Sections \ref{sec:smoothed_fields} and 
\ref{sec:testing}.
The second method selects only those PNe within an angular section
aligned along a desired position angle (the same for which the stellar
kinematics are observed).
Usually the range of angles is between 30 and 60 degrees, depending on
the number of PNe.  Then the selected PNe are folded to positive radii
(i.e., the receding side of the system) and grouped into radial bins
containing the same number of PNe. The number of objects $N_{BIN}$ per
bin ranges from 10 to 30, to reach a compromise between number of bins
and a statistically significant number of PNe in each bin.
In each bin the weighted mean velocity $V_{BIN}$ and velocity
dispersion $\sigma_{BIN}$ are computed, weights are computed from the
measurements errors on the PNe velocities. The derived values of
$V_{BIN}$ and $\sigma_{BIN}$ change by less that 10 \kms\ if
instead all weights are equal, or if the weights from
\citet{Hargreaves+94} are used. Errors are computed with the usual
formulae $\Delta V_{BIN}=\sigma_{BIN}/\sqrt{N_{BIN}}$ and $\Delta
\sigma_{BIN}=\sigma_{BIN}/\sqrt{2(N_{BIN}-1)}$.

Both methods have the advantage that we can extract the radial profile
along any direction we desire, and the disadvantage that it might be
contaminated by PNe far away from the direction we are interested, by
the use of a large kernel smoothing parameter in the first method or a
large angular range in the second. 
Especially if a nearly edge-on disk is present, both techniques
could therefore lead to systematically low rotation and systematically
high dispersions on the major axis, caused by the dilution from
off-axis velocities.  However, simple simulations show that this
effect is smaller than the error bars in the measurement, and in any
case leads to an underestimate of the significance of rotation in
Figure \ref{fig:simulations1}.

In Figure \ref{fig:radial_kinematic} we compare the (folded) velocity
and velocity dispersion radial profiles of stars and PNe. In general
the stellar and PNe kinematics agree well, with NGC 821 as the most
uncertain case. A more detailed description is given in Section
\ref{sec:individual_galaxies}.

\begin{figure*}
\vbox{ 
  \hbox{ 
    \psfig{file=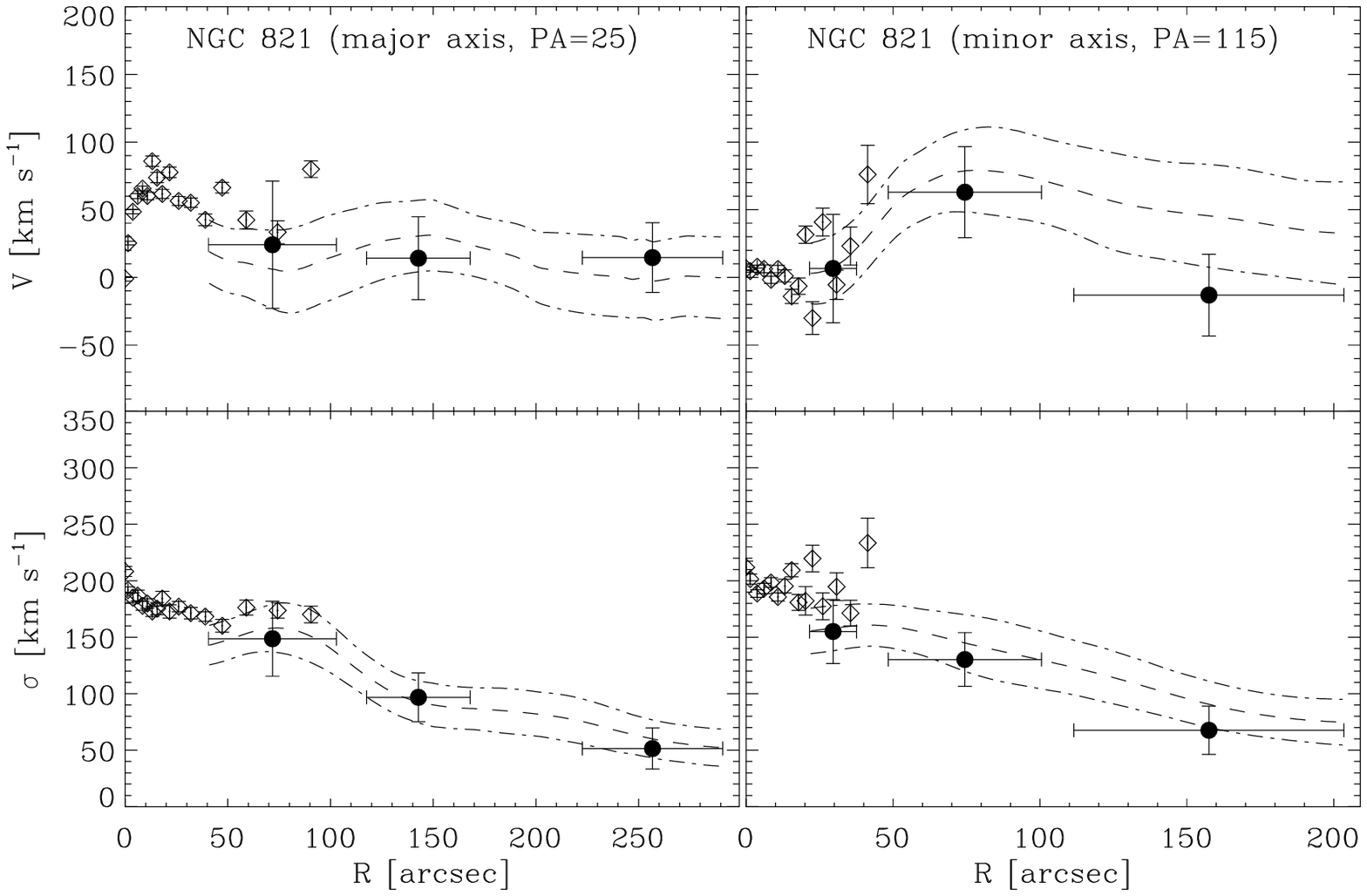,clip=,width=8.7cm}
    \psfig{file=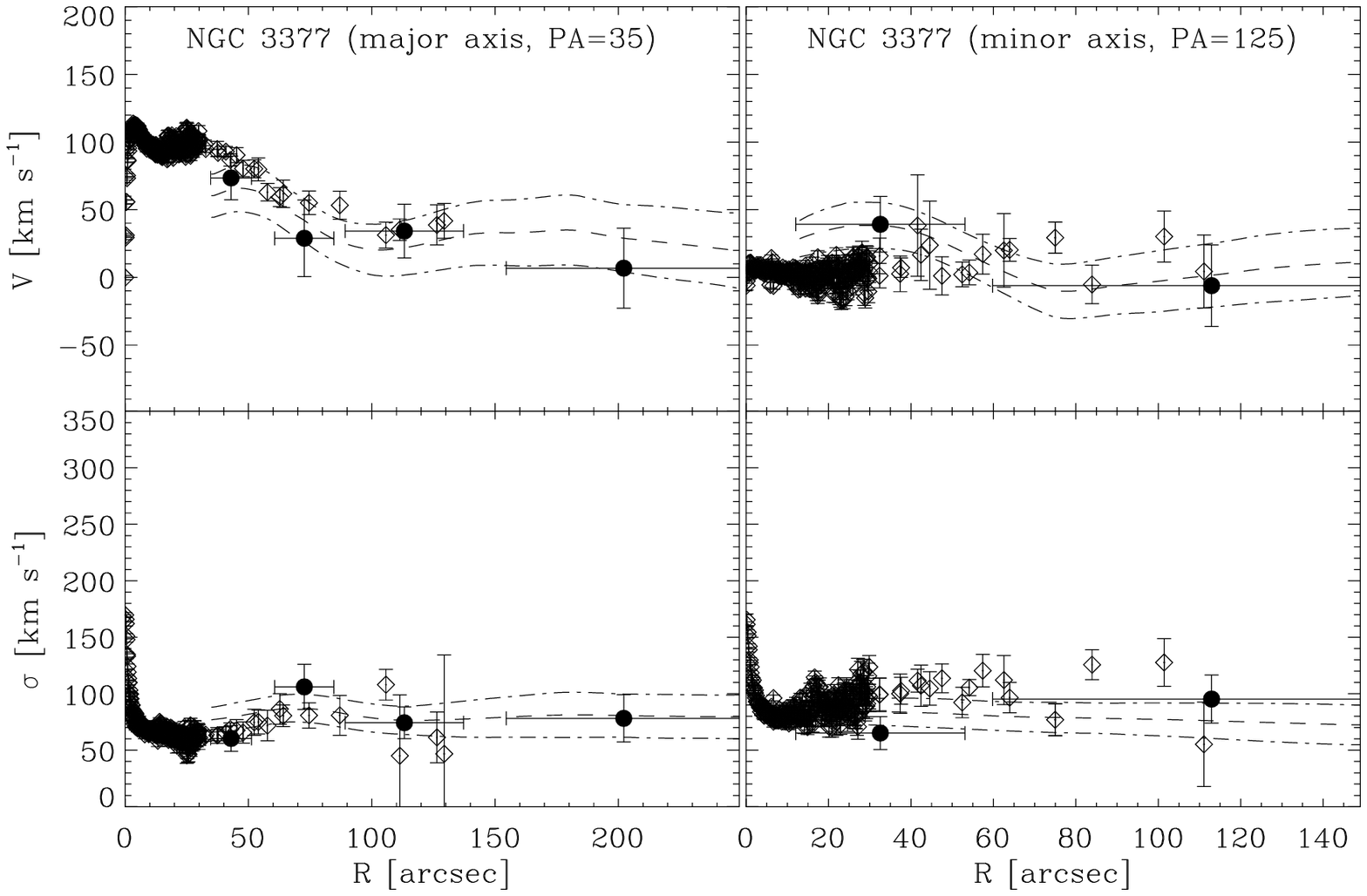,clip=,width=8.7cm} } 
\hbox{
    \psfig{file=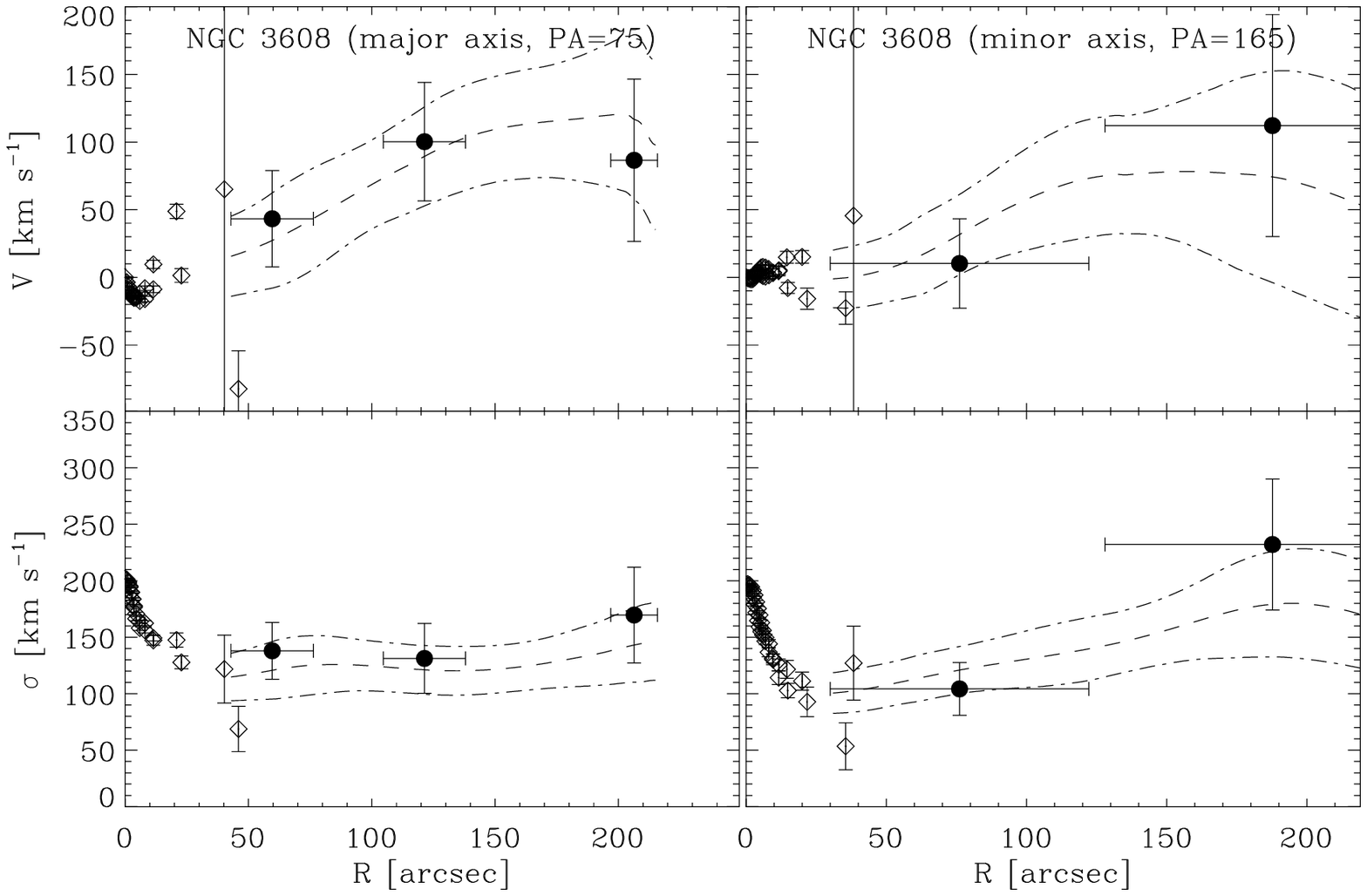,clip=,width=8.7cm}
    \psfig{file=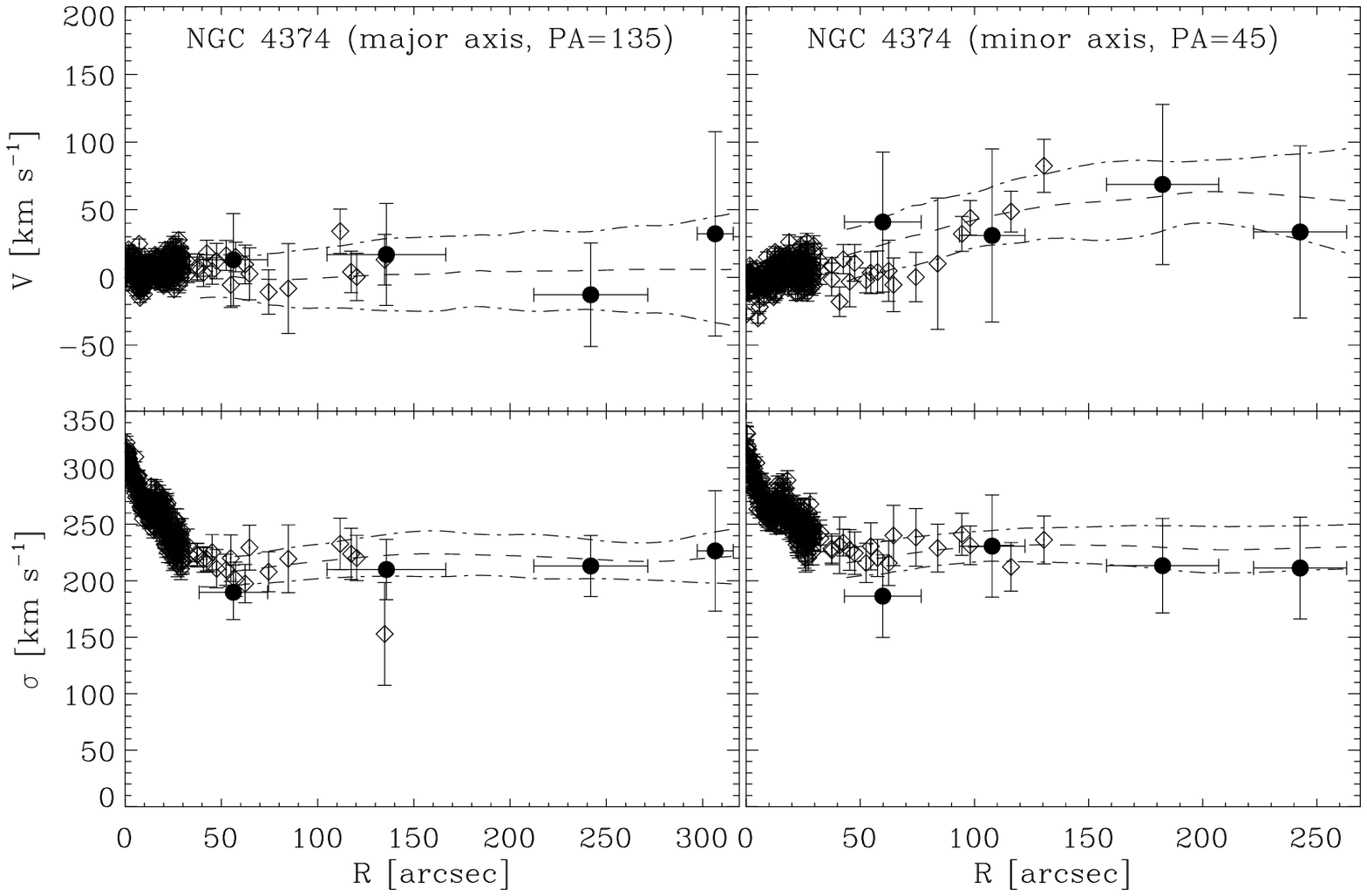,clip=,width=8.7cm} } 
\hbox{
    \psfig{file=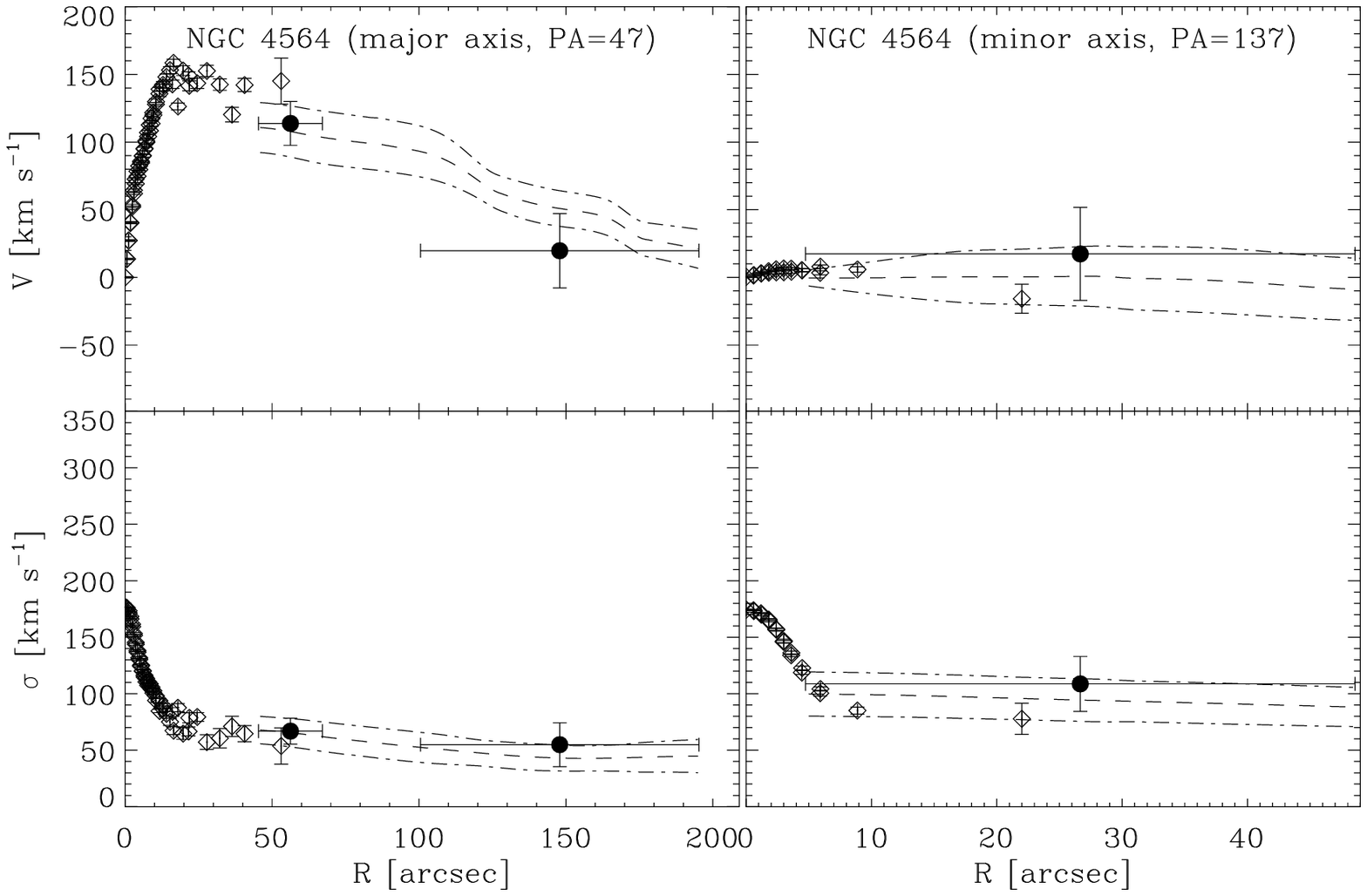,clip=,width=8.7cm}
    \psfig{file=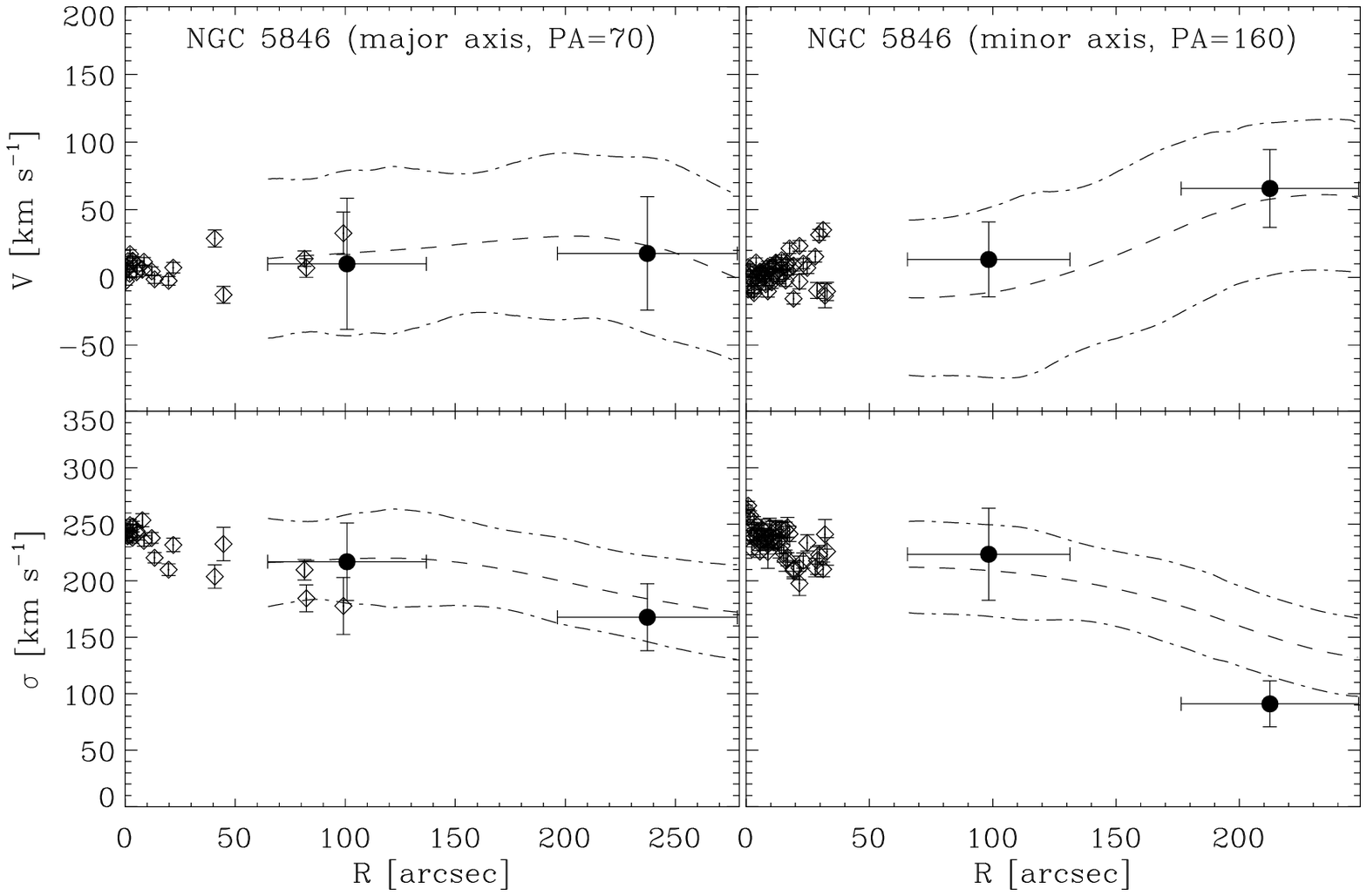,clip=,width=8.7cm} }
}
\caption{Comparison between long slit stellar and PNe kinematics.
  Each galaxy is represented by a sub-figure consisting of radial
  velocity ({\it upper panels}) and velocity dispersion profiles ({\it
    lower panels}) extracted along the photometric major ({\it left
    panels}) and minor axes ({\it right panels}). The position angles
  given are from Table \ref{tab:sample}. {\it Open diamonds} represent
  the stellar kinematics while {\it filled circles} represent the PNe
  kinematics extracted along a cone aligned with the major or minor
  axis. The {\it dashed lines} represents the kinematics extracted
  along the major or minor axis from the two-dimensional field and the
  {\it dot-dashed lines} represent the related error at the 1$-\sigma$
  level calculated from Monte Carlo simulations (see text for
  details). References for the absorption-line kinematics: NGC 821,
  \citet{Forestell+08}; NGC 3377, our VLS/FORS2 data (see Appendix
  \ref{sec:long_slit}); NGC 3608, \citet{Halliday+01}; NGC 4374, our
  VLS/FORS2 data (see Appendix \ref{sec:long_slit}); NGC 4564,
  \citet{Halliday+01}; NGC 5846, \citet[major axis]{Kronawitter+00}
  and \citet[minor axis]{Emsellem+04}. Electronic tables with the
    kinematic data plotted in this figure are available in the on-line
    version of this paper.}
\label{fig:radial_kinematic}
\end{figure*}%

\addtocounter{figure}{-1}
\begin{figure*}
\vbox{ 
  \hbox{ 
    \psfig{file=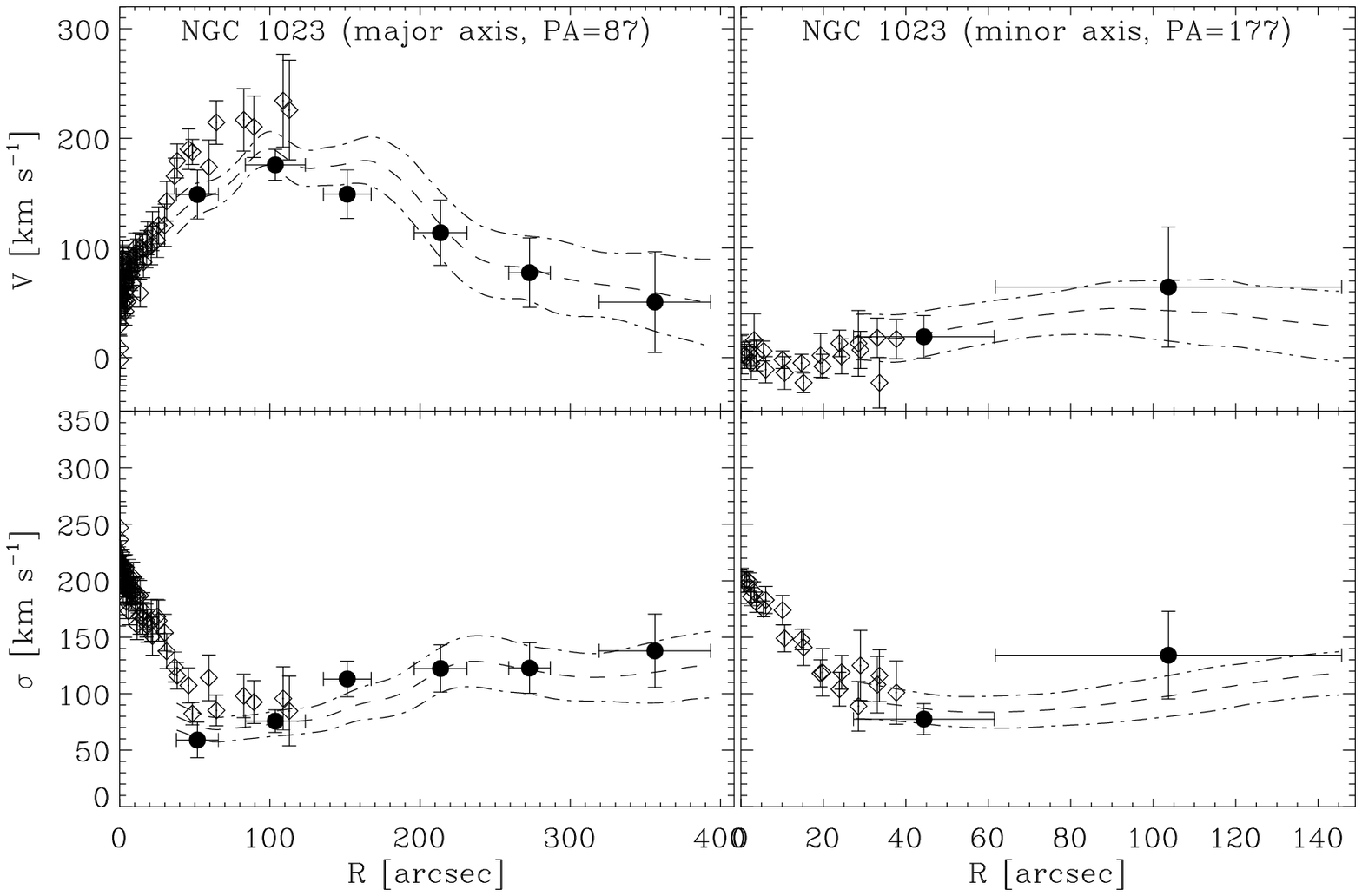,clip=,width=8.7cm}
    \psfig{file=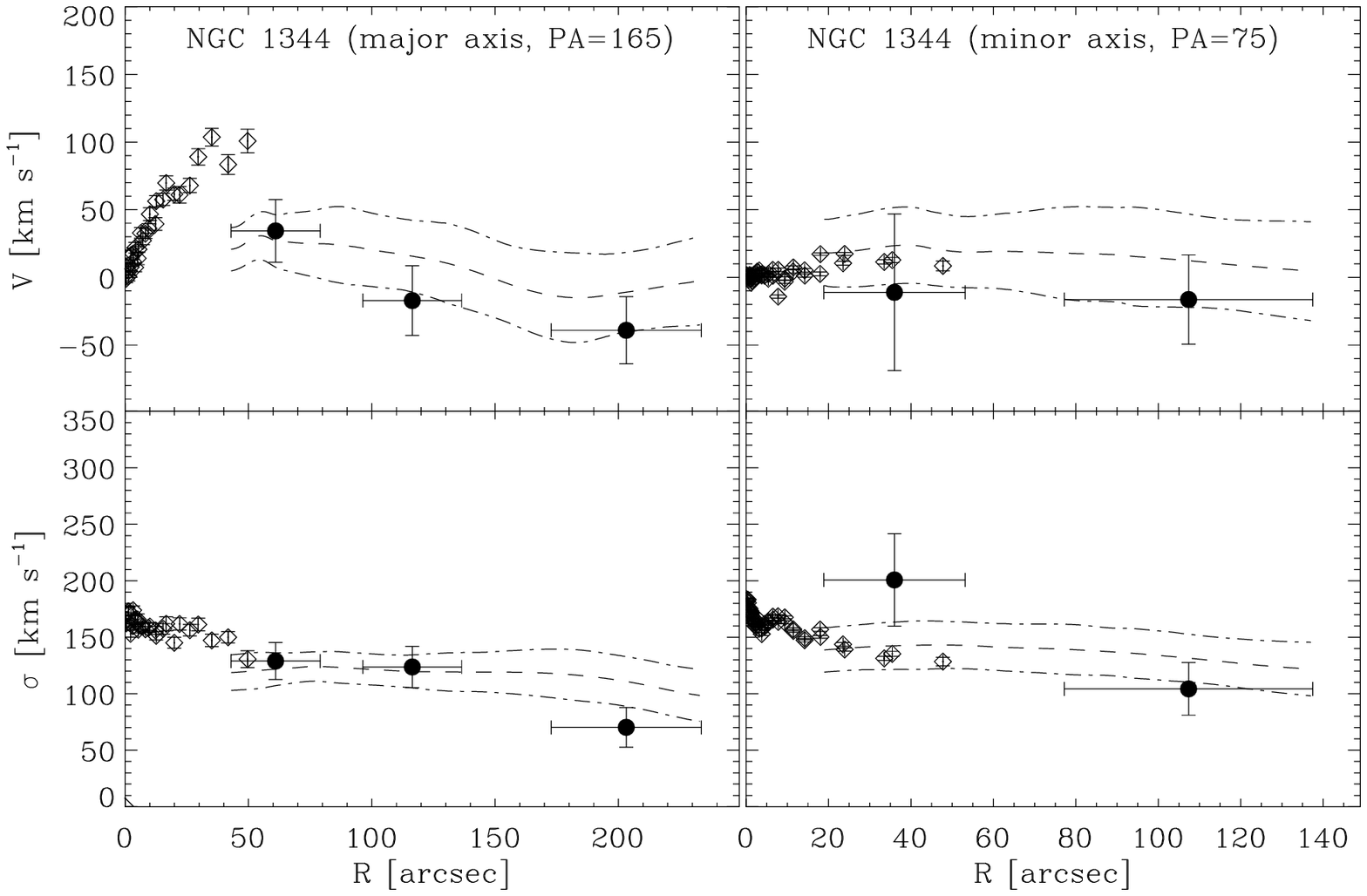,clip=,width=8.7cm} } 
\hbox{
    \psfig{file=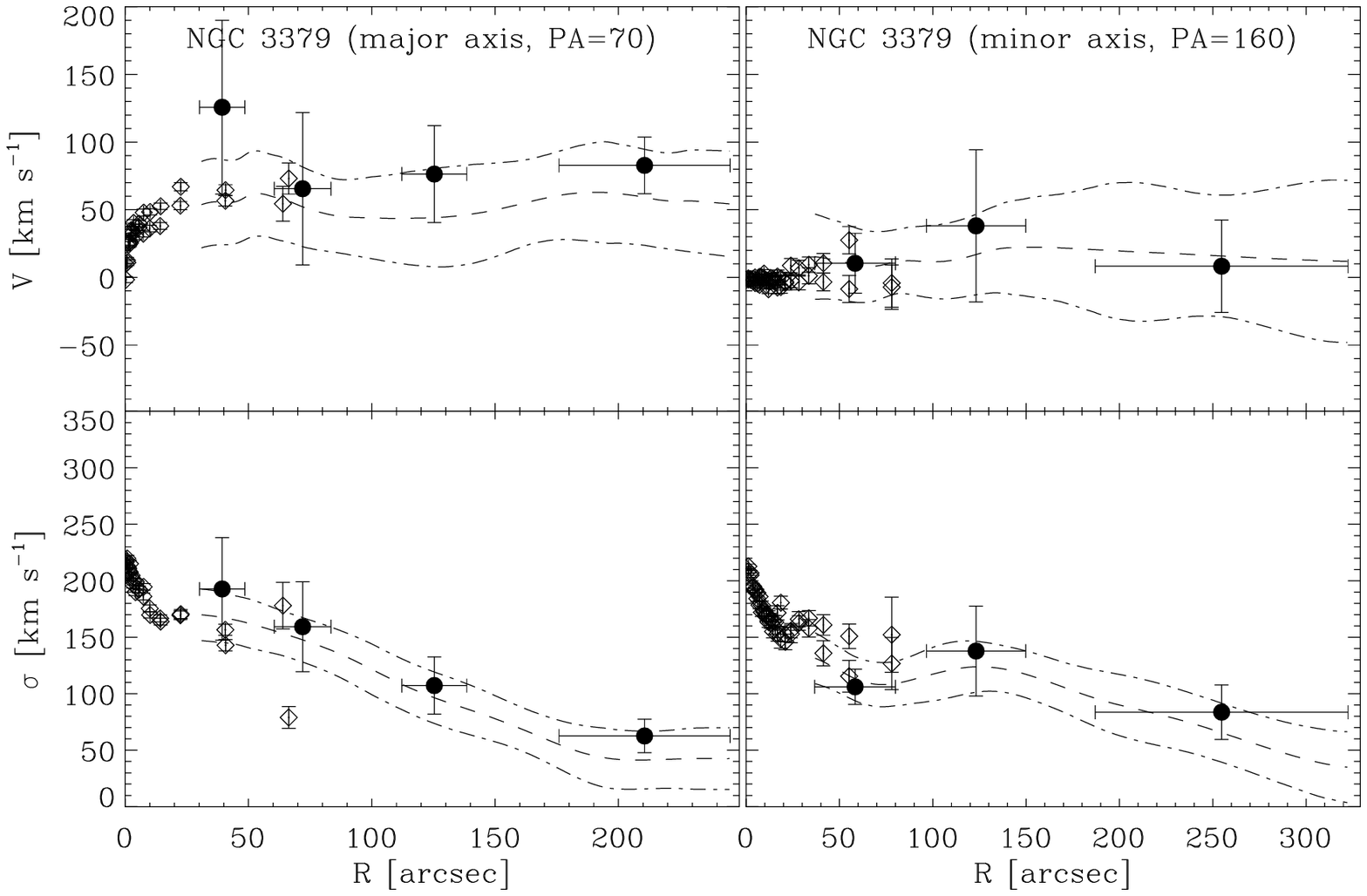,clip=,width=8.7cm}
    \psfig{file=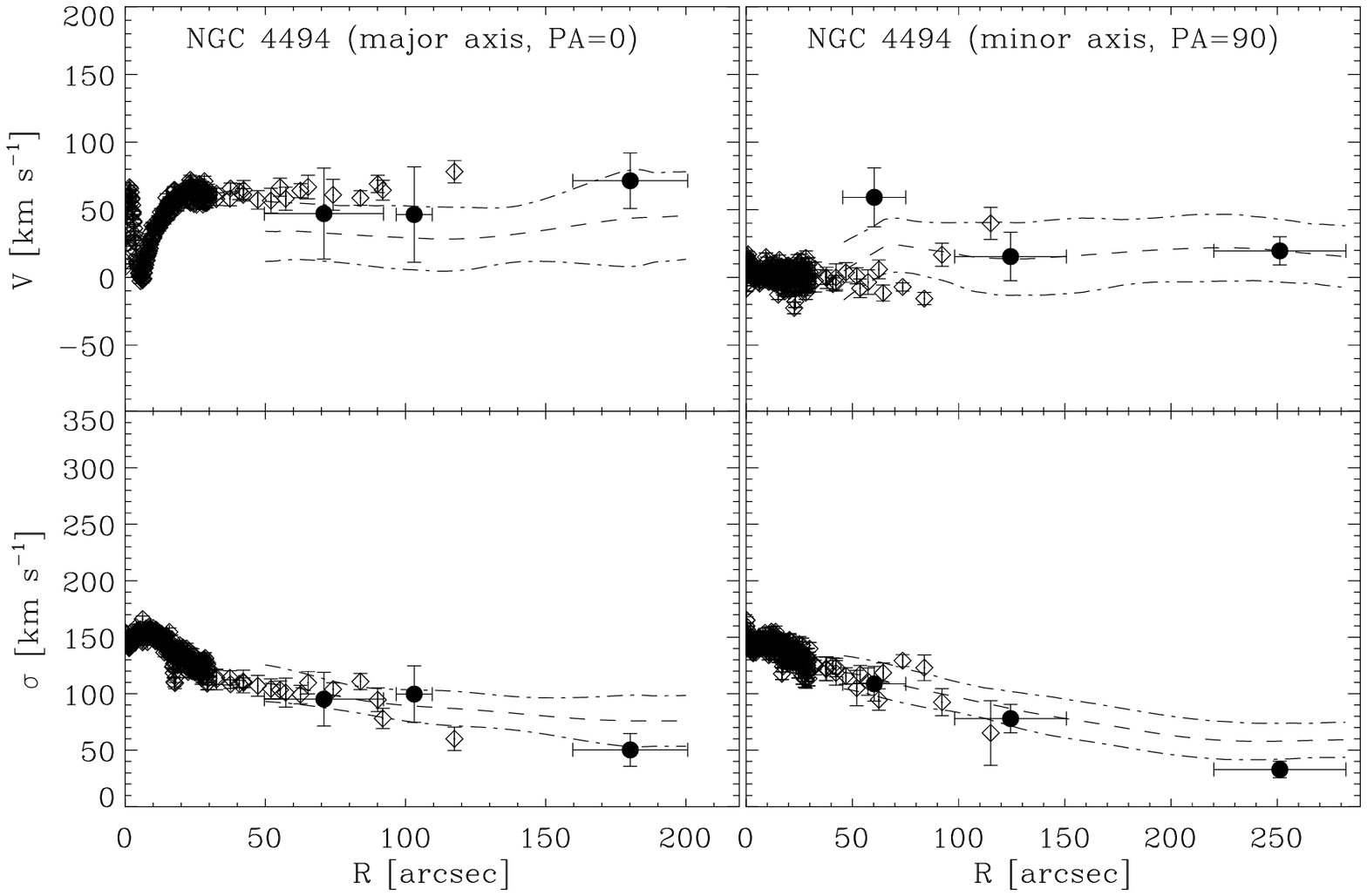,clip=,width=8.7cm} } 
\hbox{
    \psfig{file=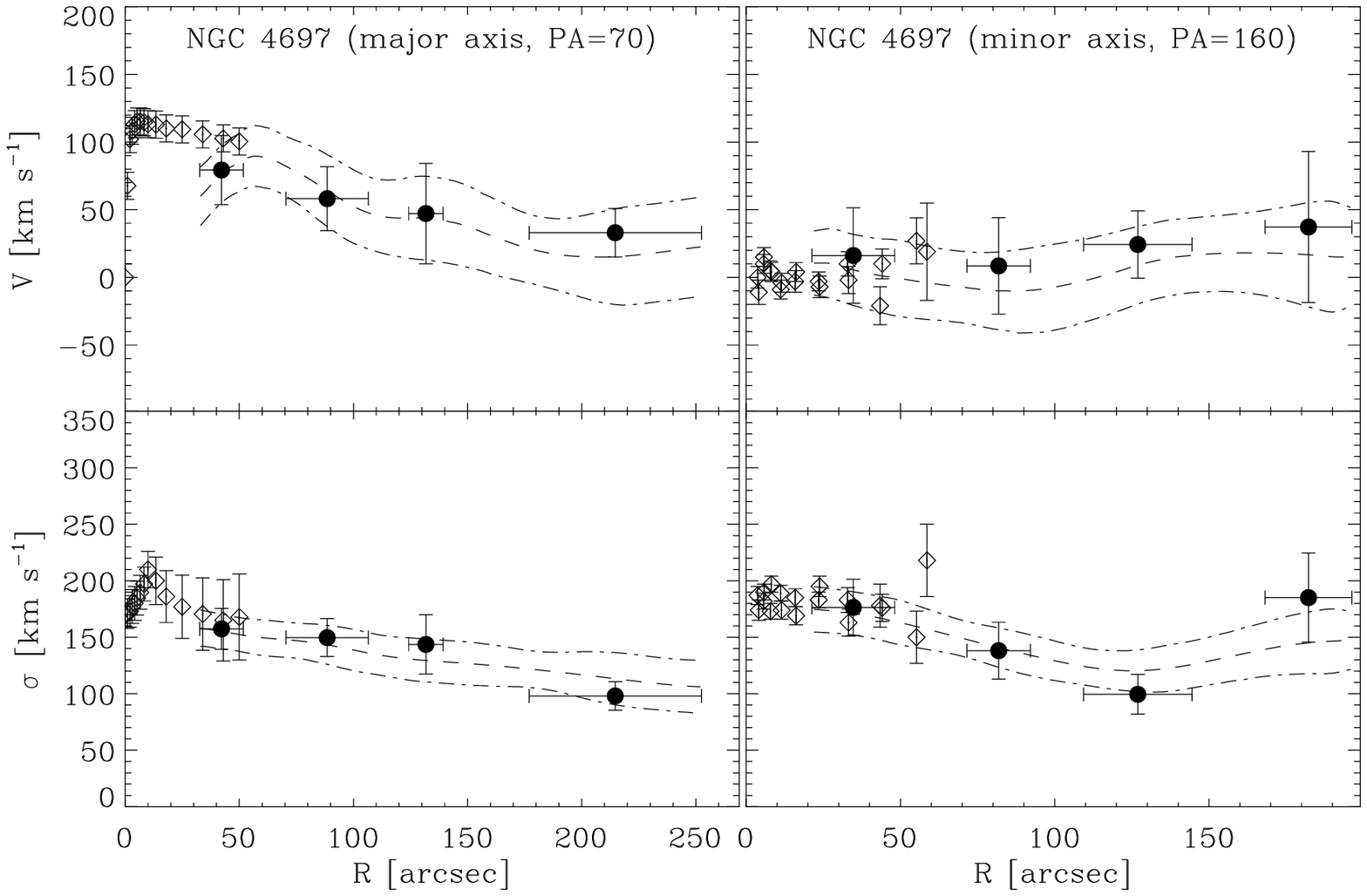,clip=,width=8.7cm}
    \psfig{file=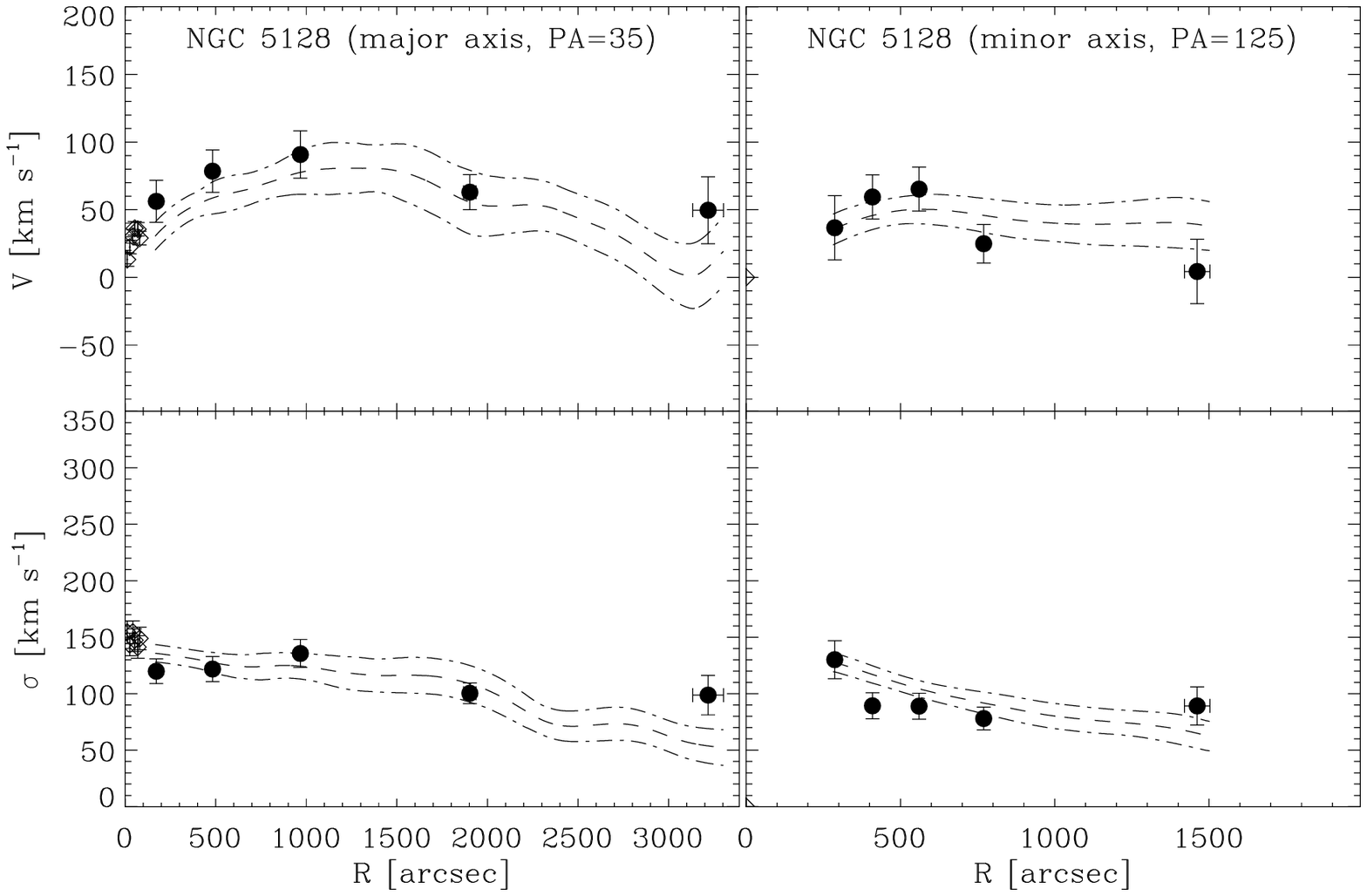,clip=,width=8.7cm} }
}
\caption{Continued. References for the absorption-line kinematics:
  NGC 1023, \citet[major axis]{Debattista+02} and \citet[minor
    axis]{Simien+97}; NGC 1344, \citet{Teodorescu+05}; NGC 3379,
  \citet{Statler+99}; NGC 4494 our VLS/FORS2 data (see Appendix
  \ref{sec:long_slit}); NGC 4697, \citet[major axis]{Dejonghe+96} and
  \citet[minor axis]{Binney+90}; NGC 5128, \citet[major axis]{Hui+95},
  minor axis not available.}
\end{figure*}%

\section{Notes on individual galaxies}
\label{sec:individual_galaxies}

\subsection{{\it Sample A}}
\subsubsection{NGC 821}
NGC 821 is a field elliptical galaxy. It is classified as an E6 in the
RC3 and NED catalogues, but we measured a mean ellipticity of 0.4.
The PNe system of NGC 821 was studied for the first time by \citet[104
  PNe]{Romanowsky+03}. The declining radial profile of the velocity
dispersion observed from the combination of stellar and PNe data was
interpreted as a signature of low dark matter concentration in this
galaxy halo, lower than the prediction by cosmological
simulations. Recently, \citet{Forestell+08} extended the stellar
kinematic measurements to $\sim 100''$, claiming a discrepancy between
stellar and PNe kinematics (the velocity dispersions measured with PNe
are lower than the ones measured with absorption lines). Using
three-integral models they derived a dark matter content that is
slightly higher than determined by \citet{Romanowsky+03}, but still
lower than predicted by cosmological simulations.

In this paper we have obtained a new PN.S catalogue with 125
detections. 
To compare their spatial
distribution with the stellar surface density profile, we used
ground-based photometric data in the $V$-band measured by
\citet{Goudfrooij+94}, combined with the HST measurements (F555W,
shifted artificially by $-0.06$ magnitudes to match the ground-based
data) given by \citet{Lauer+05}.
The extrapolation of the S\'ersic fit to the surface brightness
profile out to $\approx 300''$ agrees well within the error bars with
the number density radial profile of the PNe (Figure
\ref{fig:photom_comparison}).

The smoothed PNe two-dimensional velocity field produced in Section
\ref{sec:smoothed_fields} using the folded PNe sample shows a rotation
of $\sim 120$ \kms, with a misalignment of $56\pm25$ degrees between
the photometric major axis ($PA_{PHOT}=25$, from RC3) and the PNe
kinematic major axis ($PA_{KIN}=-31$, direction of the approaching
side). The misalignment between $PA_{PHOT}$ and $PA_{KIN}$ is
confirmed also if we use the un-folded PNe sample.  This test excludes
the possibility that the observed rotation is produced by a few background
sources erroneously identified as PNe, which skew the velocity distribution.

The comparison between PNe and stellar absorption-line kinematics is
shown in Figure \ref{fig:radial_kinematic}. Long-slit data are from
\citet{Forestell+08}, which agree with the integral-field stellar
kinematics from \citet{Emsellem+04}. The PNe velocities at $\sim 70''$
may be systematically slightly lower than the absorption-line points
but the two data sets still appear consistent within the error bars of
the PNe points and the scatter of the absorption-line data.

Along the photometric major axis, the stellar component suggests a
plateau in the rotation curve with $V \simeq 60$ \kms\ for $R<90''$,
while the PNe system shows a lower value of $\sim 15 - 20 \pm 40$
\kms\ for $R>120''$. Along the photometric minor axis the scatter in
the stellar kinematics is quite large and it is difficult to make a
clear comparison.  Nevertheless, we see an increase of the stellar
rotation in the region $20''-50''$ in good agreement with the PNe
data. The galaxy thus exhibits both major axis and minor axis rotation
and so is most likely triaxial.  
 In particular, the outer regions of NGC 821 ($R>50''$) have the
  positive side of rotation aligned with $PA=149\pm25$ (see Figure
  \ref{fig:2dfields} and Table \ref{tab:kin_angles}), while the inner
  regions have have the positive side of rotation aligned with $PA=31$
  (\citealt{Cappellari+07}, but see also Figure 4 in
  \citealt{Emsellem+04}), indicating a misalignment of 118 degrees
  between inner and outer regions.

  The fact that PNe and absorption-line kinematics show different
  misalignments with $PA_{PHOT}$, 56 and 6 degrees respectively (Table
  \ref{tab:kin_angles}), could be because these two quantities are
  computed in different radial intervals: the PNe misalignment is
  computed for $R\geq 30''-40''$, while the absorption-line
  misalignment is computed for $R<20''$ (from SAURON
  kinematics). There is a (small) region of overlap between the PNe
  kinematics and the long-slit data in Figure
  \ref{fig:radial_kinematic}, in which there is no evidence that the
  PNe and absorption line data disagree.  Both sets of data may trace
  different regions of the same velocity field, if the kinematics
  major axis changes rapidly from the centre outwards.

The combined velocity dispersion profile along both axes shows a
strong decrease, reaching $\sigma \simeq 50$ \kms\ at $250 ''$.

\subsubsection{NGC 3377}
\label{sec:individual_3377}

NGC 3377 is an E5 galaxy in the Leo I group. This galaxy has been
studied on numerous occasions. Mass models have been obtained
  over different radial ranges, from the very inner regions to
measure the supermassive black hole mass (e.g.,
\citealt{Gebhardt+03,Magorrian+98}), out to $\sim 1$ $R_e$ to
determine the mass-to-light ratio, orbital structure and dynamical
properties (e.g., \citealt{VanderMarel+07, Cappellari+06}).
The large number of PNe obtained with the PN.S (154 detections)
derived in the present paper allows us to extend the kinematic
information out to $\sim 5$ $R_e$. 

We used ground-based photometric data in the $B$-band measured by
\citet{Goudfrooij+94} and \citet{Jedrzejewski87}, combined with the
HST measurements given by \citet{Lauer+05} (F555W, shifted
artificially by $-0.08$ magnitudes to match the ground-based data). In
the small overlapping region ($60''<R<100''$) the PNe counts follow
the stellar surface brightness within the error bars. In the region
$100''<R<200''$ the PNe counts follow the S\'ersic fit extrapolation
of the stellar surface brightness. The last PNe data point at $300''$
has too large error bars to derive conclusions.

The PNe system shows a twist of the kinematic major axis, from
$PA_{KIN}=248\pm20$ at $R=60''$ up to $PA_{KIN}=167\pm20$ at $R=200''$.
The stellar surface brightness is not extended enough to check whether
or not this twisting is reproduced also in the photometry.

Our deep stellar long-slit kinematics obtained with FORS/VLT (see
Appendix \ref{sec:long_slit}) show a good agreement within errors with
the PNe velocity and velocity dispersion extracted along both major
and minor axes. Along the major axis, this galaxy shows a very steep
central rotation gradient within $5''$ together with a central
velocity dispersion $\sigma=170$ \kms\ most likely due to the
contribution of the central black hole \citep{Gebhardt+03}. The
stellar radial velocity reaches a maximum value $V\approx 115$ \kms\
within $40''$ and then it goes down to $\sim 0$ \kms\ at $210''$. On
the contrary, the velocity dispersion increases from $\approx 60$
\kms\ at $40''$ to $\approx 100$ \kms\ at $100''$ and then it goes
down again to 80 \kms. This turnover in the velocity dispersion is
observed both in the stars and the PNe.  Therefore it appears that
despite its classification as an E5, NGC 3377 shows disc-like
kinematics in its inner $\sim 70"$ (i.e., high value of the maximum
rotation velocity, low values of the velocity dispersion).

\subsubsection{NGC 3608}
\label{sec:individual_3608}

NGC 3608 is an E2 galaxy in the Leo II group.  The mass models of NGC
3608 constructed so far are confined to the inner regions and are
based on stellar long-slit \citep{VanderMarel+91, Magorrian+98,
  Gebhardt+03} or two-dimensional integral-field \citep{Cappellari+07}
kinematics.

Radial velocities for 87 PNe were derived in the present paper. It has
a close companion, NGC 3607, about 6 arcminutes to the South, whose
systemic velocity ($\sim 935$ \kms, from RC3) makes the two PNe
systems overlap in phase space. In our PNe analysis we did not
consider the objects closer to NGC 3607 than NGC 3608. This probably
still leaves  some contaminants in the PNe catalogue: in fact,
there is an over-density of detections on the side of NGC 3608 closer
to its companion (see Figure \ref{fig:spatial_distribution}). 

We used ground-based photometry by \citet{Jedrzejewski87} in $B$-band
and HST photometry by \citet{Lauer+05} (F555W, artificially shifted by
0.92 magnitudes to match the ground-based data). Unfortunately no
spatial range exists where the stellar surface brightness and PNe data
overlap, so a direct comparison is not possible. If we extrapolate the
S\'ersic fit of the luminous profile, we find a good agreement with
the PNe number denisity (much better than the $R^{1/4}+$exponential
disc fit extrapolation).

The PNe velocity field is quite noisy in the outer region and the
PNe major kinematic axis ($PA=40$\deg\ $\pm 29$) is misaligned with
the photometric one ($PA=75$). Both effects are probably caused by the
low number of detections and the possible contamination by the
companion galaxy.

Stellar kinematics ($PA=81$\deg\ and $PA=-9$\deg) are taken from
\citet{Halliday+01}. Again the stars and PNe do not overlap in radial
range, so a direct comparison between the kinematics of two systems is
not possible. Along the major axis, the stellar kinematics reveal a
kinematically decoupled core (already reported by
\citealt{Halliday+01}). Extrapolation of the stellar rotation at
$50''$ matches the PNe data well, which show an increase of the
rotation to $\approx 100$ \kms\ at $\sim 150''-200''$. Rotation is
also visible along the minor axis, but the scatter and error bars are
too large to derive general conclusions. The velocity dispersion along
the major axis decreases from the central 200 \kms\ to $\approx 100$
\kms\ at $\approx 40''-50''$ and then shows an increase to 170
\kms\ at $R=200''$.

 As pointed out before, the PNe sample of NGC 3608 might be
contaminated by the presence of PNe belonging to NGC 3607. In order to
quantify the contamination effect, we exclude all PNe on the NGC 3607
side (i.e. southern side). The remaining catalogue contains therefore
only 30 PNe on the northern side of NGC 3608, to which we apply the
point-symmetric reflection as in Section
\ref{sec:folded_catalogue}. We will refer to this folded,
northern-side catalogue as the ``N-sample''. This is a conservative
approach as it will eliminate most of the contaminants, but it will also
eliminate many PNe of NGC 3608 and the results may suffer from low
number statistics.

\begin{figure}
\vbox{
  \psfig{file=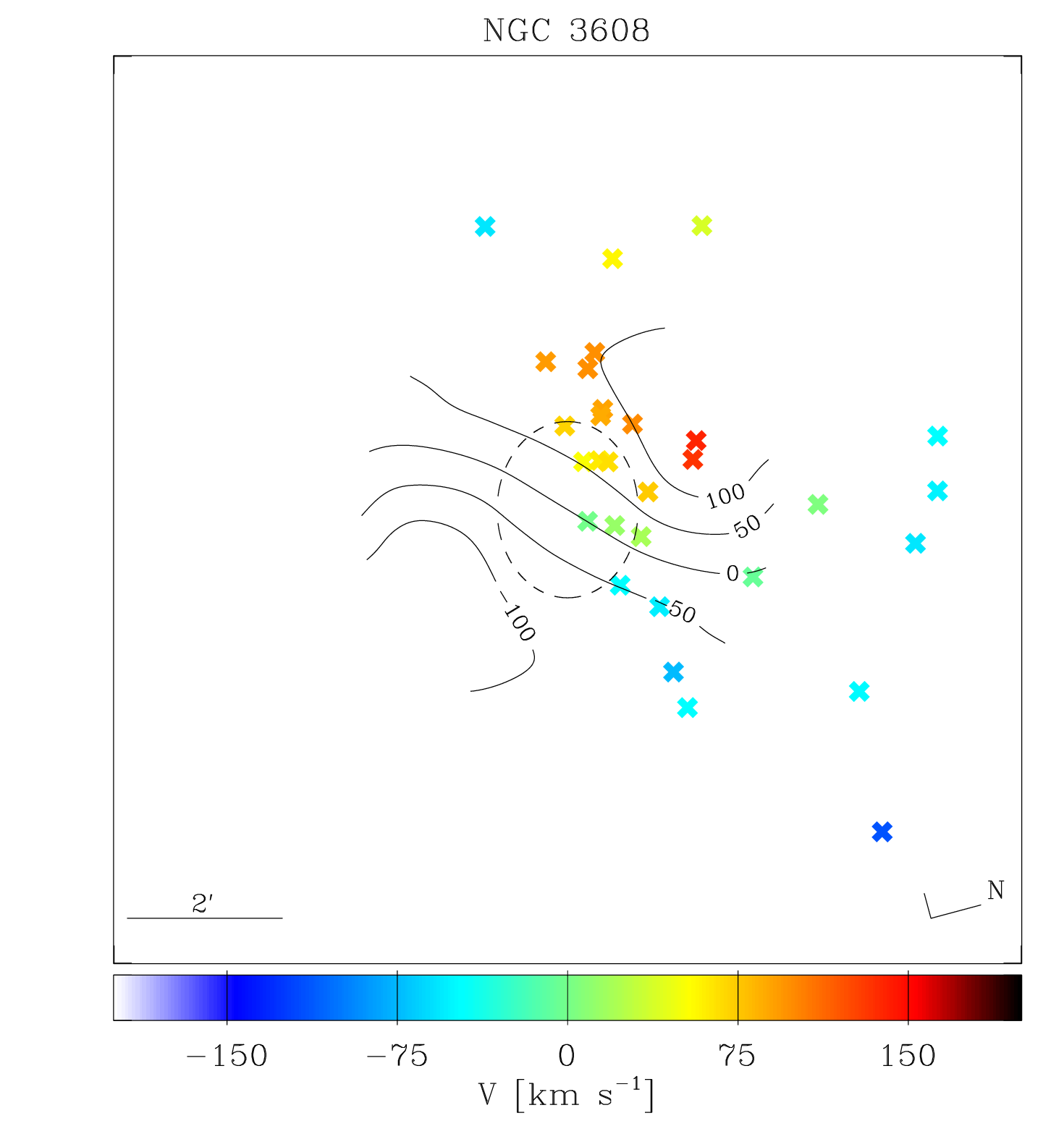,clip=,width=8.5cm}
  \psfig{file=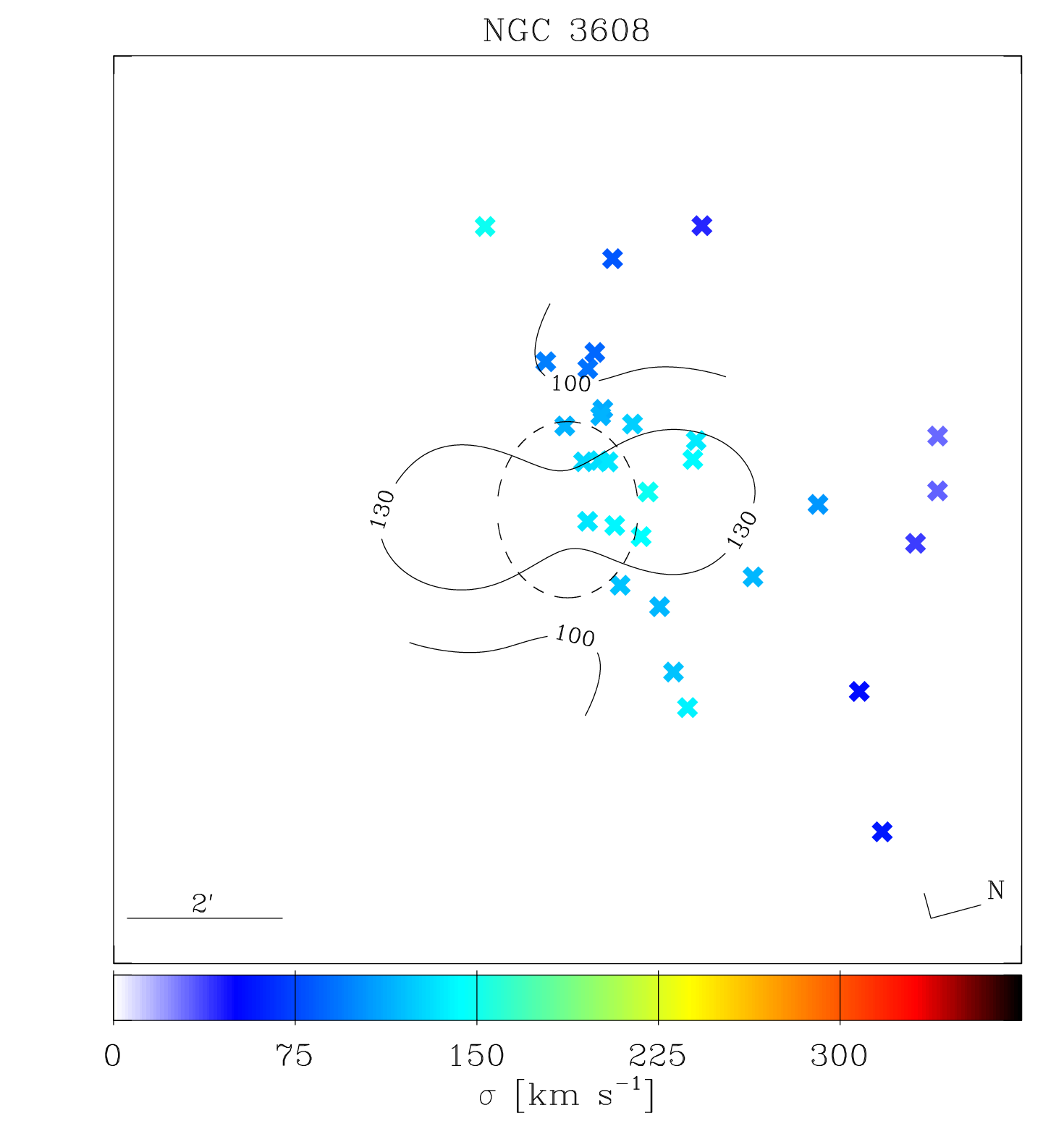,clip=,width=8.5cm}
}
\caption{ Two-dimensional velocity ({\it upper panel}) and
    velocity dispersion ({\it lower panel}) fields of NGC 3608
    obtained using the ``N-sample''.  Labels, scales and symbols are
    as in Figure \ref{fig:2dfields}.}
\label{fig:3608_Nord_2D}
\end{figure}

\begin{figure}
  \psfig{file=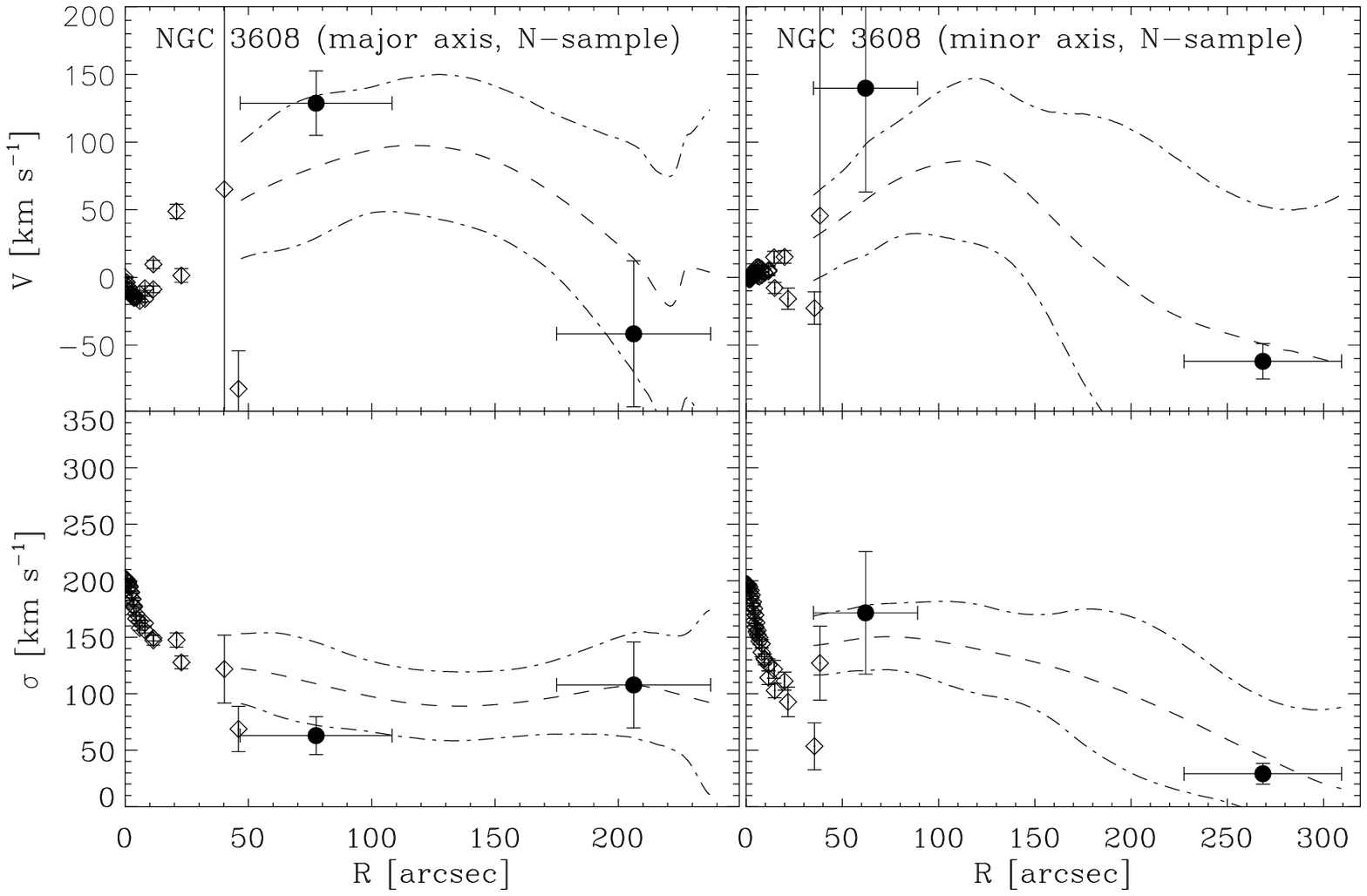,clip=,width=8.5cm}
\caption{The same as in Figure \ref{fig:radial_kinematic}, but using the NGC 3608 ``N-sample''.}
\label{fig:3608_Nord_radial}
\end{figure}

In Figures \ref{fig:3608_Nord_2D} and \ref{fig:3608_Nord_radial} we
present the two-dimensional velocity and velocity dispersion fields,
and the comparison with stellar kinematics extracted along the major
and minor axes using the ``N-sample''.
The two dimensional kinematics of the ``N-sample'' are different from 
those of the original PNe sample; the new velocity field is
characterized by more regular rotation and the new velocity dispersion
contours have a different orientation.

The new velocity field has a kinematic position angle of $44\pm33$ for
$R<100''$ and of $72\pm29$ for $R>100''$, with an average $PA_{KIN}
=75\pm22$.
Major and minor axis rotation curves are characterized by steeper
gradients: they reach the maximum of the rotation
$V_{max}^{major}=130$\kms\ at $70''$ and $V_{max}^{minor}=140$\kms\ at
$60''$ before dropping rapidly to $\sim -50$ \kms\ after $200''$.
The radial velocity dispersion profile along the major axis is
systematically lower for the ``N-sample''.

Thus the outer halo kinematics of NGC 3608 depend strongly on the
adopted PNe sample.  We therefore consider the results from the
``N-sample'' as a separate case, referring to them as NGC 3608N for
the rest of the paper.

\subsubsection{NGC 4374}
\label{sec:individual_4374}

NGC 4374 (M84) is a well-studied, bright E1 galaxy in the Virgo cluster.  Mass
models have been constructed by \citet{Kronawitter+00, Cappellari+07}
using stellar kinematics within 1 $R_e$. With our deep long-slit
observations and 457 PNe radial velocity measurements obtained in this
paper we are able to extend the kinematic information out to 6 $R_e$.

We used the extended ground-based photometry given by
\citet{Kormendy+08}, which shows a very good agreement between stellar
surface brightness and PNe number density. The data show large
variations in the position angle of the photometric major axis. Beyond
$R=80''$ the photometric position angle increases from 120 to 180
degrees. The shift in position angle may be an effect of the
  interaction with the nearby galaxy NGC 4406.

The two-dimensional velocity field of NGC 4374 indicates that the
kinematic rotation axis ($PA_{KIN}=221$) is almost orthogonal to the
photometric one ($PA_{PHOT}=135$). Simulations carried out in Section
\ref{sec:testing} clearly indicate that the rotation along that axis
is significant (see Figure \ref{fig:simulations1}). The misalignment
between kinematic and photometric major axes is seen also in the
stellar absorption-line kinematics, which agree well with the PNe
data.

\begin{figure}
\psfig{file=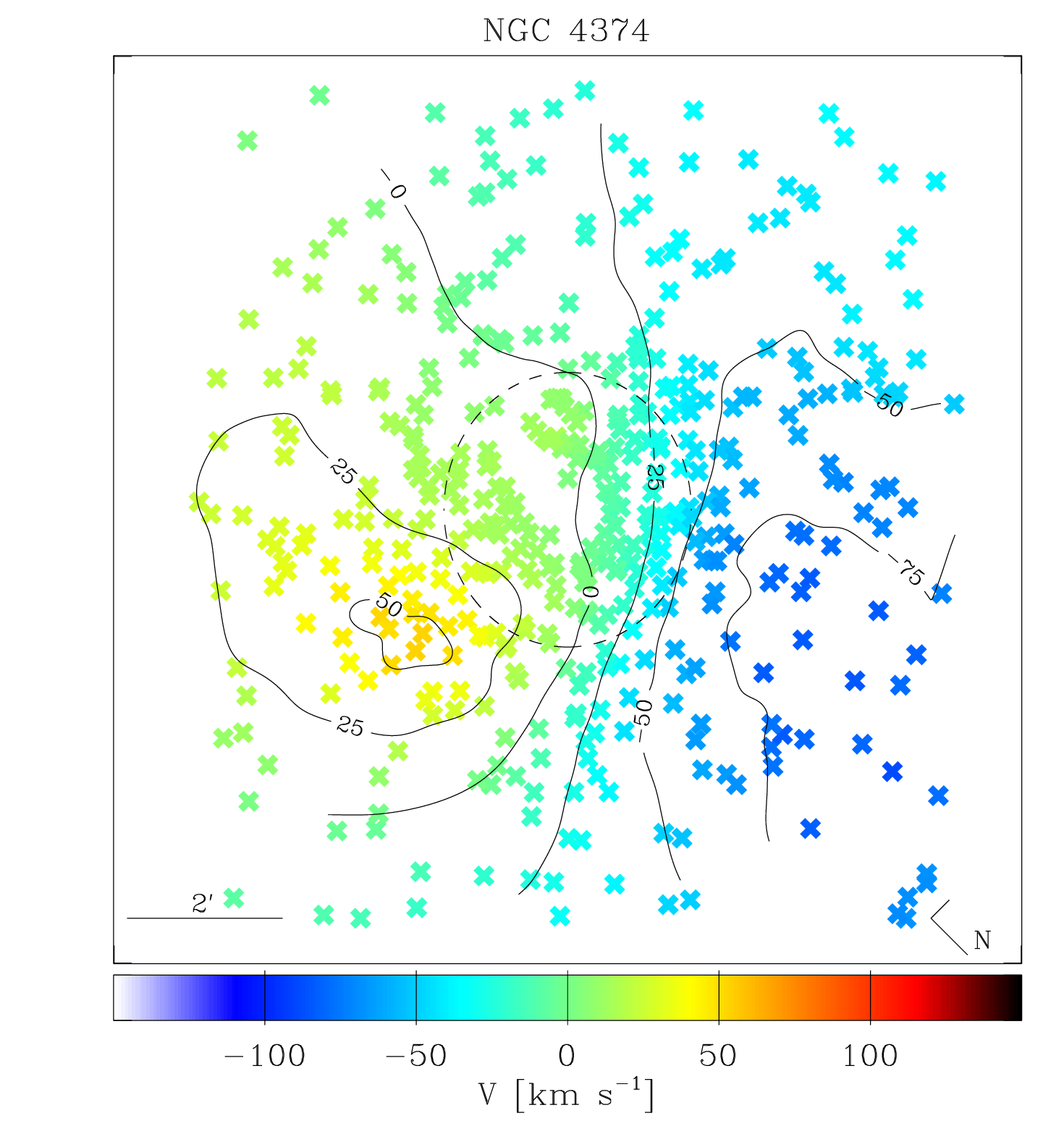,clip=,width=8.5cm}
\caption{Two-dimensional velocity field of NGC 4374 obtained using the
  adaptive kernel smoothing procedure (Section
  \ref{sec:smoothed_fields}) without imposing point symmetry on the
  PNe data set.}
\label{fig:4374_unfolded_V2d}
\end{figure}

\begin{figure}
\psfig{file=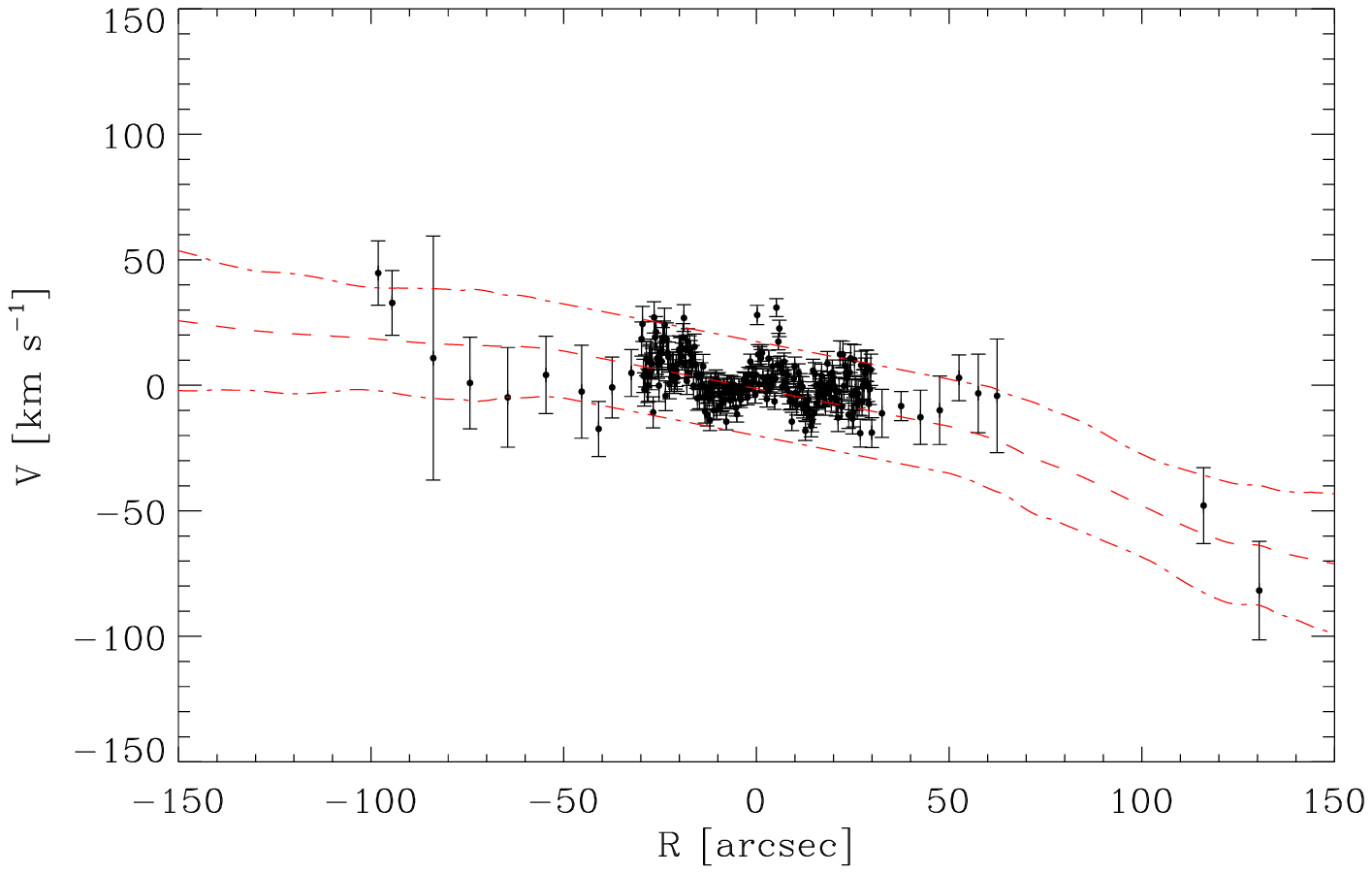,clip=,width=8.5cm}
\caption{Comparison between the stellar kinematics and the PNe
  kinematics from the ``unfolded'' data set of NGC 4374. {\it Black
    symbols:} Stellar kinematics extracted along the photometric minor
  axis, as shown in Figure \ref{fig:radial_kinematic} but without
  folding the two galaxy sides. {\it Dashed red line:} PNe kinematics
  extracted along the same axis of the stars, using the ``unfolded''
  PNe sample. {\it Red dot-dashed lines:} Error boundaries computed
  with Monte Carlo simulations.}
\label{fig:4374_unfolded_star}
\end{figure}

The mean velocity field obtained with the ``unfolded'' PNe catalogue
(Figure \ref{fig:4374_unfolded_V2d}) differs from the one calculated
with the ``folded'' catalogue.  In particular we noticed an asymmetry:
the receding side reaches $\sim +50$ \kms, while the approaching side
reaches $\sim -80$ \kms. This asymmetry in the velocity field is
confirmed also in the stellar kinematics, as shown in Figure
\ref{fig:4374_unfolded_star}.

The ``unfolded'' and ``folded'' velocity dispersion fields do not
differ significantly from each other. They both show almost constant
velocity dispersion, with small fluctuations consistent with
measurement errors.

The distortion of the velocity field in NGC 4374 is unlikely to
be caused by contamination from PNe belonging to NGC 4406 because the
two galaxies have very different systemic velocities. It could
possibly be related to the presence of a diffuse intracluster light
component in this part of the Virgo cluster \citep{Arnaboldi+96},
or to a current interaction of NGC 4374 with NGC 4406.  The angular
separation between NGC 4406 and NGC 4374 is $\sim 17' \sim 87$ kpc (at
a distance of 17.6 Mpc). NGC 4406 is on the approaching (NE) side of
NGC 4374 with a systemic velocity $\sim 1300$ \kms\ lower than NGC
4374 and it is closer by about $1.1$ Mpc (see Table \ref{tab:sample}).

We can exclude that the observed asymmetry is caused by
background Ly$\alpha$ galaxies at redshift $z\sim 3.1$ because they
are too few to generate such a major distortion of the velocity
field. In fact, according to \citet{Ciardullo+02} we would expect
$\sim 2$ Ly$\alpha$ contaminants for our field of view ($9'\times
9'$), redshift range ($\Delta z \sim 0.0087$), and completeness
magnitude (28 mag).

\subsubsection{NGC 4564}\label{sec:n4564}
NGC 4564 is an E4 galaxy in the Virgo cluster.  The number of
detections in our data (49 PNe, with 38 outside 2 $R_e$) is quite low,
and a direct comparison of their radial distribution with the stellar
surface brightness \citep{Goudfrooij+94} suffers from small number
statistics. The radial range where both data sets overlap is small,
but there is a good agreement within error bars with the extrapolation
of the S\'ersic fit to the stellar data.

The two-dimensional velocity field shows a steep central velocity
gradient parallel to the photometric major axis. The velocity
dispersion field has a central peak of $\sim 90$ \kms\ ($\sim 50''$)
and then it drops down.

Although the major axis stellar kinematics \citep{Halliday+01} are not
extended enough to have a large overlap with the PNe kinematics
extracted along the major axis, the extrapolation of the stellar
rotation curve agrees well with the PNe kinematics. The combined major
axis rotation curve reaches a plateau ($\sim 140$ \kms\ at $20''$) and
then declines from 50'' to $\sim 0$ \kms\ at $180''$. The velocity
dispersion decreases from the central 180 \kms\ to $\sim 70$ \kms\ at
$R=20''$ and then declines slowly to $\approx 30$ \kms\ at
$R=150''-200''$.  The NGC 4564 major axis rotation curve resembles
that of an S0 galaxy rather than that of an elliptical galaxy.

\subsubsection{NGC 5846}
The E0 galaxy NGC 5846 is the brightest member and the central galaxy
in the NGC 5846 group.  Mass modelling of NGC 5846 has been performed
by \citet{VanderMarel+91, Kronawitter+00, Cappellari+07}, all using
stellar kinematics within 1 effective radius.
With the PNe measurements (124 detections) obtained with our PN.S
instrument, we are able to extend the kinematic information of NGC
5846 out to 6 $R_e$ according to Table \ref{tab:sample}. A separate
paper with detailed dynamical models combining stellar kinematics, PNe
data and X-ray observations is in preparation (Das et al., in
preparation).

Stellar photometric data in $V$-band is taken from
\citet{Kronawitter+00}. PNe number density and stellar photometry
agree well within the error bars.

The two-dimensional velocity field does not show significant
rotation. The apparent rotation of $\sim$ 30 \kms\ along $PA=57$
degrees seems to be consistent with 0  within the measurement
errors (see Figure \ref{fig:simulations1}).

Long-slit stellar data along the major \citep{Kronawitter+00} and
minor \citep{Emsellem+04} axes are also consistent with no rotation,
even if their radial coverage is not sufficient to ensure a proper
comparison with PNe kinematics (Figure \ref{fig:radial_kinematic}).

The velocity dispersion along the major axis shows a slightly
declining profile, from the central $\sim$250 \kms\ down to $\sim$170
\kms\ at $250''$.

\subsection{{\it Sample B}}

\subsubsection{NGC 1023}
NGC 1023 is an S0 galaxy, the brightest member in the eponymous group.
\citet{Noordermeer+08} studied the kinematic data of 204 PNe in the S0
galaxy NGC 1023, obtained with the PN.S. The combined velocity curve
from stellar and PNe data shows a maximum value of $\sim$240 \kms\ at
$110''$ followed by a remarkable decline to $\sim$50 \kms\ at $360''$
. The peak in the radial velocity of NGC 1023 occurs approximately at
the same position as the minimum in the velocity dispersion. Outside
$110''$ the galaxy is dominated by a more pressure-supported
component, and the velocity dispersion increases up to 140 \kms.

This was interpreted as a signature of the complex evolutionary
history of the galaxy: the inner regions with high rotation and low
velocity dispersion indicate a quiescent disc formation process, while
the outer parts with declining rotation and higher velocity dispersion
indicate heating from a merger or strong interaction with the
companion galaxy.

Figure \ref{fig:photom_comparison} shows a good agreement between the
stellar surface brightness and PNe number density within the
relatively large errors. The PNe densities are different from the
original values presented in \citet{Noordermeer+08} because the
completeness correction was not performed there.

The two-dimensional PNe velocity and velocity dispersion fields are
derived using the PNe catalogue of \citet{Noordermeer+08} and
indicate rapid rotation in the central $\approx 100''$ followed by a
decline towards larger radii (Section \ref{sec:2d_literature}).

Approximate agreement is found between absorption-line and PNe
kinematics (Figure \ref{fig:radial_kinematic}) within the error
bars. The small discrepancy in the rotation curve may be caused
by the high inclination of the galaxy disk ($\sim 72^{\circ}$
according to \citealt{Noordermeer+08}). Stellar kinematics are taken
from \citet{Debattista+02} for the major axis and \citet{Simien+97}
for the minor axis.

\subsubsection{NGC 1344  (NGC 1340)}

NGC 1344 is an E5 elliptical galaxy in the Fornax cluster.  We use the
stellar long-slit kinematics and the radial velocities of the 195 PNe
in NGC 1344 from \citet{Teodorescu+05}, who estimated a dark matter
halo of $3.8\cdot 10^{11}$ $M_{\odot}$ within $160''$ (3.5 $R_e$) by
combining the two datasets. The PNe number density profile agrees well
with the stellar surface brightness \citep{Sikkema+07}; see Figure
\ref{fig:photom_comparison}.

The PNe velocity field (see Appendix \ref{sec:2d_literature})
reaches $\sim40$ \kms\ and shows a misalignment of $32\pm27$ degrees
with respect to the photometric major axis.
The absorption-line kinematics taken from \citet{Teodorescu+05} extend
to $60''$ along the major axis. The rotation velocity reaches a
maximum of $\sim 100$ \kms\ at $60''$. The PNe velocities, which are
measured outside this region show essentially no rotation (already
reported in \citealt{Teodorescu+05}). The spatial overlap between the
two systems is too small to ascertain the cause of this steep decrease
in rotation. Along the minor axis both components show no
rotation. The velocity dispersion profile declines from a central
value of $\sim 180$ \kms\ to $\sim 90$ \kms\ at $200''$.

\subsubsection{NGC 3379}

NGC 3379 (M105) is an E1 galaxy in the Leo I group, close in
projection (7') to NGC 3384.  The PNe system of NGC 3379 has been
widely studied in the literature. Radial velocities are presented by
\citet[29 PNe]{Ciardullo+93}, \citet[a preliminary sample of 109 PNe
  from the PN.S]{Romanowsky+03}, \citet[54 PNe]{Sluis+06} and by
\citet[the final sample of 191 PNe from the PN.S]{Douglas+07}.
Also the dynamical structure of NGC 3379 has been studied by several
authors, using models derived from stellar kinematics
\citep{Kronawitter+00, Cappellari+07} or from the combination of
stellar and PNe data \citep{Romanowsky+03, Samurovic+05,
Douglas+07}. The PNe kinematics at large radii show a  steep
decline in the velocity dispersion, which was previously interpreted
as evidence for a lower content of dark matter than predicted by
standard cosmological simulations.

In a recent paper, \citet{DeLorenzi+08a} presented a new
approach to the analysis of NGC 3379 using an N-particle model (NMAGIC,
\citealt{DeLorenzi+07}), combining stellar and PNe kinematics. They
generalised previous works by relaxing the assumption of spherical
symmetry and demonstrated that, due to the strong degeneracy between
radial anisotropy, intrinsic shape of the galaxy and mass
distribution, the data do allow moderately massive dark matter haloes.

The stellar surface brightness and PNe number density show good
agreement (see Figure \ref{fig:photom_comparison}, confirming the
analysis of \citealt{Douglas+07}).

In this paper we use the 191 PNe radial velocities published by
\citet{Douglas+07}. The smoothed two-dimensional fields are shown in
Appendix \ref{sec:2d_literature}. Our smoothing technique is
different from that used previously, but the general features of the
kinematics derived from the two methods are consistent (in particular,
the misalignment found between the photometric and kinematic major
axes).

Absorption-line kinematics along the major and minor axes
\citep{Statler+99} agree well with the PNe data, as already shown in
\citet{Romanowsky+03,Douglas+07}. The combined velocity dispersion
profile decreases to $\sim 60$ \kms\ at $\sim 210''$.

\subsubsection{NGC 4494}
NGC 4494 is an E1 galaxy in the NGC 4565 group.  Together with NGC 821
and NGC 3379, it is the third galaxy studied in \citet{Romanowsky+03}
in which the remarkable decline of the velocity dispersion radial
profile was interpreted in terms of a low content of dark matter. A
new detailed analysis is performed by \citet{Napolitano+08}, using
a catalogue with almost three times the number of detections obtained
with the new PN.S pipeline version described in
\citet{Douglas+07}. \citet{Napolitano+08} also present deep
long-slit data (see Figure \ref{fig:longslit}) and new deep
photometric data.

The two-dimensional fields presented in Appendix
  \ref{sec:2d_literature} are obtained using a different technique
compared to that used in \citet{Napolitano+08}. The main difference
between the two velocity fields is the presence of a kinematic
substructure rotating $\sim 30$\deg\ off the major axis. Nevertheless,
the small amplitude of this rotation ($\sim 25$ \kms) is consistent
with zero, if we take into account the measurement errors and the
errors determined by Monte Carlo simulations ($\sim 30$ \kms), as
described in Section \ref{sec:testing}.

Absorption-line and PNe kinematics show a good agreement, except at
$\sim 60''$ along the minor axis where the velocity dispersion
measured from the PNe is higher than that measured from the stars
(see also \citealt{Napolitano+08}).

\subsubsection{NGC 4697}

NGC 4697 is an E6 galaxy in the direction of Virgo.  This galaxy has
535 PNe detections by \citet{Mendez+01,Mendez+08}. They constructed a
spherical mass model using isotropic velocity dispersions and did not
require a dark matter halo. In this data set, \citet{Sambhus+06}
observed the presence of two distinct sub-populations of PNe with
different luminosity functions, spatial distributions and radial
velocities. The second population was particularly prominent in the
brightest PNe. Therefore \citet{DeLorenzi+08b} used only a sub-sample
of 351 PNe out of the Mendez catalogue for their new set of dynamical
models, which also incorporates kinematic constraints from long-slit
data. They found that a wide range of halo mass distributions was
consistent with the kinematic data for the galaxy, including some with
a dark matter content in agreement with cosmological merger simulations.

For this galaxy we could not correct the PNe data for the radial
incompleteness because of the lack of original images (see Figure
\ref{fig:photom_comparison}). This may explain why the PNe number
density profile falls below the stellar surface brightness
\citep{Goudfrooij+94,DeLorenzi+08b} in the inner regions.

We present the velocity and velocity dispersion fields of NGC 4697 in
(Appendix \ref{sec:2d_literature}) using the sub-sample defined
by \citet{DeLorenzi+08b}. A clear rotation is visible along the
photometric major axis, with a peak of $\sim 90$ \kms\ around $\sim 1$
$R_e$ and a declining profile thereafter.

Absorption-line kinematics along the major \citep{Dejonghe+96} and
minor \citep{Binney+90} axes show good agreement with the PNe
kinematics.

\subsubsection{NGC 5128}

NGC 5128 (Centaurus A) is a nearby merger remnant classified S0pec
galaxy, in the NGC 5128 group.  NGC 5128 is the early-type galaxy with the
largest number of PNe detections. It was first studied by
\citet{Hui+95}, using radial velocities of 433 PNe out to $\sim$22 kpc,
and then by \citet{Peng+04}, using radial velocities of 780 PNe out to
$\sim80$ kpc.  They found a total mass enclosed within 80 kpc of $5-6
\cdot 10^{11}$ M$_{\odot}$, depending on the particular model assumed
for the dark matter halo.

The more recent analysis of \citet{Woodley+07} included the radial
velocities of 320 globular clusters. They obtained a mass of $1.0 \cdot
10^{12}$ M$_{\odot}$ (using only PNe within 90 kpc) and $1.3 \cdot
10^{12}$ M$_{\odot}$ (using only globular clusters).

In Section \ref{sec:2d_literature} we show the two-dimensional
velocity and velocity dispersion fields of NGC 5128, which are very
similar to the ones presented by \citet{Peng+04}. The long-slit data
\citet{Hui+95} covers only the central region of the galaxy.

\subsection{{\it Sample C}}

\subsubsection{NGC 1316}
NGC 1316 (Fornax A) is a bright S0 galaxy in the Fornax cluster.
\citet{Arnaboldi+98} measured the positions and radial velocities of
43 PNe in this galaxy, computing an enclosed
mass of $2.9 \cdot 10^{11}$ M$_{\odot}$ within 16 kpc. Kinematic data
(PNe and stellar long-slit) used in our work are taken from their
paper. The relatively low number of PNe and their sparse distribution
did not allow us to construct a reliable two-dimensional velocity
field with our adaptive Gaussian kernel smoothing procedure. Also the
comparison with the stellar surface brightness suffers from low number
statistics. In addition, the stellar long-slit data
\citep{Arnaboldi+98} does not overlap sufficiently in radius with the
PNe kinematics to make a comparison. The combined velocity dispersion
radial profile is relatively flat for $R>R_e$ (see Section
\ref{sec:sigma_profiles}).

\subsubsection{NGC 1399}

NGC 1399 is an E1 galaxy in the centre of Fornax cluster.  This galaxy
has been modelled in detail by \citet{Saglia+00} using kinematic
measurements from their stellar long-slit data, and globular cluster
\citep{Kissler-Patig+98} and PNe \citep{Arnaboldi+94} radial
velocities out to $100''$. They found a total mass of $1.2-2.5 \cdot
10^{12}$ M$_{\odot}$, consistent with the results obtained from X-ray
determinations \citep{Ikebe+96}.
A more detailed analysis of the PN sample was presented in
\citet{Napolitano+02}. They found that the PN dispersion profile,
depending on the definition of outliers, is consistent with staying
approximately flat at 200\kms\ but also with increasing to 400 \kms\
outside  $\sim 3R_e$ as the globular cluster profile.

The small number of PNe (37 detections) is too low to derive reliable
two-dimensional smoothed fields with our procedure. Moreover, there is
no overlap between stellar and PNe data; therefore a comparison
between the PNe distribution and stellar surface brightness or between
PNe and stellar long-slit kinematics is not possible.

The combined velocity dispersion declines from $\sim 370$ \kms\ at the
centre to $\sim 200$ \kms\ at $170''$ ($\sim 5$ $R_e$)
\citep{Saglia+00}.

\subsubsection{NGC 3384 (NGC 3371)}

NGC 3384 is an S0 galaxy in the Leo I group, close in projection ($7'$)
to NGC 3379.  The two galaxies have comparable systemic velocities,
which makes the two PNe systems' radial velocities overlap.

Mass models of this galaxy are mostly based on stellar kinematics and
are confined to the innermost regions (e.g., \citealt{Gebhardt+03,
  Cappellari+07}). There are no large catalogues of PNe radial
velocities for this galaxy: \citet[63 PNe]{Tremblay+95}, \citet[23
  PNe]{Douglas+07} and \citet[50 PNe]{Sluis+06}. For the analysis in
this paper we use the catalogue of \citet{Tremblay+95}. A
  cross-check between the Douglas et al. and Sluis \& Williams
  catalogues is done in \citet{Douglas+07}. In \citet{Tremblay+95} PNe
  positions are not published, therefore a cross-check with that
  catalogue is not possible.

The relatively low number of PNe and their sparse distribution did not
allow us to construct a reliable two-dimensional velocity field or a
reliable number density profile. The major axis stellar long-slit data
\citep{Fisher97} did not significantly overlap in radius with the PNe
kinematics. This did not allow a comparison between the two data sets,
but the combined velocity dispersion profiles is shown in Figure
\ref{fig:slope}.

\subsubsection{NGC 4406}

NGC 4406 (M86) is an E3 galaxy in the Virgo cluster, near in
projection to NGC 4374. Surface-brightness fluctuations (Table
\ref{tab:sample}) place this galaxy at a distance of $1.1 \pm 1.4$ Mpc
closer than NGC 4374. The X-ray emitting hot gas envelopes of both
galaxies appear to be interacting (E. Churazov, private
communication). NGC 4406 has a radial velocity of about $-250$ \kms\
and is falling towards M87 from behind.
The catalogue of PNe radial velocities \citep{Arnaboldi+96} consists
only of 16 detections, but this has been sufficient to constrain the
velocity dispersion profile out to 11 kpc ($\sim 1.5R_e$).

The relatively low number of PNe and their sparse distribution did not
allow us to construct a reliable two-dimensional velocity field. Also
the comparison with the stellar surface brightness suffers from low
number statistics. 

The stellar long-slit data \citep{Bender+94} do not overlap
sufficiently with the PNe kinematics for a comparison. The combined
velocity dispersion ranges between 200 and 250 \kms\ in the central
$\sim 0.5 R_e$, and then declines to 96 \kms\ at $142''$.

\section{Sample properties: Results and discussion}
\label{sec:discussion}

In this section we analyse the general properties of the whole galaxy
sample (Table \ref{tab:sample}), combining the information from stars
and PNe. In particular, we look for correlations between kinematic
properties (such as the $V/\sigma$ ratio and the velocity
dispersion radial profile), photometric properties (such as the total
luminosity, UV emission or the shape of the isophotes) and the ratio
of PNe number to luminosity ($\alpha$ parameter). All quantities
determined for this purpose in the following sections will be
presented in Table \ref{tab:mean_values}.

\subsection{The $\alpha$ parameter}
\label{sec:alpha_parameter}

One important aspect of the comparison between PNe and the stellar
surface brightness is the evaluation of the $\alpha$ parameter
\citep{Jacoby80}, which specifies the number of PNe associated with
the amount of light emitted by the stellar population.

For the galaxies in the PN.S database we compute more
specifically $\alpha_{B,1.0}$, the ratio between the number of PNe
down to one magnitude fainter than the PNLF bright cut-off, $m^*$, and
the total stellar luminosity in the $B$-band, $L_B$, expressed in
solar units, as follows:

\begin{itemize}

\item{} We determine the number $N_c^{TOT}$ of PNe brighter than
  $m_{80\%}$ accounting for completeness correction factors $c_R$ at
  different radii from the galaxy centre. $m_{80\%}$ and $c_R$ were
  defined in Section \ref{sec:PNe_vs_sb}.
  
\item {} We integrate the fitted S\'ersic stellar surface brightness
  profile in the $B$-band over the radial range in which the PNe have
  been detected, giving us $L_B(R_{PNe})$. For NGC 4374 and
    NGC 5846 where the $V$-band profiles are available instead, we
    convert the result into the $B$-band using the total
    extinction-corrected colour $(B-V_0)_T$ as reported in the RC3
    catalogue: $(B-V_0)_T = 0.94$ for NGC 4374 and $(B-V_0)_T = 0.96$
    for NGC 5846.

\item {} We compute $\alpha_{B,m^*-m_{80\%}}$ as the ratio $N_c^{TOT}
  / L_B(R_{PNe})$. Uncertainties in the radial extent of the PNe
  detection range, in the S\'ersic fit extrapolation and in the
  completeness corrections give us an error estimate for $\alpha$.

\item {} Finally, we scale the measured $\alpha_{B,m^*-m_{80\%}}$ to
  $\alpha_{B,1.0}$ by integrating over the PNLF. We have

\begin{equation}
   \alpha_{B,1.0} = \alpha_{B,m^*-m_{80\%}}  \cdot \frac{\int_{m^*}^{m^*+1} F(m^*, m')dm'}{\int_{m^*}^{m_{80\%}} F(m^*,m')dm'} 
\end{equation}
where $F(m^*, m')$ is the analytic expression of the PNLF
\citep{Ciardullo+89}, which depends only on the cut-off magnitude
$m^*$ observed in our galaxies.
\end{itemize}

For galaxies not included in our PN.S database we use the
$\alpha_{B,1.0}$ as listed by \citet{Buzzoni+06}.

\begin{figure}
\psfig{file=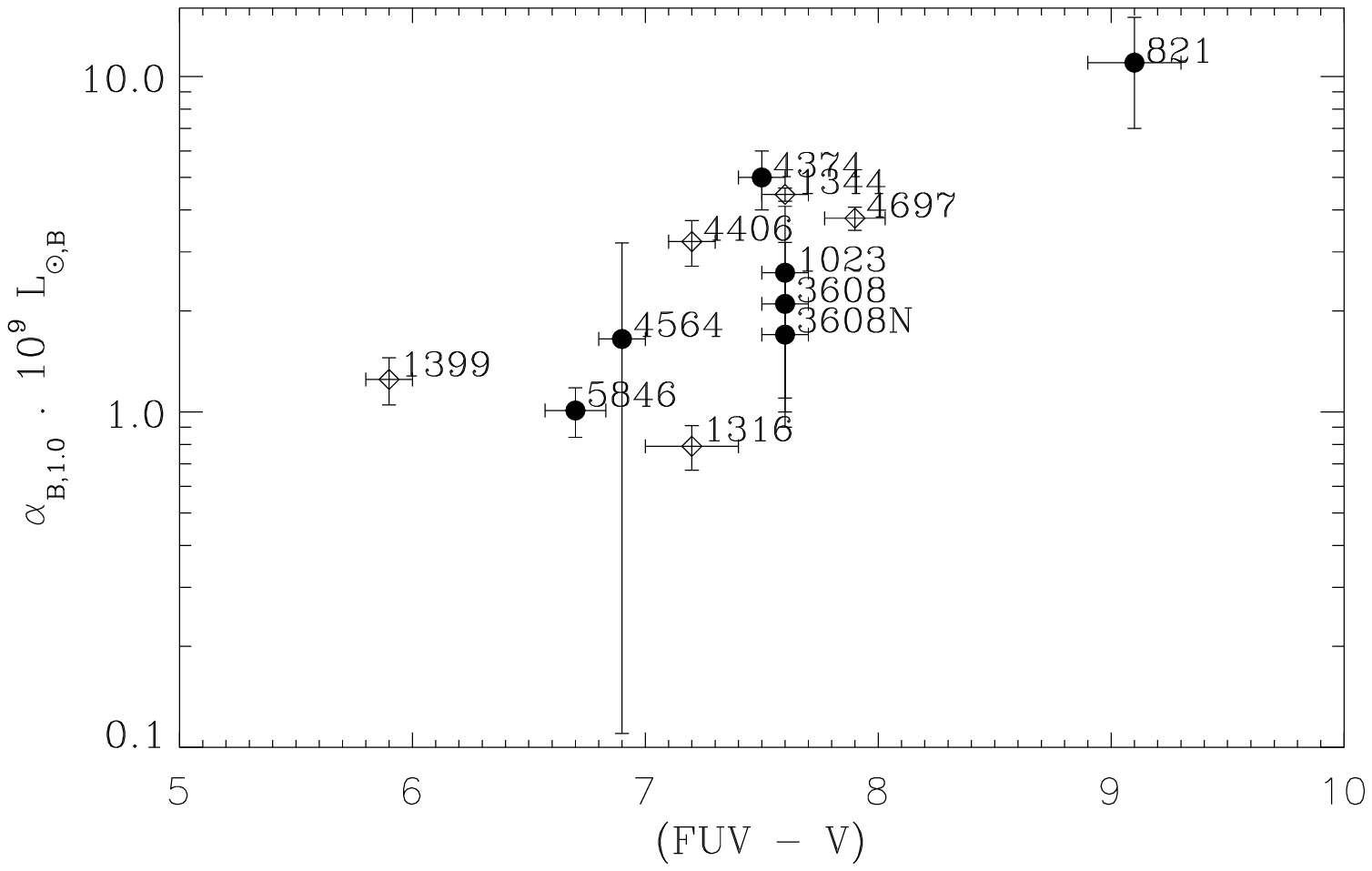,width=8.5cm, clip=}
\caption{Correlation between the $\alpha_{B,1.0}$ parameter and ${FUV-V}$
  colour, measured from total extinction-corrected magnitudes FUV
  (from GALEX) and $V$ (from RC3). {\it Filled circles:} Sample
  galaxies for which $\alpha$ is calculated from our PN.S data,
  according to the prescription given in Section
  \ref{sec:alpha_parameter}. {\it Open diamonds:} Sample galaxies for
  which $\alpha$ values are taken from \citet{Buzzoni+06}.}
\label{fig:uv_alpha}
\end{figure}

It is known that $\alpha$ is related to the UV emission in
galaxies \citep{Buzzoni+06}. We therefore compare the $\alpha_{B,1.0}$
values to the UV excess of the sample galaxies. The latter is
measured using the  total FUV (1344-1786\AA) magnitude from the
GALEX database (where available) and the extinction-corrected
total $V$ magnitude from the RC3 catalogue. GALEX FUV magnitudes are
corrected for extinction using the relation $A_{FUV}=8.376\cdot
E(B-V)$ as in
\citet{Wyder+05}. Values for the colour excess $E(B-V)$ are taken from
the NED database. The $\alpha_{B,1.0}$ parameters and the ${FUV-V}$
colours are listed in Table \ref{tab:mean_values}.
In Figure \ref{fig:uv_alpha} we plot $\alpha_{B,1.0}$ against the
GALEX ${FUV-V}$ colours. The figure shows a clear anti-correlation:
galaxies with a larger UV excess, which are the massive ellipticals,
also have a smaller number of PNe per unit luminosity,
$\alpha_{B,1.0}$.

Using stellar population models, \citet{Buzzoni+06} interpreted this trend
as a consequence of the mean post-AGB (PAGB) core mass being smaller in
massive ellipticals, which is a result of a higher rate of mass loss.
If the mean PAGB core mass falls below $M_{core} \le 0.52\, M_\odot$
one may expect a larger fraction of Horizontal-Branch (HB) stars to
follow the AGB-{\it manqu\'e} channel, i.e., the stars move directly
onto the high-temperature white dwarf cooling sequence after leaving
the HB, thus missing the AGB and PN phases entirely
\citep{Greggio+90}.  A larger fraction of AGB-{\it manqu\'e} stars is
consistent both with a lower value of $\alpha_{B,1.0}$ and a strongly
enhanced galaxy UV emission, as is observed for the more massive
elliptical galaxies in the current PNe sample.

\subsection{The average $V/\sigma$ ratio for stellar and PNe systems}
\label{sec:v_over_sigma}

The average value of the ratio $V/\sigma$ in a galaxy has
traditionally been used to describe the relative importance of
rotation and anisotropy for dynamical equilibrium and the shape. Usually, it has
been calculated from long-slit kinematics using the formula

\begin{equation}
V/\sigma = V_{max}/\sigma_{0}
\label{eqn:vs_classic}
\end{equation}
where $V_{max}$ is the maximum value of the observed radial velocity
and $\sigma_{0}$ is the average value of the velocity dispersion within
0.5 $R_e$ (e.g., \citealt{Bender+94}).

\citet{Binney05} revisited the calculation of the average $V/\sigma$,
adapting it for two-dimensional data. The new formulation is in terms
of the sky-averaged values of $V^2$ and $\sigma^2$, weighted by the
surface density. For SAURON integral-field data, this was approximated
by \citet{Cappellari+07}:

\begin{equation}
 (V/\sigma)_{stars}  = \sqrt{\frac{\langle V^2\rangle}{\langle \sigma^2\rangle}} =   \sqrt{ \frac{\sum_i F_i V_i^2}{\sum_i F_i \sigma_i^2}}
\label{eqn:vs}
\end{equation}
where the sum extends over the data points in the two-dimensional
field, on isophotes up to 1 $R_e$, depending on the radial extent of
the data. $V_i$ and $\sigma_i$ are the velocity and velocity
dispersion of the data point and $F_i$ is the flux associated with it.

``Classical'' measurements $V_{max}/\sigma_{0}$ (Equation
\ref{eqn:vs_classic}) can be rescaled to the new ``two-dimensional''
values by multiplying by 0.57 \citep{Cappellari+07}.

\begin{itemize}

\item{}{\bf {\it Stellar $V/\sigma$}}
\label{sec:stellarvs}

Stellar $(V/\sigma)_{stars}$ are taken from the SAURON survey (from
\citealt{Cappellari+07}), where available. For the remaining sample
galaxies in Table \ref{tab:sample}, we compute $V_{max}/\sigma_{0}$
from long-slit data using Equation \ref{eqn:vs_classic}, and rescale
to the two-dimensional values by multiplying with 0.57. Errors are
taken to be 0.03 for SAURON values, or are computed using error
propagation (on Equation \ref{eqn:vs_classic}) and are also
rescaled by 0.57.

\smallskip 

\item{}{\bf {\it PNe $V/\sigma$}}

We compute $V/\sigma$ for the PNe system using the smoothed
two-dimensional fields (calculated in Section
\ref{sec:smoothed_fields}), summing over the positions of the detected PNe:

\begin{landscape}
\begin{table}
\centering
\caption{Measured parameters for the sample galaxies}
\begin{tabular}{l c c c c c c c c c c c c c c c}
\hline \hline \noalign{\smallskip}
Name   &\slast      &\snorm &$\sigma_{5.0}$& $\frac{\sigma_{\rm MIN}}{\sigma_{1.0}}$ &$V_{rms,LAST}$&$V_{rms,1.0}$&$V_{rms,5.0}$&$\frac{V_{rms,MIN}}{V_{rms,1.0}}$&$V/\sigma_{PNe}$&$V/\sigma_{STARS}$&$m^*$, $m_{80\%}$ &$\alpha_{B,1.0}$  & (FUV-V)             &   $M_*/L_B$ & $M_*$\\
\noalign{\smallskip}   
 (NGC) &   (\kms)   &(\kms)&(\kms)&     &  (\kms) &(\kms)&(\kms)&   &   &(mag, mag) &($10^{-9} L_{B,\odot}$) &   (mag)         &                & & ($10^{10}$ M$_{\odot}$)\\
 (1)   &    (2)     & (3)  & (4) & (5) &(6)       &(7)&(8)&(9)&(10)& (11)     &    (12)              &  (13)           &    (14)         & (15)    &    (16)        \\
\noalign{\smallskip}
\hline
\noalign{\smallskip} 
821  &$  51\pm18$ &  167 & 76 & 0.31 &$  56\pm18$  &181& 66 &0.31&$0.31\pm0.02$&$0.26\pm0.03$ &27.4, 28.6& $11.  \pm  4.  $      &  9.1$\pm$  0.2  & $7.1^{+1.9}_{-2.0} $  & $14.7\pm4.0$\\
 1023  &$ 119\pm29$ &  110 & 79 & 0.54 &$ 123\pm29$&209& 147&0.59&$1.18\pm0.03$&$0.34\pm0.03$ &25.8, 27.3& $2.6 \pm  1.5 $       &  7.6$\pm$  0.1  &      --        &  --                 \\
 1316  &$ 152\pm31$ &  152 & 152& 1.00 &$ 179\pm31$&184& 179&0.97&$0.42\pm0.13$&$0.50\pm0.12$ &-, -      & $0.79 \pm 0.12 ^{(a)}$&  7.2$\pm$  0.2  & $3.9^{+0.8}_{-1.1} $  & $54.4\pm13.2$\\
 1344  &$ 109\pm23$ &  148 & 109& 0.73 &$ 133\pm23$&173& 142&0.77&$0.20\pm0.02$&$0.33\pm0.12$ &-, -      & $ 4.45\pm0.2 ^{(a)}  $&  7.6$\pm$  0.1  &     --         &  --   \\
 1399  &$ 198\pm47$ &  240 & 198& 0.83 &$ 213\pm47$&243& 213&0.88&$0.64\pm0.23$&$0.06\pm0.03$ &-, -      & $ 1.25\pm0.2 ^{(a)}  $&  5.9$\pm$  0.1  &$10.4^{+3.3}_{-2.7}$  & $48.1\pm13.7$\\
 3377  &$  77\pm21$ &  67  & 77 & 0.75 &$  64\pm21$&110& 64 &0.58&$0.37\pm0.03$&$0.49\pm0.03$ &25.2, 27.0& $ 6.  \pm  5.  $      &    --           &      --        &  --   \\
 3379  &$  46\pm21$ &  153 & 69 & 0.30 &$  46\pm21$&162& 77 &0.29&$0.18\pm0.02$&$0.14\pm0.03$ &25.5, 26.5& $ 24 \pm 10        $  &    --           &$ 8.4^{+2.8}_{-2.4} $  & $13.9\pm4.3$\\
 3384  &$  72\pm20$ &  63  & 72 & 0.93 &$ 146\pm20$&162& 157&0.80&$1.20\pm0.20$&$0.44\pm0.03$ &-, -      & $ 9.5\pm2.5 ^{(b)}  $ &    --            &$ 3.2^{+0.6}_{-0.6} $  & $3.7\pm0.7$\\
 3608  &$ 169\pm43$ &  130 & 169& 0.99 &$ 191\pm43$&141& 191&0.98&$0.45\pm0.03$&$0.05\pm0.03$ &27.1, 28.0& $ 2.1 \pm  1.1 $      &  7.6$\pm$  0.1  &      --        &  --   \\
 3608N &$ 107\pm33$ &  102 & 107& 0.55 &$ 134\pm33$&134& 134&1.00&$0.65\pm0.06$&$0.05\pm0.03$ &27.2, 28.0& $ 1.7 \pm  0.8 $      &  7.6$\pm$  0.1  &      --        &  --   \\
 4374  &$ 207\pm33$ &  215 & 207& 0.96 &$ 208\pm33$&216& 208&0.96&$0.16\pm0.01$&$0.03\pm0.03$ &26.9, 28.0& $ 5.0 \pm  1.0 $      &  7.5$\pm$  0.1  &      --        &  --   \\
 4406  &$  96\pm26$ &  147 & 96 & 0.65 &$ 219\pm26$&219& 219&1.00&$0.80\pm0.17$&$0.09\pm0.04$ &-, -      & $ 3.2\pm0.5 ^{(c)}  $ &  7.2$\pm$  0.1  & $8.1^{+2.4}_{-2.2} $  & $52.4\pm14.9$\\
 4494  &$  46\pm15$ &  106 & 63 & 0.44 &$  44\pm15$&122& 66 &0.36&$0.18\pm0.02$&$0.25\pm0.02$ &26.0, 27.0& $10.  \pm  3.  $      &    --           &      --        &  --   \\
 4564  &$  56\pm15$ &  70  & 58 & 0.79 &$  58\pm15$&163& 93 &0.36&$1.02\pm0.08$&$0.58\pm0.03$ &26.3, 27.8& $ 1.6 \pm  1.5 $      &  6.9$\pm$  0.1  &      --        &  --   \\
 4697  &$  92\pm20$ &  154 & 92 & 0.60 &$ 106\pm20$&180& 106&0.59&$0.32\pm0.02$&$0.41\pm0.14$ &-, -      & $ 3.8\pm0.3 ^{(a)}  $ &  7.9$\pm$  0.1   & $6.7^{+2.3}_{-1.7}$  & $15.0\pm4.5$\\
 5128  &$  69\pm69$ &  132 & 89 & 0.52 &$ 118\pm25$&147& 133&0.80&$0.48\pm0.02$&$0.13\pm0.11$ &-, -      & $12.5\pm0.6 ^{(a)}  $ &    --           &      --        &   --  \\
 5846  &$ 170\pm30$ &  210 & 170& 0.81 &$ 174\pm30$&219& 174&0.79&$0.12\pm0.01$&$0.03\pm0.03$ &27.3, 28.3& $ 1.01\pm  0.17$      &  6.7$\pm$  0.1  & $11.1^{+3.7}_{-3.2}  $  & $52\pm16$\\
 \noalign{\smallskip} 
\hline
\end{tabular}
\label{tab:mean_values}
\begin{minipage}{23.5cm}
  Notes -- 
   Col.1: Galaxy name. NGC 3608N refer to the results obtained using half of the PNe sample, 
          as explained in Section \ref{sec:individual_3608}.\\
   Col.2: Outermost value of the velocity dispersion as given by the fit. \\
   Col.3: Velocity dispersion at 1$R_e$, as given by the fit.\\
   Col.4: Velocity dispersion at 5$R_e$, as given by the fit.\\
   Col.5: Minimum value of the velocity dispersion as given by the fit, scaled by \snorm\ (from Col.3).\\
   Col.6: Outermost value of $V_{rms}$ as given by the fit. \\
   Col.7: $V_{rms}$  at 1$R_e$, as given by the fit.\\\
   Col.8: $V_{rms}$  at 5$R_e$, as given by the fit.\\
   Col.9: Minimum value of $V_{rms}$ as given by the fit, scaled by \snorm\ (from Col.7).\\
   Col.10: Average value of $V/\sigma$ for the PNe component. For the data from two-dimensional field, Equation 
          \ref{eqn:vs} was used. For the data from major axis kinematics Equation \ref{eqn:vs_classic} was used
          and values have been multiplied by 0.57 to rescale them to the two-dimensional case, as done by \citet{Cappellari+07}.\\
   Col.11: Average value of $V/\sigma$ for the stellar component, calculated with Equation \ref{eqn:vs_classic} on long-slit 
          data and corrected by the factor 0.57, or using the value reported by \citet{Cappellari+07} if available.\\
   Col.12: Bright cut-off magnitude and $m_{80\%}$ computed for the galaxies in Figure \ref{fig:photom_comparison}.\\
   Col.13: The $\alpha_{B,1.0}$ parameter computed using the PN.S data. Values marked by $^{(a)}$ are from \citet{Buzzoni+06}, 
          $^{(b)}$ from \citet{Ciardullo+89b} and $^c$ from \citet{Jacoby+90}. Total $V$ band photometry for NGC 4374 and 
          NGC 5846 has been converted into $B$ band using $(B-V)=0.94$ for NGC 4374 and $(B-V)=0.96$ for NGC 5846 
          (values taken from RC3).\\
   Col.14: Difference between the extinction-corrected magnitudes in the far UV (from GALEX) and $V$-band (from RC3). 
          The symbol $-$ means that the galaxy is not present in the GALEX database. \\
  Col.15: Mass-to-light ratio from single stellar population models listed in \citet{Napolitano+05}. The 
          symbol -- means that the measurement is not available.  \\
  Col.16: Total stellar mass computed from the mass-to-light ratio (Col.15), $B_T$ and distance from Table \ref{tab:sample}.      
\end{minipage}
\end{table}
\end{landscape}

\begin{equation}
 (V/\sigma)_{PNe} = \sqrt{\frac{\langle V^2\rangle}{\langle \sigma^2\rangle}}
  = {\sqrt{ \frac{\sum_P \frac{1}{c_R}\tilde{V}(x_P,y_P)^2}{\sum_P
        \frac{1}{c_R}\tilde{\sigma}(x_P,y_P)^2}}}
\label{eqn:vs_pne}
\end{equation}
This implicitly incorporates the weighting by the local stellar
surface density, which was shown earlier to be proportional to the PNe
number density if the completeness correction factor $c_R$
(interpolated for $x_p, y_p$) is taken into account: Equation
\ref{eqn:vs_pne} weights every region according to the
completeness-corrected number of PNe in the region (see Section
\ref{sec:PNe_vs_sb}).
Note that Equation \ref{eqn:vs_pne} is an average over the whole
PNe field, and therefore probes a much larger region of the galaxy
than the value for the stars, which was confined to within 1 $R_e$.

For a few galaxies the number of PNe detections is too small for a
reliable determination of the two-dimensional velocity fields. For
these galaxies we use the PNe mean velocities and velocity dispersions
in bins along the major axis as given in the respective paper from
which the data were taken. We then take the maximum of these mean
velocities and the average of all the velocity dispersion values to
obtain a major-axis $V/\sigma$ value analogous to the classical
formula given in Equation \ref{eqn:vs_classic}, but not confined to
within 0.5 $R_e$. These values are then rescaled to the
two-dimensional case by multiplying by 0.57.

\end{itemize}

\begin{figure}
\psfig{file=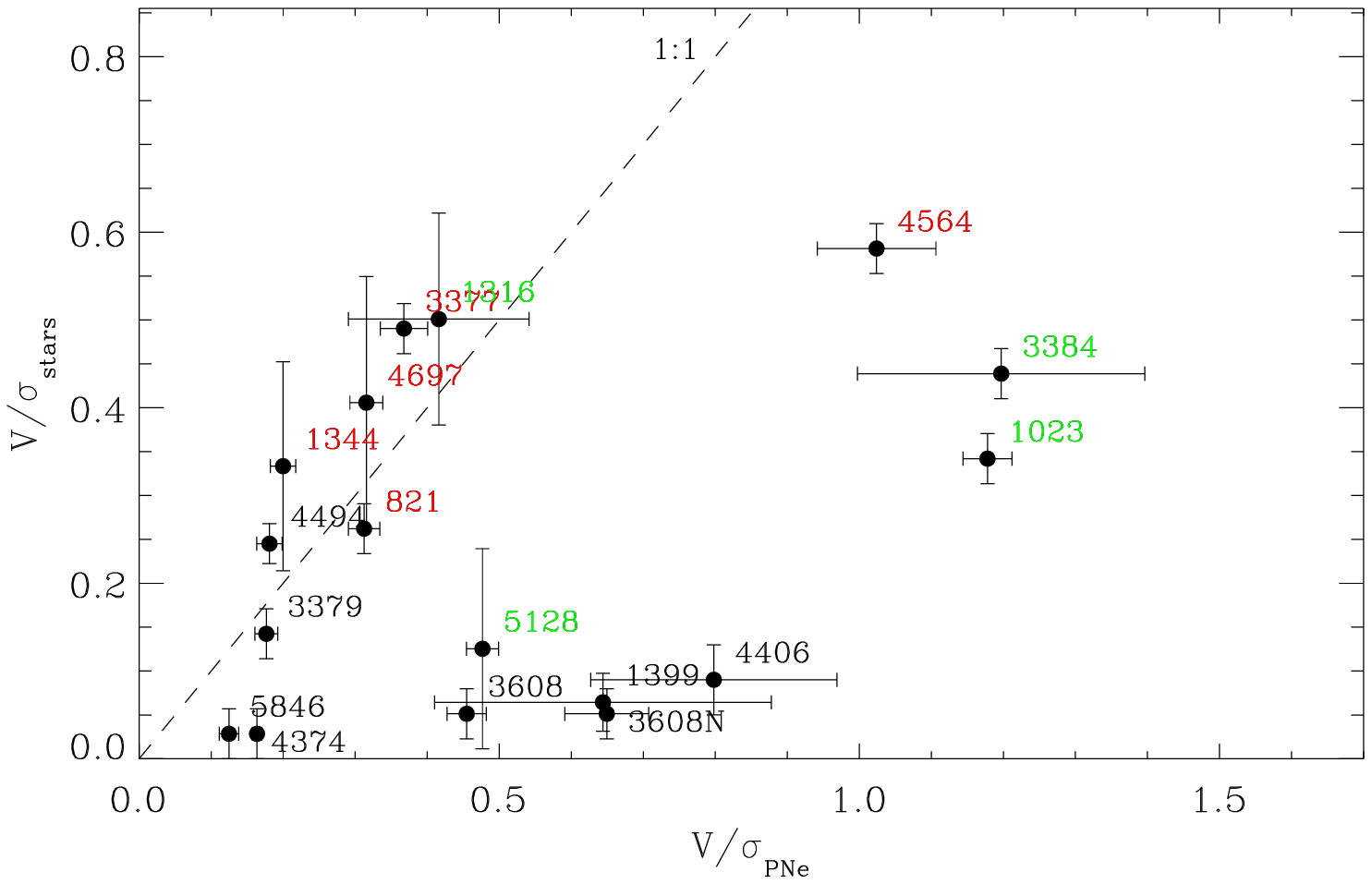,width=8.5cm}
\caption{Comparison between the two-dimensional averaged values of the $V/\sigma$
  ratio for the stellar and PNe systems. The 1:1 relation is shown by
  a {\it dashed line}. {\it Black labels} refer to round ellipticals
  ($\leq$ E3), {\it red labels} refer to flattened ($\geq$ E4) ellipticals,
  {\it green labels} refer to S0 galaxies. }
\label{fig:vsigma_comparison}
\end{figure}

In Figure \ref{fig:vsigma_comparison} we compare these stellar and PNe
$V/\sigma$, providing information on how the dynamics of a galaxy
change from the inner to the outer region. From this figure we can see
that many galaxies (nearly 50\%) have $(V/\sigma)_{PNe} > V/\sigma$,
while for the rest, the two values are similar. Figure
\ref{fig:vsigma_comparison} also shows the differences between the
distributions of flattened ($\geq $E4) ellipticals (red), round ($\leq
$E3) ellipticals (black) and S0s (green). This shows that:

(i) Three of the galaxies (NGC 1023, NGC 3384 and NGC 4564) have large
inner $V/\sigma$ and even larger $(V/\sigma)_{PNe}$ in the halo. The
first two are S0 galaxies while the third exhibits S0-like properties
(see Section \ref{sec:n4564}). In these systems the $(V/\sigma)$ rises
from the centre to the disk-dominated region, like in the disky
elliptical galaxies studied by \citet{Rix+99}.

(ii) For flattened (NGC 821, NGC 1344, NGC 3377 and NGC 4697) and
rotating, round galaxies (NGC 3379 and NGC 4494) in the sample, the
inner and outer values for $V/\sigma$ are equal to within the errors.

(iii) Galaxies with a small $V/\sigma$ in the inner parts may have
either small (NGC 4374 and NGC 5846) or larger (best exemplified by
NGC 5128 and with larger uncertainities, NGC 1399 and NGC 3608)
$V/\sigma$ in the outer parts.

\subsection{The $\lambda_R$ radial profile for stellar and PNe systems}
\label{sec:lambda_profile}

In \citet{Emsellem+07}, a kinematic classification scheme for galaxies
is proposed based on the $\lambda_R$ profile, which measures the
importance of rotation as a function of radius and is related to
the angular momentum per unit of mass within $R$:
\begin{equation}
\label{eqn:lambda_r}
\lambda_R=\frac{\sum R_i F_i \left |V_i \right|} {\sum R_i F_i \sqrt{V_i^2 + \sigma_i^2}}
\end{equation}
where the sum includes data points within $R$.
Based on the $\lambda_R$ profile within $\sim 1R_e$, galaxies can be
divided into two main groups \citep{Emsellem+07}:
 
(i) Galaxies with $\lambda_R > 0.1$ are defined as {\it fast
  rotators}. These have a small misalignment between the photometric
and kinematic major axes, and have rising $\lambda_R$ profiles.

(ii) Galaxies with $\lambda_R < 0.1$ are defined as {\it slow
  rotators}.  They exhibit a range of misalignments between the
photometric and kinematic major axes and have flat or decreasing
$\lambda_R$ profiles.

\begin{figure}
\vbox{
  \psfig{file=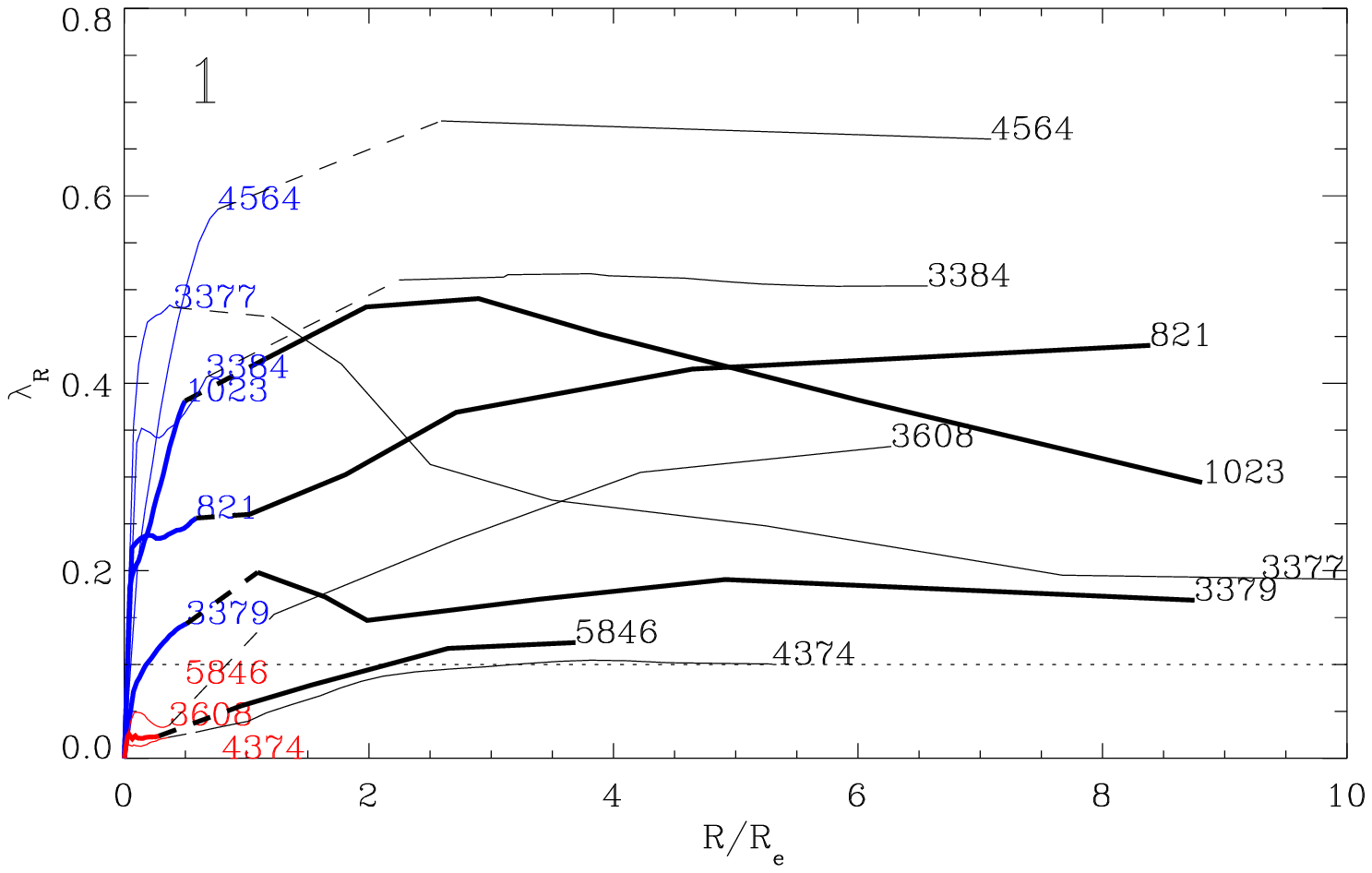,width=8.5cm}
  \psfig{file=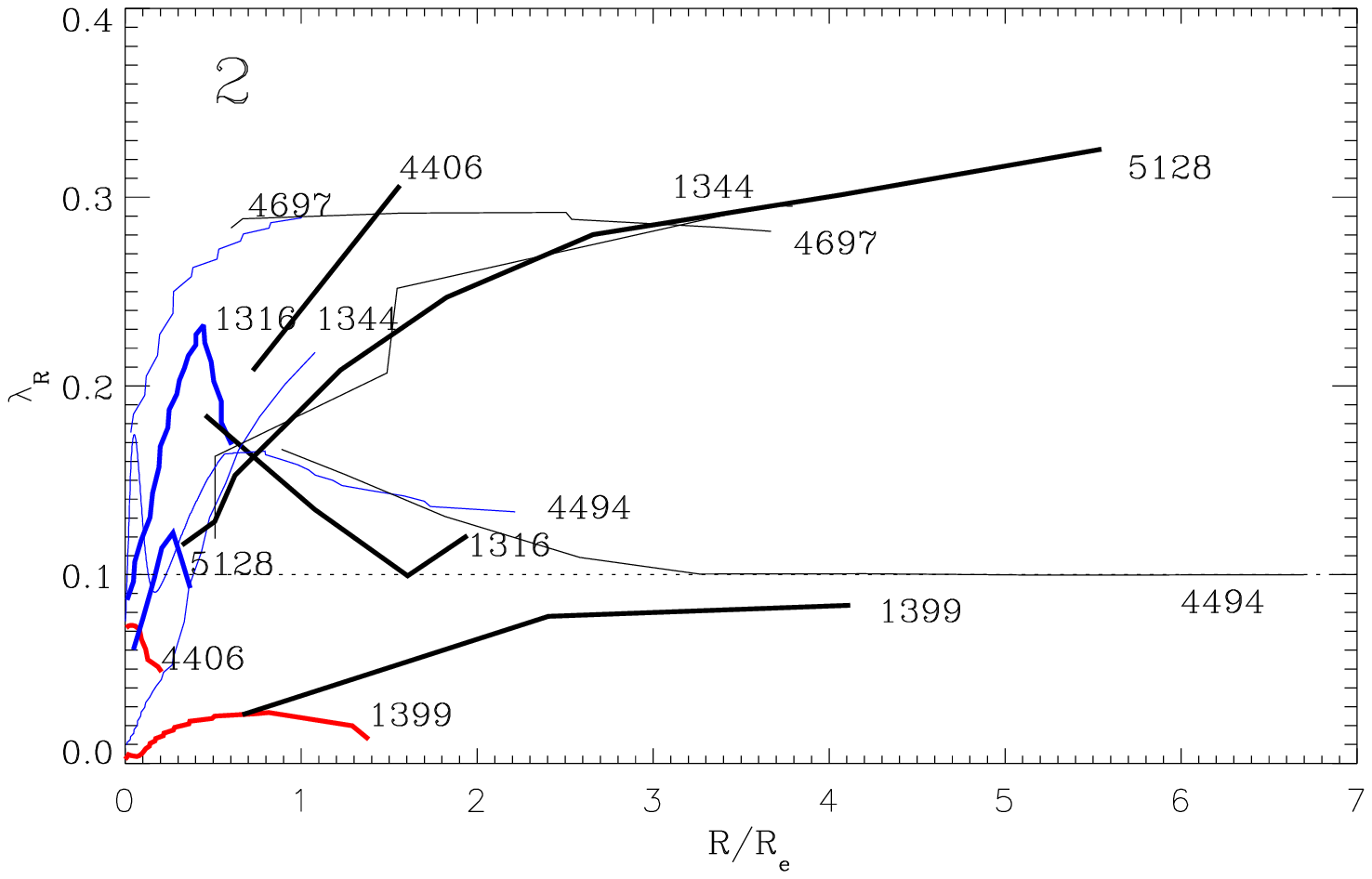,width=8.5cm}
}
\caption{{\it Panel 1:} Radial $\lambda_R$ profiles for galaxies
  common to our sample and the SAURON sample. {\it Panel 2:} Radial
  $\lambda_R$ profiles for galaxies in our sample, which are not in
  the SAURON sample. {\it Red {\rm and} blue solid lines:}
  $\lambda_R$ profiles computed from the stellar kinematics for slow
  and fast rotators respectively (those in {\it panel 1} were kindly
  provided by Eric Emsellem). {\it Black solid lines:} $\lambda_R$
  profiles extracted from the PNe kinematics. {\it Dashed black lines}
  in {\it panel 1} connect the last SAURON data point with the first
  PNe data point, to guide the eye on the plot. These lines were not
  plotted in {\it panel 2} to avoid overcrowding. The {\it dotted line}
  ($\lambda_R = 0.1$) separates fast and slow rotator regions. In
  both panels some radial $\lambda_R$ profiles are shown with thicker
  lines for clarity.}
\label{fig:lambdar}
\end{figure}

To extend their analysis to the outer haloes of elliptical galaxies
(beyond several $R_e$), we calculate the $\lambda_R$ profiles using
the PNe kinematics for the sample galaxies. For galaxies in
which the two dimensional field is available, we use Equation
\ref{eqn:lambda_r}, to sum over the velocity and velocity dispersion
field values at the positions of the PNe, using the factor $1/c_R$ as the
weight instead of the flux values $F_i$. This incorporates the
weighting by the stellar surface density as discussed in the case of
Equation \ref{eqn:vs_pne}. For galaxies in which the kinematics are
available only along the major axis, we use Equation
\ref{eqn:lambda_r} where the sum is meant to be extended only along
the major axis, $F_i$ are extracted from the extrapolation of the
stellar surface brightness and the 0.57 correction is applied.
In Figure \ref{fig:lambdar} we show the profiles derived using the
stellar kinematics (which we used to separate between fast and slow
rotators, to be consistent with previous works) and the PNe
kinematics. To simplify matters, Figure \ref{fig:lambdar} is divided
into two panels separating galaxies for which the stellar $\lambda_R$
is available from the SAURON data, from the remaining galaxies in our
sample. Several results emerge from this figure, reflecting the
changes in both the rotation and velocity dispersion profiles:

(i) In some galaxies, there is a marked change in the behaviour of
$\lambda_R$ at radii larger than 1 or 2 effective radii compared to
that in the central regions. In the case of NGC 1023, NGC 1316,
NGC 3377 and NGC 4494, after the initial increase of $\lambda_R$ (as
observed in the majority of fast rotators within $1R_e$) the profiles
drop, and they reach the ``slow rotator'' region (NGC 1316 and NGC
4494) or arrive close to it (NGC 1023, NGC 3377). According to
\citet{Krajnovic+08}, the majority of fast rotators have structures
with disk-like kinematics, which are responsible for the high values
of $\lambda_R$.  In these four fast rotators, the observed outer
decrease in $\lambda_R$ may imply that the light associated with the
disk structure fades towards larger radii.

(ii) In the galaxies of the sample, which are slow rotators according
to the SAURON classification, the $\lambda_R$ profiles grow slowly to
values of 0.1--0.3, thus entering the ``fast rotator'' regime
(NGC 3608, NGC 5846 and partially for NGC 4374). A larger increase is
visible in NGC 4406, but the small number of PNe for this galaxy makes
its $\lambda_R$ uncertain. The outward rise of $\lambda_R$ for
NGC 1399 is slower; this galaxy remains below the ``slow rotator'' 
demarcation line.

(iii) The remaining galaxies in the sample are ``fast rotators''
throughout the radial range probed. The $\lambda_R$ profile either
rises or stays almost flat throughout.

The large radial extent of the PNe data enables us to follow the
rotation properties of elliptical galaxies out into the halo,
providing new constraints on the processes involved in the formation
of these galaxies (see Section \ref{sec:comp_models}).

Although the kinematics of the stars within $R_e$ as measured by
SAURON (and most traditional long-slit studies) reflect half of the
total stellar luminosity and mass, the same is not true for the total
stellar angular momentum.  As discussed in Appendix~\ref{app:angmom},
these stars may represent only $\sim10$\% of the galaxy's angular
momentum, and an adequate global rotational picture of the galaxy
requires observations to $\sim 5 R_e$.

\subsection{The shape of the velocity dispersion and $V_{rms}$ radial profiles}
\label{sec:sigma_profiles}

In order to obtain a general overview of the shapes of the velocity
dispersion profiles $\sigma \left( R \right)$ for our sample galaxies,
we first parameterised the major axis profiles with suitable
functions. There is no universally valid function for all the observed
profiles, so we experimented with various {\it ad hoc} functions and
selected the one that best matched the observations for each
galaxy.  The fitted velocity dispersion profiles were then scaled by
the effective radius (values from Table \ref{tab:sample}) and
normalised to the value of the velocity dispersion at \fre. We did not
use the central velocity dispersion value for the normalisation (i.e.,
$\sigma\left( R=0 \right)$) because it might be artificially boosted
by the presence of a central supermassive black hole or an unresolved
central velocity gradient. We performed a similar paramerisation for
the rotation profiles to compute the $V_{rms}$ velocity profiles,
where $V_{rms}^2=\sigma^2 + V^2$.

\begin{figure}
\psfig{file=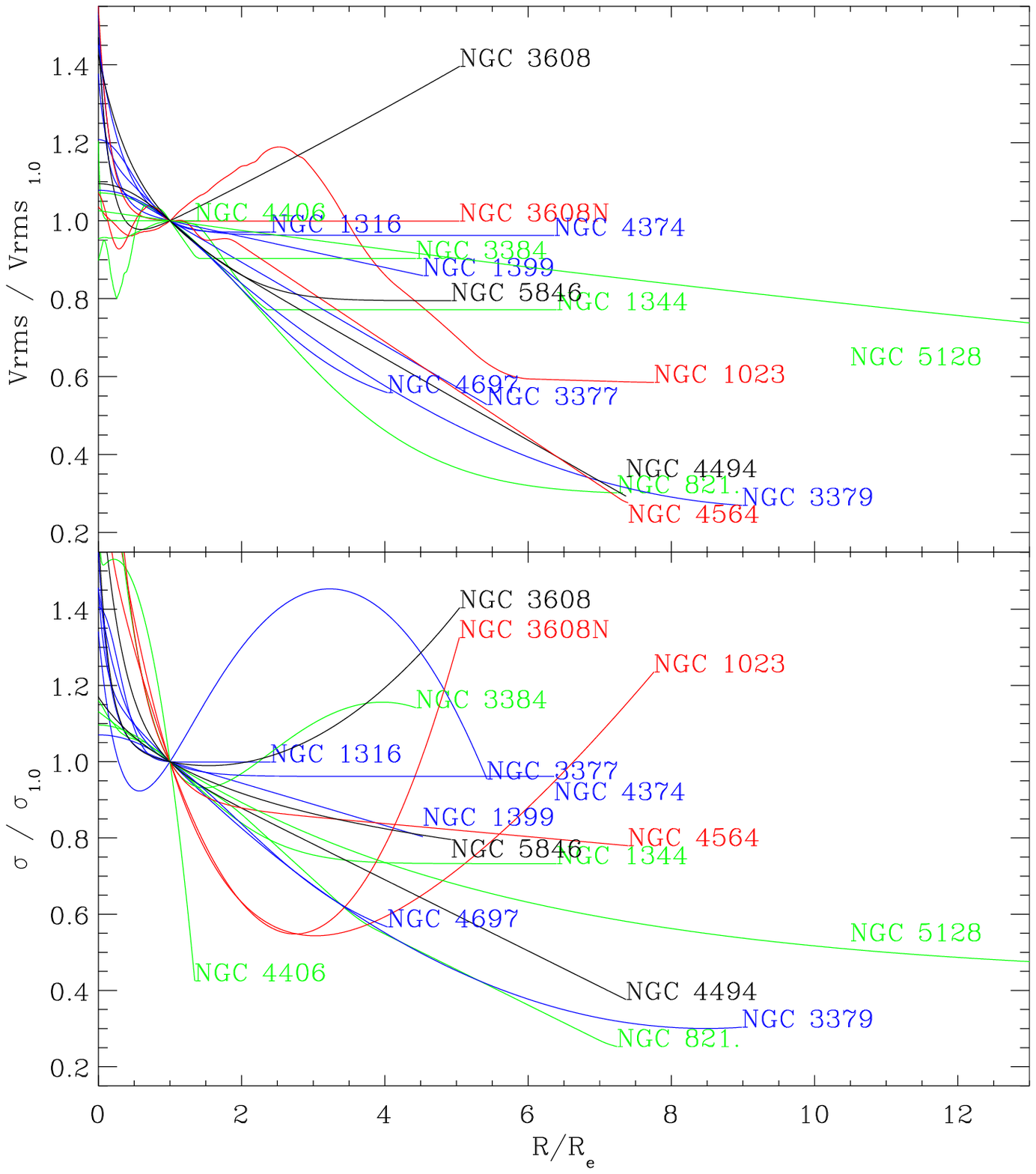,width=8.5cm}
\caption{Comparison between radial profiles of $V_{rms}$ ({\it top
    panel}) and velocity dispersion ({\it bottom panel}) of the sample
  galaxies, obtained by combining the stellar and PNe kinematics along
  the major axis. Profiles have been scaled to the effective radius
  and normalised to their value at \fre.  Colours are chosen in
    order to highlight the contrast between lines and thus better
    distinguish different profiles.  }
\label{fig:slope}
\end{figure}

According to the normalized $V_{rms}$ profiles shown in Figure
\ref{fig:slope}, galaxies fall into two main groups. The larger part
of the sample shows a slightly decreasing profile from the centre
outwards. The second group of galaxies (NGC 821, NGC 3377, NGC 3379,
NGC 4564, NGC 4494, NGC 4697) show strongly decreasing $V_{rms}$
profiles. NGC 1023 appears to be anomalous in that its $V_{rms}$
increases initially before falling steeply. This reflects the strong
rotation in the disc of the galaxy combined with a steep central drop
in the velocity dispersion profile (Figure \ref{fig:radial_kinematic},
see also \citealt{Noordermeer+08}). For NGC 3608, the apparent
increase in the $V_{rms}$ profile is doubtful because of the possible
contamination of PNe from NGC 3607 and the large errors associated
with the velocity and velocity dispersion measurements (Figure
\ref{fig:radial_kinematic}); the radial profile of NGC 3608N (see
Section \ref{sec:individual_3608}) does not show anomalies, and
belongs to the first group.

A greater variety of trends is visible in the velocity dispersion
profiles, reflecting a larger variance in the individual contributions
of ordered and random motions to the $V_ {rms}$, e.g., due to the
strong disc component in galaxies such as NGC 3377. There are galaxies
in which the velocity dispersion remains high (i.e., the drop in
velocity dispersion is less than 50\% of the normalisation value);
those, which exhibit a big drop (i.e., the last value of dispersion is
less than 50\% of the central value); and those that show an increase
towards large radii, as in the case of NGC 1023, NGC 3377 and
NGC 3608. In NGC 3608, the observed increase is again doubtful. In the
case of NGC 4406, the massive drop is based on the results derived
from a small number of PNe velocities, which show an associated
increase in rotation velocity.

\subsection{Correlations between kinematic, photometric and morphological properties of the sample galaxies}
\label{sec:gen_kin_properties}

In this section we study the relations between some physical
properties of early-type galaxies and their kinematic properties at
large radii.

\begin{figure*}
\psfig{file=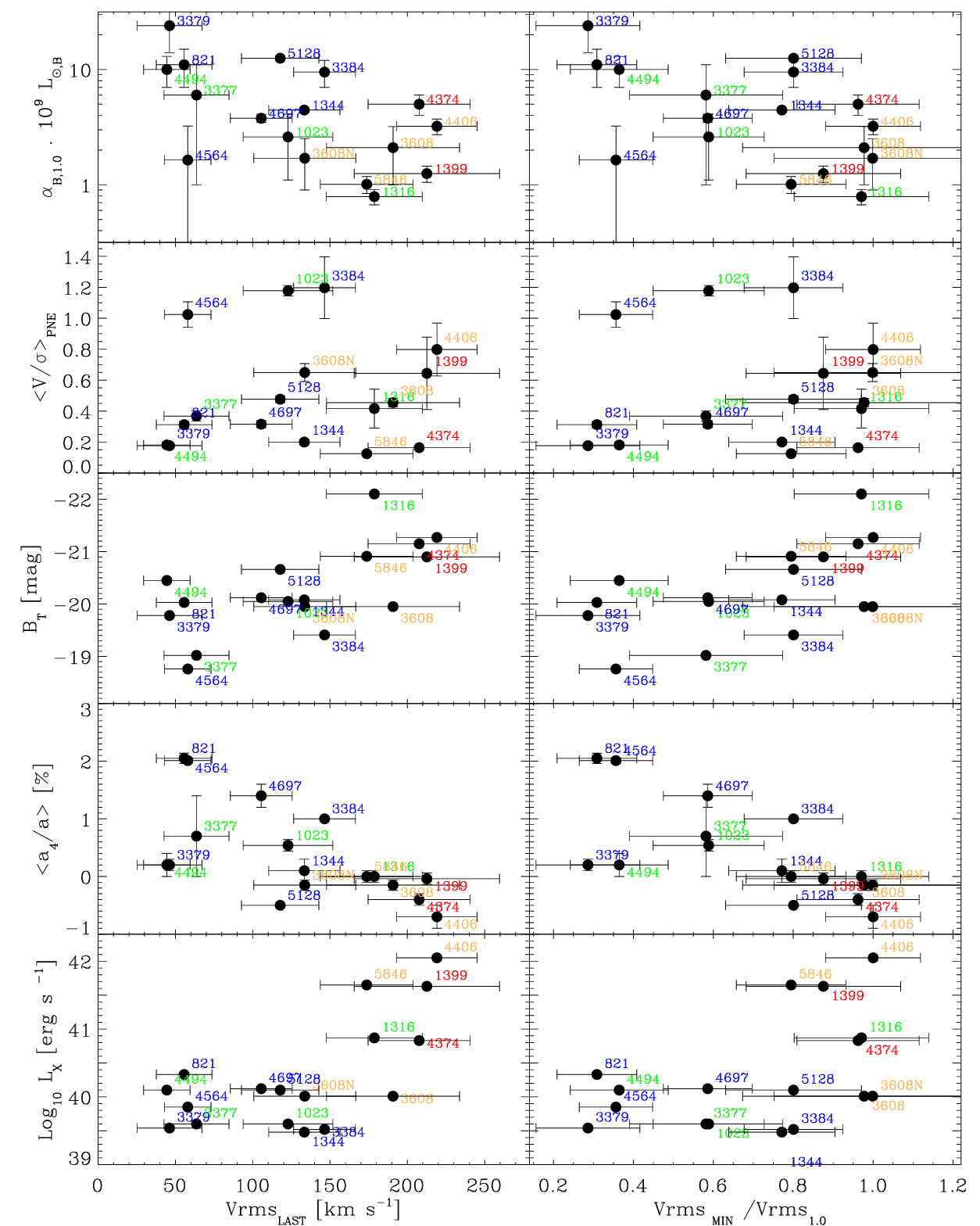,width=17.5cm}
\caption{Halo kinematics versus other physical parameters. {\it
      Left panels:} $V_{rms,LAST}$ values for the sample galaxies
      plotted versus total X-ray luminosity, $a_4$ shape coefficient,
      total extinction-corrected $B$ magnitude, mean
      $(V/\sigma)_{PNe}$ and $\alpha_{B,1.0}$. {\it Blue:} fast
      rotators, {\it green:} fast rotators with a declining
      $\lambda_R$, {\it red:} slow rotators, {\it orange:} slow
      rotators with $\lambda_R>0.1$ in the outer parts. For galaxies
      not listed in Table \ref{tab:photometry}, the $a_4$ is taken
      from \citet{Bender+89} for NGC 4406, from \citet{Goudfrooij+94}
      for NGC 1399, and from \citet{Napolitano+05} and references
      therein for the others.  {\it Right panels:} same as left
      panels, but for the normalised minimum of $V_{rms}$.}
\label{fig:comparison1_vrms}
\end{figure*}

In the the left panels of Figure \ref{fig:comparison1_vrms} we show
how the outermost values of $V_{rms}$ ($V_{rms,LAST}$, measured from
the ad-hoc fitted profiles of Figure \ref{fig:slope}) are related to
the galaxy's X-ray luminosity, total $B$-band luminosity, mean
isophotal shape parameter $<a_4>$, mean $(V/\sigma)_{PNe}$ measured
from the PNe data, and $\alpha_{B,1.0}$ parameter.
In the right panels of Figures \ref{fig:comparison1_vrms} we show similar
plots, using the minimum value of $V_{rms}$ normalized by the value at
\fre, $V_{rms,MIN}/V_{rms,1.0}$ in place of $V_{rms,LAST}$.

\begin{figure*}
\psfig{file=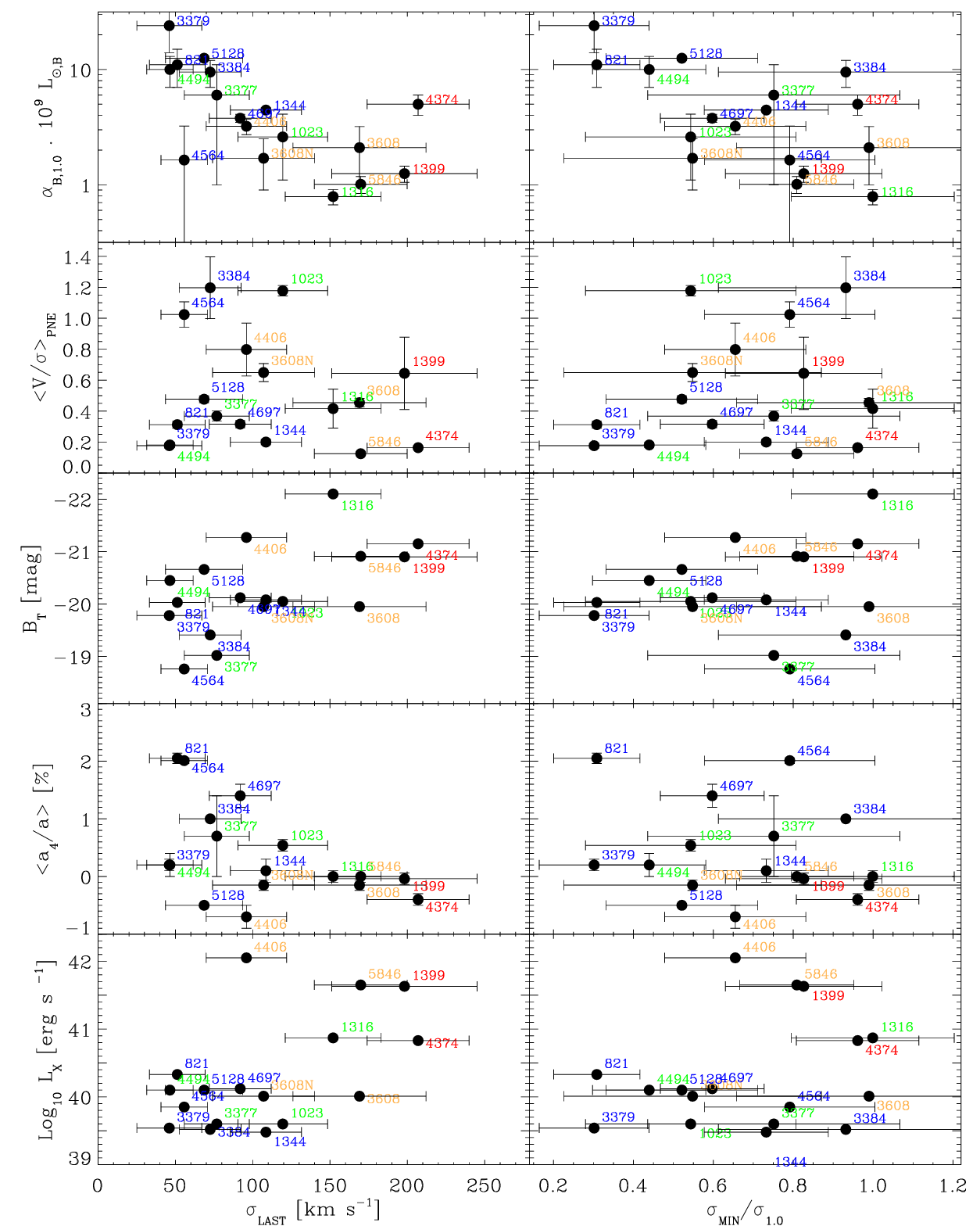,width=17.5cm}
\caption{Same as Figure \ref{fig:comparison1_vrms}, but for \slast
  ({\it left panels}) and \smin\ ({\it right panels}).}
\label{fig:comparison1}
\end{figure*}

Similarly, Figure \ref{fig:comparison1} shows corresponding plots
with the last values of the velocity dispersion (\slast, measured from
the ad-hoc fitted profiles of Figure \ref{fig:slope}) and \smin, the
minimum value of velocity dispersion normalized by the velocity
dispersion at \fre.

The outermost values of $V_{rms}$ and velocity dispersion
characterize their typical values in the galaxy halo and therefore
also the galaxy mass, while the ratios between their minima and their
values at \fre\ give a measure of how much these profiles fall towards
the outer radii.

We note that using the values of $V_{rms}$ and velocity dispersion at
a fixed radius of $5 R_e$ in these figures, instead of their outermost
values, gives very similar results. Also, plotting the physical
parameters against the logarithmic gradient of $V_{rms}$ or velocity
dispersion between $1R_e$ and $5R_e$ results in similar trends as
shown in the right parts of Figures \ref{fig:comparison1_vrms} and
\ref{fig:comparison1}.  For reference, the $V_{rms}(5R_e)$ and
$\sigma(5R_e)$ interpolated from the parametric fits are given in
Table~\ref{tab:mean_values}.

From Figures \ref{fig:comparison1_vrms} and \ref{fig:comparison1} we
notice that more luminous galaxies tend to have larger values of
$V_{rms,LAST}$ and \slast.  This is not surprising given that
luminosity, $V_{rms}$ and velocity dispersion are known to be related
to the galaxy mass.  What is new here is that we are exploring a
larger radial range, probing the relation for the galaxy halo.  
We notice also that galaxies with higher $V_{rms,LAST}$ and
\slast\ tend to have only boxy profiles ($a_4 < 0$), while galaxies
with low values of $V_{rms,LAST}$ and \slast\ have a wider range of
shapes ($-1 < a_4 < 2$).  This is also a reflection of the known trend
for massive ellipticals to be more boxy in shape (e.g.,
\citealt{Bender+89}).
 \citet{Napolitano+05} have also found the $a_4$ and other galaxy
  parameters (luminosity, stellar mass and central surface brightness
  profile) to correlate with the mass-to-light ratio gradients,
  showing that there might be a link between the galaxy structural
  parameters and the total mass of the galaxies. %

In addition, galaxies which have the highest peaks in the $\lambda_R$
parameter (NGC 1023, NGC 3384 and NGC 4564, see Figure
\ref{fig:lambdar}) have fainter $B$ magnitudes and higher values of
mean $(V/\sigma)_{PNe}$. 
Moreover, galaxies with higher $V_{rms,LAST}$ and \slast\ have smaller
$\alpha_{B,1.0}$ values (i.e., less PNe per unit luminosity). As
discussed in Section \ref{sec:alpha_parameter}, this is probably a
consequence of massive early-type systems harbouring a larger
proportion of stars on the Horizontal Branch that do not enter the PN
stage. The correlation between $\alpha_{B,1.0}$ and $\sigma$ was
already explored by \citet{Buzzoni+06}, using the central velocity
dispersion measurements.

We see in the right panels of Figures \ref{fig:comparison1_vrms} and
\ref{fig:comparison1} that these relations still hold if we replace
$V_{rms,LAST}$ with the normalized minimum of $V_{rms}$
(i. e. $V_{rms,MIN}/V_{rms,1.0}$) and if we replace \slast\ with the
normalized minimum velocity dispersion (i. e. \smin), even if the scatter is
larger \footnote{Normalization at \altfre\ or \altfree\ gives
    similar results.}. This is a consequence of more massive galaxies
having preferentially larger values of $V_{rms,LAST}$, \slast\ and
flatter profiles of $V_{rms}$ and velocity dispersion.

\begin{figure*}
  \psfig{file=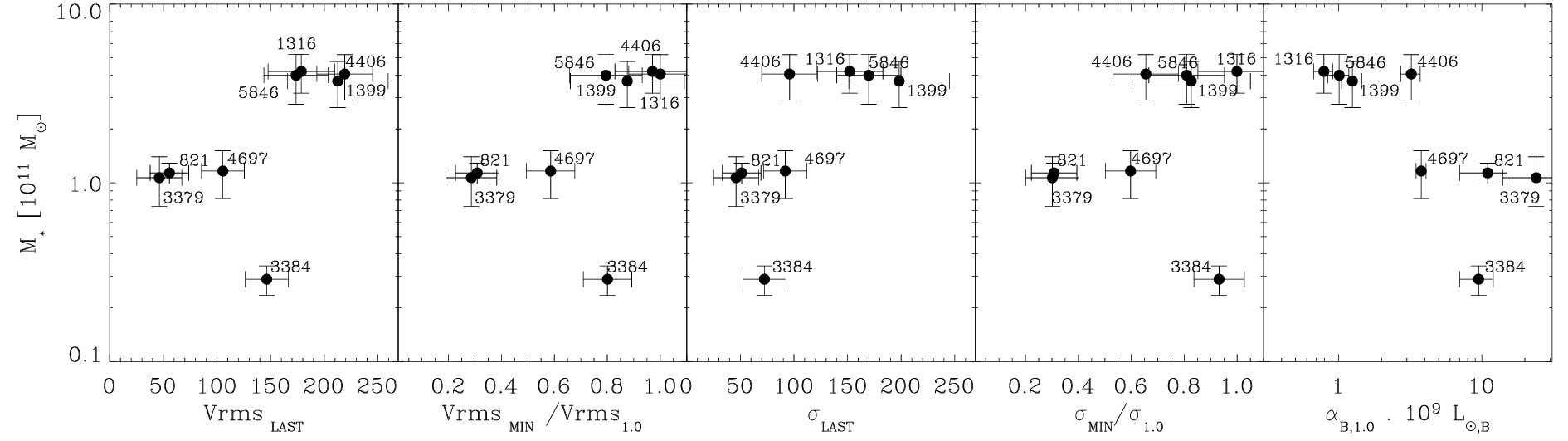,width=18.5cm}
\caption{Correlation between total stellar mass and $V_{rms}$, 
	$V_{rms}/V_{rms,1.0}$, \slast, \smin, and the $\alpha$ parameter.}
\label{fig:baryonic_mass}
\end{figure*}

In Figure \ref{fig:baryonic_mass} we plot the total stellar mass
$M_*$ computed for a subsample of galaxies as a function of
$V_{rms,LAST}$, $V_{rms,MIN}/V_{rms,1.0}$, \slast, \smin\ and the
$\alpha_{B,1.0}$ parameter. The stellar mass is computed using the
total $B$ luminosity from Table \ref{tab:sample} and the mass-to-light
ratio in the $B$-band listed in \citet{Napolitano+05} (where
available).
As expected from Figures \ref{fig:comparison1_vrms} and
\ref{fig:comparison1}, $M_*$ correlates with the $\alpha$ parameter,
$V_{rms,LAST}$, and with \slast, although with larger scatter.%%

The general message we learn from Figures
\ref{fig:comparison1_vrms}-\ref{fig:baryonic_mass} is that galaxies with
higher $V_{rms,LAST}$, and higher \slast\ and flatter profiles (i.e., higher
$V_{rms,MIN}/V_{rms,1.0}$ and higher \smin) tend to be the more
luminous, more massive galaxies, are more pressure supported at large
radii (i.e., $V/\sigma_{PNe} \leq 1$), with boxy isophotes (i.e., $a_4
< 0$) and tend to form less PNe. These results extend the picture
described by \citet{Bender+89} based on stellar kinematics within 1
$R_e$ and of \citet{Buzzoni+06}, to larger radii.%

In  Figures  \ref{fig:comparison1_vrms}  and \ref{fig:comparison1}  we
also  differentiate  between  fast  rotators,  fast  rotators  with  a
declining  $\lambda_R$ profile  in the  halo, slow  rotators  and slow
rotators with $\lambda_R>0.1$ in  the halo.  On average, fast rotators
fall on  the left side of the  plots, (i.e. they have  lower values of
$V_{rms,LAST}$,  \slast  and   more  declining  profiles)  while  slow
rotators fall  on the right side  of the plots (i.e.  they have higher
values of $V_{rms,LAST}$, \slast and flatter profiles).
This is a reflection in the halo kinematics that fast (slow) rotators
are on average less (more) massive and have more discy (boxy)
isophotes \citep{Emsellem+07}.

It is also interesting that slow rotators with $\lambda_R>0.1$ in the
outer haloes on average are located between fast and slow rotators in
Figures \ref{fig:comparison1_vrms} and \ref{fig:comparison1}, as if they
represent a link between the two classes.

\subsection{Comparison with galaxy formation models}
\label{sec:comp_models}

The results presented in Sections \ref{sec:pne_kinematics} and
\ref{sec:discussion} on the outer halo kinematics of our sample
galaxies provide new constraints for models of elliptical galaxy
formation. Most of the merger simulation papers to date compare their
remnants to data within an effective radius or so, but there are a few
predictions for the kinematics at larger radii.

Line-of-sight velocity fields for binary disc merger remnants are
published in \citet{Jesseit+07}. The progenitor galaxies in these
simulations include pure stellar discs as well as discs containing
10\% of their mass in gas. Some 3:1 mergers with gas lead to remnants
showing velocity fields with rapid rotation (e.g., their remnant
31GS19), with a peak velocity at 2-3$R_e$, similar to the case of NGC
1023. Strong misalignments such as observed in NGC 821 are more
characteristic of the 1:1 merger remnants (both dry and gas-rich,
e.g., 11C10/11S8), which also include slowly rotating remnants with
radially increasing rotational support (e.g., 11C6). Their models also
include one which shows a ring-like depression of the velocity
dispersion at $R\simeq 0.5R_e$ associated with a corresponding increase in
$h_4$ (11S2), somewhat similar to the case of NGC 3379
(see \citealt{DeLorenzi+08b}). Its velocity dispersion profile decreases by a
factor of $\sim 2$ from the centre out to $\sim 1.5R_e$. This model
originated from a merger of two disc galaxies with spin axis
perpendicular to the orbital plane.  Overall, the projected kinematics
of the merger remnants analysed by
\citet{Jesseit+07} show a variety of features seen also in our data. A
more careful comparison including several diagnostics ($V/\sigma$
values, $\lambda_R$-profiles, velocity fields, dispersion profiles)
would clearly be profitable.

Mean rotation and velocity dispersion profiles are shown by
\citet{Naab+06} for a 3:1 merger remnant from the same set of
simulations, out to $4R_e$.  The kinematics of this object are
characterised by rapid rotation and a falling dispersion profile (by a
factor $\sim 1.5$), leading to a major axis $V/\sigma$ profile
increasing to $V/\sigma\simeq 2$ at 2-4 $R_e$.  Only the S0 galaxies
and NGC 4564 in Figure \ref{fig:vsigma_comparison} reach these
values. By contrast, the $M_*$ galaxies formed in a cosmological
setting \citep[described in][]{Naab+07} rotate more slowly and are
characterised by $V/\sigma \lesssim 0.5$ with similar values in the halo
and the central parts (disregarding the counter-rotating core seen in
one of the remnants). This is more typical for the slowly rotating
galaxies in Figures \ref{fig:vsigma_comparison} and \ref{fig:lambdar}.

The kinematics of the luminous haloes of isolated galaxies are
investigated by \citet{Abadi+06}. The outer haloes of these systems are
predominantly made of stars accreted during previous merger
events. These haloes have near-spherical triaxial density
distributions, strongly increasing radial anisotropy, and
correspondingly falling line-of-sight velocity dispersion profiles (by
more than a factor of $2$ out to 0.1 times their virial radius).  Our
data probe only partially into these haloes, but it is possible that
some of the strongly falling dispersion profiles in Figure
\ref{fig:slope} are related to such accreted haloes. On the other hand,
the galaxies in our sample with slowly falling dispersion profiles
or those with twisting (NGC 3377) or misaligned rotation (NGC 821)
may not be easily reconciled with these models.

\citet{Onorbe+07} studied the properties of elliptical-like
galaxies generated in self-consistent hydro-dynamical
simulations. These galaxies resemble the characteristics of slow
rotators, i.e., are massive spheroidal systems without extended disk
components and very low (cold) gas content. Their velocity dispersion
profiles (normalized at $1R_e$) are generally flat or slightly
declining, with values of \smin\ larger than 0.7 out to $6R_e$. These
values are consistent with what we observe for the haloes of slow
rotators (see Figure \ref{fig:comparison1}). Galaxies in our sample
with more steeply declining velocity dispersion profiles are generally
fast rotators, which are not included in the \citet{Onorbe+07}
simulations.

In summary, we expect that more detailed comparisons between data 
as presented in this paper and the halo kinematics predicted by galaxy
formation models will shed new light on the merger formation
histories of elliptical galaxies.

\section{Summary}
\label{sec:summary}

We explored the outer-halo kinematics and properties of early-type
galaxies using the PNe data available for a sample of 16 objects.  Out
of the sample, 6 new catalogues with positions, radial velocities and
magnitudes of PNe are presented for the first time ({\it sample A},
namely, NGC 821, NGC 3377, NGC 3608, NGC 4374, NGC 4564 and NGC 5846).

We compared the radial distribution of PNe with the stellar surface
brightness profiles, for all galaxies in {\it sample A}, and for
galaxies in {\it sample B},  which have extended
photometry available from the literature (NGC 1023, NGC 1344, NGC
3379, NGC 4494 and NGC 4697).
We found that:

\begin{enumerate}

\item{} The PNe number density profile follows the stellar surface
  brightness profile in all the galaxies in which we have a spatial
  overlap between the stellar light and the PNe counts.

\item{} In the galaxies in which we do not have a spatial overlap, the
  PNe counts follow the extrapolation of the stellar surface
  brightness profile.

\item{} For the galaxies with a far UV magnitude measured with GALEX,
  we confirmed results from previous studies that the total number of
  PNe per unit luminosity ($\alpha_{B,1.0}$ parameter) is related to
  the UV colour excess (i.e., galaxies with higher UV emission tend to
  have fewer PNe).

\end{enumerate}

We then determined smoothed two-dimensional velocity and velocity
dispersion fields using an adaptive Gaussian kernel, for the
  galaxies in {\it sample A} and in {\it sample B} (namely NGC 1023,
  NGC 1344, NGC 3379, NGC 4494, NGC 4697 and NGC 5128). We compared
the PNe kinematics with absorption-line kinematics available in the
literature. For NGC 3377, NGC 4374 and NGC 4494 new long-slit data
were presented. The kinematic analysis showed that:

\begin{enumerate}

\item{} Rotation of the PNe system is observed in the majority of the
  studied galaxies.

\item{} There is a good agreement between the absorption-line and PNe
  kinematics along the major and minor axes. 

\end{enumerate}

We combined the PNe and absorption-line kinematics along the
photometric major axis for all 16 galaxies in our sample, extracting
the information from the two-dimensional fields where available. This
allowed us to probe their kinematics much further out than with the
use of stellar kinematics alone (usually limited to the innermost $1-2$
$R_e$). This gave the following results:

\begin{enumerate}

\item{} In several fast-rotator galaxies, kinematic twists and
misalignments are observed at large radii, which are not seen in the
SAURON data within $R_e$ (NGC 821, NGC 3377, and marginally, NGC 3379).

\item{} The average $V/\sigma$ of the stellar component (which probes
  the kinematics in the inner $0.5 - 1$ $R_e$) is equal {\bf to} or lower
  than the average value computed for the PNe (which probes the
  kinematics for $R>2$ $R_e$). This indicates that for a fraction of
  galaxies, the kinematics become increasingly supported by
  rotation in the outer parts.

\item{} The radial profiles of the $\lambda_R$ parameter (related to
  the angular momentum per unit mass) show a more complex radial
  dependence when their values in the halo region are taken into
  account. While the majority of fast rotators remain so in their
  haloes, the $\lambda_R$ profiles of slow rotators grow slowly to
  values of $0.1-0.3$, therefore requiring a slight modification to
  the classification scheme proposed by \citet{Emsellem+07} (NGC 3608,
  NGC 4374 and NGC 5846). Some fast rotators however (NGC 1316, NGC
  1023, NGC 3377 and NGC 4494), have $\lambda_R$ profiles that
  strongly decrease outwards, probably due to the presence of
  disc-like structures that dominate the kinematics within 1--3 $R_e$
  and then fade.

\item{} The normalized $V_{rms}$ profiles show that our sample
  galaxies fall into two main groups; the first group (NGC 1316, NGC
  1344, NGC 1399, NGC 3384, NGC 4374, NGC 4406, NGC 5128 and NGC 5846)
  shows a slightly declining profile from the centre outwards, the
  second group (NGC 821, NGC 3377, NGC 3379, NGC 4494, NGC 4564 and
  NGC 4697) shows a strongly declining $V_{rms}$ profile. An
  exceptional case is NGC 1023,  where $V_{rms}$ increases until
  $2.5R_e$, before falling steeply.

\item{} The radial profiles of the velocity dispersion show a variety
  of shapes. There are nearly flat profiles, in which the velocity
  dispersion falls only by a few percent from the central values;
  strongly declining profiles, in which we observe a drop of a factor
  of 2 in the velocity dispersion towards outer radii; and a few
  anomalous galaxies, which exhibit rising profiles, again related to
  the presence of a disc.

\item{} More luminous galaxies (brighter total $B$ magnitude and
  X-ray luminosity) tend to have flatter $V_{rms}$ and flatter
  velocity dispersion profiles and larger values of $V_{rms}$ and
  dispersion measured at the outermost observed point. This is related
  to the fact that more massive galaxies have nearly flat dispersion
  profiles, and generally higher values for the outer velocity
  dispersion. Moreover, there is evidence that more massive and
  luminous galaxies have on average a lower number of PNe per
  unit luminosity (smaller $\alpha$) than less massive galaxies.

\item{} Slow rotators have on average flatter $V_{rms}$ and
  flatter velocity dispersion profiles and larger values of
  $V_{rms,LAST}$ and \slast. Conversely, fast rotators have on average
  more steeply declining $V_{rms}$ and velocity dispersion profiles,
  with lower values of $V_{rms,LAST}$ and \slast.

\item{} Galaxies with high values of \slast\ or nearly flat velocity
  dispersion profiles (i.e., more massive galaxies) are preferentially
  boxy in shape ($a_4 < 0$) and have smaller values of $V/\sigma$.
  Galaxies with small values of \slast\ or declining velocity
  dispersion profiles (i.e., less massive galaxies) have a larger
  range of shapes ($-1<a_4<2$) and $V/\sigma$ ratios.

\end{enumerate}

These results show that a full picture of the kinematics and
angular momenta of elliptical galaxies requires information about
their outer velocity fields. When comparing to models of the formation
of ellipticals it is important to take into account the halo
kinematics -- the dynamical timescales in the haloes are longer and
therefore the imprint of the formation mechanisms are preserved more
strongly.

\section*{Acknowledgments}

We would like to thank Eric Emsellem for providing the $\lambda_R$
radial profiles of the SAURON data, John Kormendy for providing
surface photometry of NGC 4374 prior to publication, and
  Karl Gebhardt for useful discussion.

The minor axis absorption-line kinematics of NGC 5846 is based on data
from the SAURON archive.

This research has made use of the NASA/IPAC Extragalactic Database
(NED), which is operated by the Jet Propulsion Laboratory, California
Institute of Technology, under a contract with the National
Aeronautics and Space Administration.

PD was supported by the DFG Cluster of Excellence ``Origin and
Structure of the Universe''.
FDL was supported by the DFG Schwerpunktprogram SPP 1177 ``Witnesses
of Cosmic History''.
MRM was supported by an STFC Senior Fellowship.
NRN has been funded by CORDIS within FP6 with a Marie Curie European
Reintegration Grant, contr. n. MERG-FP6-CT-2005-014774, co-funded by
INAF.
AJR was supported by the National Science Foundation Grant
AST-0507729, and by the FONDAP centre for Astrophysics CONICYT
15010003.

\bibliography{coccato2008_R1}

\appendix
\section{Two-dimensional velocity and velocity dispersion fields of galaxies in sample B}
\label{sec:2d_literature}

Different authors adopted their own procedures and criteria to create
smoothed two-dimensional fields of the PNe, depending on the amount of
available data. Here, for a homogeneous analysis, we determine the
two-dimensional velocity and velocity dispersion fields for some of
these galaxies, adopting the procedure described in Section
\ref{sec:smoothed_fields}. We do this only for the galaxies with a
sufficient number of PNe ($>80$), namely NGC 1023, NGC 1344, NGC 3379,
NGC 4494, NGC 4697 and NGC 5128 (i.e., {\it sample B}). Parameters 
used in the kernel smoothing procedure and typical errors obtained
with Monte Carlo simulations are listed in Table
\ref{tab_app:1}. Results are shown in Figure \ref{fig:appendixA}.

\begin{table}
\centering
\caption{Typical parameters and errors for the smoothed
  two-dimensional velocity and velocity dispersion fields of the
  galaxies in {\it Sample B}.}
\begin{tabular}{l c c c c}
\hline
\hline
\noalign{\smallskip}
Name      & $A$& $B$ &  $<\Delta V>$   &   $<\Delta \sigma>$  \\
\noalign{\smallskip}                   
          &    & arcsec & \kms     & \kms   \\
 (1)      & (2)&     (3) &   (4)    &   (5)  \\
\noalign{\smallskip}                                      
\hline        
\noalign{\smallskip}  
NGC 1023  & 0.48 &  1.36   & 30 & 20  \\
NGC 1344  & 0.53 &  -2.50  & 30 & 20  \\
NGC 3379  & 0.57 &  14.92  & 30 & 30  \\
NGC 4494  & 0.48 &  11.36  & 30 & 20  \\
NGC 4697  & 0.40 & 12.80   & 40 & 20  \\
NGC 5128  & 0.08 &  18.56  & 15 & 15  \\
\noalign{\smallskip}
\hline
\end{tabular}
\label{tab_app:1}
\begin{minipage}{8cm}
  Notes -- Cols. 2 -- 3: Values of $A$ and $B$ used in the kernel
  smoothing procedure (see Equation \ref{eqn:k}) as determined from
  the simulations. Cols. 4 -- 5 typical error on the two-dimensional
  velocity and velocity dispersion fields as determined from Monte
  Carlo simulations (see Section \ref{sec:testing} for details).
\end{minipage}
\end{table}

\begin{figure*}
 \vbox{
   \hbox{
     \psfig{file=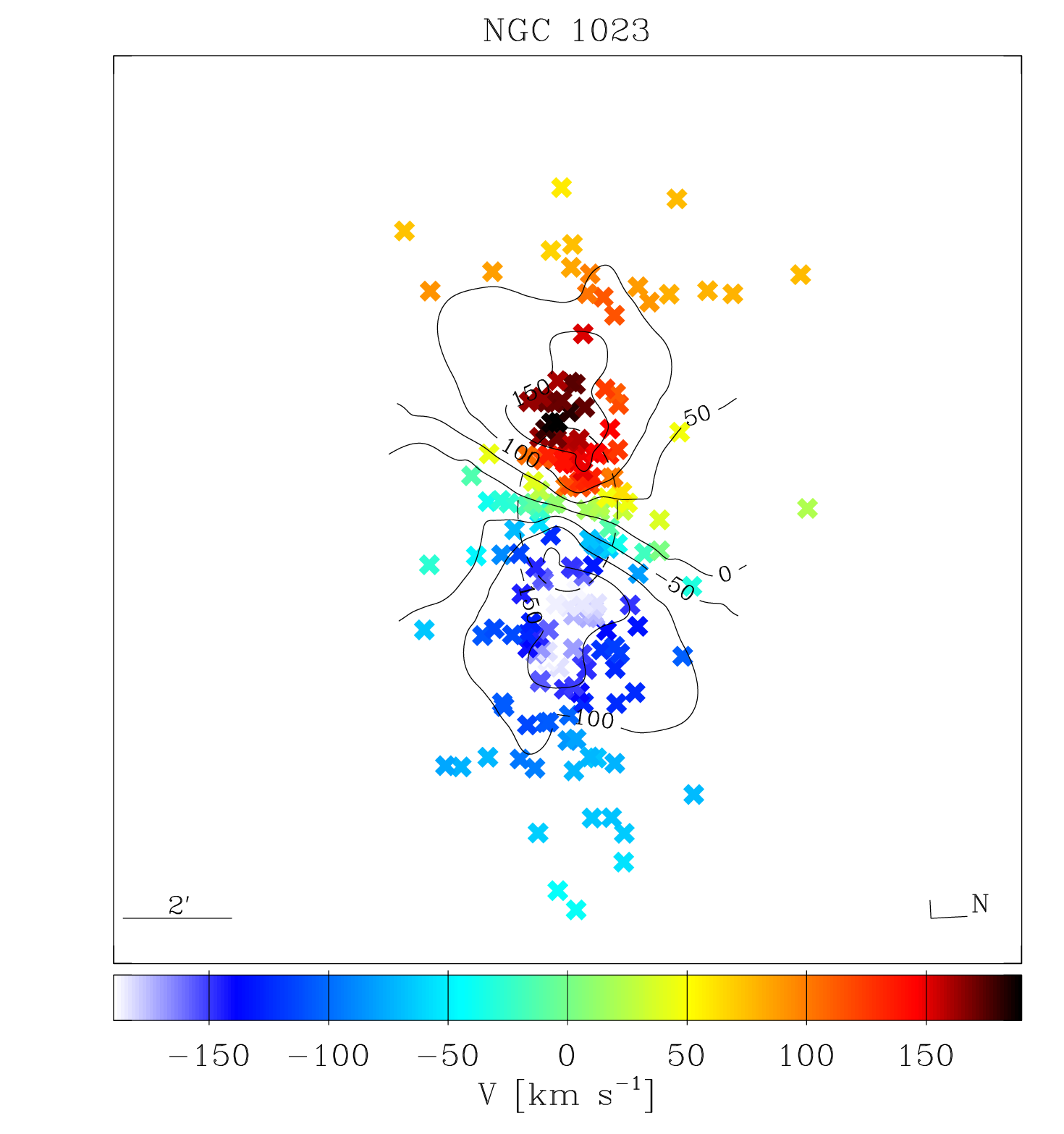,clip=,width=7.2cm}
     \psfig{file=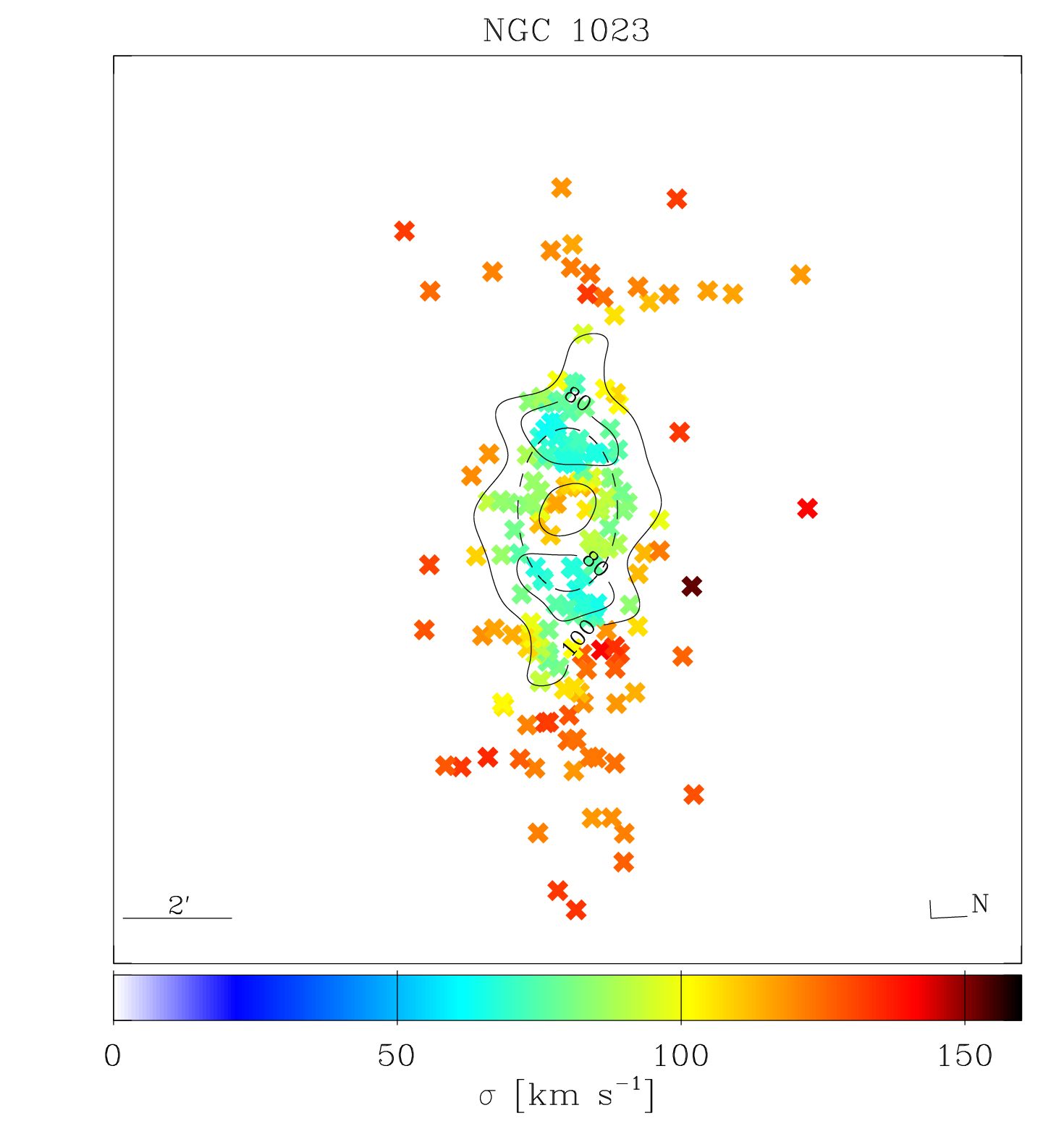,clip=,width=7.2cm}}
   \hbox{
     \psfig{file=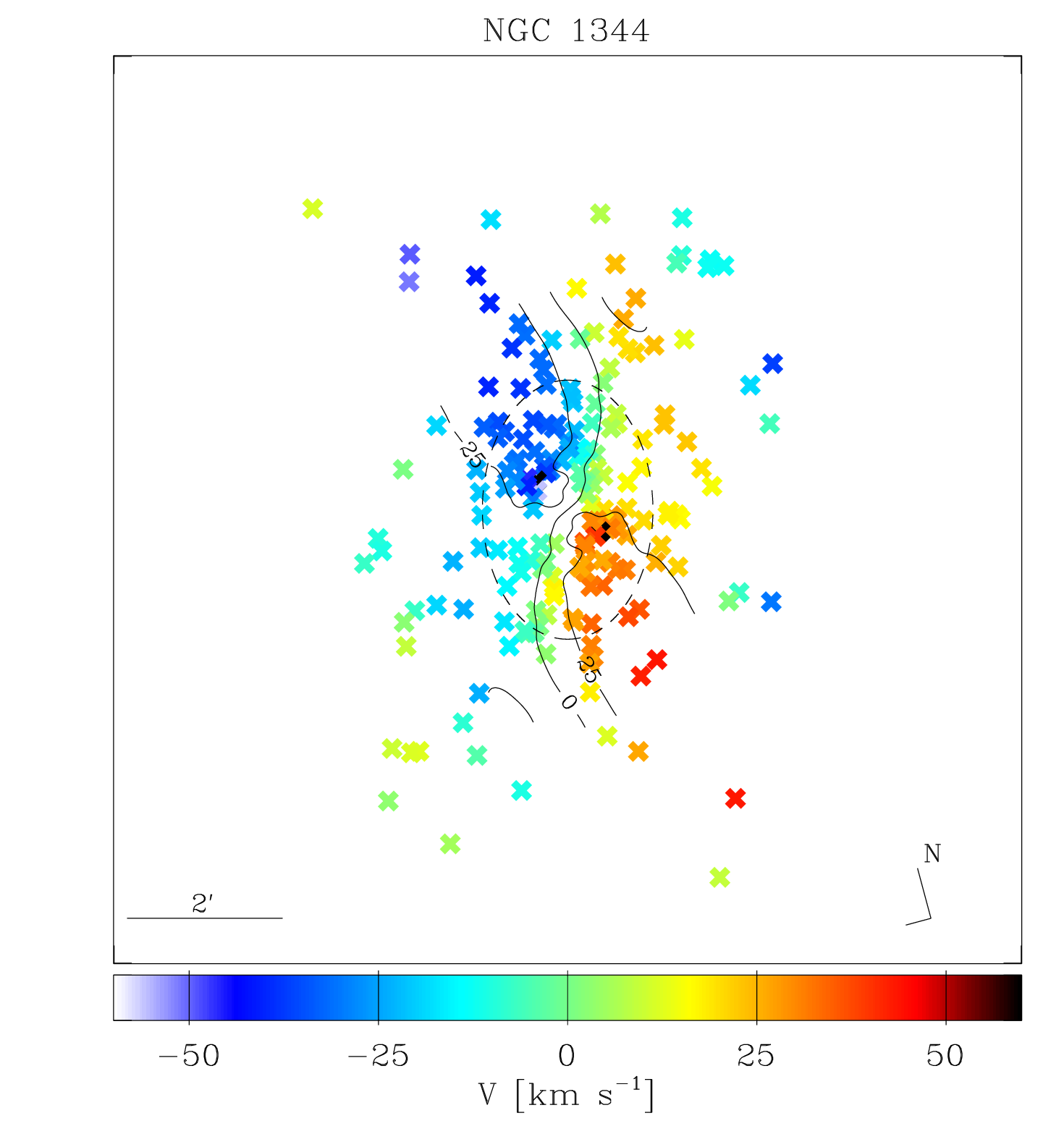,clip=,width=7.2cm}
     \psfig{file=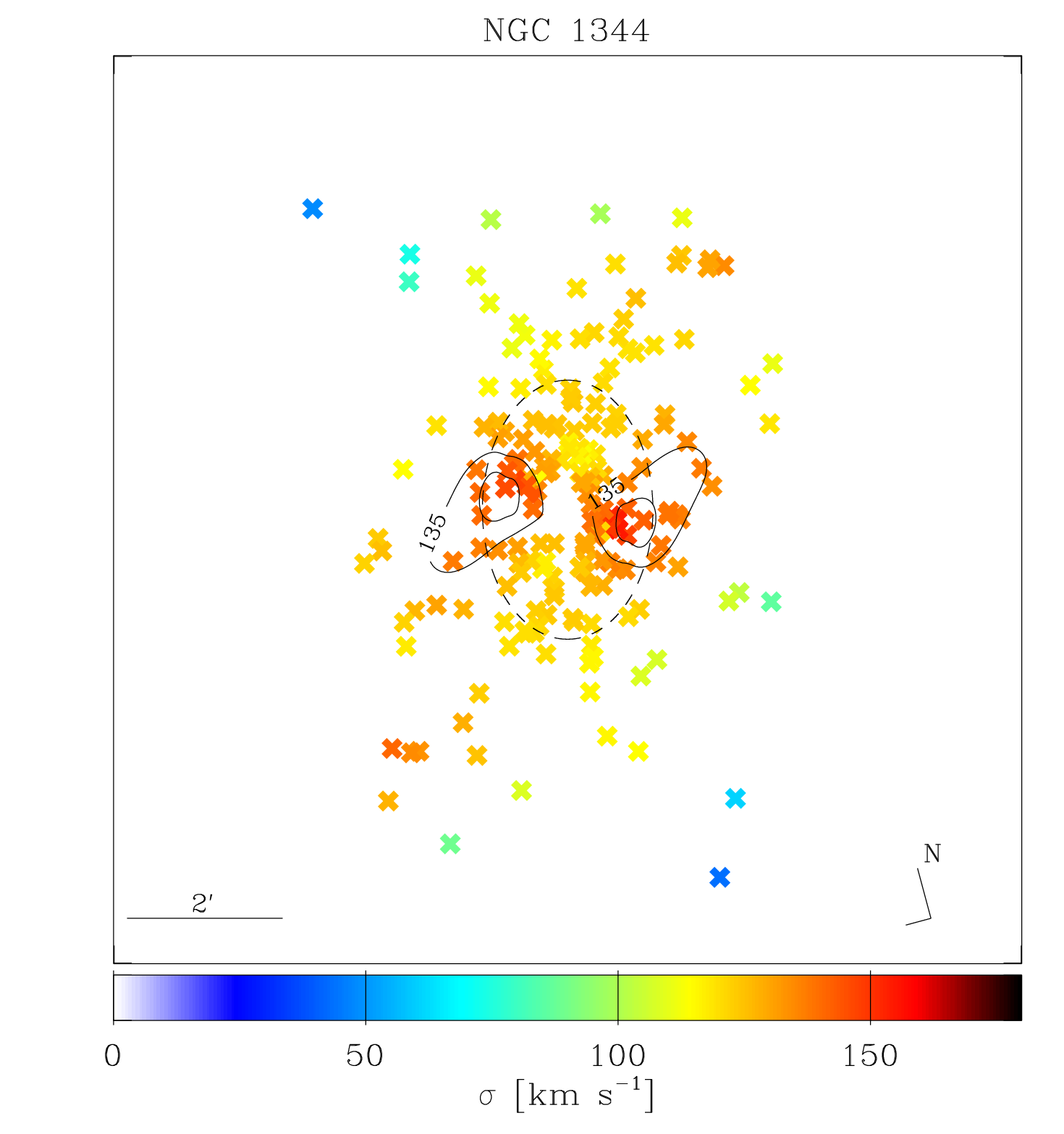,clip=,width=7.2cm}}
   \hbox{
     \psfig{file=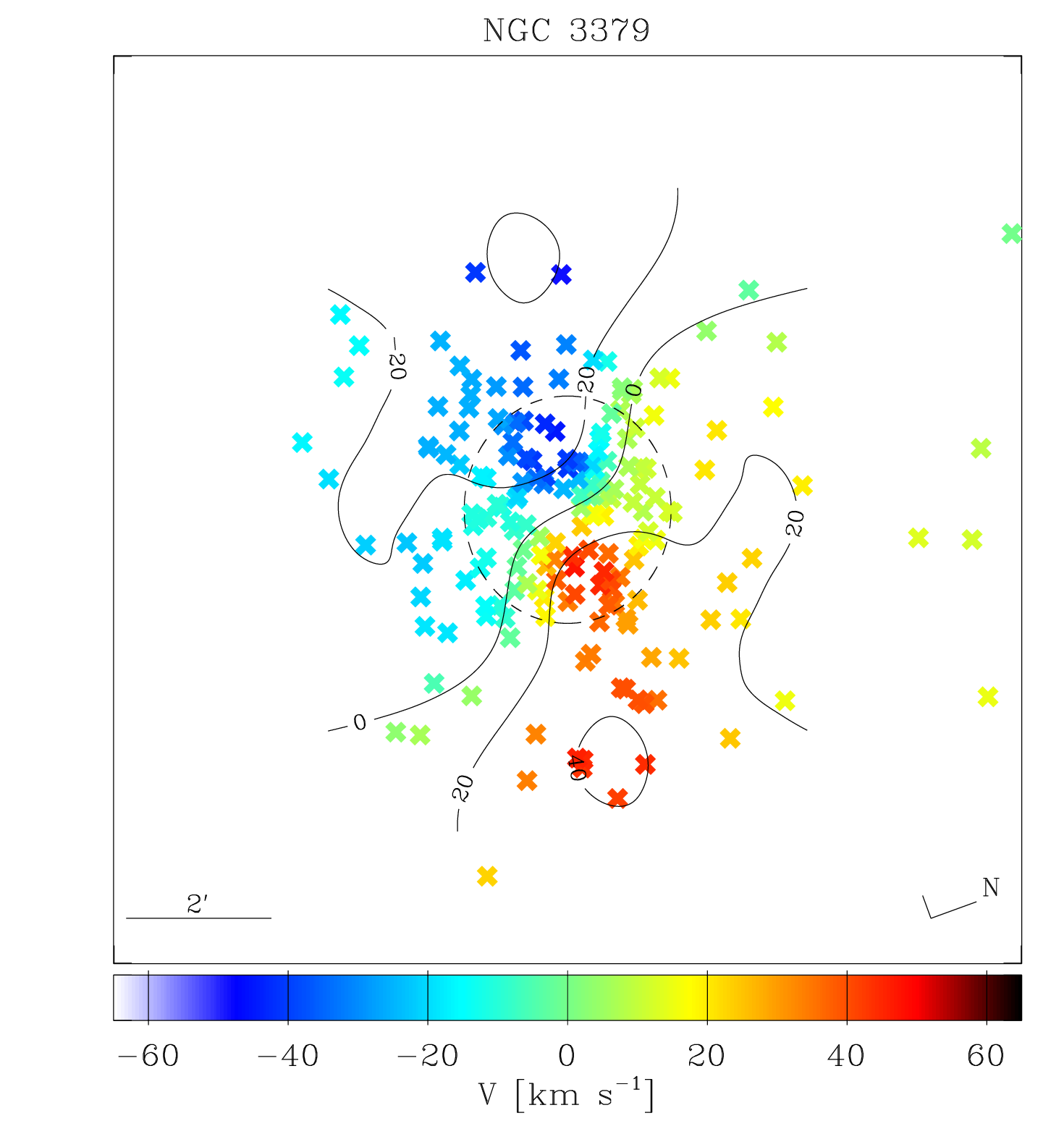,clip=,width=7.2cm}
     \psfig{file=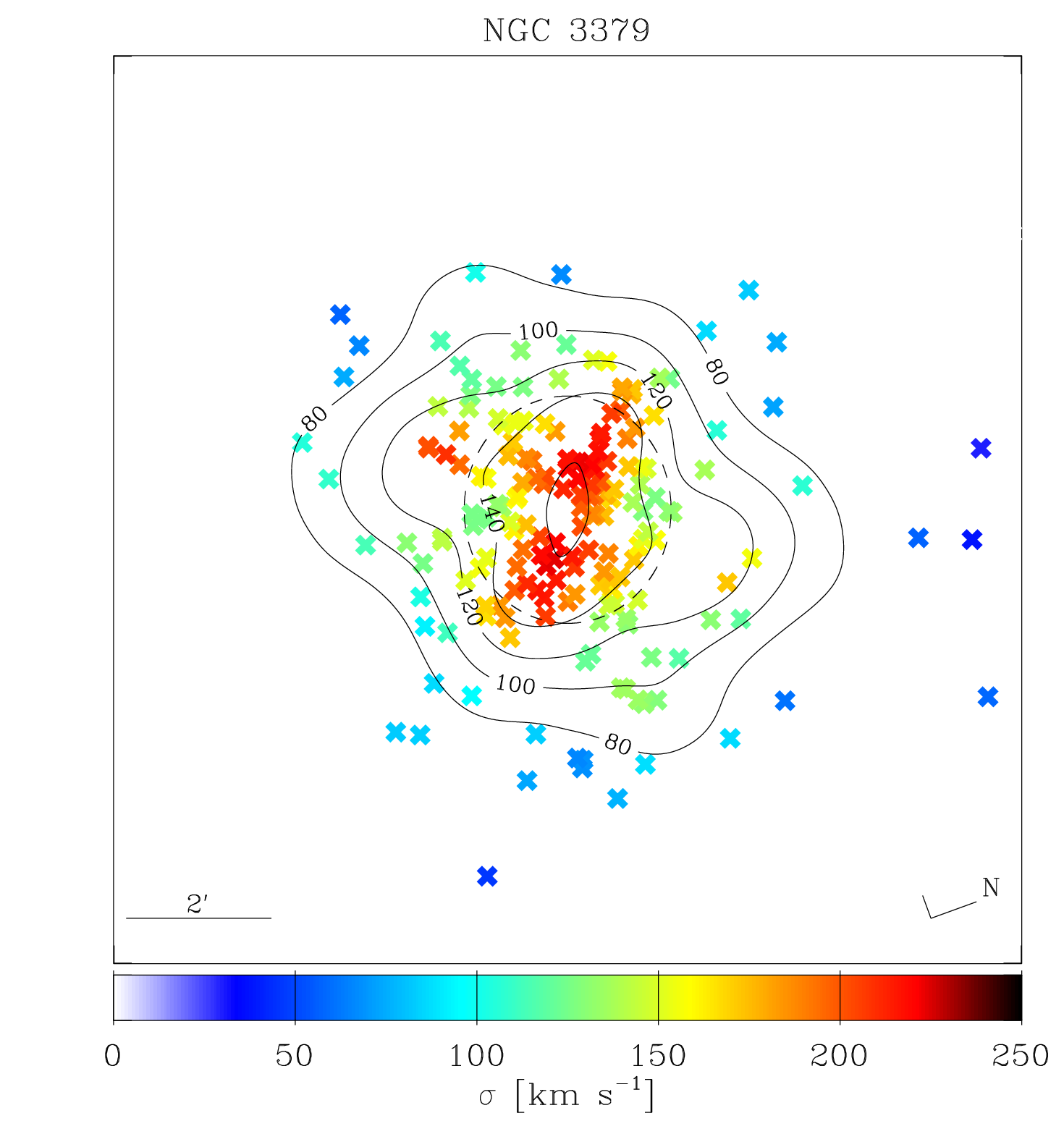,clip=,width=7.2cm}}
  }
  \caption{PNe smoothed two-dimensional velocity and velocity
    dispersion fields for galaxies in {\it sample B}. Symbols,
    orientations and scales are as in Figure \ref{fig:2dfields}.}
  \label{fig:appendixA}
\end{figure*}

\addtocounter{figure}{-1}
\begin{figure*}
 \vbox{
   \hbox{
     \psfig{file=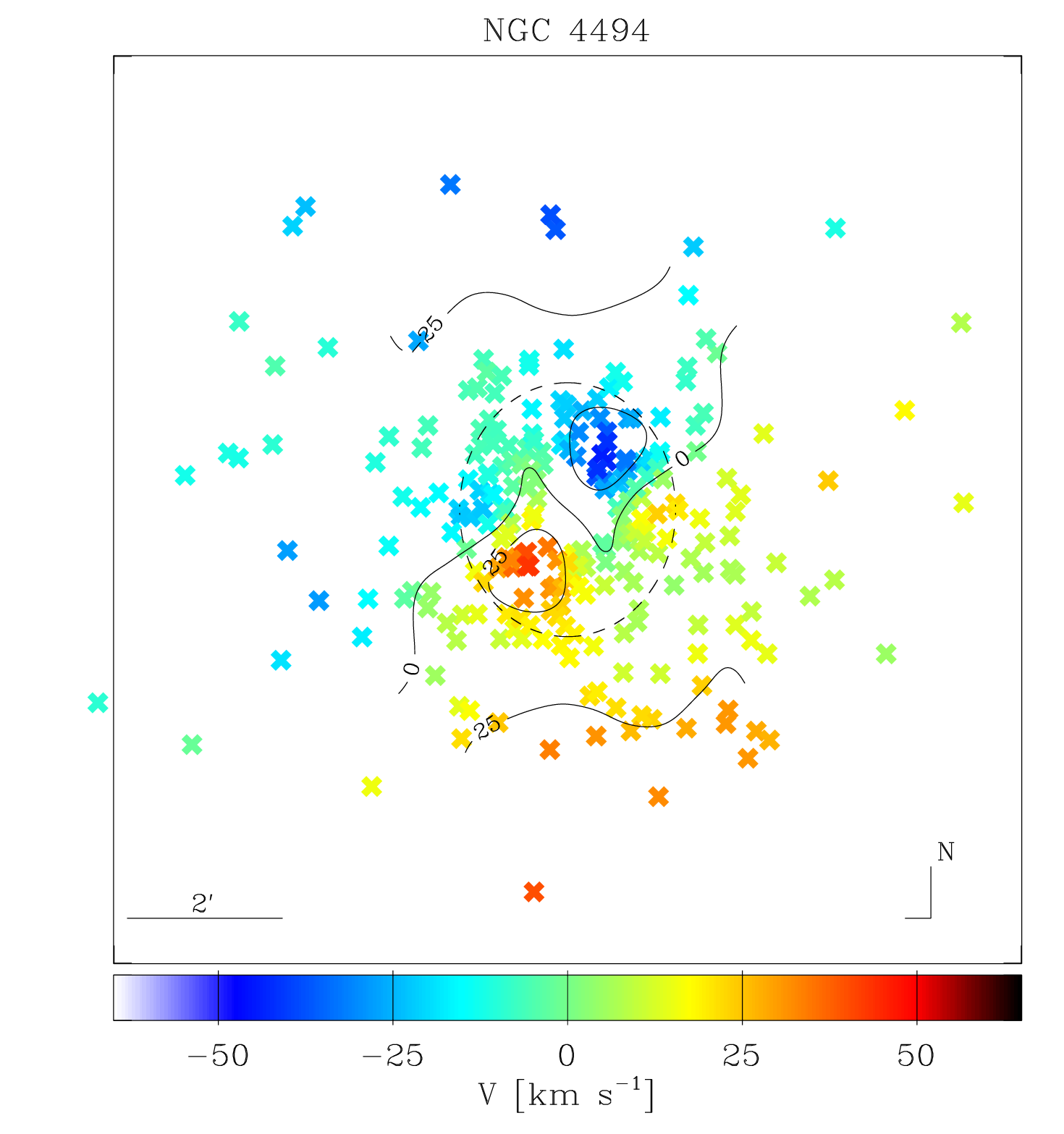,clip=,width=7.2cm}
     \psfig{file=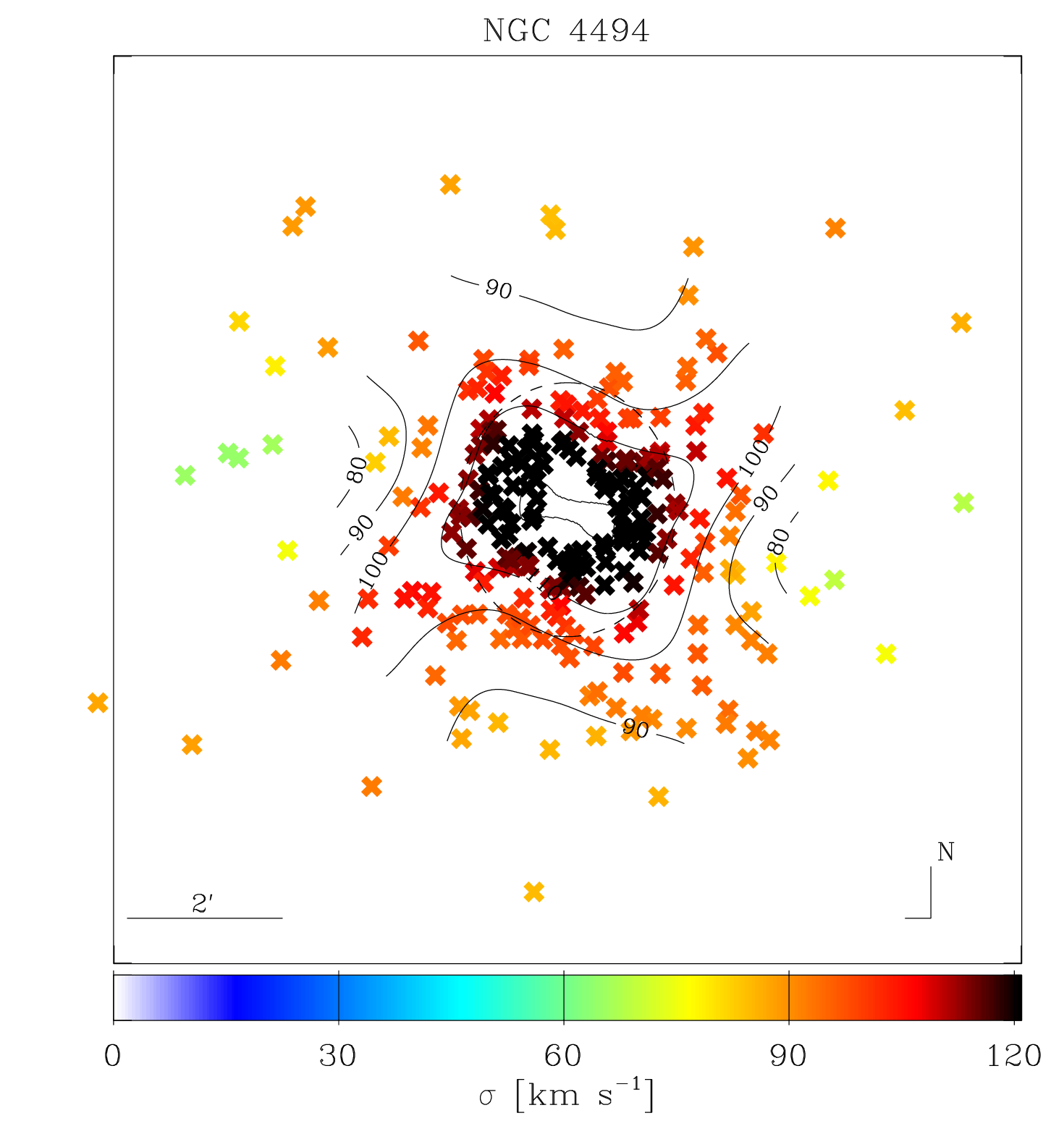,clip=,width=7.2cm}}
   \hbox{
     \psfig{file=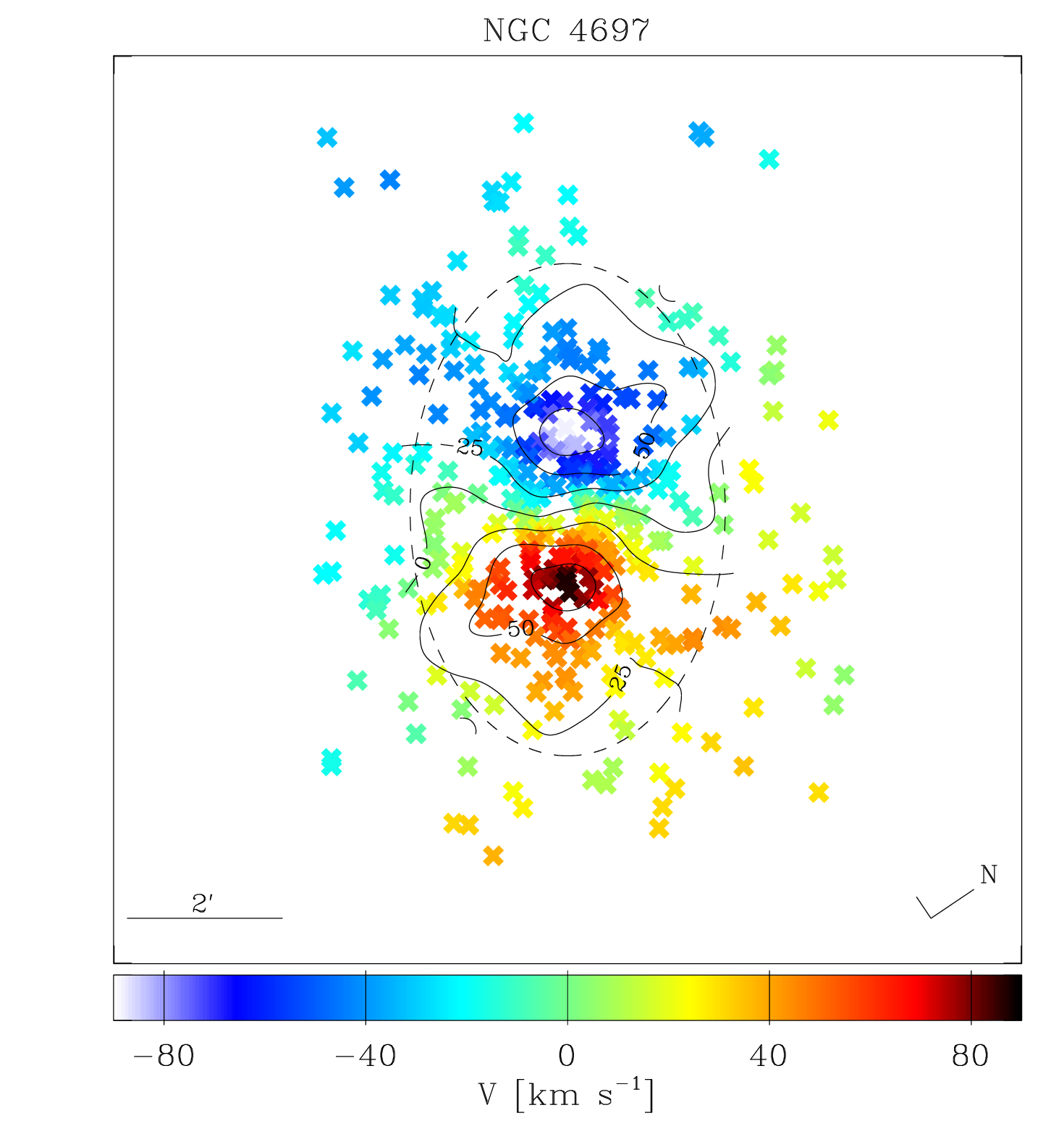,clip=,width=7.2cm}
     \psfig{file=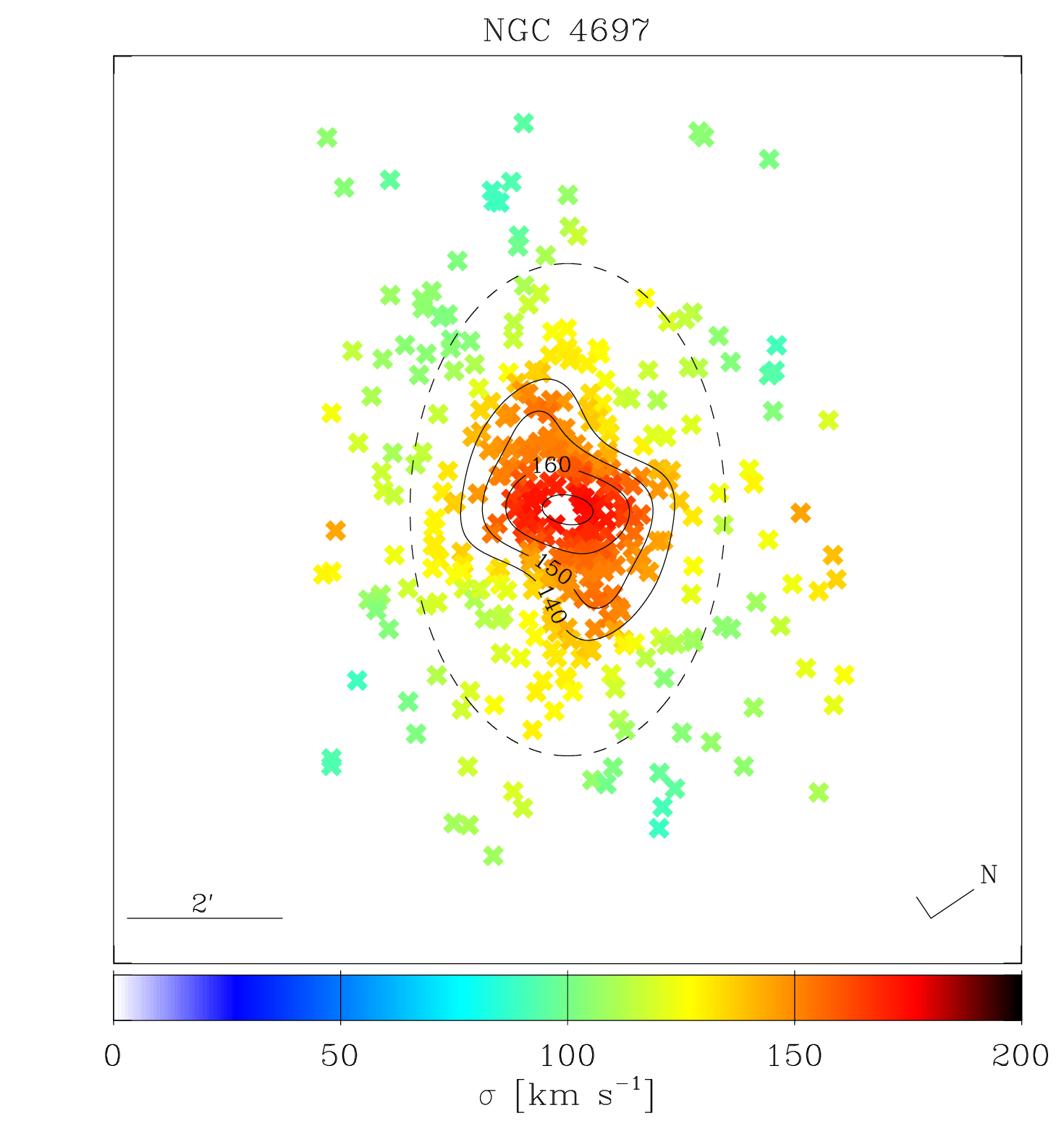,clip=,width=7.2cm}}
   \hbox{
     \psfig{file=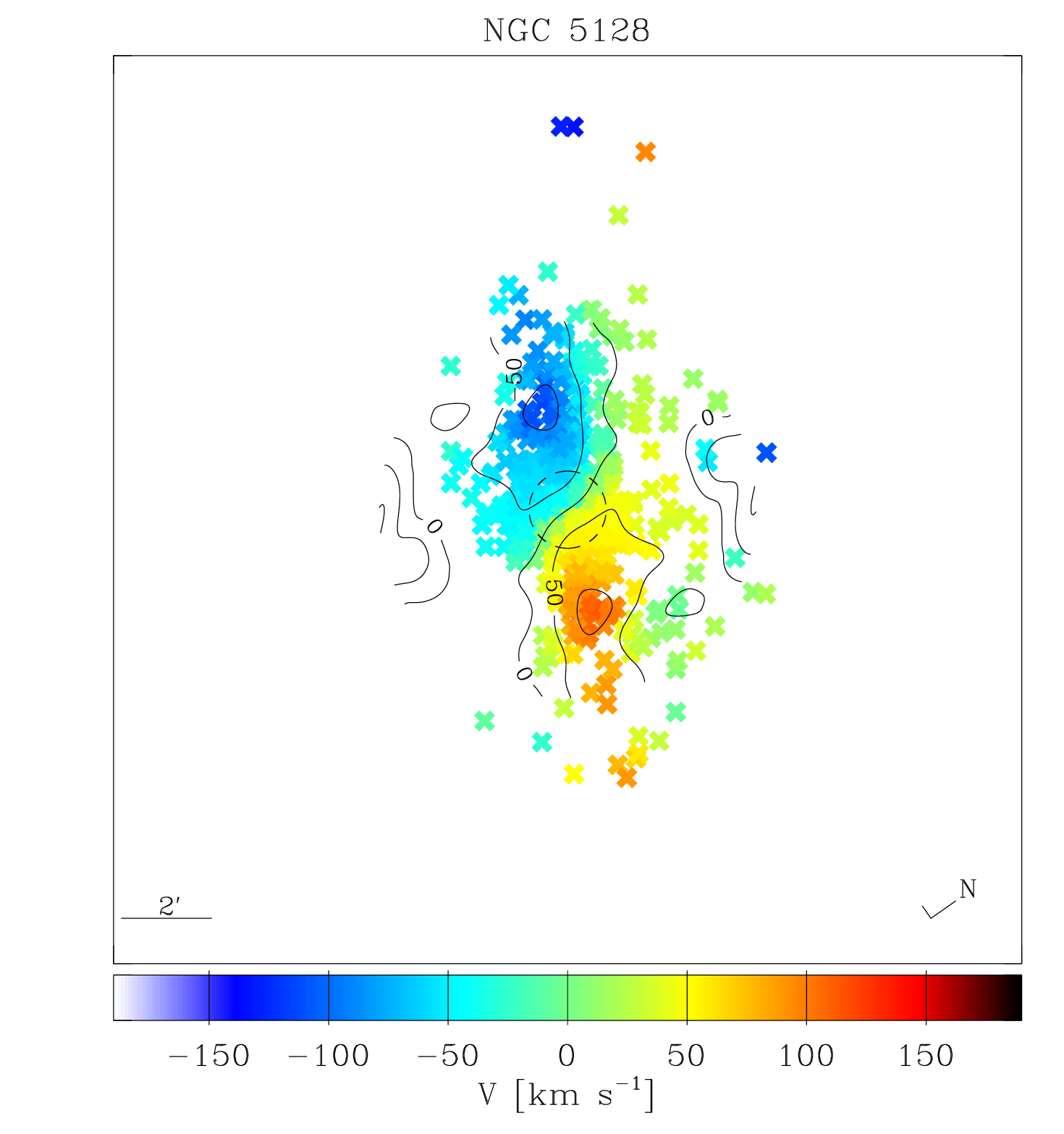,clip=,width=7.2cm}
     \psfig{file=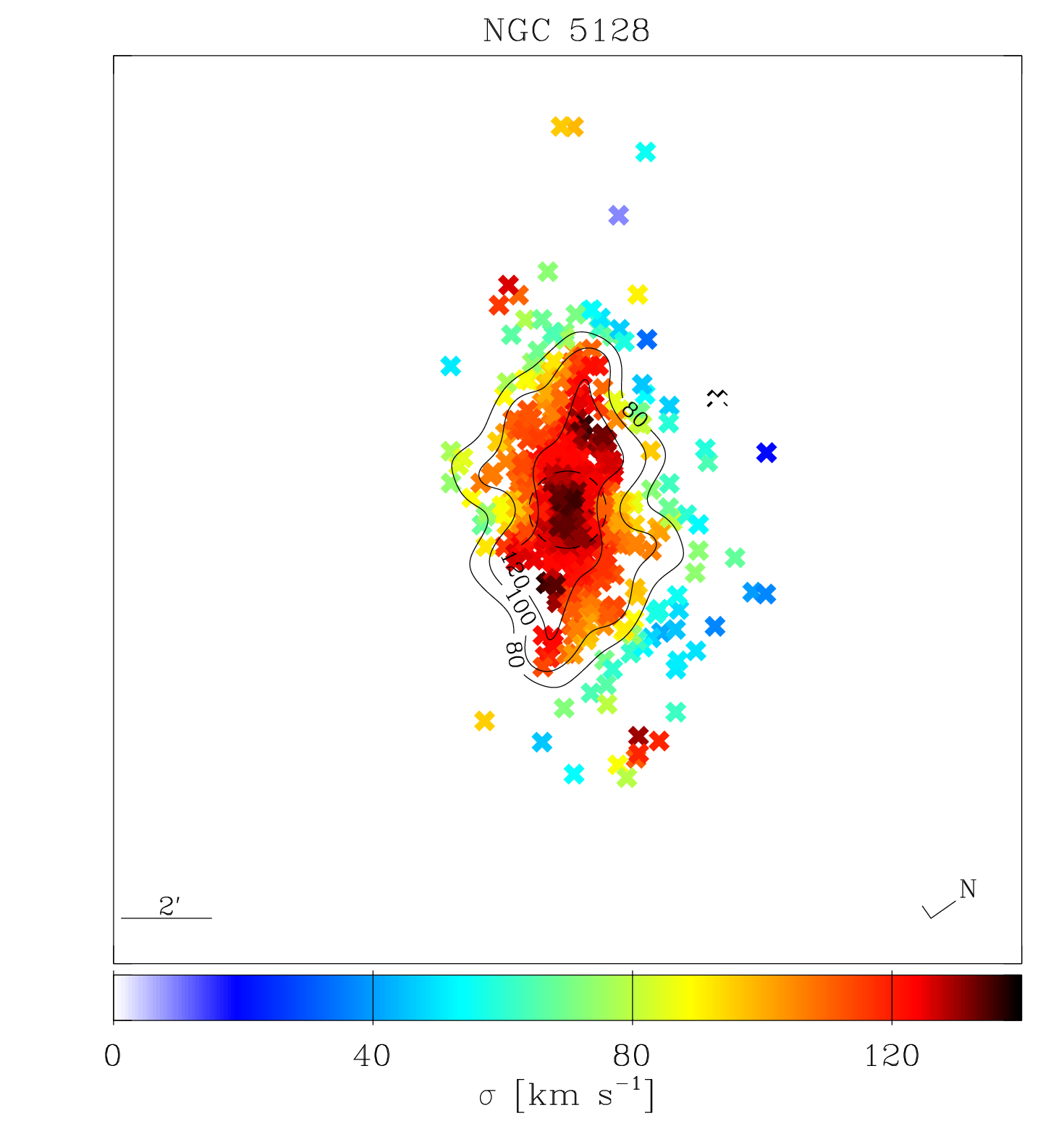,clip=,width=7.2cm}}
  }
  \caption{Continued}
\end{figure*}

\section{Unsmoothed velocity fields}
\label{app:unsmoothed}

The smoothing technique described in Section \ref{sec:smoothed_fields}
is required to measure the mean velocity and the velocity dispersion
along the line of sight of the PNe system. A single PN radial velocity
measurement randomly deviates from the local mean velocity by an
amount that depends on the local velocity dispersion. One can ask how
the unsmoothed map of PNe radial velocity measurements compares
with the smoothed two-dimensional velocity field.

\begin{figure*}
   \hbox{
     \psfig{file=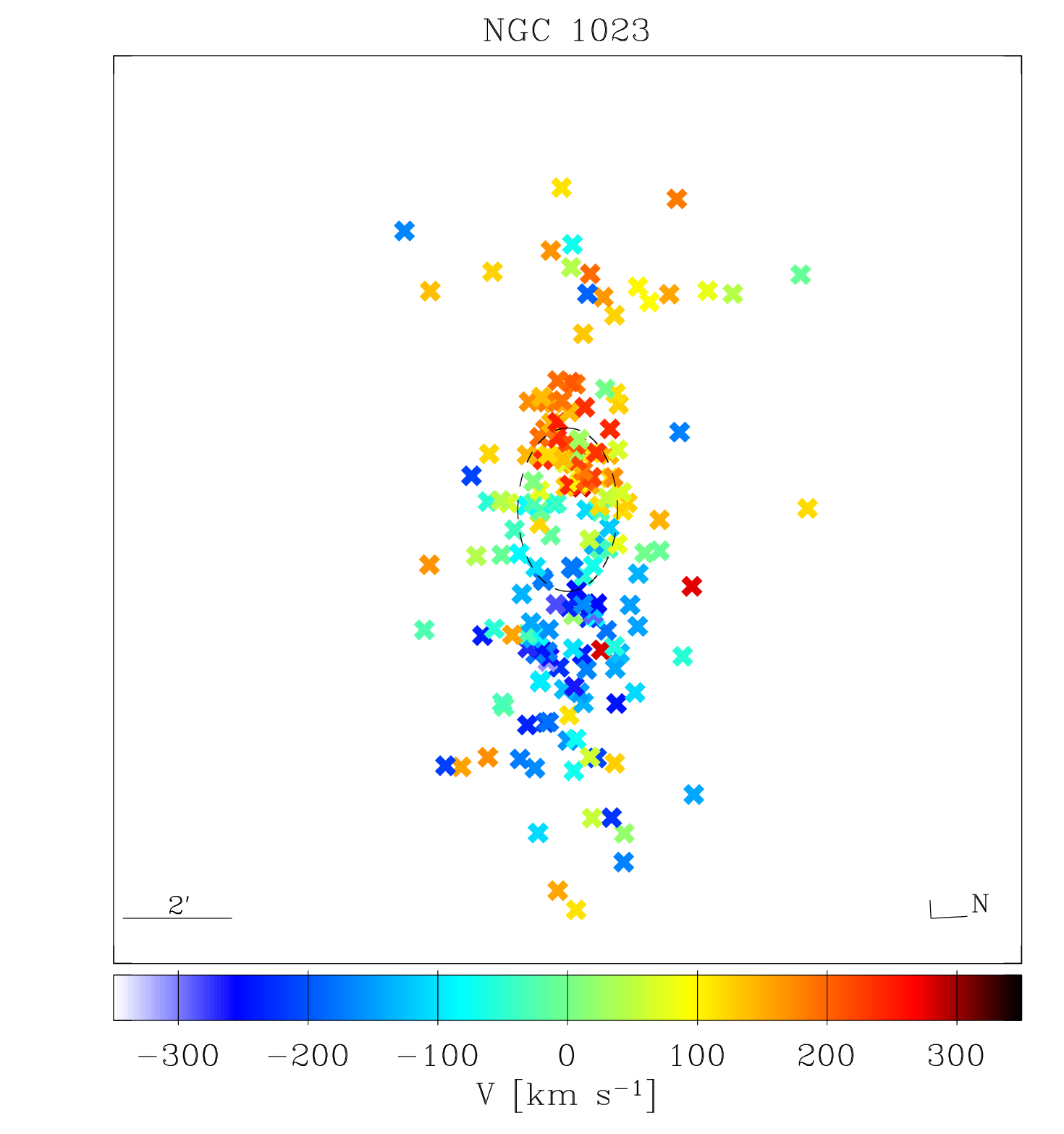,clip=,width=8.3cm}
     \psfig{file=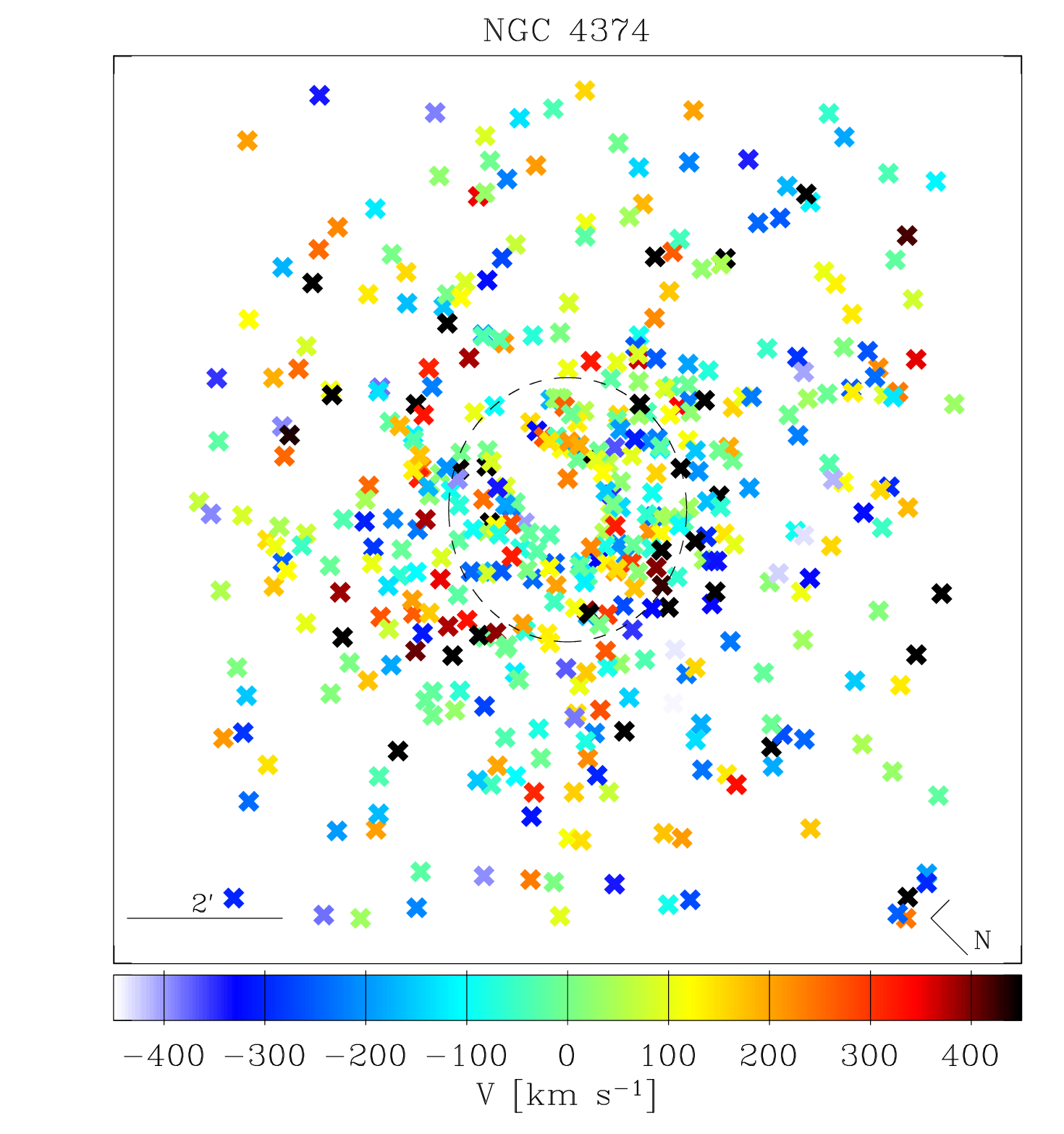,clip=,width=8.3cm}}
\caption{Two-dimensional velocity fields of the individual PNe
radial velocity measurements for NGC 1023 (left panel) and NGC 4374
(right panel), without applying the adaptive kernel smoothing.}
\label{fig:unsmoothed}
\end{figure*}

In galaxies with high $V/\sigma$, we expect that the velocity map of
individual PNe radial velocity measurements preserves the
characteristics of the mean velocity field. On the contrary, in
galaxies with low $V/\sigma$ we expect it to be chaotic and the mean
rotation to be hidden in the scatter of the velocity dispersion, which
dominates the kinematics.  As examples, we show in Figure
\ref{fig:unsmoothed} the two-dimensional maps of individual PNe radial
velocities for NGC 1023 (a system dominated by rotation) and NGC 4374
(a system dominated by velocity dispersion).

\section{Long-slit kinematics}
\label{sec:long_slit}

Long-slit observations for NGC 3377, NGC 4374 and NGC 4494 were
carried out with the Very Large Telescope (VLT) at the European
Southern Observatory (ESO) in Paranal (Chile) from December 2005 to
April 2006 (Proposal 76.B--0788A) in service mode under dark time
conditions.
The Unit Telescope 1 (Antu) was mounted with the Focal Reducer/low
dispersion Spectrograph (FORS2), which was equipped with the 1400V
Grism and the 0\farcs5 slit. Spectra were taken along both the major
and minor axes.
Basic data reduction was performed using standard
ESO-MIDAS\footnote{MIDAS is developed and maintained by the European
  Southern Observatory.} and IRAF \footnote{IRAF is distributed by
  NOAO, which is operated by AURA Inc., under contract with the
  National Science Foundation.} routines. All the spectra were
bias-subtracted, flat-field corrected by quartz lamp and twilight
exposures, cleaned of cosmic rays, and wavelength calibrated. After
calibration, the different spectra obtained for a given galaxy along
the same position angle were co-added using the centre of the
stellar-continuum radial profile as a reference. The contribution of
the sky was determined at the two edges of the resulting frames where
the galaxy light was negligible, and then subtracted.  Spectra from
adjacent rows had been binned together in order to ensure a
signal-to-noise ratio of at least 25. The instrumental FWHM measured
on the comparison spectra is $\approx 140$ \kms.
\begin{figure*}
 \vbox{
   \hbox{
     \psfig{file=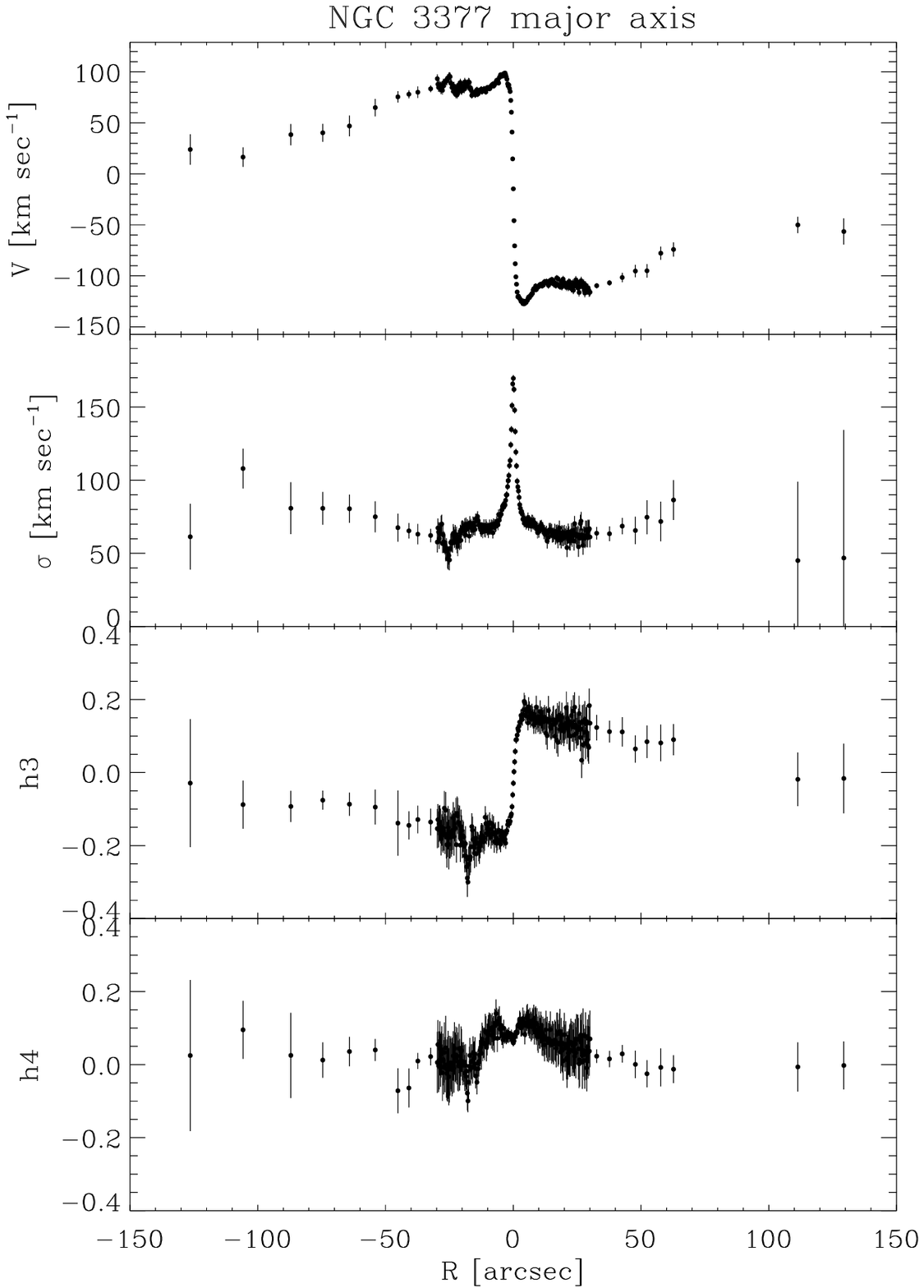,clip=,width=8.3cm}
     \psfig{file=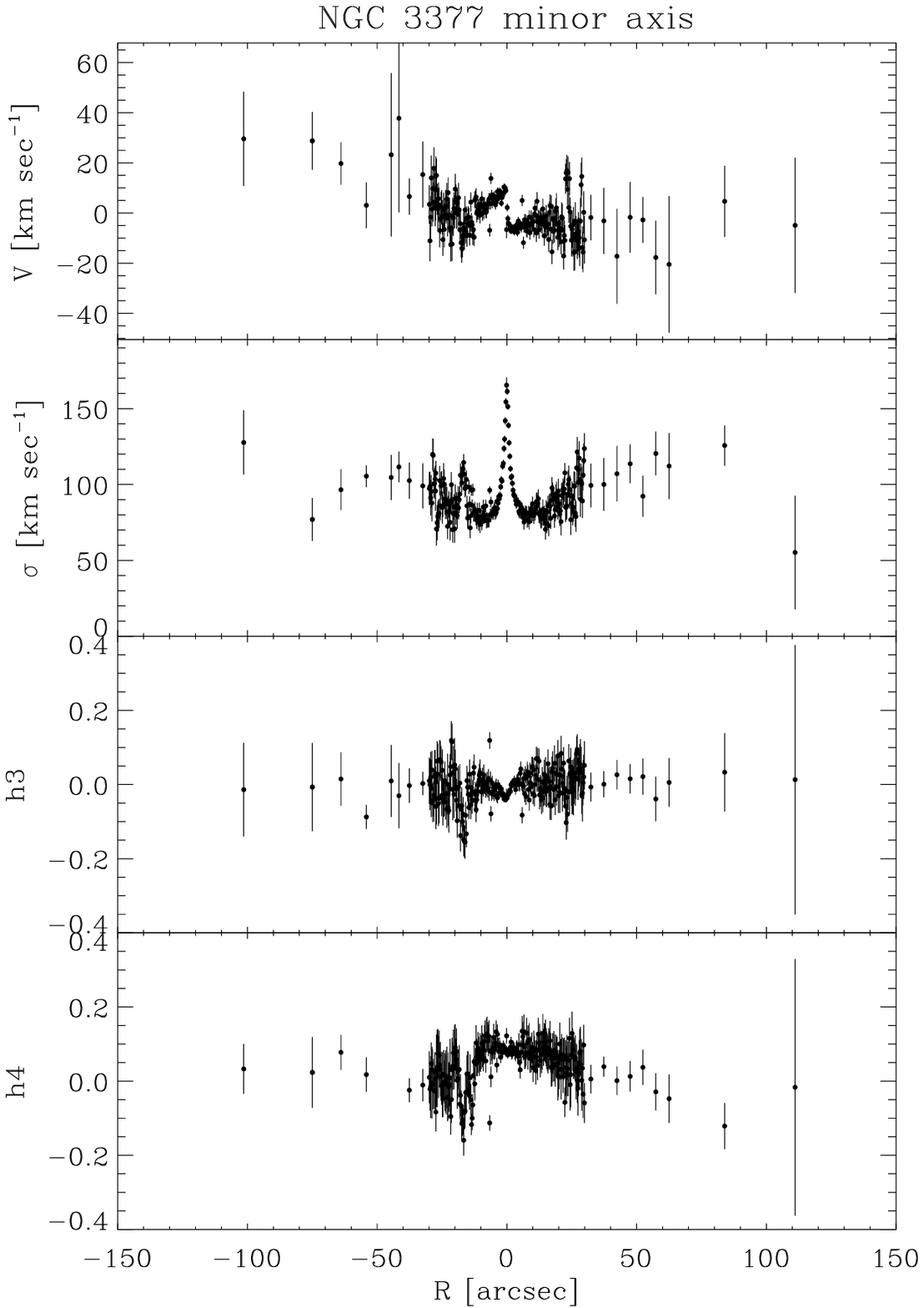,clip=,width=8.3cm}}
}
\caption{Radial velocities, velocity dispersions, $h_3$ and $h_4$
  coefficients for NGC 3377 along the major and minor axes.}
\label{fig:longslit}
\end{figure*}

The stellar kinematics (mean velocity, velocity dispersion, $h_3$ and $h_4$
Gauss-Hermite coefficients) was measured by means of the Penalized
Pixel-Fitting method by \citet{Cappellari+04}, kinematic stellar
templates were chosen from the Indo--U.S. Coud\'e Feed Spectral
Library \citep{Valdes+04} and the MILES library
\citep{Sanchez-Blazquez+06}, and then convolved with a Gaussian
function to match the instrumental FWHM.
Spectral regions with known emission lines (\hb, \oiii) were
masked and not included in the fit.

In Figure \ref{fig:longslit} we show the measured long-slit
kinematics. Table \ref{sec:long_slit}1 containing the measured
absorption-line kinematics is available as Supplementary Material in
the online version of this article.

\addtocounter{figure}{-1}
\begin{figure*}
 \vbox{
   \hbox{
     \psfig{file=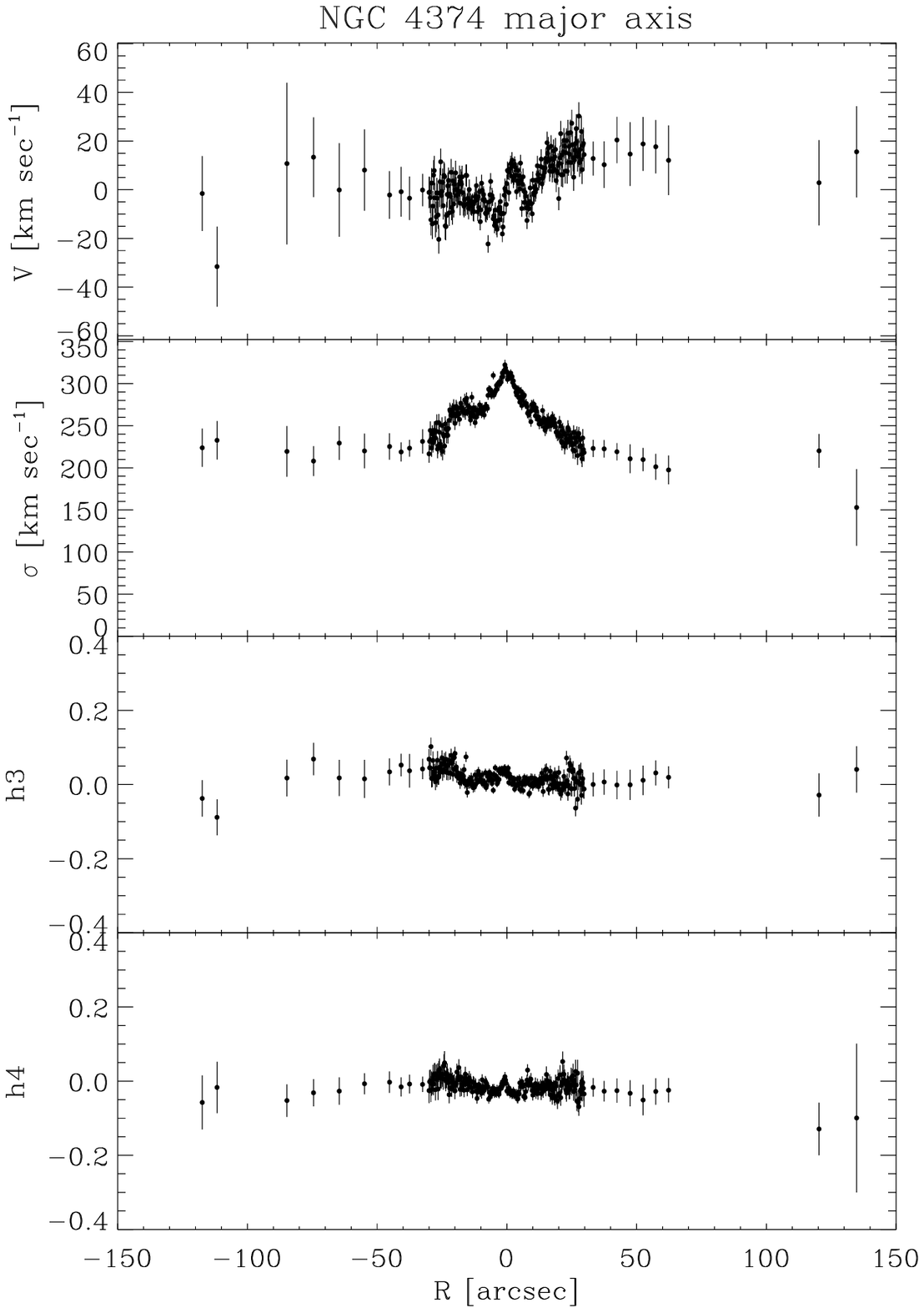,clip=,width=8.3cm}
     \psfig{file=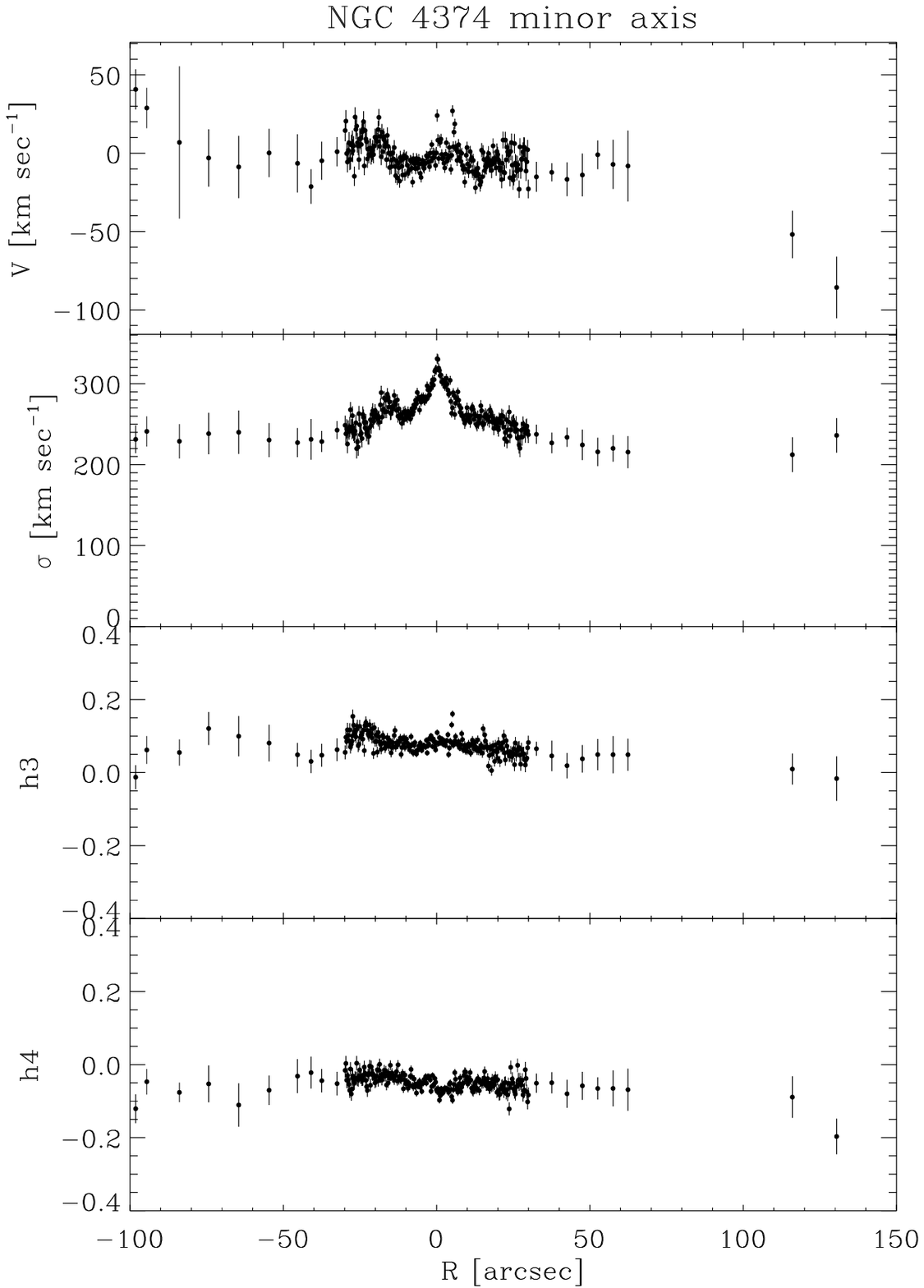,clip=,width=8.3cm}}
   \hbox{
     \psfig{file=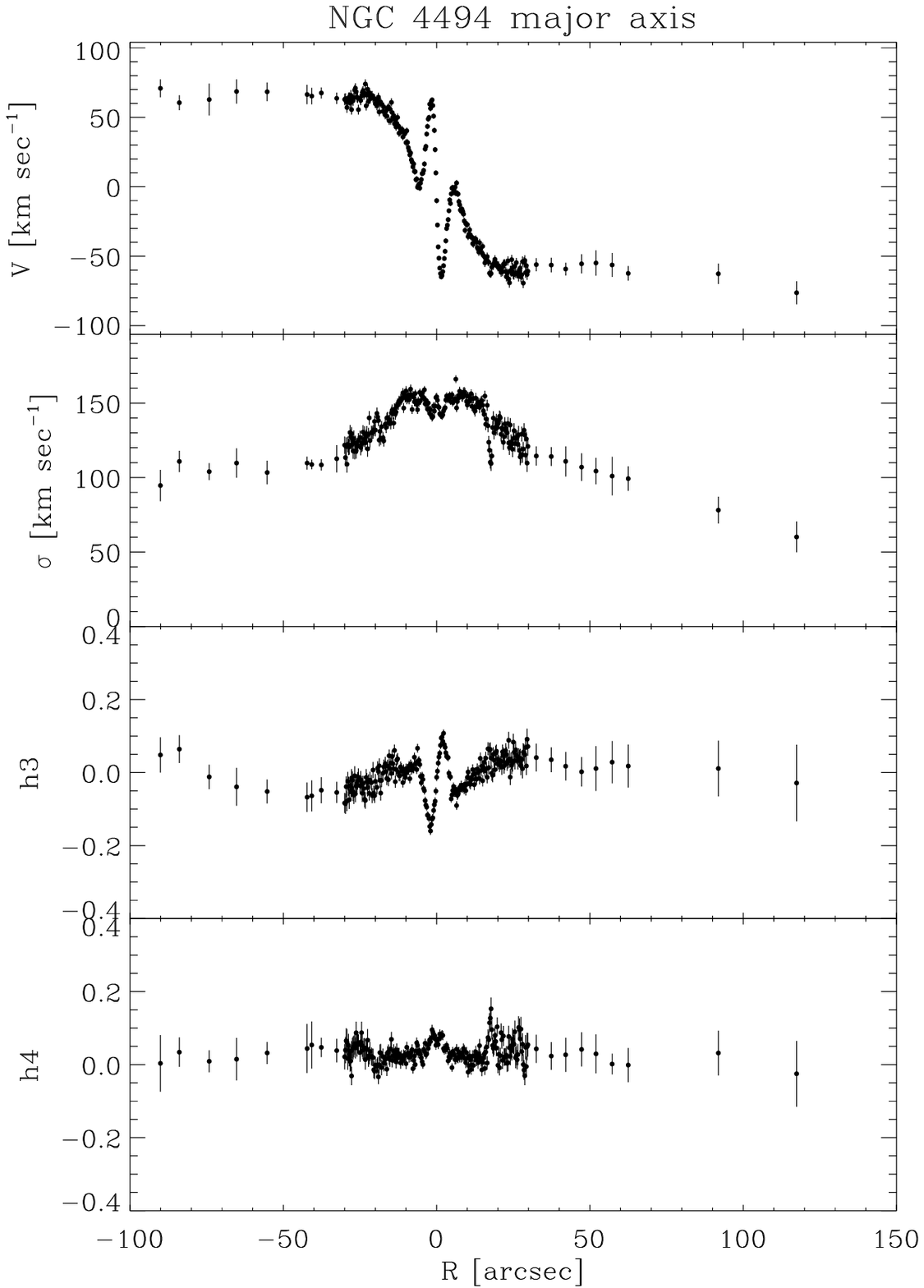,clip=,width=8.3cm}
     \psfig{file=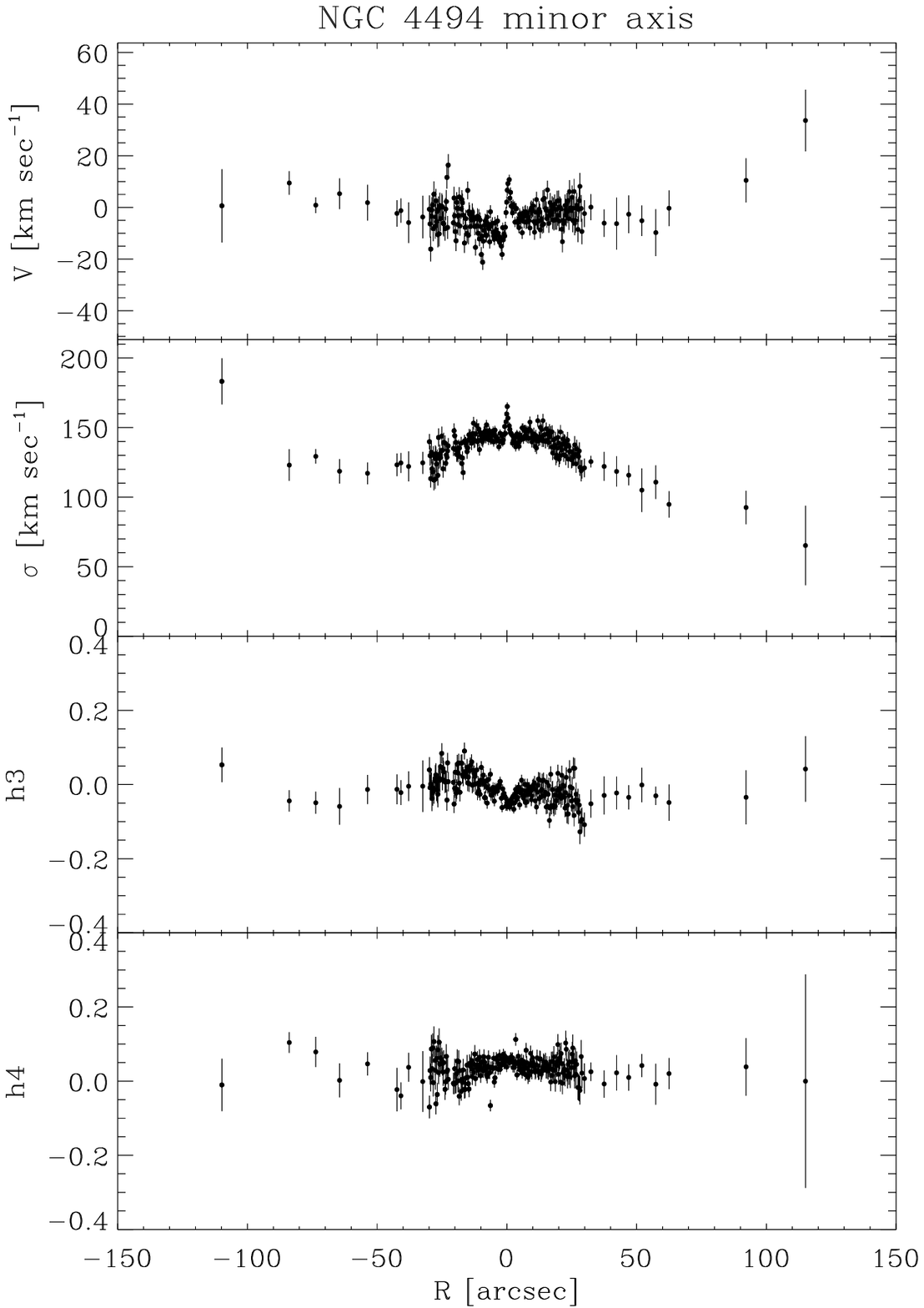,clip=,width=8.3cm}}
}
\caption{Continued, for NGC 4374 and NGC 4494.}
\end{figure*}

\section[]{Rotation and angular momentum considerations}\label{app:angmom}

Since the angular momentum of a star is weighted by its
  galactocentric radius, the outer regions of a galaxy are
  particularly important for estimating the total angular momentum.
  One cannot in general observe the three-dimensional positions and
  velocities necessary to calculate the full angular momentum, so we
  will use a two-dimensional projected proxy.  This proxy angular
  momentum can be simplified to a one-dimensional problem by adopting
  the approximation that the rotation is aligned with the photometric
  major axis, and its azimuthal dependence is related to the
  major-axis velocity by $V(\phi) = V_{\rm max} \cos(\phi)$
  (cf. \citealt{Krajnovic+08}).

The cumulative proxy angular momentum inside a radius $R$ is then:
\begin{equation}
LP(R) = \pi \int_0^R V_{\rm max}(R') \Sigma(R') R'^2 dR' ,
\end{equation}
where the stellar surface density is $\Sigma(R)$.  With the simplest
assumptions that the galaxy light follows the $R^{1/4}$ law, and that
the rotation velocity $V_{\rm max}$ and the stellar mass-to-light
ratio are both constant with radius, one can calculate that 9\% of the
total proxy angular momentum resides inside $R_e$, and 50\% inside
5.4~$R_e$.  The latter ``angular momentum effective radius'' ($\equiv
R_{LP}$), is naturally larger if $V_{\rm max}$ increases with $R$
or if the galaxy light is more extended (e.g. $R_{LP} = 12.1 R_e$
for S\'ersic index $n=6$).  It is smaller if $V_{\rm max}$ decreases
or if the light is more compact (e.g. $R_{LP} = 2.4 R_e$ for
$n=2$).

These calculations show that probing well outside $R_e$ is essential
for characterizing the angular momentum of an elliptical galaxy.  The
dimensionless cumulative angular momentum-like parameter $\lambda_R$
introduced by \citet{Emsellem+07} can also be shown to have the same
limitation, so that the global angular momentum requires
measurements to $\gtrsim 5 R_e$.

An alternative approach is to consider the radius of gyration,
$R_{\rm g}$, as the rule-of-thumb target radius for measuring the
angular momentum.  $R_{\rm g}$ is defined as the radius where a
point mass would have the same specific moment of inertia as the
extended mass profile.
Using the formula
\begin{equation}
R_{\rm g} \equiv \sqrt{\frac{I}{M}} = \left(\frac{\int_0^\infty \Sigma(R) R^3 dR}{\int_0^\infty \Sigma(R) R dR}\right)^{1/2} ,
\end{equation}
we find very similar radii to the $R_{LP}$ calculations reported above,
e.g., $R_{\rm g} = 2.1 R_e, 4.7 R_e, 10.1 R_e$ for S\'ersic $n=2,4,6$.

\end{document}